\documentclass[preprints,article,accept,moreauthors,pdftex]{mdpi} 

\firstpage{1} 
\pubvolume{xx}
\issuenum{1}
\articlenumber{5}
\pubyear{2019}
\copyrightyear{2019}
\history{Received: date; Accepted: date; Published: date}
\usepackage{bm}
\usepackage{mathrsfs}
\usepackage{verbatim}
\usepackage{bbold}
\usepackage{amssymb}

\Title{Neutrino Oscillations in Neutrino-Dominated Accretion Around Rotating Black Holes}

\Author{J.~D.~Uribe $^{1,2,\dagger}$\orcidA{}, E.~A.~Becerra-Vergara  $^{1,2,3,\dagger}$\orcidB{} and J.~A.~Rueda $^{2,4,5,6,\dagger}$\orcidC{}}% and F.~D.~Lora-Clavijo $^{3,\dagger}$\orcidD{}}

\AuthorNames{J.~D.~Uribe, E.~A.~Becerra-Vergara and J.~A.~Rueda}

\address{%
$^{1}$ \quad Dipartimento di Fisica, Sapienza Universit\`a di Roma, P.le Aldo Moro 5, I--00185 Rome, Italy\\
$^{2}$ \quad ICRANet, P.zza della Repubblica 10, I--65122 Pescara, Italy; juandavid.uribe@uniroma1.it (J.D.U.); eduar.becerra@icranet.org (E.A.B.V.); jorge.rueda@icra.it (J.A.R.)\\
$^{3}$ \quad Grupo de Investigaci\'on en Relatividad y Gravitaci\'on, Escuela de F\'isica, Universidad Industrial de Santander, A. A. 678, Bucaramanga 680002, Colombia\\
$^{4}$ \quad ICRANet-Ferrara, Dipartimento di Fisica e Scienze della Terra, Universit\`a degli Studi di Ferrara, Via Saragat 1, I--44122 Ferrara, Italy\\
$^{5}$ \quad Dipartimento di Fisica e Scienze della Terra, Universit\`a degli Studi di Ferrara, Via Saragat 1, I--44122 Ferrara, Italy\\
$^{6}$ \quad INAF, Istituto di Astrofisica e Planetologia Spaziali, Via Fosso del Cavaliere 100, 00133 Rome, Italy}

\corres{Correspondence: jorge.rueda@icra.it}

\firstnote{These authors contributed equally to this work.}
\abstract{In the binary-driven hypernova model of long gamma-ray bursts, a carbon-oxygen star explodes as a supernova in presence of a neutron star binary companion in close orbit. Hypercritical (i.e. highly super-Eddington) accretion of the ejecta matter onto the neutron star sets in, making it reach the critical mass with consequent formation of a Kerr black hole. We have recently shown that, during the accretion process onto the neutron star, fast neutrino flavour oscillations occur. Numerical simulations of the above system show that a part of the ejecta keeps bound to the newborn Kerr black hole, leading to a new process of hypercritical accretion. We here address, also for this phase of the binary-driven hypernova, the occurrence of neutrino flavour oscillations given the extreme conditions of high density \texorpdfstring{(up to $10^{12}$~g~cm$^{-3}$)}{} and temperatures (up to tens of MeV) inside this disk. We estimate the \textcolor{black}{behaviour} of the electronic and non-electronic neutrino content within the two-flavour formalism \texorpdfstring{($\nu_{e}\nu_{x}$)}{} under the action of neutrino collective effects by neutrino self-interactions. We find that \textcolor{black}{in the case of inverted mass hierarchy,} neutrino oscillations inside the disk have frequencies between \texorpdfstring{$\sim (10^{5}$--$10^{9})$~s$^{-1}$}{}, leading the disk to achieve flavour equipartition. This implies that the energy deposition rate by neutrino annihilation \texorpdfstring{($\nu + \bar{\nu} \to e^{-} + e^{+}$)}{} in the vicinity of the Kerr black hole, is smaller than previous estimates in the literature not accounting by flavour oscillations inside the disk. The exact value of the reduction factor depends on the \texorpdfstring{$\nu_{e}$}{} and \texorpdfstring{$\nu_{x}$}{} optical depths but it can be as high as \texorpdfstring{$\sim 5$}{}. The results of this work are a first step toward the analysis of neutrino oscillations in a novel astrophysical context and, as such, deserve further attention.}


\begin{document}
%%%%%%%%%%%%%%%%%%%%%%%%%%%%%%%%%%%%%%%%%%

%%%%%%%%%%%%%%%%%%%%%%%%%%%%%%%%%%%%%%%%%%
\section{Introduction}\label{intro}
%%%%%%%%%%%%%%%%%%%%%%%%%%%%%%%%%%%%%%%%%%%%%%%%%%%%%%%%%%%%%%%%
%%%%%%%%%%%%%%%%%%%%%%%%%%%%%%%%%%%%%%%%%%%%%%%%%%%%%%%%%%%%%%%%

Neutrino flavour oscillations are now an experimental fact~\cite{deSalas:2017kay} and, in recent years, its study based only on Mikheyev-Smirnov-Wolfenstein (MSW) effects~\cite{1978PhRvD..17.2369W,Mikheyev1986} has been transformed by the insight that refractive effects of neutrinos on themselves due to the neutrino self-interaction potential are essential. Their behaviour in vacuum, matter or by neutrino self-interactions have been studied in the context of early universe evolution~\cite{BARBIERI1991743,Enqvist:1990ad,Savage:1990by,Kostelecky:1993dm,Kostelecky:1993ys,KOSTELECKY199346,McKellar:1992ja,2001PhRvD..64g3006L,2002NuPhB.632..363D,2002PhRvD..66b5015W,2002PhRvD..66a3008A,2004CEJPh...2..467K}, solar and atmospheric neutrino anomalies~\cite{2003JHEP...02..009B,2003hep.ph....1072B,2003hep.ph...10012F,2004APh....21..287D,2004EPJC...33S.852G,2003PhRvD..68k3010M,2010JPhCS.203a2015D,2013ARA&A..51...21H,2017arXiv170605435V}, and core-collapse supernovae (SN)~\cite{Notzold:1987ik,1992PhLB..287..128P,Qian:1994wh,Pastor:2002we,Duan:2005cp,2005PhRvD..72d5003S,Fuller:2005ae,2006PhRvL..97x1101D,Fogli:2007bk,Duan:2007fw,Raffelt:2007yz,EstebanPretel:2007ec,EstebanPretel:2007yq,Chakraborty:2008zp,Duan:2007sh,Duan:2008eb,Dasgupta:2008my,Dasgupta:2007ws,Sawyer:2008zs,Duan:2010bg,Wu:2011yi,2014arXiv1408.2864B,2015arXiv150701434K,2016JPhCS.718f2068V,2016NCimR..39....1M,2018JPhG...45d3002H,2018PhRvD..98j3020Z} and references therein. We are here interested in astrophysical situations when neutrino self-interactions becomes more relevant than the matter potential. This implies systems in which a high density of neutrinos is present and in fact most of the literature on neutrino self-interaction dominance are concentrated on supernova neutrinos. It has been there shown how collective effects, such as synchronized and bipolar oscillations, change the flavor content of the emitted neutrinos when compared with the original content deep inside the exploding star.
This article aims to explore the problem of neutrino flavour oscillations in the case of long gamma-ray bursts (GRBs), \textcolor{black}{in particular in the context of} the binary-driven hypernova (BdHN) scenario. \textcolor{black}{Long GRBs are the most energetic and powerful cosmological transients so far observed, releasing energies of up to a few $10^{54}$~erg in just a few seconds. Most of the energy is emitted in the prompt gamma-ray emission and in the X-ray afterglow. We refer the reader to \cite{2018pgrb.book.....Z} for an excellent review on GRBs and its observational properties.}

The GRB progenitor \textcolor{black}{in the BdHN model} is a binary system composed of a carbon-oxygen star (CO$_{\rm core}$) and a companion neutron star (NS) \textcolor{black}{in tight orbit with orbital periods of the order of a few minutes \cite{2006tmgm.meet..369R, 2008mgm..conf..368R,2012A&A...548L...5I, 2012ApJ...758L...7R, 2014ApJ...793L..36F,2015ApJ...798...10R}. These binaries are expected to occur in the final stages of the evolutionary path of a binary system of two main-sequence stars of masses of the order of $10$--$15~M_ \odot$, after passing from X-ray binary phase and possibly multiple common-envelope phases (see \cite{2014ApJ...793L..36F, 2015PhRvL.115w1102F} and references therein).}

The CO$_{\rm core}$ explodes as SN \textcolor{black}{creating at its center a newborn NS ($\nu$NS), and ejecting the matter from its outermost layers. Part of the ejected matter falls back and accretes onto the $\nu$NS, while the rest continues its expansion leading to} a hypercritical accretion (i.e. highly super-Eddington) process onto the NS companion. The NS \textcolor{black}{companion} reaches the critical mass for gravitational collapse, hence forming a rotating black hole (BH). \textcolor{black}{The class of BdHN in which a BH is formed have been called as of type I, i.e. BdHN I \cite{2019ApJ...874...39W}.}

\textcolor{black}{One of the most important aspects of the BdHN model of long GRBs is that different GRB observables in different energy bands of the electromagnetic spectrum are explained by different components and physical ingredients of the system. This is summarized in Table~\ref{tab:BdHNobservables}, taken from \cite{2020ApJ...893..148R}. For a review on the BdHN model and all the physical phenomena at work, we refer the reader to \cite{2019Univ....5..110R}.}

\begin{table}[h]
\centering
\caption{\textcolor{black}{Summary of the GRB observables associated with each BdHN I component and physical phenomena. Adapted from Table 1 in \cite{2020ApJ...893..148R} with the permission of the authors. References in the table: $^a$\cite{2019ApJ...874...39W}, $^b$\cite{2014ApJ...793L..36F,2016ApJ...833..107B, 2019Univ....5..110R}, $^c$\cite{2001A&A...368..377B}, $^d$ \cite{2019ApJ...886...82R, 2019ApJ...883..191R}, $^e$\cite{2018ApJ...852...53R}, $^f$ \cite{2018ApJ...869..101R, 2019ApJ...874...39W}.}}
\textcolor{black}{
\resizebox{\textwidth}{!}{%
\begin{tabular}{clclclclclc}
\hline
\multirow{2}{*}{\textbf{BdHN component / phenomena}} &  & \multicolumn{9}{c}{\textbf{GRB observable}} \\ \cline{3-11} 
 &  & \begin{tabular}[c]{@{}c@{}}X-ray\\ precursor\end{tabular} &  & \begin{tabular}[c]{@{}c@{}}Prompt\\ (MeV)\end{tabular} &  & \begin{tabular}[c]{@{}c@{}}GeV-TeV\\ emission\end{tabular} &  & \begin{tabular}[c]{@{}c@{}}X-ray flares\\ early afterglow\end{tabular} &  & \begin{tabular}[c]{@{}c@{}}X-ray plateau\\ and late afterglow\end{tabular} \\ \cline{1-1} \cline{3-3} \cline{5-5} \cline{7-7} \cline{9-9} \cline{11-11} 
 &  &  &  &  &  &  &  &  &  &  \\
\multicolumn{1}{l}{SN breakout$^a$} &  & $\bigotimes$ &  &  &  &  &  &  &  &  \\
 &  &  &  &  &  &  &  &  &  &  \\
\multicolumn{1}{l}{Hypercrit. acc. onto the NS$^b$} &  & $\bigotimes$ &  &  &  &  &  &  &  &  \\
 &  &  &  &  &  &  &  &  &  &  \\
\multicolumn{1}{l}{\begin{tabular}[c]{@{}l@{}}$e^+e^-$: transparency\\ in low baryon load region$^c$\end{tabular}} &  &  &  & $\bigotimes$ &  &  &  &  &  &  \\
 &  &  &  &  &  &  &  &  &  &  \\
\multicolumn{1}{l}{\begin{tabular}[c]{@{}l@{}}\textit{Inner engine}: BH + $B$ + matter$^d$\end{tabular}} &  &  &  &  &  & $\bigotimes$ &  &  &  &  \\
 &  &  &  &  &  &  &  &  &  &  \\
\multicolumn{1}{l}{\begin{tabular}[c]{@{}l@{}}$e^+e^-$: transparency\\ in high baryon load region$^e$\end{tabular}} &  &  &  &  &  &  &  & $\bigotimes$ &  &  \\
 &  &  &  &  &  &  &  &  &  &  \\
\multicolumn{1}{l}{\begin{tabular}[c]{@{}l@{}}Synchrotron by $\nu$NS injected\\ particles on SN ejecta$^f$\end{tabular}} &  &  &  &  &  &  &  &  &  & $\bigotimes$ \\
 &  &  &  &  &  &  &  &  &  &  \\
\multicolumn{1}{l}{$\nu$NS pulsar-like emission$^f$} &  &  &  &  &  &  &  &  &  & $\bigotimes$ \\
 &  &  &  &  &  &  &  &  &  &  \\ \hline
\end{tabular}%
}
\label{tab:BdHNobservables}
}
\end{table}

The emission of neutrinos is a crucial ingredient since they act as the main cooling process that allows the accretion onto the NS to proceed at very high rates of up to $1~M_\odot$~s$^{-1}$ ~\cite{2014ApJ...793L..36F,2015PhRvL.115w1102F,2015ApJ...812..100B,2016ApJ...833..107B, 2019ApJ...871...14B}. In~\cite{2018ApJ...852..120B}, we studied the neutrino flavour oscillations in this hypercritical accretion process onto the NS, all the way to BH formation. We showed that, the density of neutrinos on top the NS, in the accreting ``atmosphere'', is such that neutrino self-interactions dominate the flavour evolution leading to collective effects. The latter induce in this system quick flavour conversions with a short oscillation length as small as $(0.05$--$1)$~km. Far from the NS surface the neutrino density decrease and so the matter potential and MSW resonances dominate the flavour oscillations. The main result has been that the neutrino flavour content emerging on top of the accretion zone was completely different compared to the one created at the bottom of it. In the BdHN scenario, part of the SN ejecta keeps bound to the newborn Kerr BH, forming an accretion disk onto it. In this context, the study of accretion disks and their nuances related to neutrinos is of paramount importance to shed light on this aspect of the GRB central engine. In most cases, the mass that is exchanged in close binaries has enough angular momentum so that it cannot fall radially. As a consequence, the gas will start rotating around the star or BH forming a disk. \textcolor{black}{At this point, it is worth to open a parenthesis to mention the case of short GRBs. They are widely thought to be the product of mergers of compact-object binaries; e.g. NS-NS and/or NS-BH binaries (see e.g. the pioneering works \cite{1986ApJ...308L..47G,1986ApJ...308L..43P,1989Natur.340..126E,1991ApJ...379L..17N}). It is then clear that, specially in NS-NS mergers, matter can be kept bound and circularize around the new central remnant. Also in such a case, an accretion disk will form around the more massive NS or the newborn BH (if the new central object overcomes the critical mass), and therefore the results of this work become also relevant for such physical systems.}

The magneto-hydrodynamics that describe the behaviour of accretion disks are too complex to be solved analytically and full numerical analysis are time-consuming and costly. To bypass this difficulty, different models make approximations that allow casting the physics of an accretion disk as a two- or even one-dimensional problem. These approximations can be can be pigeonholed into four categories: \textit{symmetry}, \textit{temporal evolution}, \textit{viscosity} and \textit{dynamics}. Almost all analytic models are axially symmetric. This is a sensible assumption for any physical systems that rotates. Similarly, most models are time-independent although this is a more complicated matter. A disk can evolve in time in several ways. For example, the accretion rate $\dot{M}$ depends on the external source of material which need not be constant and, at the same time, the infalling material increases the mass and angular momentum of the central object, constantly changing the gravitational potential. Additionally, strong winds and outflows can continually change the mass of the disk. Nonetheless, $\dot{M}\left(\boldsymbol{x},t\right) = \dot{M} = \mbox{constant}$ is assumed. Viscosity is another problematic approximation. For the gas to spiral down, its angular momentum needs to be reduced by shear stresses. These come from the turbulence driven by differential rotation and the electromagnetic properties of the disk~\cite{1991ApJ...376..214B,1991ApJ...376..223H,1998RvMP...70....1B,2003ARA&A..41..555B} but, again, to avoid magneto-hydrodynamical calculations, the turbulence accounted for using a phenomenological viscosity $\alpha = \mbox{constant}$, such that the kinematical viscosity takes the form $\nu \approx \alpha H c_{s}$, where $c_{s}$ is the local isothermal sound speed of the gas and $H$ is the height of the disk measured from the plane of rotation (or half-thickness). This idea was first put forward by~\cite{1973A&A....24..337S} and even though there is disagreement about the value and behaviour of the viscosity constant, and it has been criticized as inadequate~\cite{2007MNRAS.376.1740K,2008MNRAS.383..683P,2012MmSAI..83..466K,2012A&A...545A.115K}, several thriving models use this prescription. Finally, the assumptions concerning the dynamics of the disk are related to what terms are dominant in the energy conservation equation and the Navier-Stokes equation that describe the fluid (apart from the ones related to symmetry and time independence). In particular, it amounts to deciding what cooling mechanisms are important, what external potentials should be considered and what are the characteristics of the internal forces in the fluid. The specific tuning of these terms breeds one of the known models: \textit{thin disks}, \textit{slim disks}, \textit{advection-dominated accretion flows} (ADAFs), \textit{thick disks}, \textit{neutrino-dominated accretion flows} (NDAFs), \textit{convection-dominated accretion flows} (CDAFs), \textit{luminous hot accretion flows} (LHAFs), \textit{advection-dominated inflow-outflow solutions} (ADIOS) and \textit{magnetized tori}. The options are numerous and each model is full of subtleties making accretion flows around a given object an extremely rich area of research. For useful reviews and important articles with a wide range of subjects related to accretion disks see \cite{1981ARA&A..19..137P,1999agnc.book.....K,1999tbha.book.....A,2000ApJ...534..734M,2002apa..book.....F,2004adjh.conf..137B,NARAYAN2008733,2008bhad.book.....K,2009A&A...498..471Q,2012arXiv1203.6851M,2013LRR....16....1A,2014ARA&A..52..529Y,2014SSRv..183...21B,2016ASSL..440....1L,LIU20171} and references therein. 

\begin{figure}[H]
\centering
\includegraphics[width=0.65\hsize,clip]{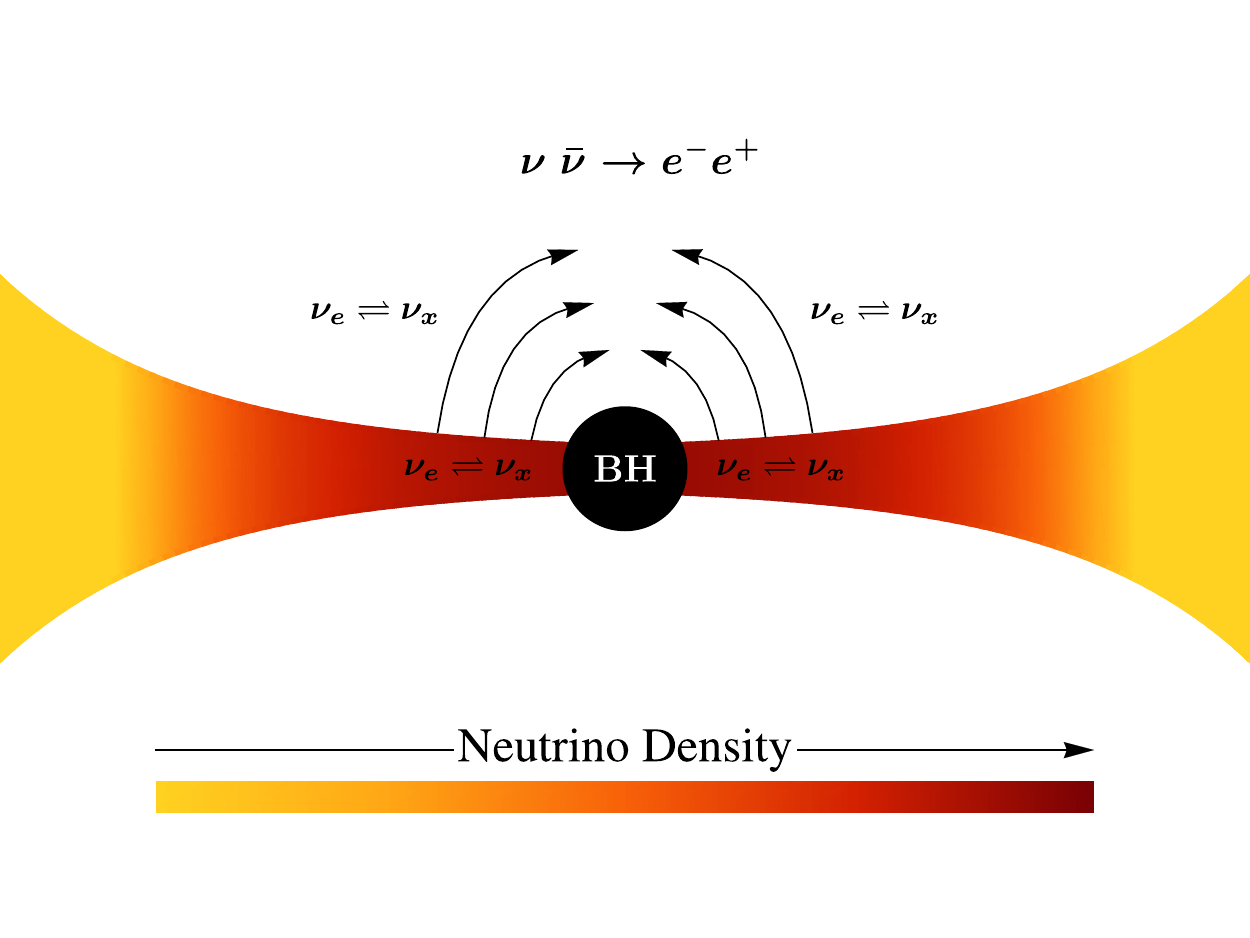}
\caption{Schematic representation of the physical system. Due to conditions of high temperature and density, neutrinos are produced in copious amounts inside the disk. Since they have a very low cross-section, neutrinos are free to escape but not before experiencing collective effects due to the several oscillation potentials. The energy deposition rate of the process $\nu + \bar{\nu} \to e^{-}\! + e^{+}$ depends on the local distribution of electronic and non-electronic (anti)neutrinos which is affected by the flavour oscillation dynamics.}
\label{fig:disk1}
\end{figure}

NDAFs are of special interest for GRBs. They are hyperaccreting slim disks, optically thick to radiation that can reach high densities $\rho \approx 10^{10}$--$10^{13}$~g~cm$^{-3}$ and high temperatures $T\approx 10^{10}$--$10^{11}$~K around the inner edge. Under these conditions, the main cooling mechanism is neutrino emission since copious amounts of (mainly electron) neutrinos and antineutrinos are created by electron-positron pair annihilation, URCA and nucleon-nucleon bremsstrahlung processes, and later emitted from the disk surface. These $\nu\bar{\nu}$ pairs might then annihilate above the disk producing an $e^{-}e^{+}$ dominated outflow. NDAFs were proposed as a feasible central engine for GRBs in~\cite{1999ApJ...518..356P} and have been studied extensively since~\cite{2001ApJ...557..949N,2002ApJ...577..311K,2002ApJ...579..706D,2005ApJ...629..341K,2005ApJ...632..421L,2006ApJ...643L..87G,2007ApJ...657..383C,2007ApJ...662.1156K,2010A&A...509A..55J,2013ApJ...766...31K,2013MNRAS.431.2362L,2013ApJS..207...23X}. In~\cite{2002ApJ...579..706D} and later in~\cite{2007ApJ...657..383C}, it was found that the inner regions of the disk can be optically thick to $\nu_{e}\bar{\nu}_{e}$ trapping them inside the disk, hinting that NDAFs may be unable to power GRBs. Yet, the system involves neutrinos propagating through dense media and, consequently, an analysis of neutrino oscillations, missing in the above literature, must be performed. Fig.~\ref{fig:disk1} represents the standard situation of the physical system of interest. The dominance of the self-interaction potential induces collective effects or decoherence. In either case, the neutrino flavour content of the disk changes. Some recent articles are starting to recognize their role in accretion disks and spherical accretion~\cite{2012PhRvD..86h5015M,2017PhRvD..95b3011F,2017PhRvD..96d3001T,2018ApJ...852..120B,PhysRevD.95.103007,2020arXiv200901843P}. \textcolor{black}{In particular, \cite{2012PhRvD..86h5015M,2020arXiv200901843P} calculate the flavour evolution of neutrinos once they are emitted from the disk, but do not take into account the oscillation behaviour inside the disk.} The energy deposition rate above a disk by neutrino-pair annihilation as a powering mechanism of GRBs in NDAFs can be affected by neutrino oscillation in two ways. The neutrino spectrum emitted at the disk surface depends not only on the disk temperature and density but also on the neutrino flavour transformations inside the disk. Also, once the neutrinos are emitted they undergo flavour transformations before being annihilated.

\textcolor{black}{Our main objective is to propose a simple model to study neutrino oscillations \emph{inside} an accretion disk and analyze its consequences. Applying the formalism of neutrino oscillations to non-symmetrical systems is difficult, so we chose a steady-state, $\alpha$-disk as a first step in the development of such a model. The generalization to more sophisticated accretion disks \cite[see e. g.][]{ Janiuk_2013,2017ApJ...837...39J, JSFI177,2019ApJ...882..163J} can be subjects of future research.}

\textcolor{black}{This article is organized as follows. We outline the features of NDAFs and discuss in detail the assumptions needed to derive the disk equations in Sec.~\ref{sec:1}. Then, in Sec.~\ref{sec:2}, we discuss the general characteristics of the equation that drives the evolution of neutrino oscillations. We use the comprehensive exposition of the accretion disk of the previous section to build a simple model that adds neutrino oscillations to NDAFs, making emphasis in how the thin disk approximation can simplify the equations of flavour evolution. In Sec.~\ref{sec:3} we set the parameters of the physical system and give some details on the initial conditions needed to solve the equations of accretion disks and neutrino oscillations. In Sec.~\ref{sec:4} we discuss the main results of our calculations and analyse in the phenomenology of neutrino oscillations in accretion disks. Finally, we present in Sec.~\ref{sec:5} the conclusions of this work. Additional technical details are presented in a series of appendices at the end.} 

%

%%%%%%%%%%%%%%%%%%%%%%%%%%%%%%%%%%%%%%%%%%%%%%%%%%%%%%%%%%%%%%%%%%%%%%%%%%%%%%%%%%%%%%%%%%%%%%%%%%%%%%%%%%%%%%%%%%%%%%%%%%%%%%%%%%%%%%%%%%%%%%%%%%%%%%%%%%%%%%%%%%%%%%%%%%%%%%%%%%%%%%%%%%%%%%
\section{Hydrodynamics}\label{sec:1}
%%%%%%%%%%%%%%%%%%%%%%%%%%%%%%%%%%%%%%%%%%%%%%%%%%%%%%%%%%%%%%%%%%%%%%%%%%%%%%%%%%%%%%%%%%%%%%%%%%%%%%%%%%%%%%%%%%%%%%%%%%%%%%%%%%%%%%%%%%%%%%%%%%%%%%%%%%%%%%%%%%%%%%%%%%%%%%%%%%%%%%%%%%%%%%

\subsection{Units, velocities and averaging}\label{subsec:1.1}

Throughout this article, we use Planck units $c=G=\hbar=k_{B}=k_{e}=1$. To describe the spacetime around a Kerr BH of mass $M$, we use the metric $g_{\mu\nu}$ in Boyer-Lindquist coordinates, with spacelike signature, and with a dimensionless spin parameter $a=J/M^2$, which can be written as:

\begin{equation} 
 ds^2 = \left( g_{tt} - \omega^2 g_{\phi\phi} \right) dt^2 + g_{\phi\phi} \left( d\phi - \omega\, dt \right)^2 + g_{rr}\,dr^2 + g_{\theta\theta}\,d\theta^2,
\label{eq:lineelement}
\end{equation}

in coordinates $(t,r,\theta,\phi)$. The covariant components $(\boldsymbol{g})_{\mu\nu}$ of the metric are

\begin{align}
g_{tt} &= -\left( 1 - \frac{2\,M\,r}{\Sigma} \right),\quad g_{rr} =\frac{\Sigma}{\Delta},\quad g_{\theta\theta} = \Sigma, \nonumber \\
g_{\phi\phi} &= \left(r^2 + M^2a^2 + \frac{2\, M^3 a^2 r }{\Sigma} \sin^2\theta\right)  \sin^2\theta, \quad g_{t\phi} = -\frac{2\,M^2\,a\, r}{\Sigma} \sin^2\theta,
\label{eq:metric}
\end{align}

and its determinant is $g=-\Sigma^2 \sin^2 \theta$, with the well known functions $\Sigma = r^2+M^2a^2 \cos^2\theta$ and $\Delta = r^2-2Mr+M^2a^2$. We denote the coordinate frame by CF. Note that these coordinates can be used by an observer on an asymptotic rest frame. The angular velocity of the \emph{locally non-rotating frame} (LNRF) is

\begin{align}
\omega &= -\frac{g_{t\phi}}{g_{\phi\phi}} = \frac{2\,a\,M^{2}}{\left(r^3 + M^2a^2r + 2M^3a^2\right)},
 \label{eq:omega}
\end{align}

and in Eq.~(\ref{eq:metric}) it can be seen explicitly that if an observer has an angular velocity $\omega = d\phi/dt$, it would not measure any differences between the $\pm \phi$ directions. The LNRF is defined by orthonormality and the coordinate change $\phi_{\textrm{LNRF}} = \tilde{\phi} = \phi - \omega\, t$ \cite{1970ApJ...162...71B,1972ApJ...178..347B}. We assume that the disk lies on the equatorial plane of the BH $(\theta = \pi/2)$. This way we represent the average movement of the fluid by geodesic circular orbits with angular velocity $\Omega=d\phi/dt=u^{\phi}/u^{t}$ plus a radial velocity so that the local rest frame (LRF) of the fluid is obtained by performing, first, an azimuthal Lorentz boost with velocity $\beta^{\hat{\phi}}$ to a co-rotating frame (CRF) \cite{1998ApJ...498..313G}, and then a radial Lorentz boost with velocity $\beta^{\tilde{r}}$. Clearly, the metric on the LNRF, CRF and LRF is ${\textrm{diag}}(-1,1,1,1)$. The expression for the angular velocity of circular orbits is obtained by setting $\dot{r}=\ddot{r}=0$ in the $r$-component of the geodesic equation

\begin{align}
\Omega^{\pm} &= \pm\frac{\sqrt{M}}{\left( r^{3/2} \pm M^{3/2}a \right)},
 \label{eq:Omega}
\end{align}

where $(+)$ is for prograde orbits and $(-)$ is for retrograde orbits. We will limit our calculations to prograde movement with $0\leq a \leq 1$ but extension to retrograde orbits is straightforward. Finally, we can get the components of the 4-velocity of the fluid by transforming $\boldsymbol{u}_{\textrm{LRF}}=(1,0,0,0)$ back to the CF

\begin{equation}
  u^{\mu}\! =\! \left( \frac{\gamma_{\tilde{r}}\gamma_{\hat{\phi}}}{\sqrt{\omega^2 g_{\phi\phi}\!-g_{tt}}}, \frac{\gamma_{\tilde{r}}\beta^{\tilde{r}}}{\sqrt{g_{rr}}}, 0,\frac{\gamma_{\tilde{r}}\gamma_{\hat{\phi}}\Omega}{\sqrt{\omega^2 g_{\phi\phi}\!-g_{tt}}}\right),
       \label{eq:fourvel}
\end{equation}

leaving $\beta^{\tilde{r}}$ to be determined by the conservation laws. In Eq.~(\ref{eq:fourvel}) we have replaced $\beta^{\hat{\phi}}$ with Eq.~(\ref{eq:betaphi}). A discussion on the explicit form of the transformations and some miscellaneous results are given in Appendix~\ref{app:christoffel}. We will also assume that the disk is in a steady-state. This statement requires some analysis. There are two main ways in which it can be false:

First, as matter falls into the BH, its values $M$ and $a$ change \cite{1970Natur.226...64B,1974ApJ...191..507T}, effectively changing the spacetime around it. For the spacetime to remain the same \textcolor{black}{(i.e. for $M$ and $a$ to stay constant)} we require $\Omega^{-1} \ll t_{\textrm{acc}} = \Delta M_0/\dot{M}_{\textrm{acc}}$, where $\Delta M_{0}$ is the total mass of the disk and $\dot{M}_{\textrm{acc}}$ is the accretion rate. The characteristic accretion time must be bigger than the dynamical time of the disk so that flow changes due to flow dynamics are more important than flow changes due to spacetime changes. Equivalent versions of this condition that appear throughout disk accretion articles are $t_{\textrm{dym}} \ll t_{\textrm{visc}}$ and

\begin{equation}
\beta^{r} \ll \beta^{\phi} < 1,
    \label{eq:appr1}
\end{equation}

where it is understood that the accretion rate obeys $\dot{M}_{\textrm{acc}} \approx \Delta M_{0}/t_{\textrm{acc}}$. To put this numbers into perspective, consider a solar mass BH $(M = 1 M_\odot)$ and a disk with mass between $\Delta M_{0} = (1-10)M_\odot$. For accretion rates up to $\dot{M}_{\textrm{acc}} = 1 M_\odot / {\textrm{s}}$ the characteristic accretion time is $t_{\textrm{acc}} \lesssim (1-10)$~${\textrm{s}}$, while $\Omega^{-1} \sim (10^{-5}-10^{-1})$~${\textrm{s}}$ between $r=r_{\textrm{ISCO}}$ and $r=2000 M_{\odot}$. Consequently, a wide range of astrophysical system satisfy this condition and it is equivalent to claiming that both $\boldsymbol{\partial}_t$ and $\boldsymbol{\partial}_\phi$ are Killing fields.

Second, at any point inside the disk, any field $\boldsymbol{\psi}(t,r,\theta,\phi)$ that reports a property of the gas may variate in time due to the turbulent motion of the flow. So, to assume that any field is time-independent and smooth enough in $r$ for its flow to be described by Eq.~(\ref{eq:fourvel}) means replacing such field by its average over an \emph{appropriate} spacetime volume. The same process allows to choose a \emph{natural} set of variables that split the hydrodynamics into $r$-component equations and $\theta$-component equations. The averaging process has been explained in \cite{1973blho.conf..343N,1974ApJ...191..499P,1998ApJ...498..313G}. We include the analysis here and try to explain it in a self-consistent manner. The turbulent motion is characterized by the eddies. The azimuthal extension of the largest eddies can be $2\pi$, like waves crashing around an island, but their linear measure cannot be larger than the thickness of the disk, and, as measured by an observer on the CRF, their velocity is of the order of $\beta^{\tilde{r}}$ so that their period along the $r$ component is $\Delta \tilde{t} \approx ({\textrm{Thickness}})/\beta^{\tilde{r}}$ \cite[e.g. $\S 33$][]{1959flme.book.....L}. If we denote by $H$ the average half-thickness of the disk as measured by this observer at $r$ over the time $\Delta \tilde{t}$, then the appropriate volume $\mathcal{V}$ is composed by the points $(t,r,\theta,\phi)$ such that $t\in[t^{*}-\Delta t/2,t^{*}+\Delta t/2]$, $\theta\in[\theta_{\textrm{min}},\theta_{\textrm{max}}]$ and $\phi\in[0,2\pi)$, where we have transformed $\Delta \tilde{t}$ and $\Delta \tilde{r}$ back to the CF using Eqs.~(\ref{eq:deltas}) as approximations. The values $\theta_{\textrm{min}}$ and $\theta_{\textrm{max}}$ correspond to the upper and lower faces of the disk, respectively. Then, the average takes the form

\begin{equation}
\boldsymbol{\psi}\left(t,r,\theta,\phi\right) \mapsto\boldsymbol{\psi}\left( r,\theta \right) = \langle \boldsymbol{\psi}\left(t,r,\theta,\phi\right) \rangle = \frac{ \int_{t^{*}-\Delta t/2}^{t^{*}+\Delta t/2} \int_{0}^{2\pi} \boldsymbol{\psi}\left(r,t,\theta,\phi\right) \sqrt{\frac{-g}{g_{rr}g_{\theta\theta}}} dt d\phi}{\int_{t^{*}-\Delta t/2}^{t^{*}+\Delta t/2} \int_{0}^{2\pi} \sqrt{\frac{-g}{g_{rr}g_{\theta\theta}}} dt d\phi }.
    \label{eq:average}
\end{equation}

The steady-state condition is achieved by requiring that the Lie derivative of the averaged quantity along the Killing field $\boldsymbol{\partial}_t$ vanishes: $\mathcal{L}_{\boldsymbol{\partial}_t}\langle \boldsymbol{\psi} \rangle = 0$. Note that the thickness measurement performed by the observer already has an error $\sim M^2a^2H^{3}/6r^{4}$ since it extends the Lorentz frame beyond the local neighbourhood but, if we assume that the disk is thin $(H/r \ll 1)$, and we do, this error remains small. At the same time, we can take all metric components evaluated at the equator and use Eq.~(\ref{eq:fourvel}) as the representative average velocity. Under these conditions, we have $\theta_{\textrm{max}}-\theta_{\textrm{min}} \approx 2H/r$ and the term $\sqrt{-g/g_{rr}}$ in Eq.~(\ref{eq:average}) cancels out. It becomes clear that an extra $\theta$ integral is what separates the radial and polar variables. In other words, the $r$-component variables are the \emph{vertically integrated} fields

\begin{equation}
\boldsymbol{\psi}\left(r,\theta\right) \mapsto \boldsymbol{\psi}\left(r\right) =\!\! \int^{\theta_{\textrm{max}}}_{\theta_{\textrm{min}}}\!\!\boldsymbol{\psi}\left(r,\theta\right)\sqrt{g_{\theta\theta}}d\theta.
\label{eq:vertav}    
\end{equation}

The vertical equations of motion can be obtained by setting up Newtonian (with relativistic corrections) equations for the field $\boldsymbol{\psi}\left(r,\theta\right)$ at each value of $r$ \cite[see e.g.][]{1973blho.conf..343N,Abramowicz_1996,Abramowicz_1997,LIU20171}. 

%%%%%%%%%%%%%%%%%%%%%%%%%%%%%%%%%%%%%%%%%%%%%%%%%%%%%%%%%%%%%%%%%%%%%%%%%%%%%%%%%%%%%%%%%%%%%%%%%%%%%%%%%%%%%%%%%%%
\subsection{Conservation Laws}\label{subsec:1.2}
%%%%%%%%%%%%%%%%%%%%%%%%%%%%%%%%%%%%%%%%%%%%%%%%%%%%%%%%%%%%%%%%%%%%%%%%%%%%%%%%%%%%%%%%%%%%%%%%%%%%%%%%%%%%%%%%%%%

The equations of evolution of the fluid are contained in the conservation laws $\nabla_{\mu}T^{\mu\nu} = 0$ and $\nabla_{\mu}(\rho u^{\mu}) = 0$. The most general stress-energy tensor for a Navier-Stokes viscous fluid with heat transfer is \cite{1973grav.book.....M,1984oup..book.....M}

\begin{align}
\boldsymbol{T} = & \overbrace{\left( \rho + U + P \right)\boldsymbol{u}\otimes\boldsymbol{u} + P\boldsymbol{g}}^{\mbox{Ideal Fluid}}\,+\, \overbrace{\left( - 2\eta\boldsymbol{\sigma} - \zeta\left(\boldsymbol{\nabla}\cdot\boldsymbol{u}\right)\boldsymbol{P}\right)}^{\mbox{Viscous Stress}}\,+\, \overbrace{\boldsymbol{q}\otimes\boldsymbol{u} + \boldsymbol{u}\otimes\boldsymbol{q}}^{\mbox{Heat flux}},
\label{eq:stress-energy}
\end{align}

where $\rho$, $P$, $U$, $\zeta$, $\eta$, $\boldsymbol{q}$, $\boldsymbol{P}$ and $\boldsymbol{\sigma}$ are the rest-mass energy density, pressure, internal energy density, dynamic viscosity, bulk viscosity, heat-flux 4-vector, projection tensor and shear tensor, respectively, and thermodynamic quantities are measured on the LRF. We do not consider electromagnetic contributions and ignore the causality problems associated with the equations derived from this stress-energy tensor since we are not interested in phenomena close to the horizon \cite{1998ApJ...498..313G}. Before deriving the equations of motion and to add a simple model of neutrino oscillations to the dynamics of disk accretion we must make some extra assumption. We will assume that the $\theta$ \textcolor{black}{integral in Eq.~(\ref{eq:vertav}) can be approximated by}

\begin{equation}
\color{black}\int_{\theta_{\textrm{min}}}^{\theta_{\textrm{max}}}\!\!\color{black}\boldsymbol{\psi}\sqrt{g_{\theta\theta}}d\theta \approx \boldsymbol{\psi}r\left(\theta_{\textrm{max}}-\theta_{\textrm{min}}\right) \approx 2H\boldsymbol{\psi},
    \label{eq:vertaverage}
\end{equation}

for any field $\boldsymbol{\psi}$. Also, we use Stokes' hypothesis ($\zeta = 0$). Since we are treating the disk as a thin fluid in differential rotation, we will assume that, on average, the only non-zero component of the shearing stress on the CRF is $\sigma_{\tilde{r}\tilde{\phi}}$ (there are torques only on the $\phi$ direction), and $q_{\tilde{\theta}}$ is the only non-zero component of the energy flux (on average the flux is vertical). By $u^{\mu}\sigma_{\mu\nu}=0$ and Eq.~(\ref{eq:shear}) we have 

\begin{equation}
\sigma_{r\phi} =\frac{\gamma^{3}_{\hat{\phi}}}{2}\frac{g_{\phi\phi}}{\sqrt{\omega^2 g_{\phi\phi}-g_{tt}}}\partial_{r}\Omega\, ,\;\;\sigma_{rt} = -\Omega\sigma_{r\phi}.
    \label{eq:shear2}
\end{equation}

Finally, the turbulent viscosity is estimated to be $\sim l \Delta u$ where $l$ is the size of the turbulent eddies and $\Delta u$ is the average velocity difference between points in the disk separated by a distance $l$. By the same arguments in \cite[$\S 33$][]{1959flme.book.....L} and in Sec.~\ref{subsec:1.2}, $l$ can be at most equal to $2H$ and $\Delta u$ can be at most equal to the isothermal sound speed $c_{s}=\sqrt{\partial P/\partial\rho}$ or else the flow would develop shocks \cite{2002apa..book.....F}. The particular form of $c_{s}$ can be calculated from Eq.~(\ref{eq:press}). This way we get

\begin{equation}
\eta = \Pi\nu_{\textrm{turb}} = 2\alpha \Pi H c_{s},
    \label{eq:alpha}
\end{equation}

with $\alpha \leq 1$ and $\Pi = \rho + U + P$. In a nutshell, this is the popular $\alpha$-prescription put forward by \cite{1973A&A....24..337S}. As we mentioned at the end of Sec.~\ref{subsec:1.1}, on the CRF for a fixed value of $r$, the polar equation takes the form of Euler's equation for a fluid at rest where the acceleration is given by the tidal gravitational acceleration. Namely, the $\theta$ component of the fluid's path-lines relative acceleration in the $\theta$ direction is

\begin{equation}
\frac{1}{r}\partial_{\theta}P \approx \rho r\cos\theta\left[\boldsymbol{R}\left(\boldsymbol{u},\boldsymbol{\partial}_{\tilde{\theta}},\boldsymbol{u}\right)\cdot\boldsymbol{\partial}_{\tilde{\theta}}\right]_{_{\theta=\pi/2}},
\label{eq:vertdiffeq}
\end{equation}

with $\boldsymbol{R}$ the Riemann curvature tensor. With $u^{\tilde{\mu}}\approx(1,0,0,0)$, Eq.~(\ref{eq:vertaverage}), Eq.~(\ref{eq:riemanncurv}) and assuming that there is no significant compression of the fluid under the action of the tidal force, integration of this equation yields the relation up to second order in $\pi/2-\theta$

\begin{equation}
P=\frac{1}{2}\rho \! \left. {R^{\,\tilde{\theta}}}_{\tilde{t}\tilde{\theta}\tilde{t}}\, \right|_{_{\theta=\pi/2}} \left(H^2 - r^2\left(\frac{\pi}{2} - \theta\right)^2\right),
    \label{eq:verteq}
\end{equation}

where we used the condition $P=0$ at the disk's surface. Hence, the average pressure inside the disk is \cite[cf.][]{Abramowicz_1997,LIU20171,2007ApJ...657..383C}

\begin{equation}
P=\frac{1}{3}\rho H^{2}\! \left. {R^{\,\tilde{\theta}}}_{\tilde{t}\tilde{\theta}\tilde{t}}\, \right|_{_{\theta=\pi/2}}.
    \label{eq:press}
\end{equation}

The equation of mass conservation is obtained by directly inserting into Eq.~(\ref{eq:mass}) the averaged density and integrating vertically to obtain

\begin{equation}
2Hr\rho u^{r}= {\textrm{constant}} = - \frac{\dot{M}}{2\pi},
\label{eq:masscon}
\end{equation}

where the term $2 H r \rho u^{r}$ is identified as the average inward mass flux through a cylindrical surface of radius $r$ per unit azimuthal angle and thus must be equal to the accretion rate divided by $2\pi$. The same process applied to Eq.~(\ref{eq:fineq}) yields the energy conservation equation

\begin{equation}
u^{r}\left[\partial_{r}\left(HU\right)-\frac{U + P}{\rho}\partial_{r}\left(H\rho\right)\right] =2\eta H\sigma^{r\phi}\sigma_{r\phi} - H\epsilon,
\label{eq:energycon}
\end{equation}

where factors proportional to $H/r$ were ignored and we assume $\Pi\approx\rho$ to integrate the second term on the left-hand side. $\epsilon$ is the average energy density measured on the LRF (see the discussion around Eq.~(\ref{eq:vertflux})). The first term on the right hand side is the viscous heating rate $F_{\text{heat}}$ and the second term is the cooling rate $F_{\text{cool}}$. The last constitutive equation is obtained by \textcolor{black}{applying the \emph{zero torque at the last stable orbit} condition. These relations are calculated in Appendix~\ref{app:christoffel}. We just replace the density in Eq.~(\ref{eq:masscon}) using Eq.~(\ref{eq:pol}) obtaining}

\begin{equation}
u^{r}=-\frac{4\alpha H c_{s}\sigma^{r}_{\phi}}{Mf\left(x,x^{*}\right)}.
\label{eq:eqforvel}
\end{equation}
% 

%%%%%%%%%%%%%%%%%%%%%%%%%%%%%%%%%%%%%%%%%%%%%%%%%%%%%%%%%%%%%%%%%%%%%%%%%%%%%%%%%%%%%%%%%%%%%%%%%%%
\subsection{Equations of State}\label{subsec:1.3}
%%%%%%%%%%%%%%%%%%%%%%%%%%%%%%%%%%%%%%%%%%%%%%%%%%%%%%%%%%%%%%%%%%%%%%%%%%%%%%%%%%%%%%%%%%%%%%%%%%% 

We consider that the main contribution to the rest-mass energy density of the disk is made up of neutrons, protons and ions. This way $\rho = \rho_{\textrm{B}} = n_{B}m_{B}$ with baryon number density $n_{B}$ and baryon mass $m_{B}$ equal to the atomic unit mass. The disk's baryonic mass obeys Maxwell-Boltzmann statistics and its precise composition is determined by the Nuclear Statistical Equilibrium (NSE). We denote the mass fraction of an ion $i$ by $X_{i}=\rho_{i}/\rho_{B}$ (if $i= p \textrm{ or } n$ then we are referring to proton or neutrons) and it can be calculated by the Saha equation \cite{1965MmRAS..69...21C,2007ApJ...656..313C}

\begin{equation}\label{eq:saha}
X_{i} = \frac{A_{i}m_{B}}{\rho}  G_{i}\left(\frac{TA_{i}m_{B}}{2\pi}\right)^{3/2} \exp\left(\frac{Z_{i}\left(\mu_{p} + \mu^{C}_{p}\right) + N_{i}\mu_{n} - \mu^{C}_{i} + B_{i}}{T}\right),
\end{equation}

with the constraints:

\begin{equation}\label{eq:sahacons}
\sum _{i}X_{i} = 1,\quad \sum_{i}Z_{i}Y_{i} = Y_{e}.
\end{equation}

In these equations $T$, $A_{i}$, $N_{i}$, $Z_{i}$, $Y_{e}$, $Y_{i}$, $G_{i}$, $\mu_{i}$ and $B_{i}$ are the temperature, atomic number, neutron number, proton number, electron fraction (electron abundance per baryon), ion abundance per baryon, nuclear partition function, chemical potential (including the nuclear rest-mass energy) and ion binding energy. The $\mu^{C}_{i}$ are the Coulomb corrections for the NSE state in a dense plasma (see Appendix~\ref{app:Coulomb}). The binding energy data for a large collection of nuclei can be found in \cite{2018IJMPE..2750015M} and the temperature-dependent partition functions are found in \cite{2000ADNDT..75....1R,2003ApJS..147..403R}. Even though we take into account Coulomb corrections in NSE we assume that the baryonic mass can be described by an ideal gas\footnote{Since bulk viscosity effects appear as a consequence of correlations between ion velocities due to Coulomb interactions and of large relaxation times to reach local equilibrium, the NSE and ideal gas assumptions imply that imposing Stokes' hypothesis becomes de rigueur \cite{1965itpg.book.....V,1984oup..book.....M,Buresti2015}}\textsuperscript{,}\footnote{\textcolor{black}{We will consider accretion rates of up to 1$M_{\odot}$~s$^{-1}$. These disks reach densities of $10^{13}$~g~cm$^{-3}$. Baryons can be lightly degenerate at these densities but we will still assume that the baryonic mass can be described by an ideal gas.}} and

\begin{equation}
P_{B} = \sum_{i}P_{i} = n_{B}T\sum_{i}\frac{X_{i}}{A_{i}}, \;\; U_{B} = \frac{3}{2}P_{B}.
    \label{eq:baryonpressureenergy}
\end{equation}

The disk also contains photons, electrons, positrons, neutrinos and antineutrinos. As it is usual in neutrino oscillations analysis, we distinguish only between electron (anti)neutrinos $\nu_{e},(\bar{\nu}_e)$ and $x$ (anti)neutrinos $\nu_{x}(\bar{\nu}_{x})$, where $x=\mu+\tau$ is the superposition of muon neutrinos and tau neutrinos. Photons obey the usual relations

\begin{equation}
P_{\gamma} = \frac{\pi^{2}T^{4}}{45} , \;\; U_{\gamma} = 3P_{\gamma},
    \label{eq:radiationpressureenergy}
\end{equation}

while, for electrons and positrons we have

\begin{subequations}
\begin{align}
n_{e^{\pm}} &= \frac{\sqrt{2}}{\pi^{2}}\xi^{3/2}\left[\mathcal{F}_{1/2,0}\left(\xi,\eta_{e^{\pm}}\right) + \xi \mathcal{F}_{3/2,0}\left(\xi,\eta_{e^{\pm}}\right)\right], \label{eq:elecnumb} \\
U_{e^{\pm}} &= \frac{\sqrt{2}}{\pi^{2}}\xi^{5/2}\left[\mathcal{F}_{3/2,0}\left(\xi,\eta_{e^{\pm}}\right) + \xi \mathcal{F}_{5/2,0}\left(\xi,\eta_{e^{\pm}}\right)\right], \label{eq:elecener} \\
P_{e^{\pm}} &= \frac{2\sqrt{2}}{3\pi^{2}}\xi^{5/2}\left[\mathcal{F}_{3/2,0}\left(\xi,\eta_{e^{\pm}}\right) + \frac{\xi}{2} \mathcal{F}_{5/2,0}\left(\xi,\eta_{e^{\pm}}\right)\right], \label{eq:elecpress}
\end{align}\label{eq:fermion}\end{subequations}

with $\xi=T/m_{e}$ and written in terms of the generalized Fermi functions

\begin{equation}
\mathcal{F}_{k,\ell}\left(y,\eta\right) = \int\limits_{\ell}^{\infty}\frac{x^{k}\sqrt{1+xy/2}}{\exp\left(x-\eta\right) + 1}dx.
    \label{eq:generalfermi}
\end{equation}

In these equations $\eta_{e^{\pm}}=\left(\mu_{e^{\pm}} - m_{e}\right)/T$ is the electron (positron) degeneracy parameter without rest-mass contributions (not to be confused with $\eta$ in Sec.~(\ref{subsec:1.2})). Since electrons and positrons are in equilibrium with photons due to the pair creation and annihilation processes $(e^{-}\!+e^{+}\! \to 2\gamma)$ we know that their chemical potentials are related by $\mu_{e^{+}} = -\mu_{e^{-}}$, which implies $\eta_{e^{+}} = -\eta_{e^{-}} - 2/\xi$. From the charge neutrality condition and we obtain 

\begin{equation}
n_{B}Y_{e} = n_{e^{-}} - n_{e^{+}}.
    \label{eq:chargeneutrality}
\end{equation}

For neutrinos, the story is more complicated. In the absence of oscillations and if the disk is hot and dense enough for neutrinos to be trapped within it and in thermal equilibrium, $n_{\nu},U_{\nu},P_{\nu}$ can be calculated with Fermi-Dirac statistics using the same temperature $T$

\begin{subequations}
\begin{align}
n^{\textrm{trapped}}_{\nu\left(\bar{\nu}\right)} &= \frac{T^{3}}{\pi^{2}}\mathcal{F}_{2,0}\left(\eta_{\nu\left(\bar{\nu}\right)}\right), \label{eq:numbpost} \\
U^{\textrm{trapped}}_{\nu\left(\bar{\nu}\right)} &= \frac{T^{4}}{\pi^{2}}\mathcal{F}_{3,0}\left(\eta_{\nu\left(\bar{\nu}\right)}\right), \label{eq:enpost} \\
P^{\textrm{trapped}}_{\nu\left(\bar{\nu}\right)} &= \frac{U^{\textrm{trapped}}_{\nu\left(\bar{\nu}\right)}}{3}, \label{eq:presspost}
\end{align}
\label{eq:neutrinospost}\end{subequations}

where it is understood that $\mathcal{F}(\eta)=\mathcal{F}(y\! =\! 0,\eta)$ with $\eta_{\nu\left(\bar{\nu}\right)}=\mu_{\nu\left(\bar{\nu}\right)}/T$ and the ultra-relativistic approximation $m_{\nu} \ll 1$ for any neutrino flavour is used. If thermal equilibrium is has not been achieved, Eq.~(\ref{eq:neutrinospost}) cannot be used. Nevertheless, at any point in the disk and for a given value of $T$ and $\rho$, (anti)neutrinos are being created through several processes. The processes we take into account are \emph{pair annihilation} $e^{-}+e^{+}\! \to\! \nu+\bar{\nu}$, \emph{electron or positron capture by nucleons} $p + e^{-}\! \to n + \nu_{e} \textrm{ or } n + e^{+}\! \to p + \bar{\nu}_{e}$, \emph{electron capture by ions} $A + e^{-}\! \to A' + \nu_{e}$, plasmon decay $\tilde{\gamma}\!\to\nu+\bar{\nu}$ and \emph{nucleon-nucleon bremsstrahlung} $n_{1}+n_{2}\to n_{3}+n_{4} + \nu + \bar{\nu}$. The emission rates can be found in Appendix~\ref{app:emisscross}. The chemical equilibrium for these processes determines the values of $\eta_{\nu\left(\bar{\nu}\right)}$. In particular,

\begin{subequations}
\begin{align}
\eta_{\nu_{e}} &= \eta_{e^{-}} + \ln\left(\frac{X_{p}}{X_{n}}\right) + \frac{1 - \mathbb{Q}}{\xi}, \\
\eta_{\bar{\nu}_{e}} &= - \eta_{\nu_{e}}, \\
\eta_{\nu_{x}} &= \eta_{\bar{\nu}_{x}} = 0,
\end{align}\label{eq:neutrinoschempot}\end{subequations}

satisfy all equations. Here, $\mathbb{Q}= (m_{n}-m_{p})/m_{e} \approx 2.531$. Once the (anti)neutrino number and energy emission rates $(R_{i},Q_{i})$ are calculated for each process $i$, the (anti)neutrino thermodynamic quantities are given by

\begin{subequations}
\begin{align}
n^{\textrm{free}}_{\nu\left(\bar{\nu}\right)} &= \color{black} H\sum_{i}R_{i,\nu\left(\bar{\nu}\right)}, \label{eq:numbpre} \\
U^{\textrm{free}}_{\nu\left(\bar{\nu}\right)} &= H\sum_{i}Q_{i,\nu\left(\bar{\nu}\right)},  \label{eq:enpre} \\
P^{\textrm{free}}_{\nu\left(\bar{\nu}\right)} &= \frac{U^{\textrm{free}}_{\nu\left(\bar{\nu}\right)}}{3}. \label{eq:presspre}
\end{align}
\label{eq:neutrinospre}\end{subequations}

Remember we are using Planck units so in these expressions there should be an $H/c$ instead of just an $H$. The transition for each (anti)neutrino flavour between both regimes occurs when Eq.~(\ref{eq:enpost}) and Eq.~(\ref{eq:enpre}) are equal and it can be simulated by defining the parameter

\begin{equation}
w_{\nu\left(\bar{\nu}\right)} = \frac{U^{\textrm{free}}_{\nu\left(\bar{\nu}\right)}}{U^{\textrm{free}}_{\nu\left(\bar{\nu}\right)} + U^{\textrm{trapped}}_{\nu\left(\bar{\nu}\right)}}.
    \label{eq:disttran}
\end{equation}

With this equation, the (anti)neutrino average energy can be defined as

\begin{equation}
\langle E_{\nu\left(\bar{\nu}\right)} \rangle = \left(1 - w_{\nu\left(\bar{\nu}\right)}\right)\frac{U^{\textrm{free}}_{\nu\left(\bar{\nu}\right)}}{n^{\textrm{free}}_{\nu\left(\bar{\nu}\right)}} + w_{\nu\left(\bar{\nu}\right)}\frac{U^{\textrm{trapped}}_{\nu\left(\bar{\nu}\right)}}{n^{\textrm{trapped}}_{\nu\left(\bar{\nu}\right)}}. 
    \label{eq:neutrinoaverageenergy}
\end{equation}

and the approximated number and energy density are 

\begin{subequations}
\begin{align}
n_{\nu\left(\bar{\nu}\right)} &= 
     \begin{cases}
     &\!\!\!\!\!\! n^{\textrm{free}}_{\nu\left(\bar{\nu}\right)},\;\;\;\; {\textrm{if}}\; w_{\nu\left(\bar{\nu}\right)} < 1/2. \\
     &\!\!\!\!\!\! n^{\textrm{trapped}}_{\nu\left(\bar{\nu}\right)}\!\!\!\!\!\!,\;\;\;\; {\textrm{if}}\; w_{\nu\left(\bar{\nu}\right)} \geq 1/2. \\ 
     \end{cases} \label{eq:numbreal} \\
U_{\nu\left(\bar{\nu}\right)} &= 
     \begin{cases}
     &\!\!\!\!\!\! U^{\textrm{free}}_{\nu\left(\bar{\nu}\right)},\;\;\;\; {\textrm{if}}\; w_{\nu\left(\bar{\nu}\right)} < 1/2. \\
     &\!\!\!\!\!\! U^{\textrm{trapped}}_{\nu\left(\bar{\nu}\right)}\!\!\!\!\!\!,\;\;\;\; {\textrm{if}}\; w_{\nu\left(\bar{\nu}\right)} \geq 1/2. \\ 
     \end{cases} \label{eq:enerreal} \\
P_{\nu\left(\bar{\nu}\right)} &= \frac{U_{\nu\left(\bar{\nu}\right)}}{3}. \label{eq:pressreal}
\end{align}     
\label{eq:neutrinosreal}\end{subequations}

Note that both Eq.~(\ref{eq:presspre}) and Eq.~(\ref{eq:pressreal}) are approximations since they are derived from equilibrium distributions, but they help make the transition smooth. Besides, the neutrino pressure before thermal equilibrium is negligible. This method was presented in \cite{2007ApJ...657..383C} where it was used only for electron (anti)neutrinos. The total (anti)neutrino number and energy flux through one the disk's faces can be approximated by

\begin{subequations}
\begin{gather}
\dot{n}_{\nu_{j}\left(\bar{\nu}_{j}\right)}=\!\sum_{j\in\left\{e,x\right\}} \frac{n_{\nu_{j}\left(\bar{\nu}_{j}\right)}}{1+\tau_{\nu_{j}\left(\bar{\nu}_{j}\right)}}, \\ F_{\nu_{j}\left(\bar{\nu}_{j}\right)}=\!\sum_{j\in\left\{e,x\right\}} \frac{U_{\nu_{j}\left(\bar{\nu}_{j}\right)}}{1+\tau_{\nu_{j}\left(\bar{\nu}_{j}\right)}},
\end{gather}\label{eq:neutrinofluxpost}\end{subequations}

where $\tau_{\nu_{i}}$ is the total optical depth for the (anti)neutrino $\nu_{i}\left(\bar{\nu}_{i}\right)$. Collecting all the expressions we write the total internal energy and total pressure

\begin{subequations}
\begin{gather}
U = \sum_{j\in\left\{e,x\right\}}\left(U_{\nu_{j}} + U_{\bar{\nu}_{j}}\right) + U_{B} + U_{e^{-}} + U_{e^{+}} + U_{\gamma}\;\; {\textrm{}},  \\
P = \sum_{j\in\left\{e,x\right\}}\left(P_{\nu_{j}} + P_{\bar{\nu}_{j}}\right) + P_{B} + P_{e^{-}} + P_{e^{+}} + P_{\gamma}\;\; {\textrm{}}.
\end{gather}\label{eq:energypressure}\end{subequations}

The (anti)neutrino energy flux through the disk faces contributes to the cooling term in the energy conservation equation but it is not the only one. Another important energy sink is \emph{photodisintegration} of ions. To calculate it we proceed as follows. The energy spent to knocking off a nucleon of an ion $i$ is equal to the binding energy per nucleon $B_{i}/A_{i}$. Now, consider a fluid element of volume $V$ whose moving walls are attached to the fluid so that no baryons flow in or out. The total energy of photodisintegration contained within this volume is the sum over $i$ of (energy per nucleon of ion $i$)$\times$(\# of freed nucleons of ion $i$ inside $V$). This can be written as $\sum_{i}(B_{i}/A_{i})n_{f,i}V$, or, alternatively, $n_{B}V\sum_{i}(B_{i}/A_{i})X_{f,i}$. If we approximate $B_{i}/A_{i}$ by the average binding energy per nucleon $\bar{B}$ (which is a good approximation save for a couple of light ions) the expression becomes $n_{B}V\bar{B}\sum_{i}X_{f,i} = n_{B}V\bar{B}X_{f} = n_{B}V\bar{B}(X_{p} + X_{n})$. \textcolor{black}{We place the value of $\bar{B}$ in Sec.~\ref{sec:3}.}

The rate of change of this energy on the LRF, denoting the proper time by $\lambda$, is

\begin{equation}
\frac{d}{d\lambda}\left[n_{B}V\bar{B}\left(X_{p}+X_{n}\right)\right] =  n_{B}V\bar{B}\frac{d}{d\lambda}\left(X_{p}+X_{n}\right).
    \label{eq:photodis1}
\end{equation}

The derivative of $n_{B}V$ vanishes by baryon conservation. Transforming back to CF and taking the average we find the energy density per unit time used in disintegration of ions

\begin{equation}
\epsilon_{\textrm{ions}} = n_{B}\bar{B}u^{r}H\partial_{r}\left(X_{p}+X_{n}\right).
    \label{eq:photodis2}
\end{equation}

The average energy density measured on the LRF $\epsilon$ appearing in Eq.~(\ref{eq:energycon}) is

\begin{equation}
\epsilon = \epsilon_{\textrm{ions}} + \frac{1}{H}\sum_{i\in\left\{e,x\right\}}\left( F_{\nu_{i}} + F_{\bar{\nu}_{i}}\right).
    \label{eq:finalepsilon}
\end{equation}

Finally, a similar argument allows us to obtain the equation of lepton number conservation. For any lepton $\ell$, the total lepton number density is $\sum_{\ell \in \{e,\mu,\tau\}} \left(n_{\ell}-n_{\bar{\ell}}+n_{\nu_{\ell}}-n_{\bar{\nu}_{\ell}} \right)$. So, with Eq.~(\ref{eq:chargeneutrality}), calculating the rate of change as before, using Gauss' theorem and taking the average we get

\begin{equation}
u^{r}H\left[n_{B}\partial_{r}Y_{e} + \partial_{r} \!\! \sum_{\ell \in \{e,x\}} \! \left(n_{\nu_{\ell}}\! - n_{\bar{\nu}_{\ell}}\right)\right] = \! \sum_{\ell \in \{e,x\}} \! \left( \dot{n}_{\bar{\nu}_{\ell}}\! - \dot{n}_{\nu_{\ell}} \right),
\label{eq:leptonconservation}
\end{equation}

where the right hand side represents the flux of lepton number through the disk's surface.

%%%%%%%%%%%%%%%%%%%%%%%%%%%%%%%%%%%%%%%%%%%%%%%%%%%%%%%%%%%%%%%%%%%%%%%%%%%%%%%%%%%%%%%%%%%%%%%%%%%%%%%%%%%%%%%%%% 
\section{Neutrino Oscillations}\label{sec:2}
%%%%%%%%%%%%%%%%%%%%%%%%%%%%%%%%%%%%%%%%%%%%%%%%%%%%%%%%%%%%%%%%%%%%%%%%%%%%%%%%%%%%%%%%%%%%%%%%%%%%%%%%%%%%%%%%%%

To study the flavour evolution of neutrinos within a particular system, a Hamiltonian governing neutrino oscillation must be set up. The relative strength of the potentials appearing in such Hamiltonian depends on four elements: geometry, mass content, neutrino content and neutrino mass hierarchy. Geometry refers to the nature of net neutrino fluxes and possible gravitational effects. Mass and neutrino content refers to the distribution of leptons of each flavour $(e,\mu,\tau)$ present in the medium. Finally, mass hierarchy refers to the relative values of the masses $m_{1},m_{2},m_{3}$ for each neutrino mass eigenstates (see Table~\ref{tab:massvalues}). We dedicate this section to a detailed derivation of the equations of flavour evolution for a neutrino dominated accretion disk. To maintain consistency with traditional literature of neutrino oscillations we will reuse some symbols appearing in \textcolor{black}{previous} sections. To avoid confusion we point out that the symbols in this section are independent of \textcolor{black}{previous} sections unless we explicitly draw a comparison.

\begin{table}[H]
\caption{Mixing and squared mass differences as they appear in \cite{PhysRevD.98.030001}. Error values in parenthesis are shown in 3$\sigma$ interval. The squared mass difference is defined as $\Delta m^2 = m^{2}_{3}-\left(m^{2}_{2}+m^{2}_{1}\right)/2$ and its sign depends on the hierarchy $m_{1}<m_{2}<m_{3}$ or $m_{3}<m_{1}<m_{2}$.}
\centering
\begin{tabular}{l}
\toprule
$ \Delta m^{2}_{21}  = 7.37\,(6.93-7.96)\times 10^{-5}$ eV$^2 $ \\
$|\Delta m^{2}| = 2.56\,(2.45-2.69) \times 10^{-3}$ eV$^2$ Normal Hierarchy \\
$|\Delta m^{2}| = 2.54 \,(2.42-2.66) \times 10^{-3}$ eV$^2$ Inverted Hierarchy \\ 
$\sin^2\theta_{12} = 0.297\,(0.250-0.354)$ \\
$\sin^2\theta_{23} (\Delta m^2 > 0) = 0.425\,(0.381-0.615)$ \\
$\sin^2\theta_{23} (\Delta m^2 < 0) = 0.589\,(0.383-0.637)$ \\
$\sin^2\theta_{13} (\Delta m^2 > 0) = 0.0215\,(0.0190-0.0240)$ \\
$\sin^2 \theta_{13} (\Delta m^2 < 0) = 0.0216\,(0.0190-0.0242)$ \\
\bottomrule
\end{tabular}
\label{tab:massvalues}
\end{table}
%

%%%%%%%%%%%%%%%%%%%%%%%%%%%%%%%%%%%%%%%%%%%%%%%%%%%%%%%%%%%%%%%%
%%%%%%%%%%%%%%%%%%%%%%%%%%%%%%%%%%%%%%%%%%%%%%%%%%%%%%%%%%%%%%%%
\subsection{Equations of Oscillation}\label{subsec:2.1}
%%%%%%%%%%%%%%%%%%%%%%%%%%%%%%%%%%%%%%%%%%%%%%%%%%%%%%%%%%%%%%%%
%%%%%%%%%%%%%%%%%%%%%%%%%%%%%%%%%%%%%%%%%%%%%%%%%%%%%%%%%%%%%%%%

The equations that govern the evolution of an ensemble of mixed neutrinos are the Boltzmann collision equations

\begin{subequations}
\begin{gather}
i\dot{\rho}_{\mathbf{p},t} = C \left(\rho_{\mathbf{p},t}\right),\\
i\dot{\bar{\rho}}_{\mathbf{p},t} = C \left(\bar{\rho}_{\mathbf{p},t}\right).
\end{gather}\label{eq:Liouville1}\end{subequations}

The collision terms should include the vacuum oscillation plus all possible scattering interactions that neutrinos undergo through their propagation. For free streaming neutrinos, only the vacuum term and the forward-scattering interactions are taken into account so that the equations become

\begin{subequations}
\begin{gather}
i\dot{\rho}_{\mathbf{p},t} =\left[\mathsf{H}_{\mathbf{p},t},\rho_{\mathbf{p},t}\right], \\
i\dot{\bar{\rho}}_{\mathbf{p},t} = \left[\mathsf{\bar{H}}_{\mathbf{p},t},\bar{\rho}_{\mathbf{p},t}\right].
\end{gather}\label{eq:Liouville}\end{subequations}

Here, $\mathsf{H}_{\mathbf{p},t}$ ($\bar{\mathsf{H}}_{\mathbf{p},t}$) is the oscillation Hamiltonian for (anti)neutrinos and $\rho_{\mathbf{p},t}$ ($\bar{\rho}_{\mathbf{p},t}$) is the matrix of occupation numbers: $(\rho_{\mathbf{p},t})_{ij}=\langle a^{\dagger}_{j}a_{i}\rangle_{\mathbf{p},t}$ for neutrinos and ($(\bar{\rho}_{\mathbf{p},t})_{ij}=\langle \bar{a}^{\dagger}_{i}\bar{a}_{j}\rangle_{\mathbf{p},t}$ for antineutrinos), for each momentum $\mathbf{p}$ and flavours $i,j$. The diagonal elements are the distribution functions $f_{\nu_{i}\left(\bar{\nu}_{i}\right)}\left(\mathbf{p}\right)$ such that their integration over the momentum space gives the neutrino number density $n_{\nu_{i}}$ of a determined flavour $i$ at time $t$. The off-diagonal elements provide information about the \emph{overlapping} between the two neutrino flavours. Taking into account the current-current nature of the weak interaction in the standard model, the Hamiltonian for each equation is \cite{Dolgov:1980cq,Sigl:1992fn,Hannestad:2006nj}

\begin{subequations}
\begin{align}
\mathsf{H}_{\mathbf{p},t}&=\Omega_{\mathbf{p},t}+\sqrt{2}G_{F}\!\!\int\!\!\left( l_{\mathbf{q},t}-\bar{l}_{\mathbf{q},t}\right)\left( 1-\mathbf{v}_{\mathbf{q},t}\cdot\mathbf{v}_{\mathbf{p},t} \right)\frac{d^3\mathbf{q}}{\left(2\pi\right)^3} +\sqrt{2}G_{F}\!\!\int\!\!\left( \rho_{\mathbf{q},t}-\bar{\rho}_{\mathbf{q},t}\right)\left( 1-\mathbf{v}_{\mathbf{q},t}\cdot\mathbf{v}_{\mathbf{p},t} \right)\frac{d^3\mathbf{q}}{\left(2\pi\right)^{3}},\\
\mathsf{\bar{H}}_{\mathbf{p},t}&=-\Omega_{\mathbf{p},t}+\sqrt{2}G_{F}\!\!\int\!\!\left( l_{\mathbf{q},t}-\bar{l}_{\mathbf{q},t}\right)\left( 1-\mathbf{v}_{\mathbf{q},t}\cdot\mathbf{v}_{\mathbf{p},t} \right)\frac{d^3\mathbf{q}}{\left(2\pi\right)^3}  +\sqrt{2}G_{F}\!\!\int\!\!\left( \rho_{\mathbf{q},t}-\bar{\rho}_{\mathbf{q},t}\right)\left( 1-\mathbf{v}_{\mathbf{q},t}\cdot\mathbf{v}_{\mathbf{p},t} \right)\frac{d^3\mathbf{q}}{\left(2\pi\right)^{3}}.
\end{align}\label{eq:FullHam}\end{subequations}

where $G_{F}$ is the Fermi coupling constant, $\Omega_{\mathbf{p},t}$ is the matrix of vacuum oscillation frequencies, $l_{\mathbf{p},t}$ and $\bar{l}_{\mathbf{p},t}$ are matrices of occupation numbers for charged leptons built in a similar way to the neutrino matrices, and $\mathbf{v}_{\mathbf{p},t}=\mathbf{p}/ p$ is the velocity of a particle with momentum $\mathbf{p}$ (either neutrino or charged lepton). As stated before, we will only consider two neutrino flavours: $e$ and $x=\mu+\tau$. Three-flavour oscillations can be approximated by two-flavour oscillations as a result of the strong hierarchy of the squared mass differences $\vert \Delta m^{2}_{13} \vert \approx \vert \Delta m^{2}_{23} \vert \gg \vert \Delta m^{2}_{12} \vert$. In this case, only the smallest mixing angle $\theta_{13}$ is considered. We will drop the suffix for the rest of the discussion. Consequently, the relevant oscillations are $\nu_e \rightleftharpoons \nu_x$ and $\bar\nu_e \rightleftharpoons \bar\nu_x$, and each term in the Hamiltonian governing oscillations becomes a 2~$\times$~2 Hermitian matrix. Now, consider an observer on the LRF (which is almost identical to the CRF due to Eq.~(\ref{eq:appr1}) at a point $r$. In its spatial local frame, the unit vectors  $\hat{x},\hat{y},\hat{z}$ are parallel to the unit vectors $\hat{r},\hat{\theta},\hat{\phi}$ of the CF, respectively. Solving Eq.~(\ref{eq:Liouville}) in this coordinate system would yield matrices $\rho,\bar{\rho}$ as functions of time $t$. However, in our specific physical system, both the matter density and the neutrino density vary with the radial distance from the BH. This means that the equations of oscillations must be written in a way that makes explicit the spatial dependence, i.e. in terms of the coordinates $x,y,z$. For a collimated ray of neutrinos, the expression $dt=dr$ would be good enough, but for radiating extended sources or neutrino gases the situation is more complicated.

In Eq.~(\ref{eq:Liouville}) we must replace the matrices of occupation numbers by the space-dependent Wigner functions $\rho_{\mathbf{p,x},t}$ (and $\bar{\rho}_{\mathbf{p,x},t}$) and the total time derivative by the Liouville operator \cite{Cardall:2007zw,Strack:2005ux}

\begin{equation}
\dot{\rho}_{\mathbf{p,x},t}\quad=\overbrace{\frac{\partial \rho_{\mathbf{p,x},t}}{\partial t}}^{\mbox{Explicit Time}}\!\!\!\!\!\! + \; \overbrace{\mathbf{v}_\mathbf{p} \cdot \nabla_{\mathbf{x}}\,\rho_{\mathbf{p,x},t}}^{\mbox{Drift}} \; + \!\!\!\! \overbrace{\dot{\mathbf{p}}\cdot\nabla_{\mathbf{p}}\,\rho_{\mathbf{p,x},t}}^{\mbox{External Forces}}.
	\label{eq:Loperator}
\end{equation}

In this context, $\mathbf{x}$ represents a vector in the LRF. In the most general case, finding $\rho_{\mathbf{p,x},t}$ and $\bar{\rho}_{\mathbf{p,x},t}$ means solving a 7D neutrino transport problem in the variables $x,y,z,p_x,p_y,p_z,t$. Since our objective is to construct a simple model of neutrino oscillations inside the disk, to obtain the specific form of Eq.~(\ref{eq:Liouville}) we must simplify the equations by imposing on it conditions that are consistent with the assumptions made in Sec.~\ref{sec:1}. 

\begin{itemize}[leftmargin=*,labelsep=5.8mm]
    \item 
    Due to axial symmetry, the neutrino density is constant along the $\mathbf{z}$ direction. Moreover, since neutrinos follow null geodesics, we can set $\dot{p}_{z} \approx \dot{p}_{\phi}=0$.
    \item 
    Within the thin disk approximation (as represented by Eq.~(\ref{eq:vertaverage})) the neutrino and matter densities are constant along the $\mathbf{y}$ direction and the momentum change due to curvature along this direction can be neglected, that is, $\dot{p}_{y} \approx 0$.
    \item 
    In the LRF, the normalized radial momentum of a neutrino can be written as $p_{x} = \pm r/\sqrt{r^2-2Mr+M^2a^2}$. Hence, the typical scale of the change of momentum with radius is $\Delta r_{p_{x},\text{eff}} = \left\vert d\ln p_{x}/dr  \right\vert^{-1} = (r/M)\left(r^2 -2Mr+M^2a^2\right)/\left(Ma^2 - r\right)$, which obeys $\Delta r_{p_{x},\text{eff}} > r_{s}$ for $r > 2 r_{\text{in}}$. This means we can assume $\dot{p}_{x} \approx 0$ up to regions very close to the inner edge of the disk. 
    \item 
    We define an effective distance $\Delta r_{\rho,\text{eff}} = \left\vert d\ln\left( Y_{e}n_{B} \right)/dr\right\vert^{-1}$. For all the systems we evaluated we found that is comparable to the height of the disk $(\Delta r_{\rho,\text{eff}}\sim 2-5$~$r_{s}$). This means that at any point of the disk we can calculate neutrino oscillations in a small regions assuming that both the electron density and neutrino densities are constant.  
    \item 
    We neglect energy and momentum transport between different regions of the disk by neutrinos that are recaptured by the disk due to curvature. This assumption is reasonable except for regions very close to the BH but is consistent with the thin disk model \cite[see e.g.][]{1974ApJ...191..499P}. We also assume initially that the neutrino content of neighbouring regions of the disk (different values of $r$) do not affect each other. As a consequence of the results discussed above, we assume that at any point inside the disk and at any instant of time an observer can describe both the charged leptons and neutrinos as isotropic gases around small enough regions of the disk. This assumption is considerably restrictive but we will generalize it in Sec.~\ref{sec:4}.
\end{itemize}

The purpose of these approximations is twofold. On one hand, we can reduce the problem considerably since they allow us to add the neutrino oscillations to a steady-state disk model by simply studying the behaviour of neutrinos at each point of the disk using the constant values of density and temperature at that point. We will see in Sec.~\ref{sec:4}, that this assumption would correspond \textcolor{black}{to a} transient state of an accretion disk since, very fast, neighbouring regions of the disk start interacting. On the other hand, the approximations allow us to simplify the equations of oscillation considering that all but the first term in Eq.~(\ref{eq:Loperator}) vanish, leaving only a time derivative. In addition, both terms of the form $\mathbf{v}_{\mathbf{q},t}\cdot\mathbf{v}_{\mathbf{p},t}$ in Eq.~(\ref{eq:FullHam}) average to zero so that $\rho_{\mathbf{p,x},t} = \rho_{p,t}$ and $\bar{\rho}_{\mathbf{p,x},t} = \bar{\rho}_{p,t}$. We are now in a position to derive the simplified equations of oscillation for this particular model. Let us first present the relevant equations for neutrinos. Due to the similarity between $\mathsf{H}_{p,t}$ and $\mathsf{\bar{H}}_{p,t}$, the corresponding equations for antineutrinos can be obtained analogously. For simplicity, we will drop the suffix $t$ since the time dependence is now obvious. In the two-flavour approximation, $\rho_{p}$ is a $2\times 2$ Hermitian matrix and can be expanded in terms of the Pauli matrices $\sigma_i$ and a polarization vector $\mathsf{P}_{p} = \left(\mathsf{P}^{x},\mathsf{P}^{y},\mathsf{P}^{z}\right)$ in the neutrino flavour space, such that

\begin{equation}
\small
\rho_{p}=\begin{pmatrix}
  \rho_{ee} & \rho_{ex}\\
  \rho_{xe} & \rho_{xx}\\
   \end{pmatrix}
   =
 \frac{1}{2}\left(f_{p}\boldsymbol{1} +\mathsf{P}_p \cdot \vec \sigma\right),
	\label{eq:expansion_of_rho}
\end{equation}

where $f_{p}={\textrm{Tr}}[\rho_{p}]=f_{\nu_e}(p)+f_{\nu_x}(p)$ is the sum of the distribution functions for $\nu_e$ and $\nu_x$. Note that the $z$ component of the polarization vector obeys

\begin{equation}
\mathsf{P}^{z}_{p} = f_{\nu_e}(p)-f_{\nu_x}(p).
\label{eq:pzeta}
\end{equation}

Hence, this component tracks the fractional flavour composition of the system. Appropriately normalizing $\rho_{p}$ allows to define a survival and mixing probability

\begin{subequations}
\begin{gather}
P_{p,\nu_{e} \to \nu_{e}} = \frac{1}{2}\left( 1 + \mathsf{P}^{z}_{p} \right),\\
P_{p,\nu_{e} \to \nu_{x}} = \frac{1}{2}\left( 1 - \mathsf{P}^{z}_{p}\right).
\end{gather}\label{eq:survprobability}\end{subequations}

The Hamiltonian can be written as a sum of three interaction terms:

\begin{equation}
\mathsf{H} = \mathsf{H}_{\mbox{\footnotesize{vacuum}}} + \mathsf{H}_{\mbox{\footnotesize{matter}}} + \mathsf{H}_{\nu\nu}.
\label{eq:neutrinohamiltonian}
\end{equation}

The first term is the Hamiltonian in vacuum~\cite{Qian:1994wh}:

\begin{equation}
\mathsf{H}_{\mbox{\footnotesize{vacuum}}} = \frac{\omega_p}{2} \begin{pmatrix}
  -\cos 2\theta & \sin 2\theta\\
  \sin 2\theta & \cos 2\theta \\
  \end{pmatrix} =\frac{\omega_p}{2} \mathbf{B}\cdot \vec{\sigma},
\label{eq:Hvacuum}
\end{equation}

where $\omega_p = \Delta m^2/2p$, $\mathbf{B}=(\sin2\theta,0,-\cos 2 \theta)$ and $\theta$ is the smallest neutrino mixing angle in vacuum. The other two terms in Eqs.~(\ref{eq:FullHam}) are special since they make the evolution equations non-linear. Since we are considering that the electrons inside the form an isotropic gas, the vector $\mathbf{v}_{\mathbf{q}}$ in the first integral is distributed uniformly on the unit sphere and the factor $\mathbf{v}_\mathbf{q}\cdot\mathbf{v}_p$ averages to zero. After integrating the matter Hamiltonian is given by

\begin{equation}
\mathsf{H}_{\mbox{\footnotesize{matter}}} =
\frac{\lambda}{2}\left(
 \begin{array}{cc}
  1 & 0\\
  0 & -1 \\
   \end{array}\right)
   =\frac{\lambda}{2} \mathbf{L} \cdot \vec{\sigma},
	\label{eq:Hmatter}
\end{equation}

where $\lambda = \sqrt{2}G_{F}\left(n_{e^-} - n_{e^+}\right)$ is the charged current matter potential and $\mathbf{L}=(0,0,1)$. Similarly, the same product disappears in the last term and after integrating we get

\begin{equation}
\mathsf{H}_{\nu\nu} = \sqrt{2}G_{F}\left[\mathsf{P}-\bar{\mathsf{P}} \right]\cdot \vec{\sigma}.
\label{eq:Hnunu}
\end{equation}

Clearly, $\mathsf{P} = \int \mathsf{P}_p\, d{\mathbf{p}}/(2 \pi)^3$. Introducing every Hamiltonian term in Eqs.~(\ref{eq:Liouville}), and using the commutation relations of the Pauli matrices, we find the equations of oscillation for neutrinos and antineutrinos for each momentum mode $p$:

\begin{subequations}
\begin{gather}
\dot{\mathsf{P}}_{p} = \left[ \omega_p \mathbf{B} + \lambda \mathbf{L} + \sqrt{2}G_{F}\left(\mathsf{P}-\bar{\mathsf{P}} \right) \right]\times \mathsf{P}_{p}, \\
\dot{\bar{\mathsf{P}}}_{p} = \left[ -\omega_p \mathbf{B} + \lambda \mathbf{L} + \sqrt{2}G_{F}\left(\mathsf{P}-\bar{\mathsf{P}} \right) \right]\times \bar{\mathsf{P}}_{p},
\end{gather}\label{eq:Hnu}\end{subequations}

where we have assumed that the total neutrino distribution remains constant, $\dot{f}_{p} = 0$. This shows how the polarization vectors can be normalized. Performing the transformation $\mathsf{P}_{p}/f_{p} \mapsto \mathsf{P}_{p}$ and $\bar{\mathsf{P}}_{p}/\bar{f}_{p} \mapsto \bar{\mathsf{P}}_{p}$ and, multiplying and dividing the last term by the total neutrino density Eqs.~(\ref{eq:Hnu}) become

\begin{subequations}
\begin{gather}
\dot{\mathsf{P}}_{p} = \left[ \omega_p \mathbf{B} + \lambda \mathbf{L} + \mu\mathbf{D} \right]\times \mathsf{P}_{p}, \label{eq:Hnu2.1} \\
\dot{\bar{\mathsf{P}}}_{p} = \left[ -\omega_p \mathbf{B} + \lambda \mathbf{L} + \mu\mathbf{D} \right]\times \bar{\mathsf{P}}_{p}, \label{eq:Hnu2.2}\\
\mathbf{D} = \frac{1}{n_{\nu_e}\!+n_{\nu_x}}\int\left(f_{q}\mathsf{P}_{q}-\bar{f}_{q}\bar{\mathsf{P}}_{q} \right)\frac{d{\mathbf{q}}}{(2 \pi)^3}.
\end{gather}\label{eq:Hnu2}\end{subequations}

This is the traditional form of the equations in terms of the vacuum, matter and self-interaction potentials $\omega_p$, $\lambda$ and $\mu$ with

\begin{equation}
\mu = \sqrt{2}G_{F}\!\! \sum_{i\in \{e,x\}}n_{\nu_i}.
\label{eq:SelfPotential}
\end{equation}

Different normalization schemes are possible \cite[see e.g.][]{Hannestad:2006nj,EstebanPretel:2007ec,Dasgupta:2008cu,2016NCimR..39....1M}. By assuming that we can solve the equations of oscillation with constant potentials $\lambda$ and $\mu$ simplifies the problem even further. Following \cite{2006PhRvD..74l3004D}, with the vector transformation (a rotation around the $z$ axis of flavour space)

\begin{equation}
\small
R_{z}=\begin{pmatrix}
  \cos\left(\lambda t\right) & \sin\left(\lambda t\right) & 0\\
  -\sin\left(\lambda t\right) & \cos\left(\lambda t\right) & 0\\
  0 & 0 & 1
   \end{pmatrix},
	\label{eq:transflavour}
\end{equation}

Eqs.~(\ref{eq:Hnu2}) become

\begin{subequations}
\begin{gather}
\dot{\mathsf{P}}_{p} = \left[ \omega_p \mathbf{B} + \mu\mathbf{D} \right]\times \mathsf{P}_{p}, \label{eq:Hnu3.1} \\
\dot{\bar{\mathsf{P}}}_{p} = \left[ -\omega_p \mathbf{B} + \mu\mathbf{D} \right]\times \bar{\mathsf{P}}_{p}, \label{eq:Hnu3.2}
\end{gather}\label{eq:Hnu3}\end{subequations}

eliminating the $\lambda$ potential, but making $\mathbf{B}$ time dependent. Defining the vector $\mathbf{S}_p = \mathsf{P}_p + \bar{\mathsf{P}}_p$ and, adding and subtracting Eq.~(\ref{eq:Hnu3.1}) and Eq.~(\ref{eq:Hnu3.2}) we get 

\begin{subequations}
\begin{gather}
\dot{\mathbf{S}}_{p} = \omega_p \mathbf{B}\times \mathbf{D}_p + \mu\mathbf{D}\times\mathbf{S}_p \approx \mu\mathbf{D}\times\mathbf{S}_p, \\
\dot{\mathbf{D}}_{p} = \omega_p \mathbf{B}\times \mathbf{S}_p + \mu\mathbf{D}\times\mathbf{D}_p \approx \mu\mathbf{D}\times\mathbf{D}_p.\label{eq:Hnu4.2}
\end{gather}\label{eq:Hnu4}\end{subequations}

The last approximation is true if we assume that the self-interaction potential is larger than the vacuum potential $\omega_p /\mu \ll 1$. We will show later that this is the case for thin disks (see Fig.~\ref{fig:Potentials}). The first equation implies that all the vectors $\mathbf{S}_p$ and their integral $\mathbf{S}$ evolve in the same way, suggesting the relation $\mathbf{S}_{p} = \left(f_{p}+\bar{f}_{p}\right)\mathbf{S}$. By replacing in Eq.~(\ref{eq:Hnu4.2}) and integrating

\begin{subequations}
\begin{gather}
\dot{\mathbf{S}} = \mu\mathbf{D}\times\mathbf{S},\\
\dot{\mathbf{D}} = \langle \omega \rangle\mathbf{B}\times\mathbf{S}.
\end{gather}\label{eq:Hnufin}\end{subequations}

where $\langle\omega\rangle = \int \omega_p\left(f_{p}+\bar{f}_{p}\right)d{\mathbf{p}}/(2 \pi)^3$ is the average vacuum oscillation potential. The fact that in our model the equations of oscillations can be written in this way has an important consequence. Usually, as it is done in supernovae neutrino oscillations, to solve Eq.~(\ref{eq:Hnu2}) we would need the neutrino distributions throughout the disk. If neutrinos are trapped, their distribution is given by Eq.~(\ref{eq:neutrinospost}). If neutrinos are free, their temperature is not the same as the disk's temperature. Nonetheless, we can approximate the neutrino distribution in this regime by a Fermi-Dirac distribution with the same chemical potential as defined by Eq.~(\ref{eq:neutrinoschempot}) but with an effective temperature $T^{\textrm{eff}}_{\nu}$. This temperature can be obtained by solving the equation $\langle E_{\nu} \rangle = U\left(T^{\text{eff}}_{\nu},\eta_{\nu}\right)/n\left(T^{\text{eff}}_{\nu},\eta_{\nu}\right)$ which gives

\begin{subequations}
\begin{equation}
T^{\text{eff}}_{\nu_{x},\bar{\nu}_{x}} = \langle E_{\nu_{x},\bar{\nu}_{x}} \rangle \frac{180\,\zeta(3)}{7 \pi^4},
\end{equation}
\begin{equation}
T^{\text{eff}}_{\nu_{e},\bar{\nu}_{e}} = \frac{\langle E_{\nu_{e},\bar{\nu}_{e}} \rangle}{3} \frac{\text{Li}_{3}\left(-\exp\left({\eta_{\nu_{e},\bar{\nu}_{e}}}\right)\right)}{\text{Li}_{4}\left(-\exp\left({\eta_{\nu_{e},\bar{\nu}_{e}}}\right)\right)},
\end{equation}
\label{eq:efftempneu}\end{subequations}

where $\zeta(3)$ is Ap\'ery's constant ($\zeta$ is the Riemann zeta function) and $\text{Li}_{s}\!\left(z\right)$ is Jonqui\`ere's function. For convenience and considering the range of values that the degeneracy parameter reaches (see Sec.~\ref{sec:5}), we approximate the effective temperature of electron neutrinos and antineutrinos with the expressions

\begin{subequations}
\begin{equation}
T^{\text{eff}}_{\nu_{e}} = \frac{\langle E_{\nu_{e}} \rangle}{3}\left( a\eta_{\nu_{e}}^{2} + b\eta_{\nu_{e}} + c \right),
\end{equation}
\begin{equation}
T^{\text{eff}}_{\bar{\nu}_{e}} = \frac{\langle E_{\bar{\nu}_{e}} \rangle}{3}.
\end{equation}\label{eq:efftempneu2}\end{subequations}

with constants $a = 0.0024$, $b = -0.085$, $c = 0.97$. However, Eq.~(\ref{eq:Hnufin}) allow us to consider just one momentum mode, and the rest of the spectrum behaves in the same way.

%%%%%%%%%%%%%%%%%%%%%%%%%%%%%%%%%%%%%%%%%%%%%%%%%%%%%%%%%%%%%%%%%%%%%%%%%%%%%%%%%%%%%%%%%%%%%%%%%%%%%%%%%%%%%%%%%% 
\section{Initial Conditions and Integration}\label{sec:3}
%%%%%%%%%%%%%%%%%%%%%%%%%%%%%%%%%%%%%%%%%%%%%%%%%%%%%%%%%%%%%%%%%%%%%%%%%%%%%%%%%%%%%%%%%%%%%%%%%%%%%%%%%%%%%%%%%%

In the absence of oscillations, we can use Eqs.~(\ref{eq:energycon}),  (\ref{eq:press}) and (\ref{eq:leptonconservation}) to solve for the set of functions $\eta_{e^{-}}\!\left( r \right)$, $\xi\left( r \right)$, $Y_{e}\left( r \right)$ using as input parameters the accretion rate $\dot{M}$, the dimensionless spin parameter $a$, the viscosity parameter $\alpha$ and the BH mass $M$. From \cite{2007ApJ...657..383C,LIU20171} we learn that neutrino dominated disks require accretion between $0.01~M_{\odot}$~ s$^{-1}$ and $1~M_{\odot}$~s$^{-1}$ (this accretion rate range vary depending on the value of $\alpha$). For accretion rates smaller than the lower value, the neutrino cooling is not efficient and, for rates larger than the upper value, the neutrinos are trapped within the flow. We also limit ourselves to the above accretion rate range since it is consistent with the one expected to occur in a BdHN (see e.g. \cite{2014ApJ...793L..36F,2016ApJ...833..107B,2019ApJ...871...14B}). \textcolor{black}{We also know that high spin parameter, high accretion rate, high BH mass and low viscosity parameter produces disks with higher density and higher temperature. This can be explained using the fact that several variables of the disk, like pressure, density and height are proportional to a positive power of $M$ and a positive power of the quotient $\dot{M}/\alpha$. To avoid this semi-degeneracy in the system, reduce the parameter space and considering that we want to focus on the study of the oscillation dynamics inside the disk, we fix the BH mass at $M = 3 M_{\odot}$, the viscosity parameter at $\alpha = 0.01$ and the spin parameter at $a=0.95$ while changing the accretion rate. These values also allow us to compare our results with earlier disk models.} Eqs.~(\ref{eq:energycon}) and (\ref{eq:leptonconservation}) are first order ordinary differential equations and since we perform the integration from an external (far away) radius $r_{\text{out}}$ up to the innermost stable circular orbit $r_{\text{in}}$ we must provide two boundary conditions at $r_{\text{out}}$. Following the induced gravitational collapse (IGC) paradigm of GRBs associated with type Ib/c supernovae we assume that at the external edge of the disk, the infalling matter is composed mainly by the ions present in the material ejected from an explosion of a carbon-oxygen core, that is, mainly oxygen and electrons. This fixes the electron fraction $Y_{e}\left(r_{\text{out}}\right) = 0.5$. \textcolor{black}{We can also calculate the average binding energy per nucleon that appears in Eq.~(\ref{eq:photodis1}) using the data in \cite{2018IJMPE..2750015M}. To establish the NSE we consider H2, H3, HE3, HE4, LI6, LI7, BE7, BE9, BE10, B10, B11, C11, C12, C13, C14, N13, N14, N15, O14, O15,  O16, O17, O18 and obtain the value of the average binding energy per nucleon $\bar{B}=6.35$~MeV.} The second boundary condition can be obtained by the relation $\left(T\eta + m_{B}\right)\sqrt{g_{tt}} =$ constant \cite{1934rtc..book.....T,Klein1949,RevModPhys.21.531}, with $\eta$ the degeneracy parameter of the fluid. If we require the potentials to vanishes at infinity and invoking Euler's theorem we arrive at the relation in the weak field limit

\begin{equation}
\frac{M}{r_{\text{out}}} = \left . \frac{\rho + U + P - TS}{\rho}\right\vert_{r=r_{\text{out}}}.
\label{eq:klein1}
\end{equation}

For a classical gas composed of ions and electrons this relation becomes

\begin{equation}
\frac{M}{r_{\text{out}}} \lesssim  \left . \frac{U}{\rho} \right\vert_{r=r_{\text{out}}}.
\label{eq:klein2}
\end{equation}

\textcolor{
red}{That is, the virial specific energy must be smaller or comparable to the energy per baryon.} Eq.~(\ref{eq:klein2}) can be used together with Eqs.~(\ref{eq:press}) and (\ref{eq:energypressure}) to solve for $\eta_{e^{-}}\!\left( r_{\text{out}} \right)$, $\xi\left( r_{\text{out}} \right)$. The value of $r_{\text{out}}$ is chosen to be at most the circularization radius of the accreting material as described in \cite{2015ApJ...812..100B,2016ApJ...833..107B}. We can estimate this radius by solving for $r$ in the expression of the angular momentum per unit mass for a equatorial circular orbits. So using Eq.~(\ref{eq:fourvel}) we need to solve

\begin{equation}
u_{\phi} = M\frac{x^2 - 2x + a^2}{x^{3/2}\sqrt{ x^3 - 3x + 2a }} \sim 3\times 10^{7}\,\text{cm},
\label{eq:angularmomentumcond1}
\end{equation}

where $x=\sqrt{r/M}$ which yields $r_{\rm out} \sim 1800 r_s$ and the expression is in geometric units. Finally, for the initial conditions to be accepted, they are evaluated by the gravitational instability condition \cite{1978AcA....28...91P}

\begin{equation}
\sqrt{\left. {R^{\,\tilde{\theta}}}_{\tilde{t}\tilde{\theta}\tilde{t}}\, \right|_{_{\theta=\pi/2}}}\Omega \geq 2\sqrt{3}\pi\rho.
\label{eq:angularmomentumcond2}
\end{equation}

Integration of the equations proceeds as follows, with the initial conditions we solve Eq.~(\ref{eq:leptonconservation}) to obtain the electron fraction in the next integration point. With the new value of the electron fraction we solve the differential-algebraic system of Eqs.~(\ref{eq:energycon}) and (\ref{eq:press}) at this new point. This process continues until the innermost stable circular orbit $r_{\text{in}}$ is reached.
To add the dynamics of neutrino oscillations we proceed same as before but at each point of integration, once the values of $Y_e$, $\eta$ and $\xi$ are found, we solve Eq.~(\ref{eq:Hnu2}) for the average momentum mode to obtain the survival probabilities as a function of time. We then calculate the new neutrino and antineutrino distributions with the conservation of total number density and the relations

\begin{subequations}
\begin{gather}
n^{\text{new}}_{\nu_{e}}\left(t\right)=P_{\nu_e\to \nu_e}\left(t\right)n_{\nu_e} +\left[1-P_{\nu_e\to \nu_e}(t)\right]n_{\nu_x}, \\
n^{\text{new}}_{\nu_{x}}\left(t\right)=P_{\nu_x\to \nu_x}\left(t\right)n_{\nu_x} +\left[1-P_{\nu_x\to \nu_x}(t)\right]n_{\nu_e}.
\end{gather}\label{eq:newdistr1}\end{subequations}

\begin{figure}[H]
\centering
\includegraphics[width=0.75\hsize,clip]{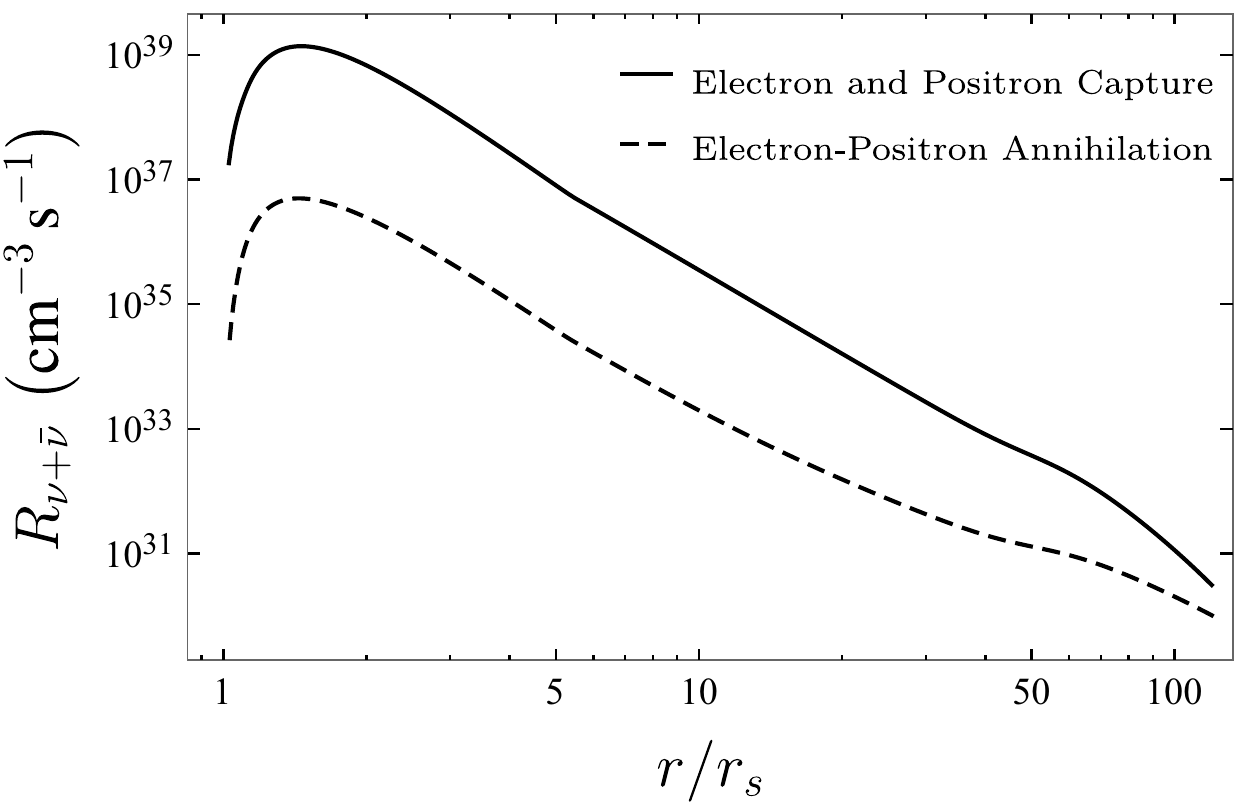}
\caption{Total number emissivity for electron and positron capture ($p + e^{-}\!\to n + \nu_{e}$, $n + e^{+}\!\to p + \bar{\nu}_{e}$) and electron-positron annihilation ($e^{-}\!+e^{+}\! \to \nu + \bar{\nu}$) for accretion disks with $\dot{M} = 0.1M_{\odot}$ s$ ^{-1}$ between the inner radius and the ignition radius.}
\label{fig:emissivities}
\end{figure}

Since the disk is assumed to be in a steady-state, we then perform a time average of Eq.~(\ref{eq:newdistr1}) as discussed in Sec.~\ref{sec:1}. With the new distributions, we can calculate the new neutrino and antineutrino average energies and use them to re-integrate the disk equations.
Neutrino emission within neutrino-cooled disks is dominated by electron and positron capture which only produces electron (anti)neutrinos. The second most important process is electron-positron annihilation but it is several orders of magnitude smaller. In Fig.~\ref{fig:emissivities} we show the total number emissivity for these two processes for an accretion rate of $\dot{M} = 0.1M_{\odot}$ s $^{-1}$. Other cases behave similarly. Moreover, although the degeneracy parameter suppresses the positron density, a high degeneracy limit does not occur in the disk and the degeneracy is kept low at values between $\sim (0.2$--$3)$, as shown in Fig.~\ref{fig:Disks}. The reason for this is the effect of high degeneracy on neutrino cooling. Higher degeneracy leads to a lower density of positrons which suppresses the neutrino production and emission, which in turn leads to a lower cooling rate, higher temperature, lower degeneracy and higher positron density. This equilibrium leads, via the lepton number conservation Eq.~(\ref{eq:leptonconservation}), to a balance between electronic and non-electronic neutrino densities within the inner regions of the disk. Given this fact, to solve the equations of oscillations, we can approximate the initial conditions of the polarization vectors with

\begin{equation}
\mathsf{P} = \bar{\mathsf{P}} \approx (0,0,1).
\label{eq:initialconditionsp}
\end{equation}
% 

%%%%%%%%%%%%%%%%%%%%%%%%%%%%%%%%%%%%%%%%%%%%%%%%%%%%%%%%%%%%%%%%%%%%%%%%%%%%%%%%%%%%%%%%%%%%%%%%%%%%%%%%%%%%%%%%%% 
\section{Results and Analysis}\label{sec:4}
%%%%%%%%%%%%%%%%%%%%%%%%%%%%%%%%%%%%%%%%%%%%%%%%%%%%%%%%%%%%%%%%%%%%%%%%%%%%%%%%%%%%%%%%%%%%%%%%%%%%%%%%%%%%%%%%%%

\begin{figure}
\centering
\includegraphics[width=0.4\hsize,clip]{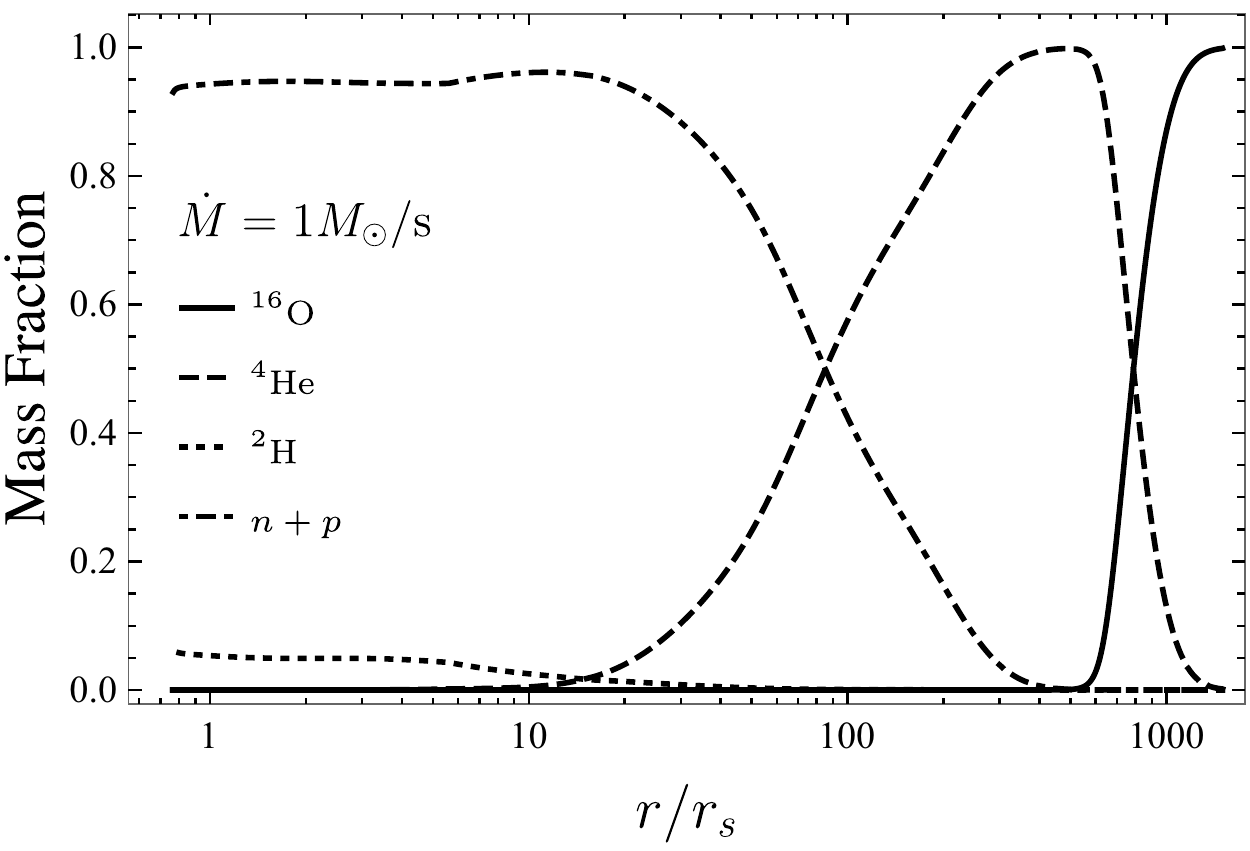}
\includegraphics[width=0.4\hsize,clip]{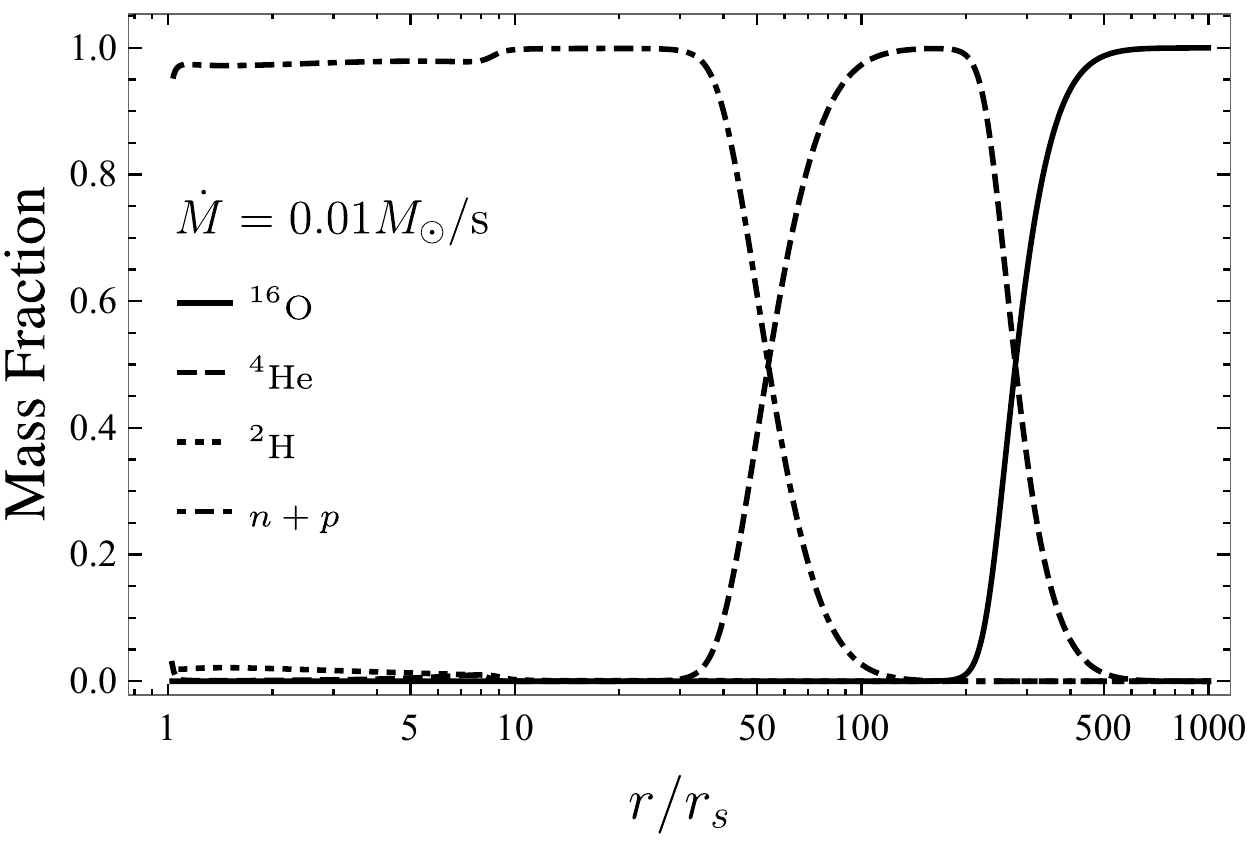}\\
\includegraphics[width=0.4\hsize,clip]{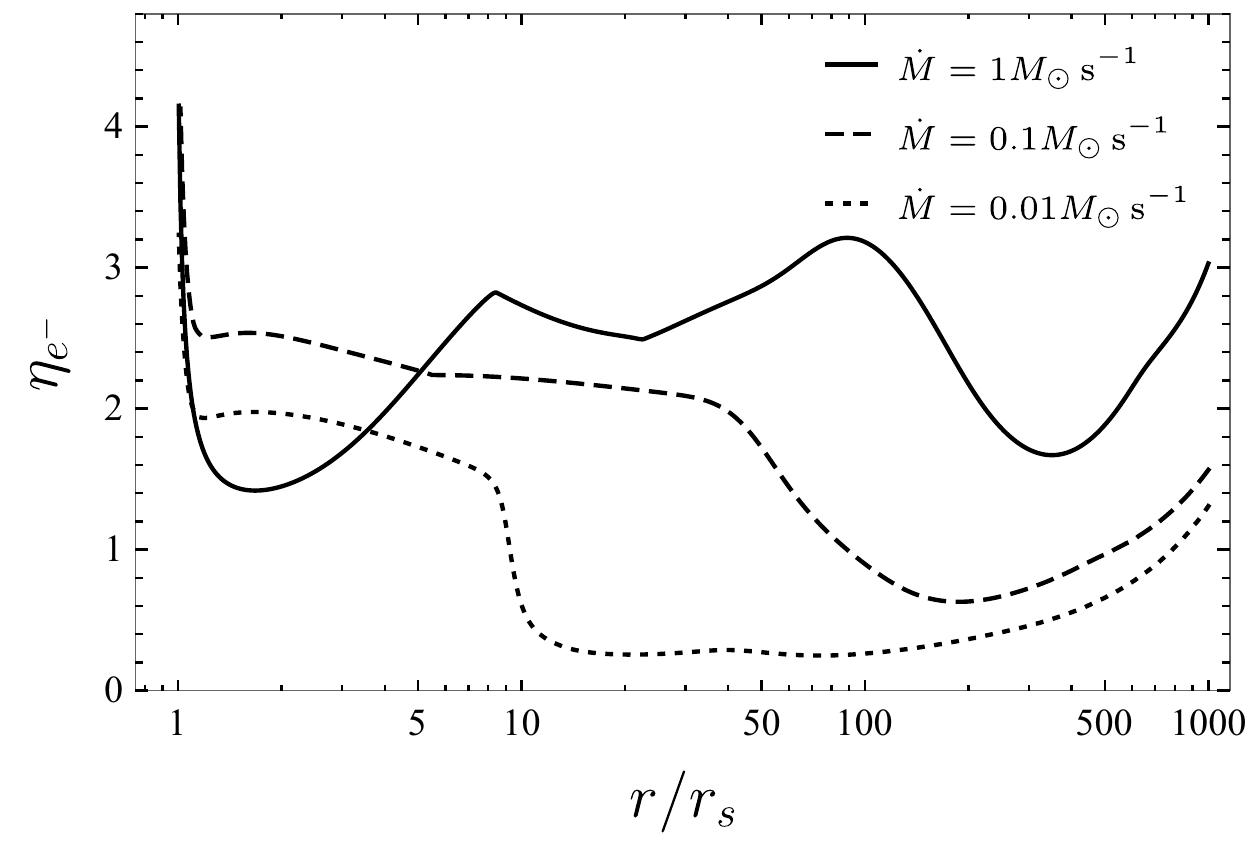}
\includegraphics[width=0.4\hsize,clip]{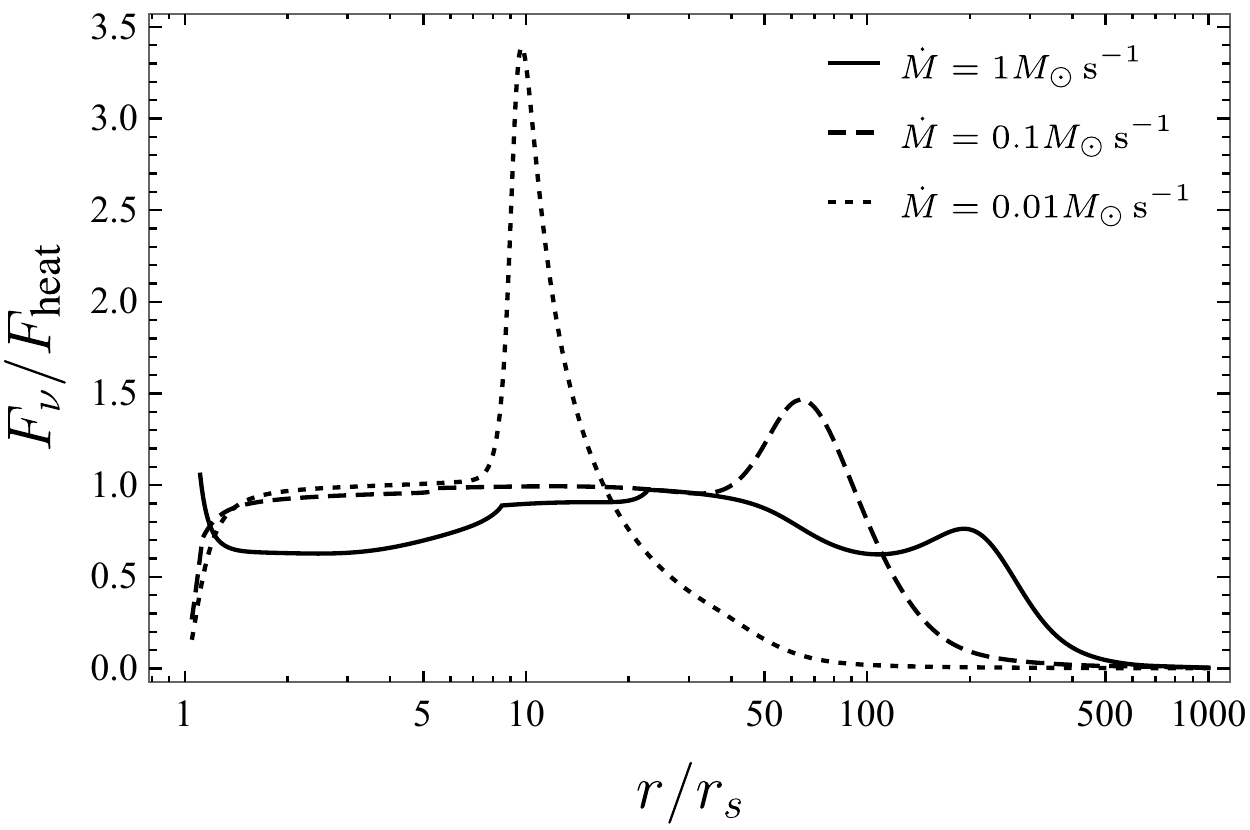}\\
\includegraphics[width=0.4\hsize,clip]{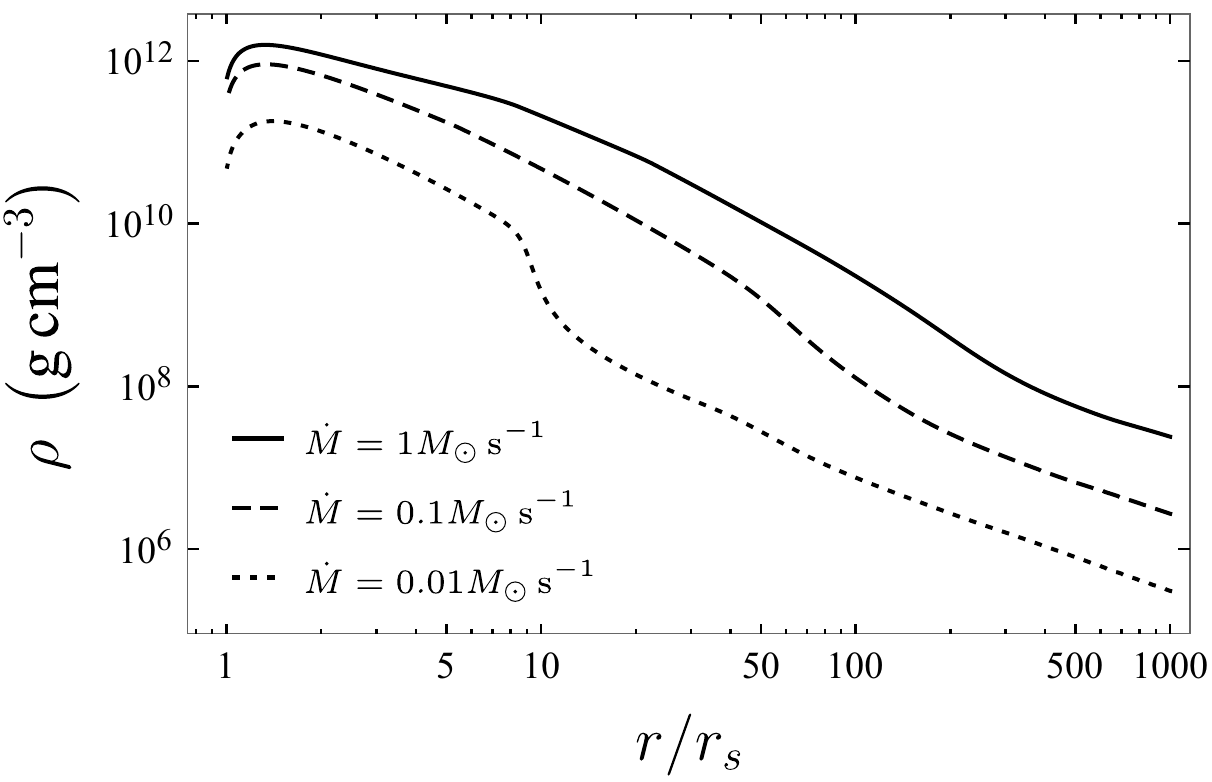}
\includegraphics[width=0.4\hsize,clip]{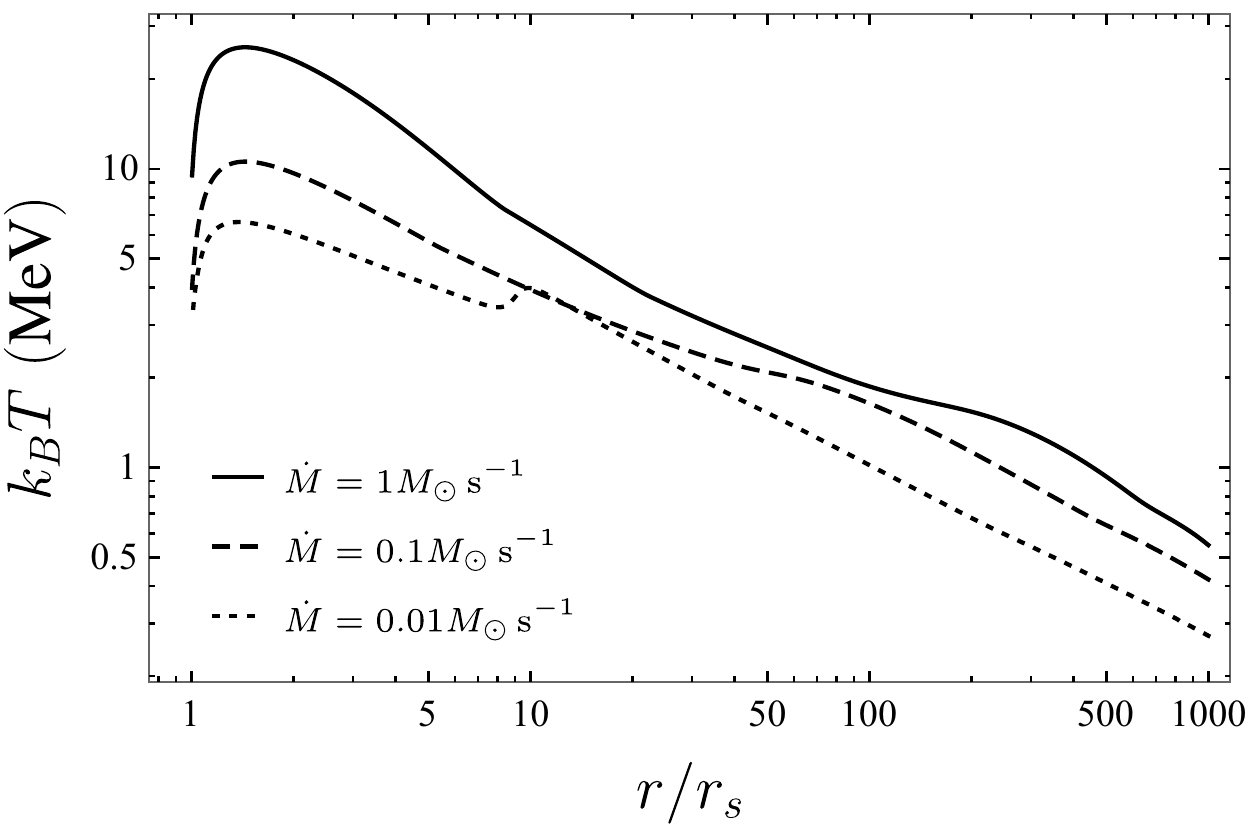}\\
\includegraphics[width=0.4\hsize,clip]{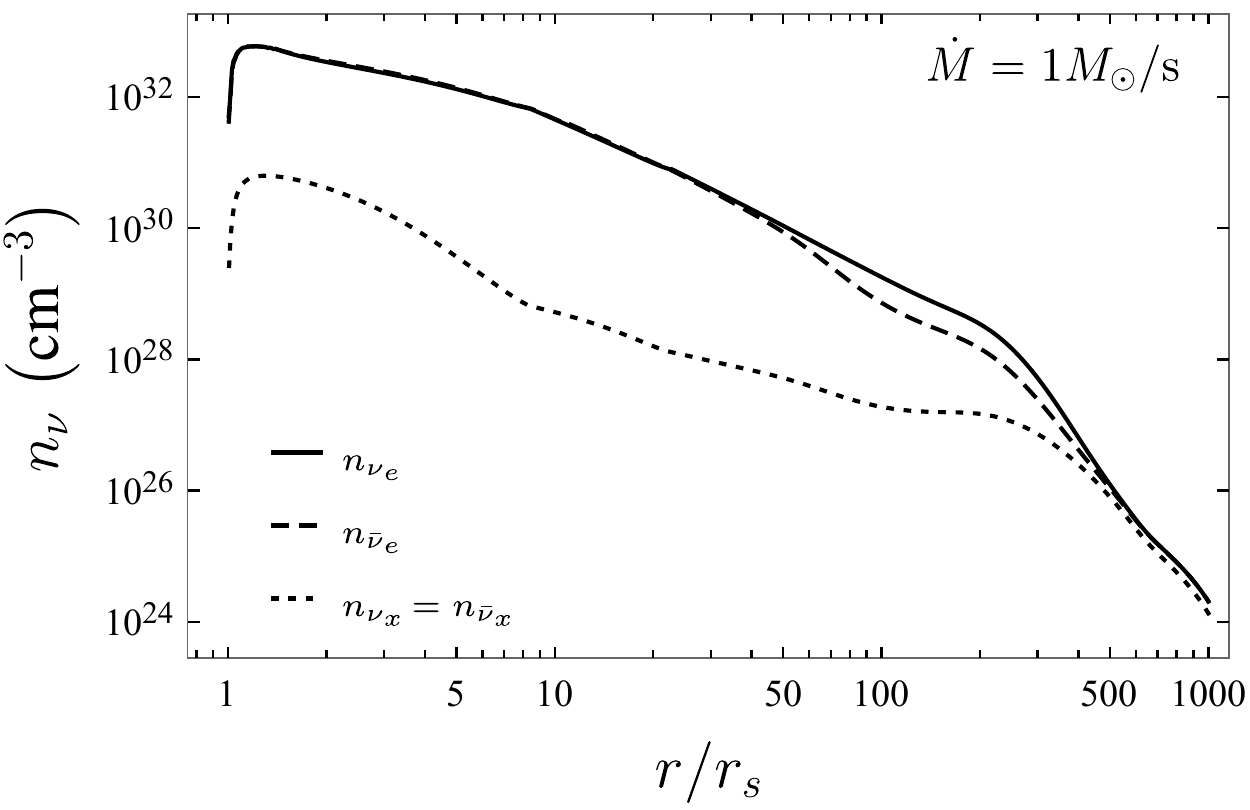}
\includegraphics[width=0.4\hsize,clip]{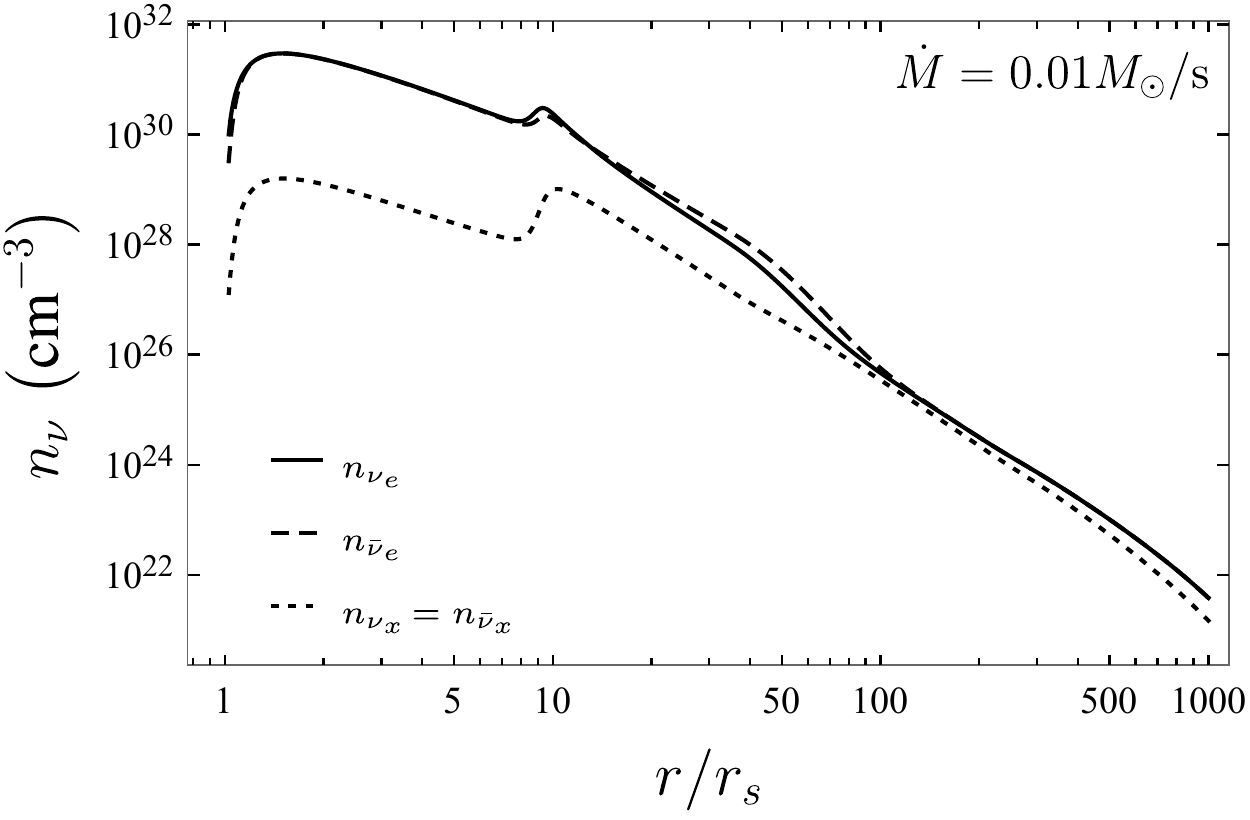}\\
\includegraphics[width=0.4\hsize,clip]{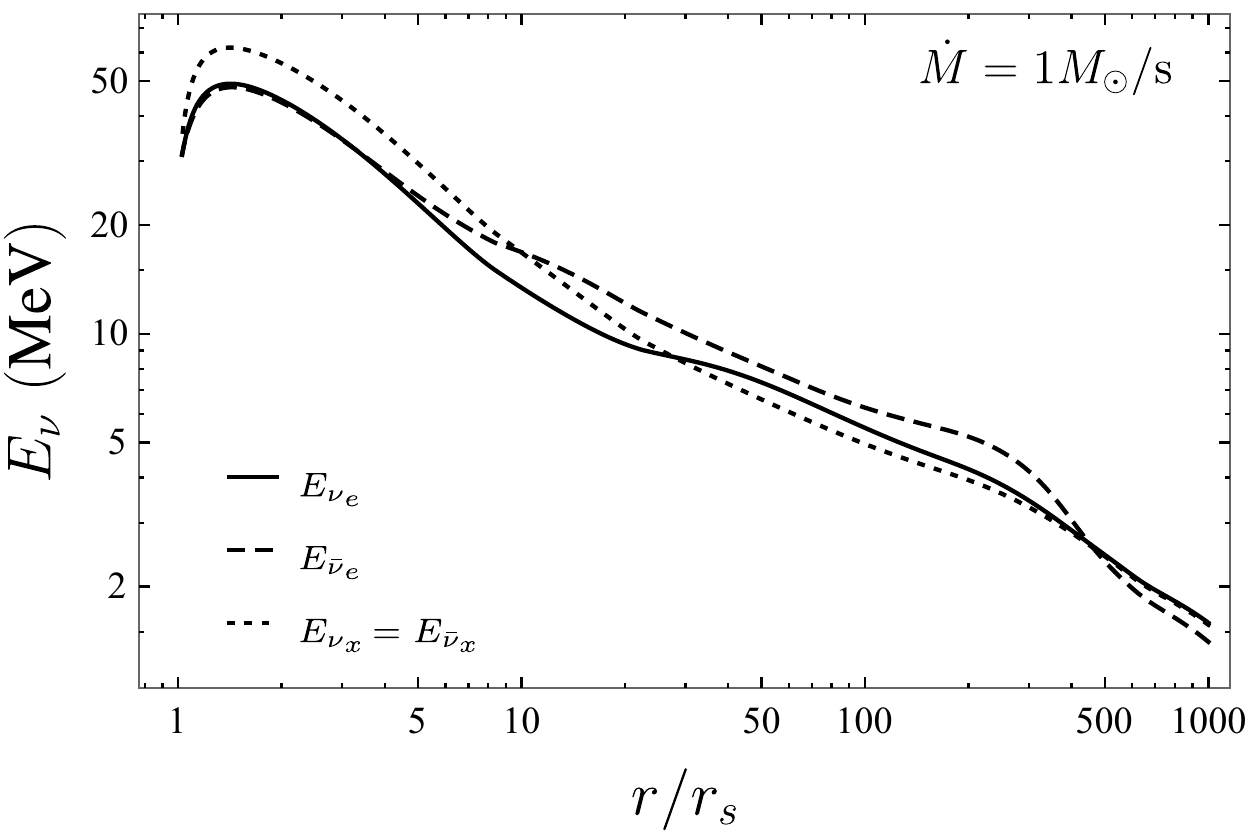}
\includegraphics[width=0.4\hsize,clip]{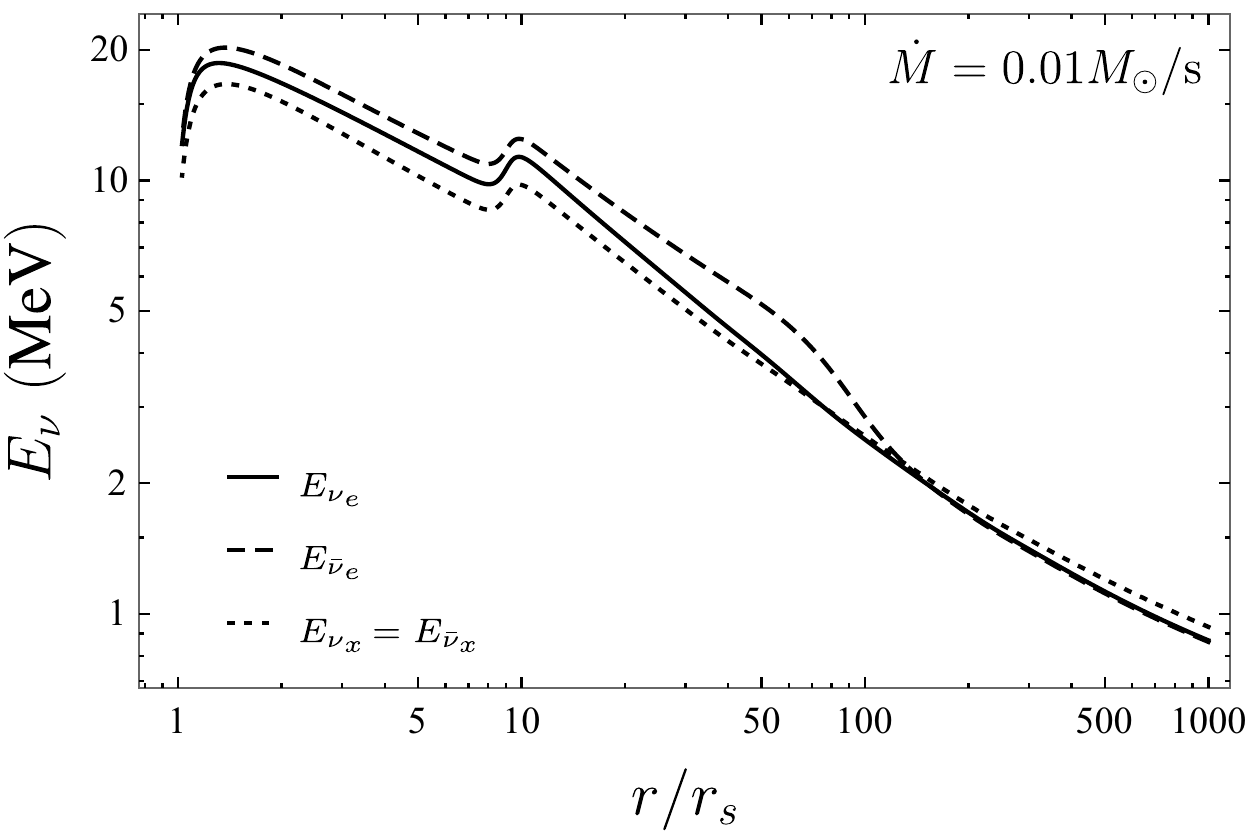}
\caption{Properties of accretion disks in the absence of oscillations with $M=3M_{\odot}$, $\alpha = 0.01$, $a = 0.95$. (\textbf{a}) and (\textbf{b}) are the Mass Fraction inside the disk. \textcolor{black}{We have plotted only the ones that appreciably change.} (\textbf{c}) is the electron degeneracy parameter. (\textbf{d}) is the comparison between the neutrino cooling flux $F_{\nu}$ and the viscous heating $F_{\text{heat}}$. (\textbf{e}) is the baryon density. (\textbf{f}) is the temperature. (\textbf{g}) and (\textbf{h}) are the neutrino number density. (\textbf{i}) and (\textbf{j}) are the average neutrino energies.} 
\label{fig:Disks}
\end{figure}

\begin{figure}
\centering
\includegraphics[width=0.49\hsize,clip]{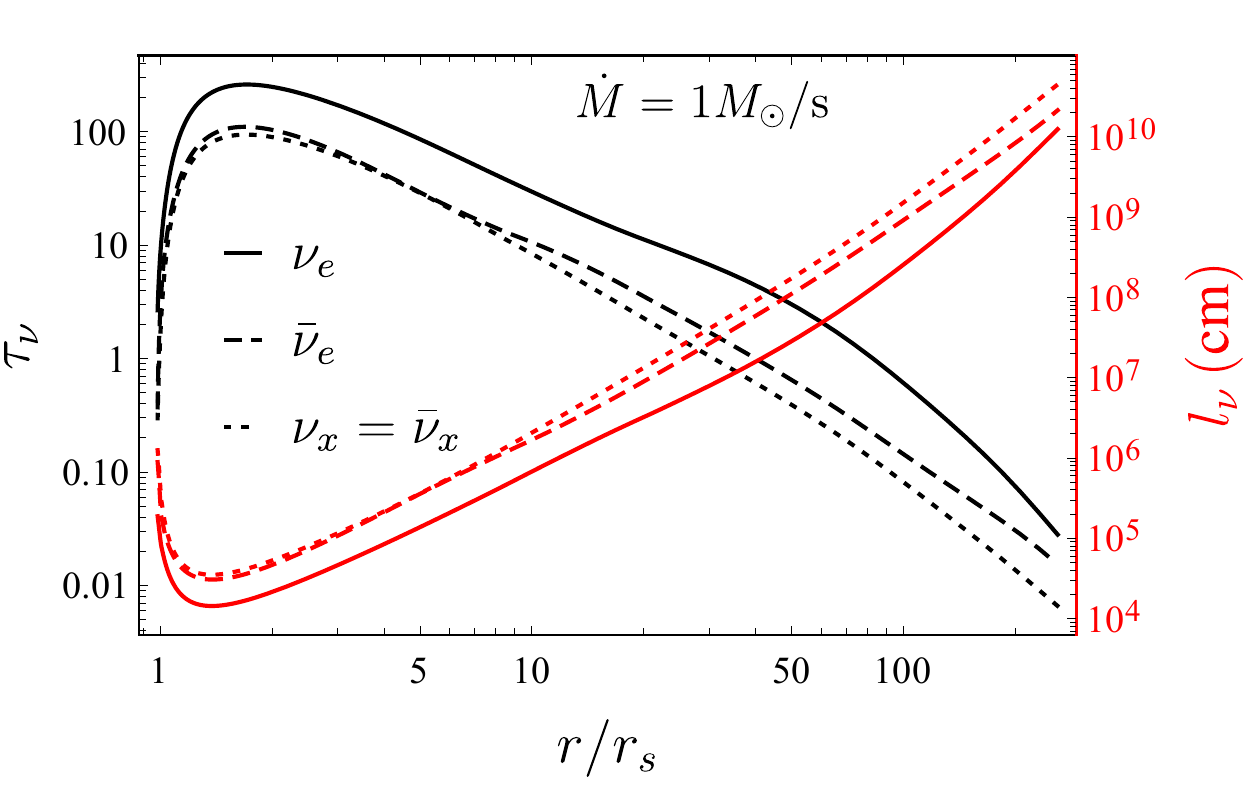}
\includegraphics[width=0.49\hsize,clip]{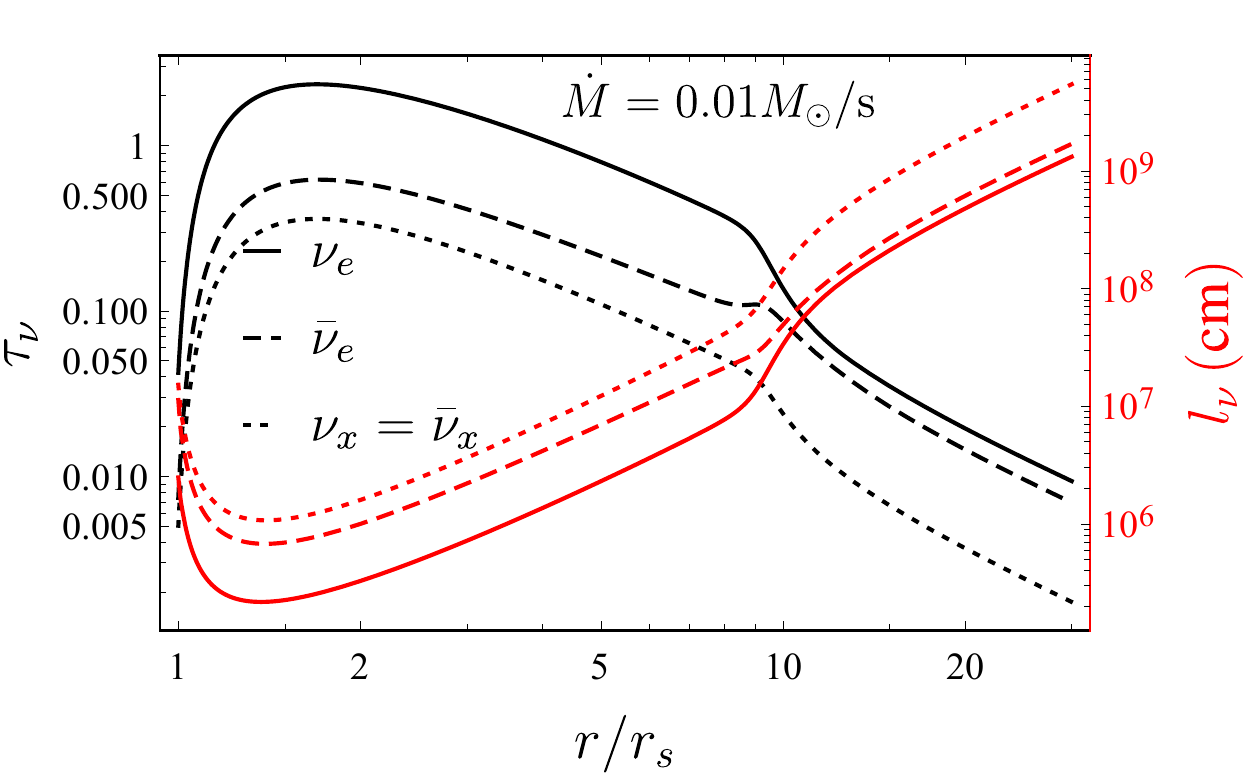}\caption{Total optical depth (left scale) and mean free path (right scale) for neutrinos and antineutrinos of both flavours between the inner radius and the ignition radius for accretion disks with  (\textbf{a}) $\dot{M} = 1M_{\odot}$~s$^{-1}$ and  (\textbf{b}) $0.01 M_{\odot}$~s$^{-1}$.} 
\label{fig:opticaldepth}
\end{figure}

In Figs.~\ref{fig:Disks} and \ref{fig:opticaldepth}, we present the main features of accretion disks for the parameters  $M=3M_{\odot}$, $\alpha = 0.01$, $a = 0.95$, and two selected accretion rates $\dot{M} = 1M_{\odot}$ s$ ^{-1}$, and $\dot{M} = 0.01M_{\odot}$ s$ ^{-1}$. It exhibits the usual properties of thin accretions disks. High accretion rate disks have higher density, temperature and electron degeneracy. Also, for high accretion rates, the cooling due to photodisintegration and neutrino emission kicks in at larger radii. For all cases, as the disk heats up, the number of free nucleons starts to increase enabling the photodisintegration cooling at $r\sim(100$--$300)r_{s}$. Only the disintegration of alpha particles is important and the nucleon content of the infalling matter is of little consequence for the dynamics of the disk. When the disk reaches temperatures $\sim$ 1.3 MeV, the electron capture switches on, the neutrino emission becomes significant and the physics of the disk is dictated by the energy equilibrium between $F_{\text{heat}}$ and $F_{\nu}$. The radius at which neutrino cooling becomes significant (called ignition radius $\color{black}r_{\text{ign}}$) is defined by the condition $F_{\nu} \sim F_{\text{heat}}/2$. For the low accretion rate $\dot{M}=0.01M_{\odot}$ s$ ^{-1}$, the photodisintegration cooling finishes before the neutrino cooling becomes significant, this leads to fast heating of the disk. Then the increase in temperature triggers a strong neutrino emission that carries away the excess heat generating a sharp spike in $F_{\nu}$ surpassing $F_{\text{heat}}$ by a factor of $\sim$ 3.5. This behaviour is also present in the systems studied in \cite{2007ApJ...657..383C}, but there it appears for fixed accretion rates and high viscosity ($\alpha = 0.1$). This demonstrates the semi-degeneracy mentioned in Sec.~\ref{sec:4}. The evolution of the fluid can be tracked accurately through the degeneracy parameter. At the outer radius, $\eta_{e^{-}}$ starts to decrease as the temperature of the fluid rises. Once neutrino cooling becomes significant, it starts to increase until the disk reaches the local balance between heating and cooling. At this point, $\eta_{e^{-}}$ stops rising and is maintained (approximately) at a constant value. Very close to $r_{\text{in}}$, the zero torque condition of the disk becomes important and the viscous heating is reduced drastically. This is reflected in a sharp decrease in the fluid's temperature and increase in the degeneracy parameter. For the high accretion rate and additional effect has to be taken into account. Due to high $\nu_{e}$ optical depth, neutrino cooling is less efficient, leading to an increase in temperature and a second dip in the degeneracy parameter. This dip is not observed in low accretion rates because $\tau_{\nu_{e}}$ does not reach significant values.

\begin{figure}
\centering
\includegraphics[width=0.49\hsize,clip]{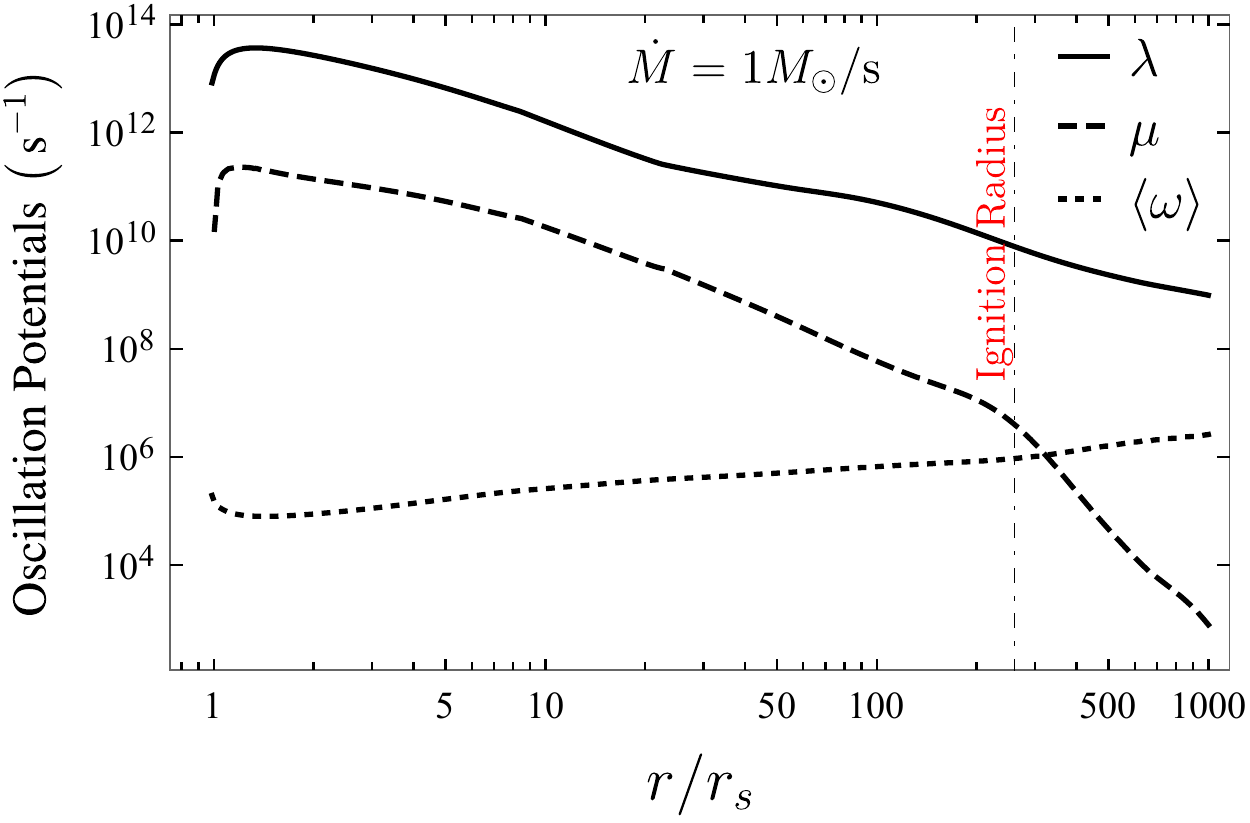}
\includegraphics[width=0.49\hsize,clip]{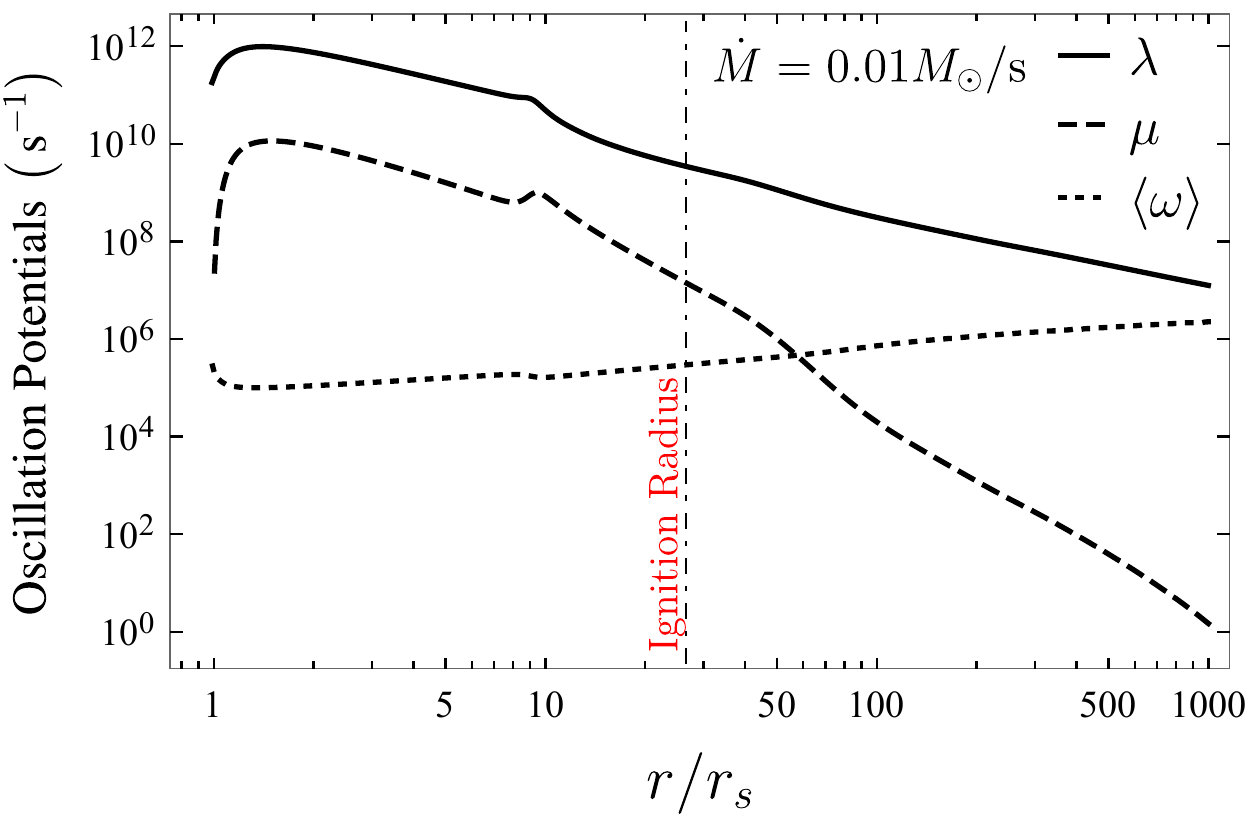}\caption{Oscillation potentials as functions of $r$ with $M=3M_{\odot}$, $\alpha = 0.01$, $a = 0.95$ for accretion rates (\textbf{a}) $\dot{M} = 1M_{\odot}$ s$ ^{-1}$ and (\textbf{b}) $\dot{M} = 0.01M_{\odot}$ s$ ^{-1}$, respectively. The vertical line represents the position of the ignition radius.} 
\label{fig:Potentials}
\end{figure}

With the information in Fig.~\ref{fig:Disks} we can obtain the oscillation potentials which we plot in Fig.~\ref{fig:Potentials}. Since the physics of the disk for $r < r_{\text{ign}}$ is independent of the initial conditions at the external radius and for $r>r_{\text{ign}}$ the neutrino emission is negligible, the impact of neutrino oscillations is important only inside $r_\text{{ign}}$.

We can see that the discussion at the end of Sec.~\ref{subsec:2.1} is justified since, for $r_{\text{in}}<r<r_{\text{ign}}$, the potentials obey the relation

\begin{equation}
\langle\omega\rangle\ll\mu\ll\lambda.
    \label{eq:potorder}
\end{equation}

Generally, the full dynamics of neutrino oscillations is a rather complex interplay between the three potentials, yet it is possible to understand the neutrino response in the disk using some numerical and algebraic results obtained in~\cite{Hannestad:2006nj,Fogli:2007bk,EstebanPretel:2007ec} and references therein. Specifically, we know that if $\mu \gg \langle\omega\rangle$, as long as the MSW condition $\lambda \simeq \langle\omega\rangle$ is not met (precisely our case), collective effects should dominate the neutrino evolution even if $\lambda \gg \mu$. On the other hand, if $\mu \lesssim \langle\omega\rangle$, the neutrino evolution is driven by the relative values between the matter and vacuum potentials (not our case). With Eq.~(\ref{eq:Hnufin}) we can build a very useful analogy. These equations are analogous to the equations of motion of a simple mechanical pendulum with a vector position given by $\mathbf{S}$, precessing around with angular momentum $\mathbf{D}$, subjected to a gravitational force $\langle\omega\rangle\mu\mathbf{B}$ with mass $\mu^{-1}$. Using Eq.~(\ref{eq:initialconditionsp}) obtains the expression $\vert\mathbf{S}\vert = S \approx 2 + O(\langle\omega\rangle / \mu)$. Calculating $\partial_{t}(\mathbf{S}\cdot\mathbf{S})$ it can be checked that this value is conserved up to fluctuations of order $\langle\omega\rangle/\mu$. The analogous angular momentum is $\mathbf{D}=\mathsf{P}-\bar{\mathsf{P}} = 0$. Thus, the pendulum moves initially in a plane defined by $\mathbf{B}$ and the $z$-axis, i.e., the plane $xz$. Then, it is possible to define an angle $\varphi$ between $\mathbf{S}$ and the $z$-axis such that

\begin{equation}
\mathbf{S}=S\left(\sin\varphi,0,\cos\varphi\right).
\label{eq:definiton_phi}
\end{equation}

The only non-zero component of $\mathbf{D}$ is $y$-component. From Eq.~(\ref{eq:Hnufin}) we find

\begin{subequations}
\begin{gather}
\dot{\varphi} = \mu D,\\
\dot{D} = - \langle\omega\rangle S \cos (\varphi + 2\theta).
\end{gather}\label{eq:evolution_phi1}\end{subequations}

These equations can be equivalently written as

\begin{equation}
\ddot{\varphi} = - k^2 \sin (2\theta + \varphi),
\label{eq:motion_phi}
\end{equation}

where we have introduced the inverse characteristic time $k$ by

\begin{equation}
k^2=\langle\omega\rangle \mu S,
\label{eq:definition_k}
\end{equation}

which is related to the anharmonic oscillations of the pendulum. The role of the matter potential $\lambda$ is to logarithmically extend the oscillation length by the relation \cite{Hannestad:2006nj}

\begin{equation}
\tau = -k^{-1}\ln\left[\frac{k}{\theta\left(k^{2}+\lambda^{2}\right)^{1/2}}\left(1 + \frac{\langle\omega\rangle}{S\mu}\right)\right].
\label{deffinition tau}
\end{equation}

The total oscillation time can then be approximated by the period of an harmonic pendulum plus the logarithmic extension

\begin{equation}
t_{\text{osc}} = \frac{2\pi}{k} + \tau.
\label{eq:tot_osc_time}
\end{equation}

The initial conditions of Eq.~(\ref{eq:initialconditionsp}) imply

\begin{equation}
\varphi\left(t=0\right) = \arcsin \left( \frac{\langle\omega\rangle}{S\mu}\sin 2\theta \right),
\label{eq:initial_phi}
\end{equation}

so that $\varphi$ is a small angle. The potential energy for a simple pendulum is

\begin{equation}
V\left(\varphi\right) = k^2 \left[ 1 - \cos\left(\varphi + 2\theta\right) \right] \approx k^2\left(\varphi + 2\theta\right)^2.
\label{eq:potential_phi}
\end{equation}

If $k^2>0$, which is true for the normal hierarchy $\Delta m^{2}  > 0$, we expect small oscillations around the initial position since the system begins in a stable position of the potential. The magnitude of flavour conversions is of the order $\sim \langle \omega \rangle/S\mu \ll 1$. We stress that normal hierarchy does not mean an absence of oscillations but rather \textit{imperceptible} oscillations in $\mathsf{P}_z$. No strong flavour oscillations are expected. On the contrary, for the inverted hierarchy $\Delta m^{2} < 0$, $k^{2}<0$ and the initial $\varphi$ indicates that the system begins in an unstable position and we expect very large anharmonic oscillations. $\mathsf{P}^z$ (as well as $\bar{\mathsf{P}}^z$) oscillates between two different maxima passing through a minimum $-\mathsf{P}^z$ ($-\bar{\mathsf{P}}^z$) several times. This implies total flavour conversion: all electronic neutrinos (antineutrinos) are converted into non-electronic neutrinos (antineutrinos) and vice-versa. This has been called bipolar oscillations in the literature~\cite{Duan:2010bg}. If the initial condition are not symmetric as in Eq.~(\ref{eq:initialconditionsp}), the asymmetry is measured by a constant $\varsigma = \bar{\mathsf{P}}^z/\mathsf{P}^z$ if $\bar{\mathsf{P}}^z < \mathsf{P}^z$ or $\varsigma = \mathsf{P}^z/\bar{\mathsf{P}}^z$ if $\bar{\mathsf{P}}^z > \mathsf{P}^z$ so that $0<\varsigma<1$. Bipolar oscillations are present in an asymmetric system as long as the relation 

\begin{equation}
\frac{\mu}{\vert\langle\omega\rangle\vert} < 4\frac{1+\varsigma}{\left(1-\varsigma\right)^2},
\label{eq:asymmetry}
\end{equation}

is obeyed \cite{Hannestad:2006nj}. If this condition is not met, instead of bipolar oscillation we get synchronised oscillations. Since we are considering constant potentials, synchronised oscillations are equivalent to the normal hierarchy case. From Fig.~\ref{fig:Potentials} we can conclude that in the normal hierarchy case, neutrino oscillations have no effects on neutrino-cooled disks under the assumptions we have made. On the other hand, in the inverted hierarchy case, we expect extremely fast flavour conversions with periods of order $t_{\text{osc}} \sim (10^{-9} - 10^{-5})$ s for high accretion rates and $t_{\text{osc}} \sim (10^{-8} - 10^{-5})$ s for low accretion rates, between the respective $r_{\text{in}}$ and $r_{\text{ign}}$.

\begin{figure}
\centering
\includegraphics[width=0.49\hsize,clip]{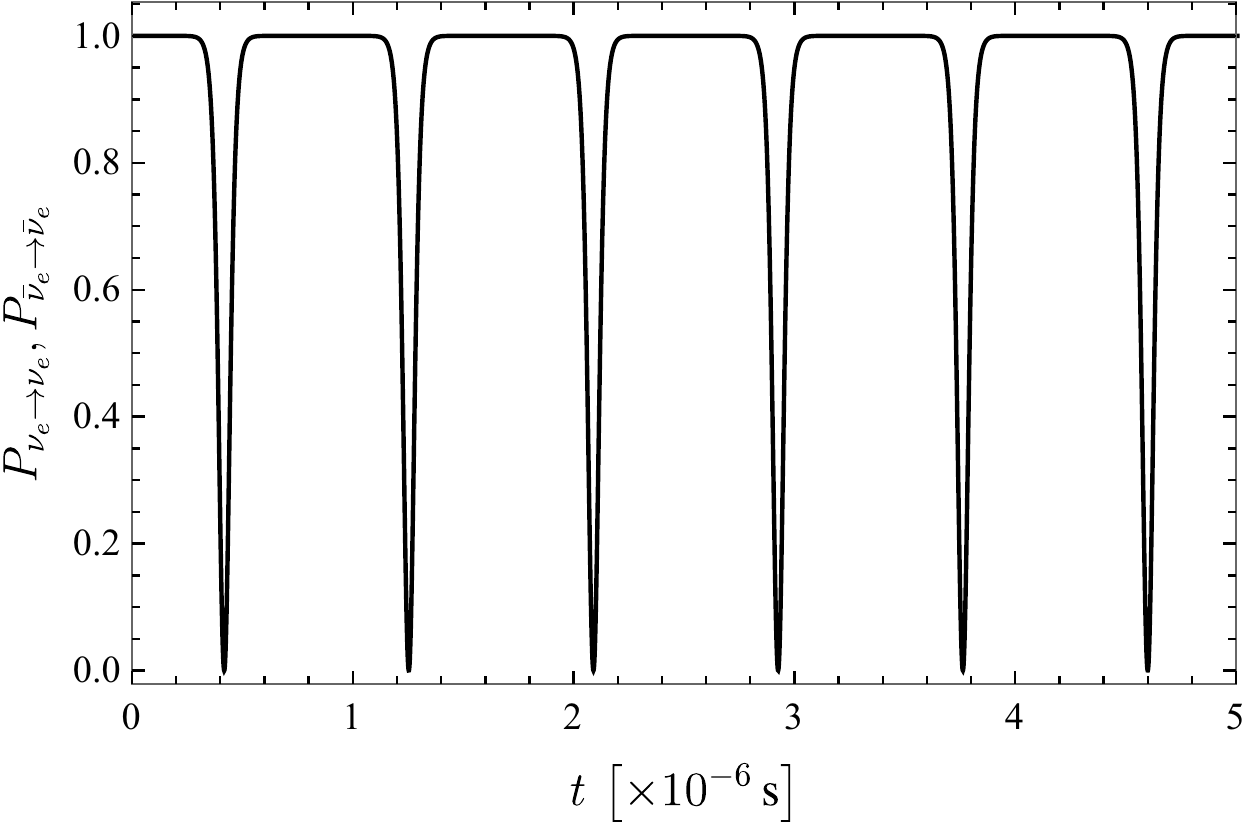}
\includegraphics[width=0.49\hsize,clip]{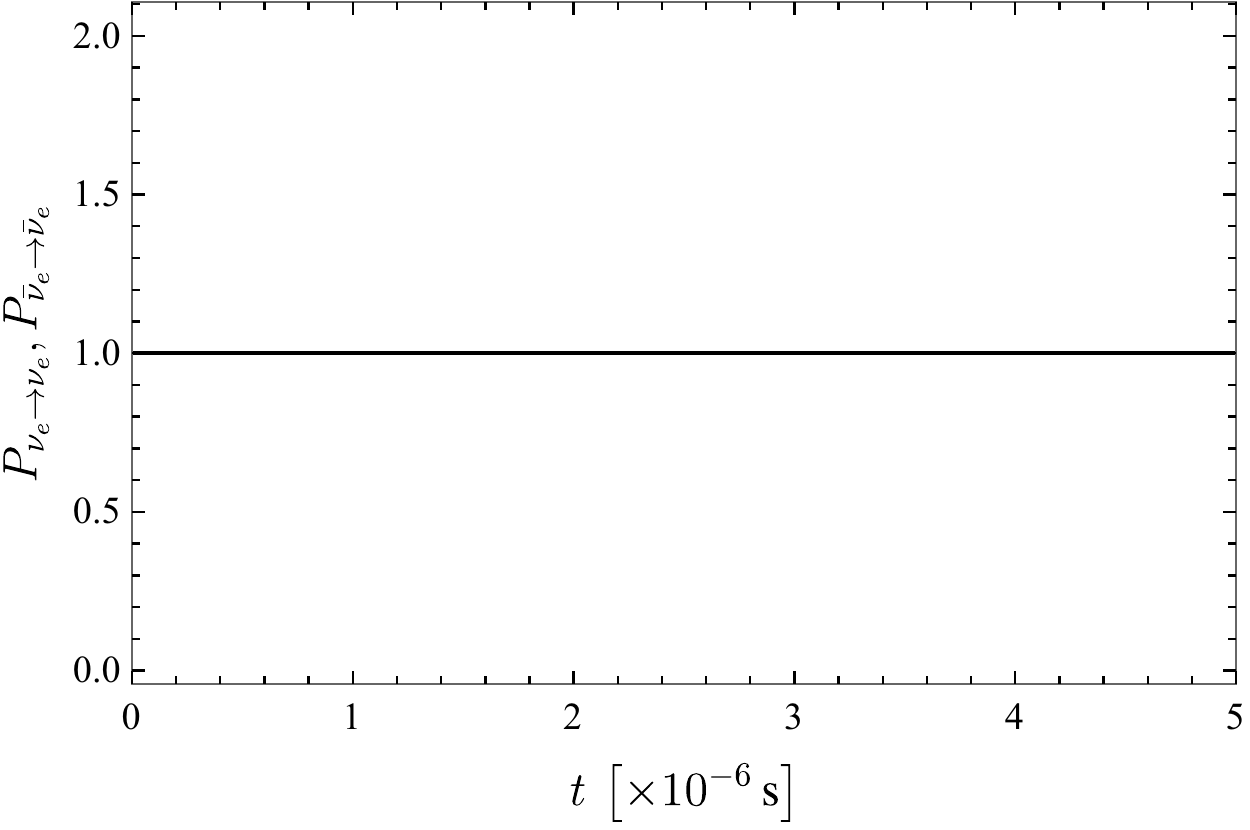}
\caption{Survival provability for electron neutrinos and antineutrinos for the accretion disk with $\dot{M}=0.1M_{\odot}$ s$ ^{-1}$ at $r=10r_{s}$. \textcolor{black}{The survival probabilities for neutrinos and antineutrinos in both plots coincide.} (\textbf{a}) Corresponds to inverted hierarchy and (\textbf{b}) Corresponds to normal hierarchy.}
\label{fig:SurvProbI}
\end{figure}

For the purpose of illustration we solve the equations of oscillations for the $\dot{M}=0.1M_{\odot}$ s$^{-1}$ case at $r=10r_{s}$. The electronic (anti)neutrino survival probability at this point is shown in Fig.~\ref{fig:SurvProbI} for inverted hierarchy and normal hierarchy, respectively. On both plots, there is no difference between the neutrino and antineutrino survival probabilities. This should be expected since for this values of $r$ the matter and self-interaction potentials are much larger than the vacuum potential, and there is virtually no difference between Eq.~(\ref{eq:Hnu2.1}) and Eq.~(\ref{eq:Hnu2.2}). Also, as mentioned before, note that the (anti)neutrino flavour proportions remain virtually unchanged for normal hierarchy while the neutrino flavour proportions change drastically \textcolor{black}{for the inverted hierarchy case}. The characteristic oscillation time of the survival probability in inverted hierarchy found on the plot is

\begin{equation}
t_{\text{osc}} \approx 8.4\times 10^{-7}\, \text{s},
\label{eq:osc_leng_plot}
\end{equation}

which agree with the ones given by Eq.~(\ref{eq:tot_osc_time}) up to a factor of order one. Such a small value suggests extremely quick $\nu_e\bar{\nu}_e \to \nu_x \bar{\nu}_x$ oscillations. A similar effect occurs for regions of the disk inside the ignition radius for all three accretion rates. In this example, the time average of the survival probabilities yield the values $\langle P_{\nu_{e} \to \nu_{e}} \rangle = \langle P_{\bar{\nu}_{e} \to \bar{\nu}_{e}} \rangle = 0.92$. With this number, Eq.~(\ref{eq:newdistr1}), and Eq.~(\ref{eq:efftempneu2}), the (anti)neutrino spectrum for both flavours can be constructed. But, more importantly, this means that the local observer at that point in the disk measures, on average, an electron (anti)neutrino loss of around $8\%$ which is represented by an excess of non-electronic (anti)neutrinos.

\begin{figure}
\centering
\includegraphics[width=0.95\hsize,clip]{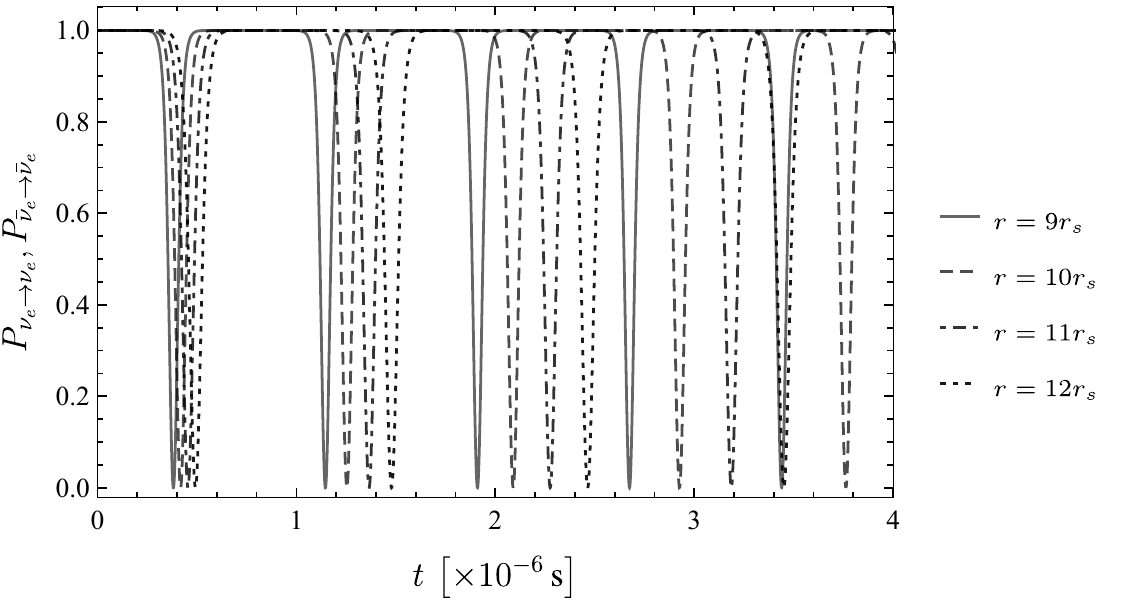}
\caption{Survival provability for electron neutrinos and antineutrinos for the accretion disk with $\dot{M}=0.1M_{\odot}$ s$ ^{-1}$ at $r=9r_{s},10r_{s},11r_{s},12r_{s}$.}
\label{fig:SurvSev}
\end{figure}

In Sec.~\ref{subsec:2.1} we proposed to calculate neutrino oscillations assuming that small neighbouring regions of the disk are independent and that neutrinos can be viewed as isotropic gases in those regions. However, this cannot be considered a steady-state of the disk. To see this consider Fig.~\ref{fig:opticaldepth}. The maximum value of the neutrino optical depth is of the order of $10^3$ for the highest accretion rate, meaning that the time that takes neutrinos to travel a distance of one Schwarzschild inside the disk radius obeys

\begin{equation}
t_{r_{s}} \ll \text{Max}\left(\tau_{\nu}\right)r_{s} \approx 10^{-2}\, \text{s},
\label{eq:travel_time}
\end{equation}

which is lower than the accretion time of the disk as discussed in Sec.~\ref{sec:1} but higher than the oscillation time. Different sections of the disk are not independent since they, very quickly, share (anti)neutrinos created with a non-vanishing momentum along the radial direction. Furthermore, the oscillation pattern between neighbouring regions of the disk is not identical. In Fig.~\ref{fig:SurvSev} we show the survival probability as a function of time for different (but close) values of $r$ for $\dot{M}=0.1M_{\odot}$ s$ ^{-1}$. The superposition between neutrinos with different oscillation histories has several consequences: (1) It breaks the isotropy of the gas because close to the BH, neutrinos are more energetic and their density is higher producing a radially directed net flux, meaning that the factor $\mathbf{v}_{\mathbf{q},t}\cdot\mathbf{v}_{\mathbf{p},t}$ does not average to zero. This implies that realistic equations of oscillations include a multi-angle term and a radially decaying neutrino flux similar to the situation in SN neutrinos. (2) It constantly changes the neutrino content \textcolor{black}{at} any value of $r$ independently of the neutrino collective evolution given by the values of the oscillation potentials at that point. This picture plus the asymmetry that electron and non-electron neutrinos experience through the matter environment (electron (anti)neutrinos can interact through $n + \nu_{e}\!\to p + e^{-}$ and $p + \bar{\nu}_{e}\!\to n + e^{+}$), suggests that the disk achieves complete flavour equipartition (decoherence). We can identify two competing causes, namely, quantum decoherence and kinematic decoherence.

Quantum decoherence is the product of collisions among the neutrinos or with a thermal background medium can be understood as follows \cite{Raffelt:1996wa}. From Appx.~\ref{app:subemisscross1} we know that different (anti)neutrino flavours posses different cross-sections and scattering rates $\Gamma_{{\nu}_{i},\bar{\nu}_{i}}$. In particular, we have $\Gamma_{\nu_x} \approx \Gamma_{\bar{\nu}_x}<\Gamma_{\bar{\nu}_e}<\Gamma_{\nu_e}$. An initial electron (anti)neutrino created at a point $r$ will begin to oscillate into $\nu_{x}(\bar{\nu}_{x})$. The probability of finding it in one of the two flavors evolves as previously discussed. However, in each interaction $n + \nu_{e}\!\to p + e^{-}$, the electron neutrino component of the superposition is absorbed, while the $\nu_{x}$ component remains unaffected. Thus, after the interaction the two flavors can no longer interfere. This allows the remaining $\nu_{x}$ oscillate and develop a new coherent $\nu_{e}$ component which is made incoherent in the next interaction. The process will come into equilibrium only when there are equal numbers of electronic and non-electronic neutrinos. That is, the continuous emission and absorption of electronic (anti)neutrinos generates a non-electronic (anti)neutrinos with an average probability of $\langle P_{\nu_e \to \nu_e} \rangle$ in each interaction and once the densities of flavours are equal, the oscillation dynamic stops. An initial system composed of $\nu_e,\bar{\nu}_e$ turns into an equal mixture of $\nu_e,\bar{\nu}_e$ and $\nu_x,\bar{\nu}_x$, reflected as an exponential damping of oscillations. For the particular case in which non-electronic neutrinos can be considered as sterile (do not interact with the medium), the relaxation time of this process can be approximated as \cite{1982PhLB..116..464H,PhysRevD.36.2273}

\begin{equation}
t_{Q} = \frac{1}{2l_{\nu\bar{\nu}}\langle\omega\rangle^{2}\sin^{2}2\theta} + \frac{2l_{\nu\bar{\nu}}\lambda^{2}}{\langle\omega\rangle^{2}\sin^{2}2\theta},
    \label{eq:t_q_osc}
\end{equation}

where $l_{\nu\bar{\nu}}$ represents the (anti)neutrino mean free path.

Kinematic decoherence is the result of a non-vanishing flux term such that at any point, (anti)neutrinos travelling in different directions, do not experience the same self-interaction potential due to the multi-angle term in the integral of Eq.~(\ref{eq:FullHam}). Different trajectories do not oscillate in the same way, leading to a de-phasing and a decay of the average $\langle P_{\nu \to \nu} \rangle$ and thus to the equipartition of the overall flavour content. The phenomenon is similar to an ensemble of spins in an inhomogeneous magnetic field. In \cite{Raffelt:2007yz} it is shown that for asymmetric $\nu\bar{\nu}$ gas, even an infinitesimal anisotropy triggers an exponential evolution towards equipartition, and in \cite{EstebanPretel:2007ec} it was shown that if the symmetry between neutrinos and antineutrinos is not broken beyond the limit of 25\%, kinematic decoherence is still the main effect of neutrino oscillations. As a direct consequence of the $\nu\bar{\nu}$ symmetry present within the ignition radius of accretion disks (see Fig.~\ref{fig:Disks}), equipartition among different neutrino flavours is expected. This multi-angle term keeps the order of the characteristic time $t_{\text{osc}}$ of Eq.~(\ref{eq:tot_osc_time}), unchanged and kinematic decoherence happens within a few oscillation cycles. The oscillation time gets smaller closer to the BH due to the $1/\mu^{1/2}$ dependence. Therefore, we expect that neutrinos emitted within the ignition radius will be equally distributed among both flavours in about few microseconds. Once the neutrinos reach this maximally mixed state, no further changes are expected. We emphasize that kinematic decoherence does not mean quantum decoherence. Figs.~\ref{fig:SurvProbI} and Fig.~\ref{fig:SurvSev} clearly show the typical oscillation pattern which happens only if quantum coherence is still acting on the neutrino system. Kinematics decoherence, differently to quantum decoherence, is just the result of averaging over the neutrino intensities resulting from quick flavour conversion. Therefore, neutrinos are yet able to quantum oscillate if appropriate conditions are satisfied. 
Simple inspection of Eq.~(\ref{eq:tot_osc_time}) and Eq.~(\ref{eq:t_q_osc}) with Fig.~\ref{fig:opticaldepth} yields $t_{\text{osc}} \ll t_{Q}$. Clearly the equipartition time is dominated by kinematic decoherence. These two effects are independent of the neutrino mass hierarchy and neutrino flavour equipartition is achieved for both hierarchies. Within the disk dynamic, this is equivalent to imposing the condition $\langle P_{\nu_e \to \nu_e} \rangle= \langle P_{\bar{\nu}_e \to \bar{\nu}_e} \rangle = 0.5$.

\begin{figure}
\centering
\includegraphics[width=0.49\hsize,clip]{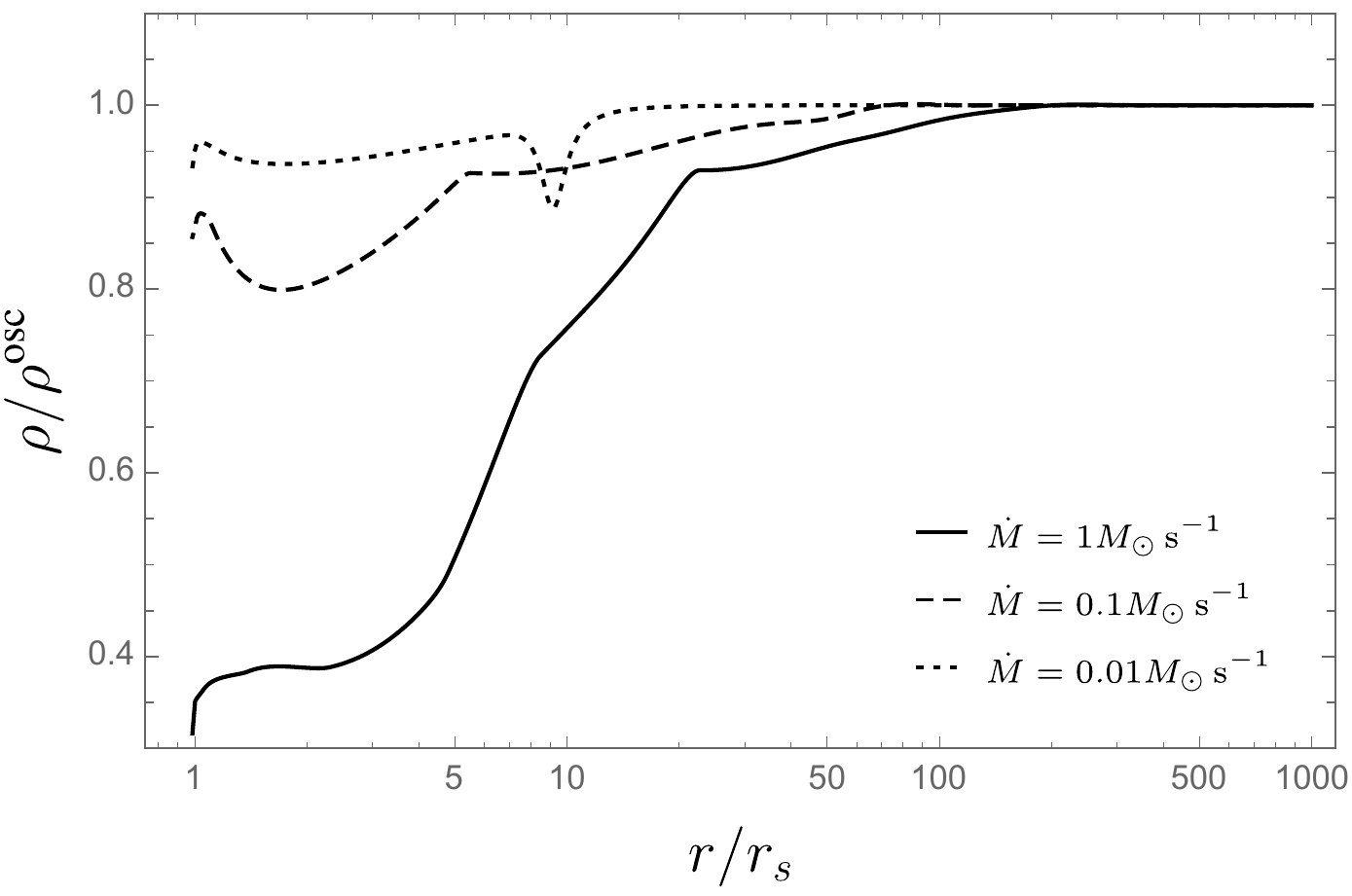}\includegraphics[width=0.49\hsize,clip]{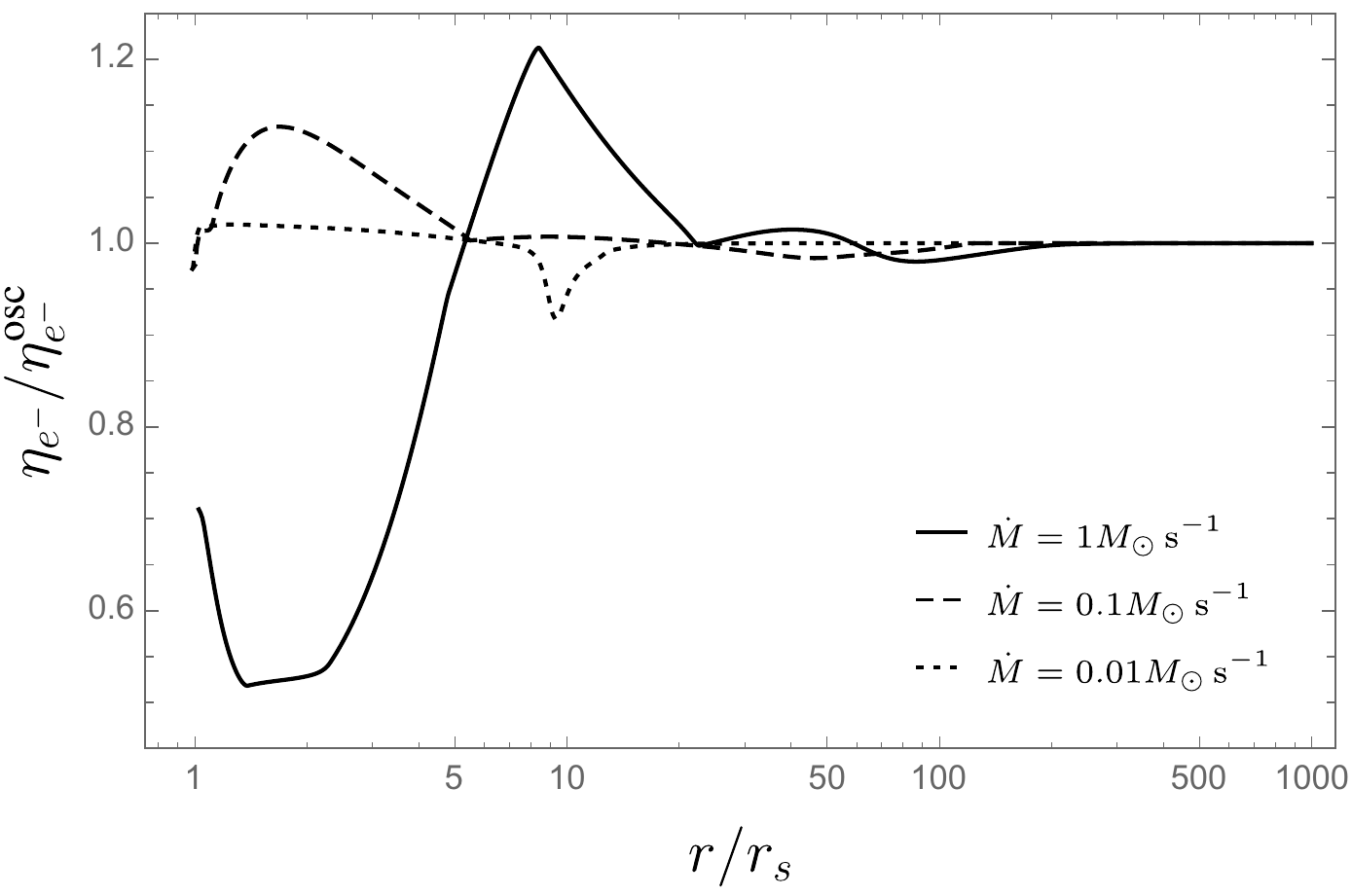}\\
\includegraphics[width=0.49\hsize,clip]{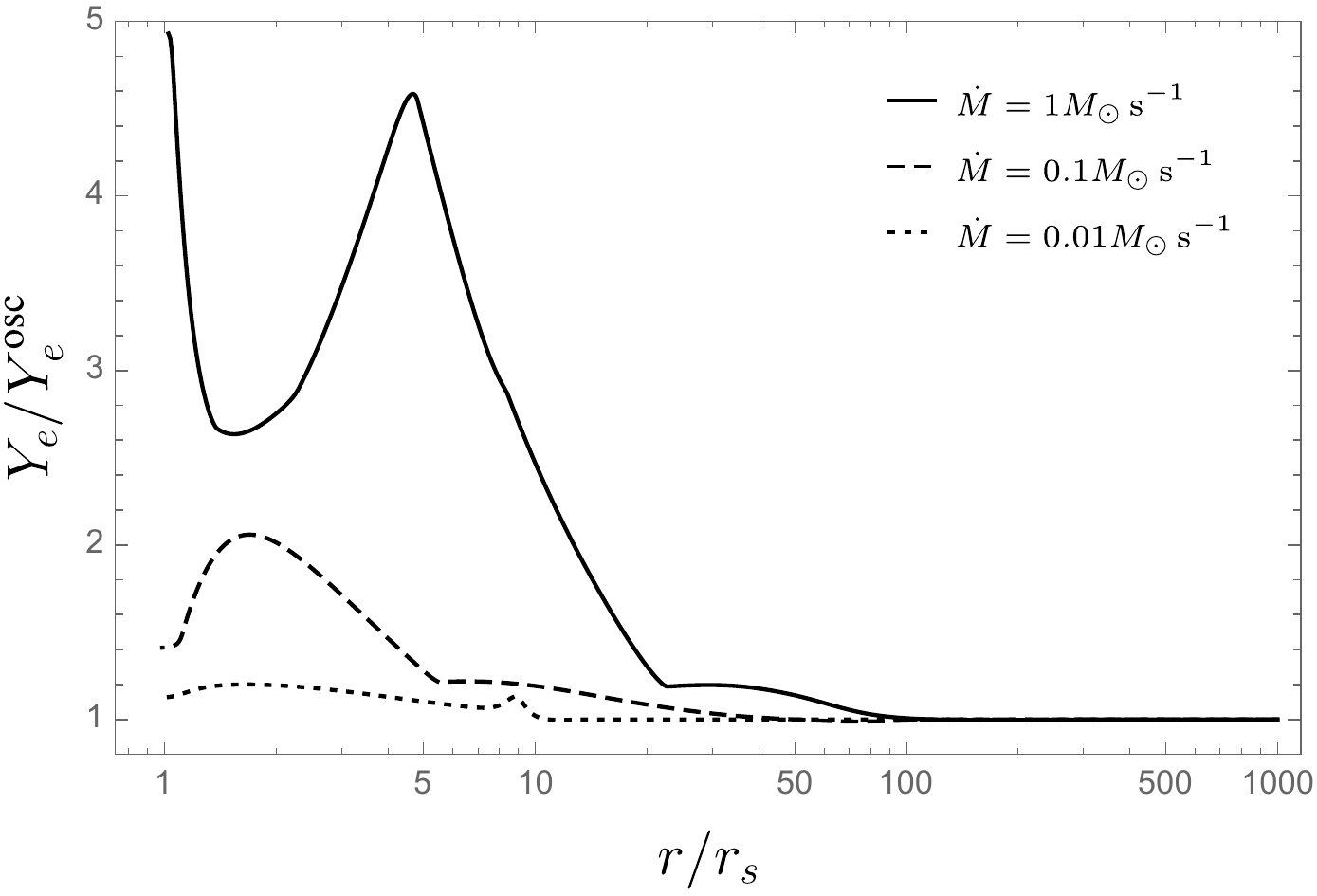}\includegraphics[width=0.49\hsize,clip]{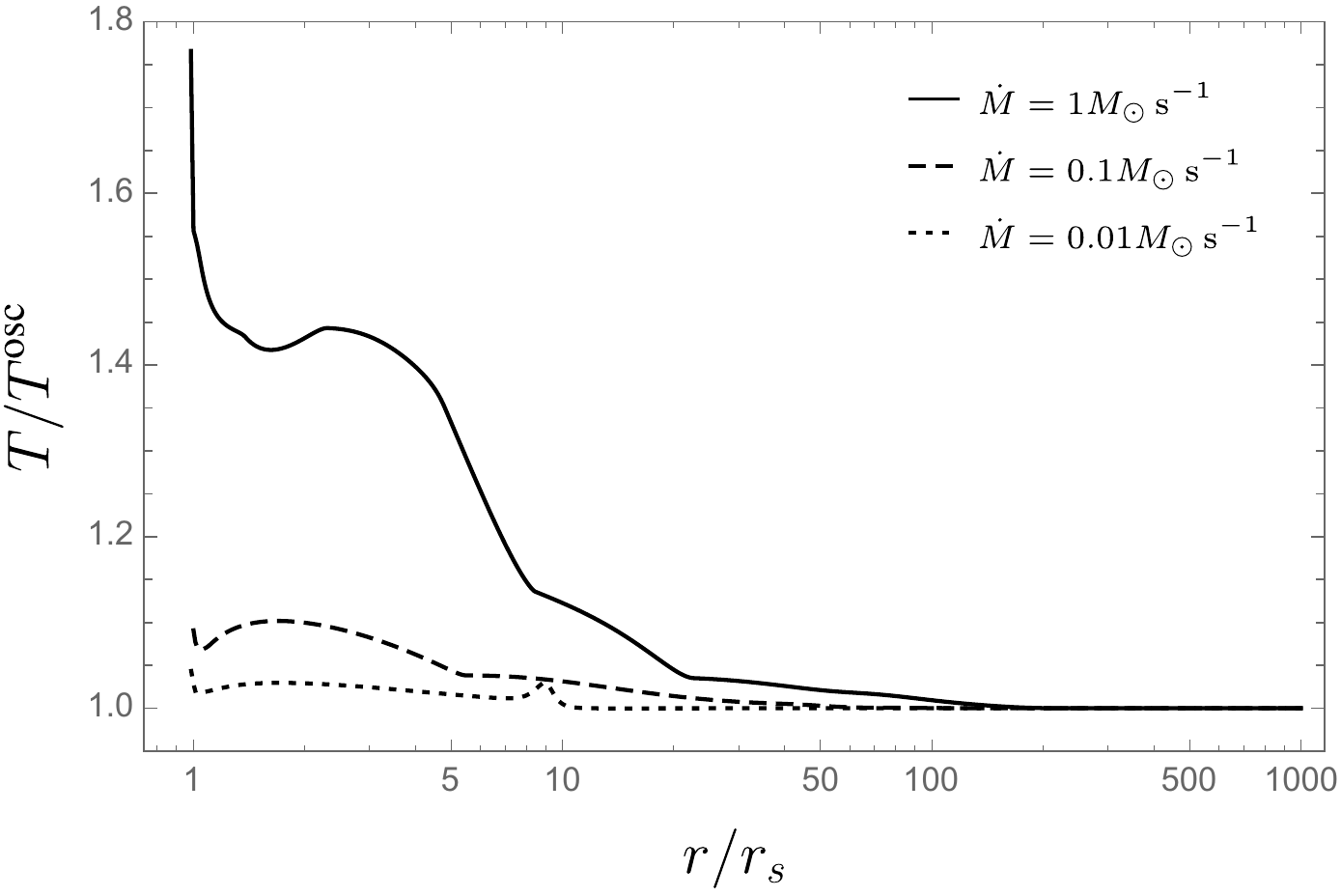}
\caption{Comparison between the main variables describing thin disks with and without neutrino flavour equipartition for each accretion rate considered. \textcolor{black}{Here $\rho^{\text{osc}},\eta^{\text{osc}}_{e^{-}},Y^{\text{osc}}_{e},T^{\text{osc}}$ are the density, electron degeneracy, electron fraction and temperature of a disk with flavour equipartition.} Together with Fig.~\ref{fig:Disks}, these plots completely describe the profile of a disk under flavour equipartition. (\textbf{a}) is the ratio between baryon densities. (\textbf{b}) is the ratio between degeneracy parameters. (\textbf{c}) is the ratio between electron fractions. (\textbf{d}) is the ratio between temperatures.}
\label{fig:comp}
\end{figure}

Figure~\ref{fig:comp} shows a comparison between disks with and without neutrino flavour equipartition for the three accretion rates considered. The role of equipartition is to increase the disk's density, reduce the temperature and electron fraction, and further stabilize the electron degeneracy for regions inside the ignition radius. The effect is mild for low accretion rates and very pronounced for high accretion rates. This result is in agreement with our understanding of the dynamics of the disk and can be explained in the following way. In low accretion systems the neutrino optical depth for all flavors is $\tau_{\nu\bar{\nu}} \lesssim 1$ and the differences between the cooling fluxes, as given by Eq.~(\ref{eq:neutrinofluxpost}) are small. Hence, when the initial (mainly electron flavour) is redistributed among both flavours, the total neutrino cooling remains virtually unchanged and the disk evolves as if equipartition had never occurred save the new emission flavour content. On the other hand, when accretion rates are high, the optical depth obeys $\tau_{\nu_{x}}\approx\tau_{\bar{\nu}_{x}}\lesssim\tau_{\bar{\nu}_{e}} < \tau_{\nu_{e}} \sim 10^3$. The $\nu_{e}$ cooling is heavily suppressed while the others are less so. When flavours are redistributed, the \textit{new} $\nu_{x}$ particles \textcolor{black}{are} free to escape, enhancing the total cooling and reducing the temperature. As the temperature decreases, so do the electron and positron densities leading to a lower electron fraction. The net impact of flavour equipartition is to make the disk evolution less sensitive to $\nu_{e}$ opacity and, thus, increase the total cooling efficiency. As a consequence, once the fluid reaches a balance between $F^{+}$ and $F_{\nu}$, this state is kept without being affected by high optical depths and $\eta_{e^{-}}$ stays at a constant value until the fluid reaches the zero torque condition close to $r_{\text{in}}$. Note that for every case, inside the ignition radius, we \textcolor{black}{find $\tau_{\nu_{x}}\approx\tau_{\bar{\nu}_{x}}\lesssim\tau_{\bar{\nu}_{e}} < \tau_{\nu_{e}}$} so that equipartition enhances, mainly, neutrino cooling $F_{\nu}$ (and not antineutrino cooling $F_{\bar{\nu}}$). The quotient between neutrino cooling with and without equipartition can be estimated with

\begin{equation}
\frac{F^{\text{eq}}_{\nu}}{F_{\nu}} \approx \frac{1}{2}\left(1 + \frac{\langle E_{\nu_{x}} \rangle}{\langle E_{\nu_{e}} \rangle}\frac{1 + \tau_{\nu_{e}}}{1 + \tau_{\nu_{x}}}\right). 
    \label{eq:fluxcomp}
\end{equation}

This relation exhibits the right limits. From Fig.~\ref{fig:Disks} we see that $\langle E_{\nu_{e}} \rangle \approx \langle E_{\nu_{x}} \rangle$. Hence, If $1 \gg \tau_{\nu_{e}} > \tau_{\nu_{x}}$, then $F^{\text{eq}}_{\nu} = F_{\nu}$ and the equipartition is unnoticeable. But if $1 < \tau_{\nu_{x}} < \tau_{\nu_{e}}$ then $F^{\text{eq}}_{\nu}/F_{\nu} > 1$. In our simulations, this fraction reaches values of 1.9 for $\dot{M} = 1M_{\odot}$ s$ ^{-1}$ to 2.5 for $\dot{M} = 0.01M_{\odot}$ s$ ^{-1}$.

The disk variables at each point do not change beyond a factor of order 5 in the most obvious case. However, these changes can be important for cumulative quantities, e.g. the total neutrino luminosity and the total energy deposition rate into electron-positron pairs due to neutrino antineutrino annihilation. To see this we perform a Newtonian calculation of these luminosities following \cite{1991A&A...244..378J,1997A&A...319..122R,1999ApJ...518..356P,2003MNRAS.345.1077R,2012MNRAS.419..713K,2013ApJS..207...23X,LIU20171}, and references therein. The neutrino luminosity is calculated by integrating the neutrino cooling flux throughout both faces of the disk:

\begin{equation}\label{eq:luminosity}
L_{\nu_i} = 4\pi \int_{r_{\text{in}}}^{r_{\text{out}}}C_{\text{cap}}F_{\nu_{i}}rdr.
\end{equation}
\begin{table}
\caption{Comparison of total neutrino luminosities $L_{\nu}$ and annihilation luminosities $L_{\nu\bar{\nu}}$ between disk with and without flavour equipartition. All luminosities are reported in $\text{MeV}^{\phantom{2}}$s$^{-1}$.}
\centering
\resizebox{\textwidth}{!}{\begin{tabular}{ c c c c c c c | c  c  c  c  c  c }
\cmidrule{2-13}
 & \multicolumn{6}{c|}{\textbf{Without oscillations}} & \multicolumn{6}{c}{\textbf{With oscillations (flavour equipartition)}} \\
\cmidrule{2-13}
 & $L_{\nu_{e}}$ & $L_{\bar{\nu}_{e}}$ & $L_{\nu_{x}}$ & $L_{\bar{\nu}_{x}}$ & $L_{\nu_{e}\bar{\nu}_{e}}$ & $L_{\nu_{x}\bar{\nu}_{x}}$ & $L_{\nu_{e}}$ & $L_{\bar{\nu}_{e}}$ & $L_{\nu_{x}}$ & $L_{\bar{\nu}_{x}}$ & $L_{\nu_{e}\bar{\nu}_{e}}$ & $L_{\nu_{x}\bar{\nu}_{x}}$ \\
\midrule
\multicolumn{1}{c|}{1 $M_{\odot}^{\phantom{2}}$ s$^{-1}$} & $6.46\times 10^{58}$ & $7.33\times 10^{58}$ & $1.17\times 10^{58}$ & $1.17\times 10^{58}$ & $1.25\times 10^{57}$ & $1.05\times 10^{55}$ & $1.87\times 10^{58}$ & $4.37\times 10^{58}$ & $7.55\times 10^{58}$ & $5.44\times 10^{58}$ & $1.85\times 10^{56}$ & $2.31\times 10^{56}$   \\
\midrule
\multicolumn{1}{c|}{0.1 $M_{\odot}^{\phantom{2}}$ s$^{-1}$} & $9.19\times 10^{57}$ & $1.08\times 10^{58}$ & $8.06\times 10^{55}$ & $8.06\times 10^{55}$ & $1.62\times 10^{55}$ & $1.27\times 10^{50}$ & $2.47\times 10^{57}$ & $4.89\times 10^{57}$ & $7.75\times 10^{57}$ & $5.27\times 10^{57}$ & $1.78\times 10^{54}$ & $1.64\times 10^{54}$ \\
\midrule
\multicolumn{1}{c|}{0.01 $M_{\odot}^{\phantom{2}}$ s$^{-1}$} & $1.05\times 10^{57}$ & $1.12\times 10^{57}$ & $2.43\times 10^{55}$ & $2.43\times 10^{55}$ & $1.78\times 10^{53}$ & $8.68\times 10^{48}$ & $4.29\times 10^{56}$ & $5.48\times 10^{56}$ & $6.71\times 10^{56}$ & $5.70\times 10^{56}$ & $3.53\times 10^{52}$ & $1.23\times 10^{52}$ \\
\bottomrule
\end{tabular}}
\label{tab:Luminosities}
\end{table}

The factor $0<C_{\text{cap}}<1$ is a function of the radius (called \textit{capture function} in \cite{1974ApJ...191..507T}) that accounts for the proportion of neutrinos that are re-captured by the BH and, thus, do not contribute to the total luminosity. For a BH with $M=3M_{\odot}$ and $a=0.95$, the numerical value of the capture function as a function of the dimensionless distance $x=r/r_{s}$ is well fitted by

\begin{equation}
C_{\text{cap}}\left(x\right) = \left(1+\frac{0.3348}{x^{3/2}}\right)^{-1},
    \label{eq:Ccap}
\end{equation}

with a relative error smaller than $0.02\%$. To calculate the energy deposition rate, the disk is modeled as a grid of cells in the equatorial plane. Each cell $k$ has a specific value of differential neutrino luminosity $\Delta\ell^{k}_{\nu_{i}} = F^{k}_{\nu_{i}}r_{k}\Delta r_{k}\Delta\phi_{k}$ and average neutrino energy $\langle E_{\nu_i} \rangle^{k}$. If a neutrino of flavour $i$ is emitted from the cell $k$ and an antineutrino is emitted from the cell $k^{\prime}$, and, before interacting at a point $\mathbf{r}$ above the disk, each travels a distance $r_{k}$ and $r_{k^{\prime}}$, then, their contribution to the energy deposition rate at $\mathbf{r}$ is (see Appx.~\ref{app:nu_barnu_ann} for details)

\begin{align}
\Delta Q_{\nu_{i}\bar{\nu}_{i}kk^{\prime}} &= A_{1,i}\frac{\Delta\ell^{k}_{\nu_{i}}}{r^2_{k}}\frac{\Delta\ell^{k^{\prime}}_{\bar{\nu}_{i}}}{r^2_{k^{\prime}}}\left( \langle E_{\nu_i} \rangle^{k} + \langle E_{\bar{\nu}_i} \rangle^{k^{\prime}}\right)\left(1 - \frac{\mathbf{r}_{k}\cdot \mathbf{r}_{k^{\prime}}}{r_{k}r_{k^{\prime}}}\right)^2 \nonumber \\
& + A_{2,i}\frac{\Delta\ell^{k}_{\nu_{i}}}{r^2_{k}}\frac{\Delta\ell^{k^{\prime}}_{\bar{\nu}_{i}}}{r^2_{k^{\prime}}}\left( \frac{\langle E_{\nu_i} \rangle^{k} + \langle E_{\bar{\nu}_i} \rangle^{k^{\prime}}}{\langle E_{\nu_i} \rangle^{k} \langle E_{\bar{\nu}_i} \rangle^{k^{\prime}}}\right)\left(1 - \frac{\mathbf{r}_{k}\cdot \mathbf{r}_{k^{\prime}}}{r_{k}r_{k^{\prime}}}\right).
    \label{eq:differentialdepp}
\end{align}

The total neutrino annihilation luminosity is the sum over all pairs of cells integrated in space

\begin{equation}
L_{\nu_{i}\bar{\nu}_{i}} = 4\pi \int\limits_{\mathcal{A}} \sum_{k,k^{\prime}}\Delta Q_{\nu_{i}\bar{\nu}_{i}kk^{\prime}}d^{3}\mathbf{r},
    \label{eq:ann_lum_tot}
\end{equation}

where $\mathcal{A}$ is the entire space above (or below) the disk.

\begin{figure}[!hbtp]
\centering
\includegraphics[width=0.5\hsize,clip]{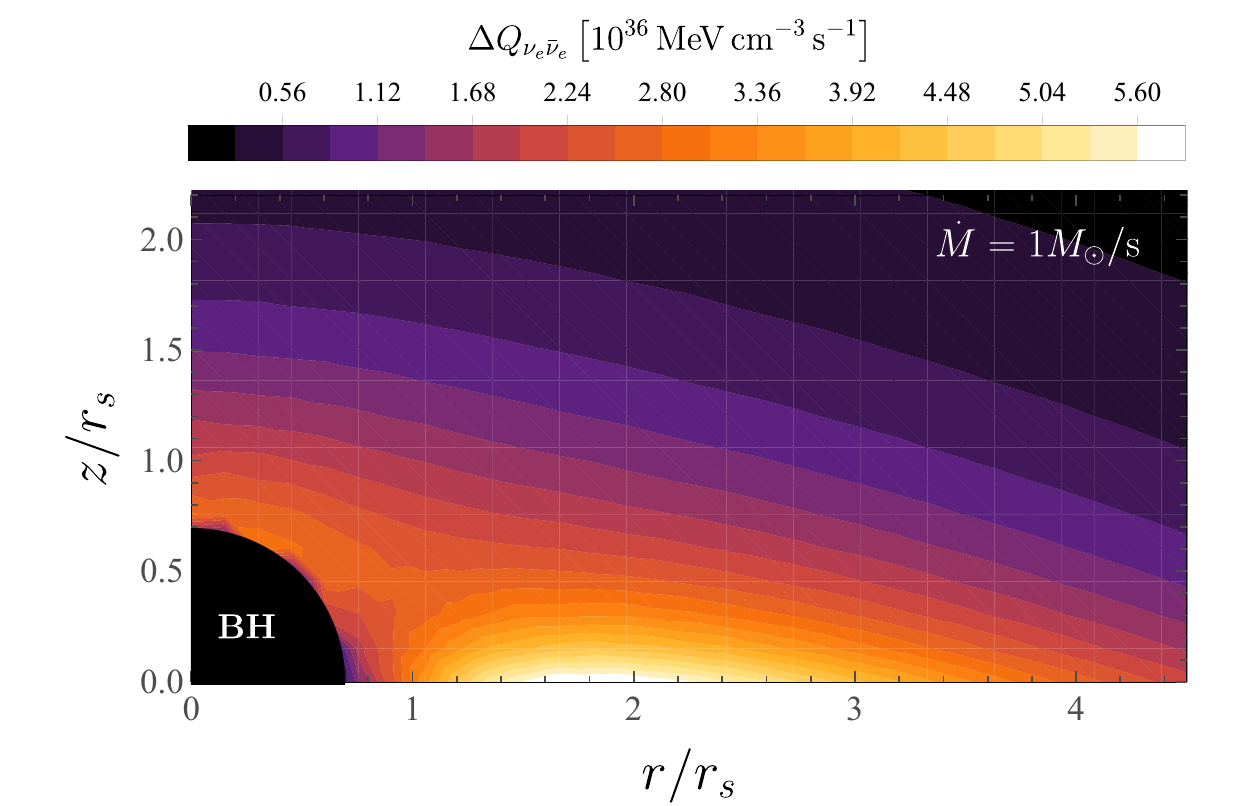}\includegraphics[width=0.5\hsize,clip]{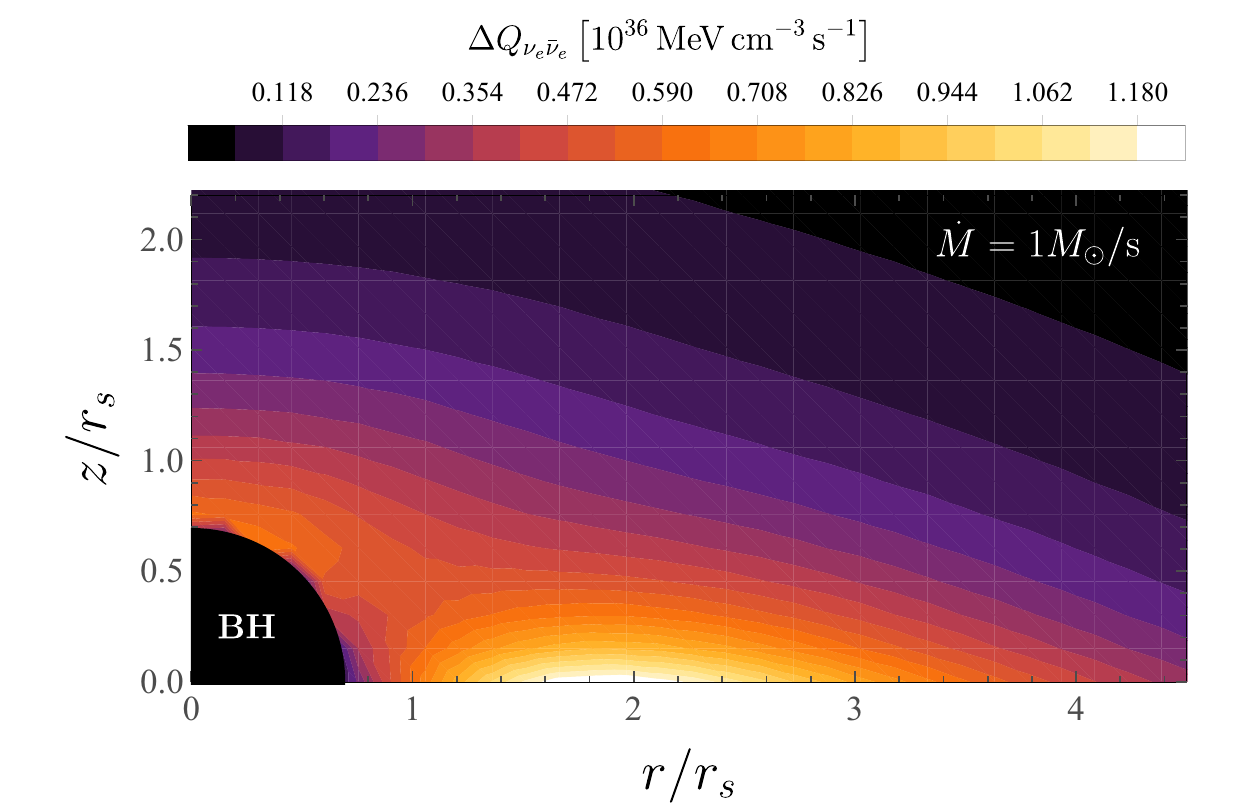}\\
\includegraphics[width=0.5\hsize,clip]{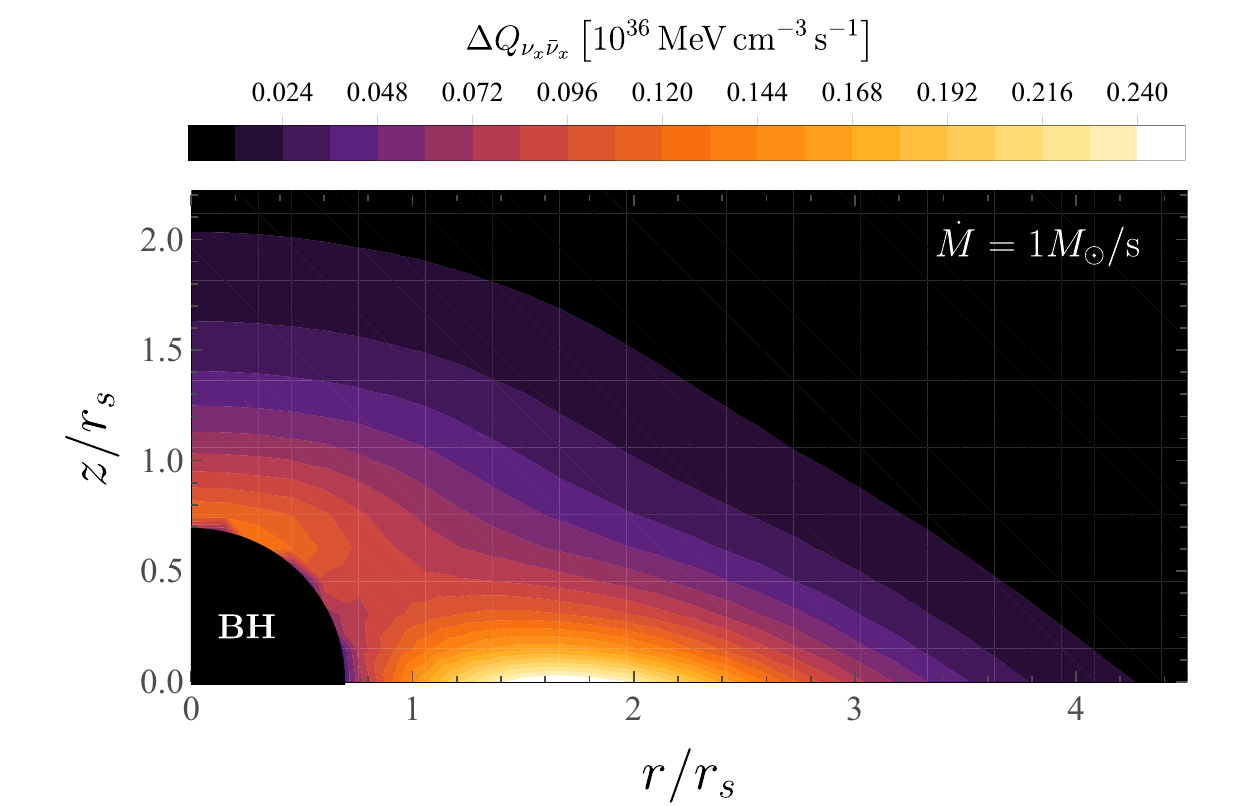}\includegraphics[width=0.5\hsize,clip]{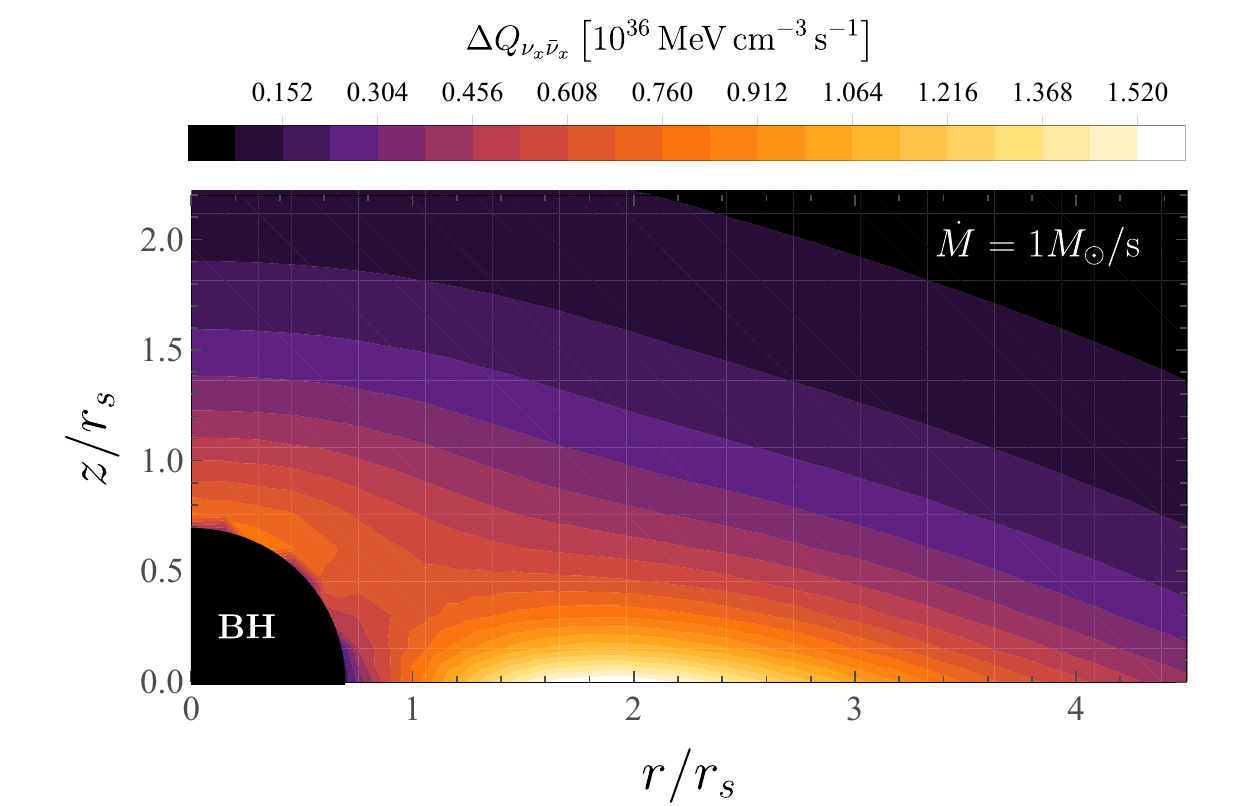}\\
\includegraphics[width=0.5\hsize,clip]{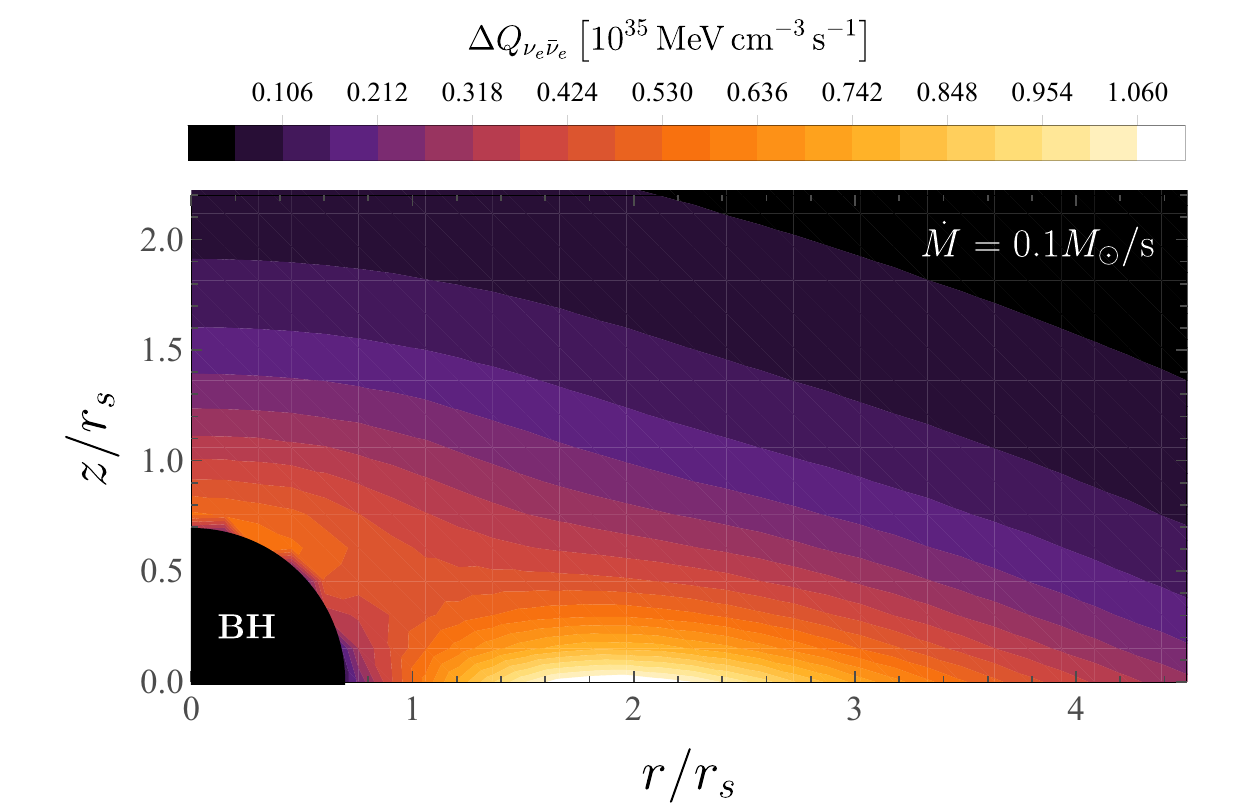}\includegraphics[width=0.5\hsize,clip]{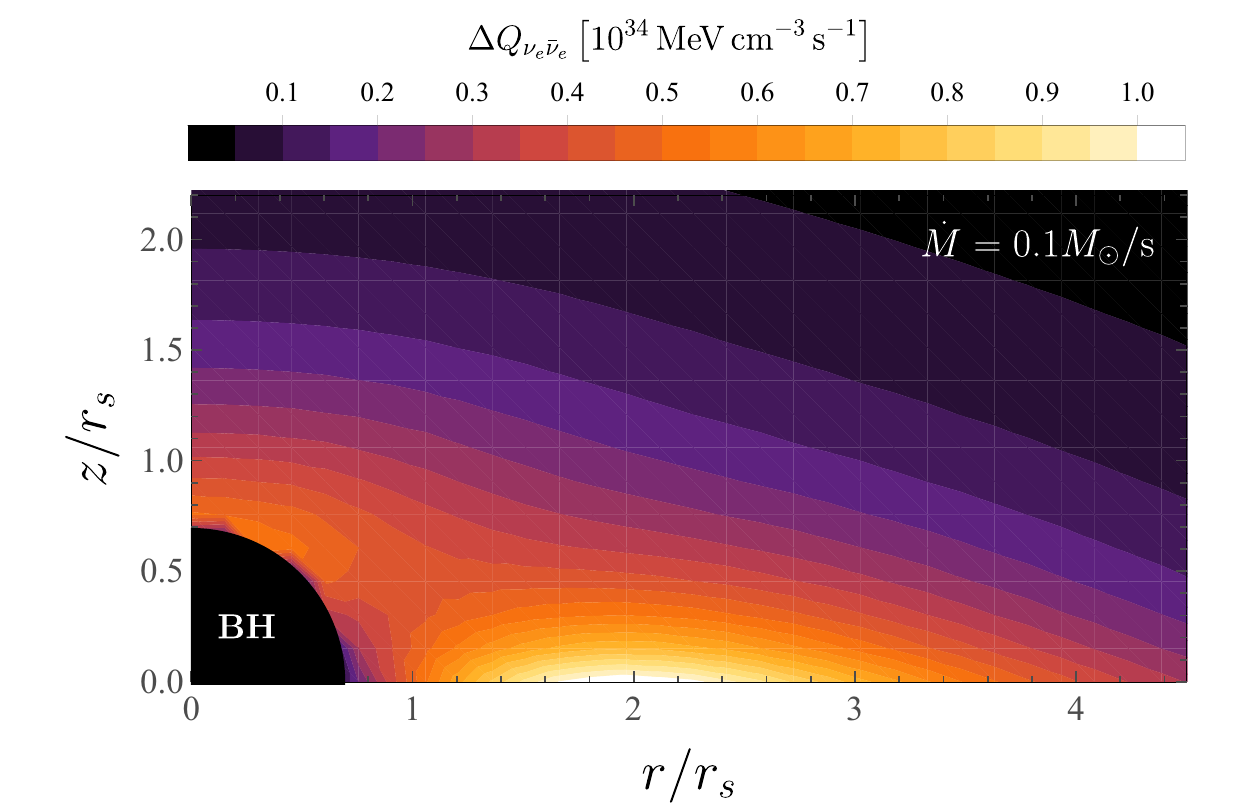}\\
\includegraphics[width=0.5\hsize,clip]{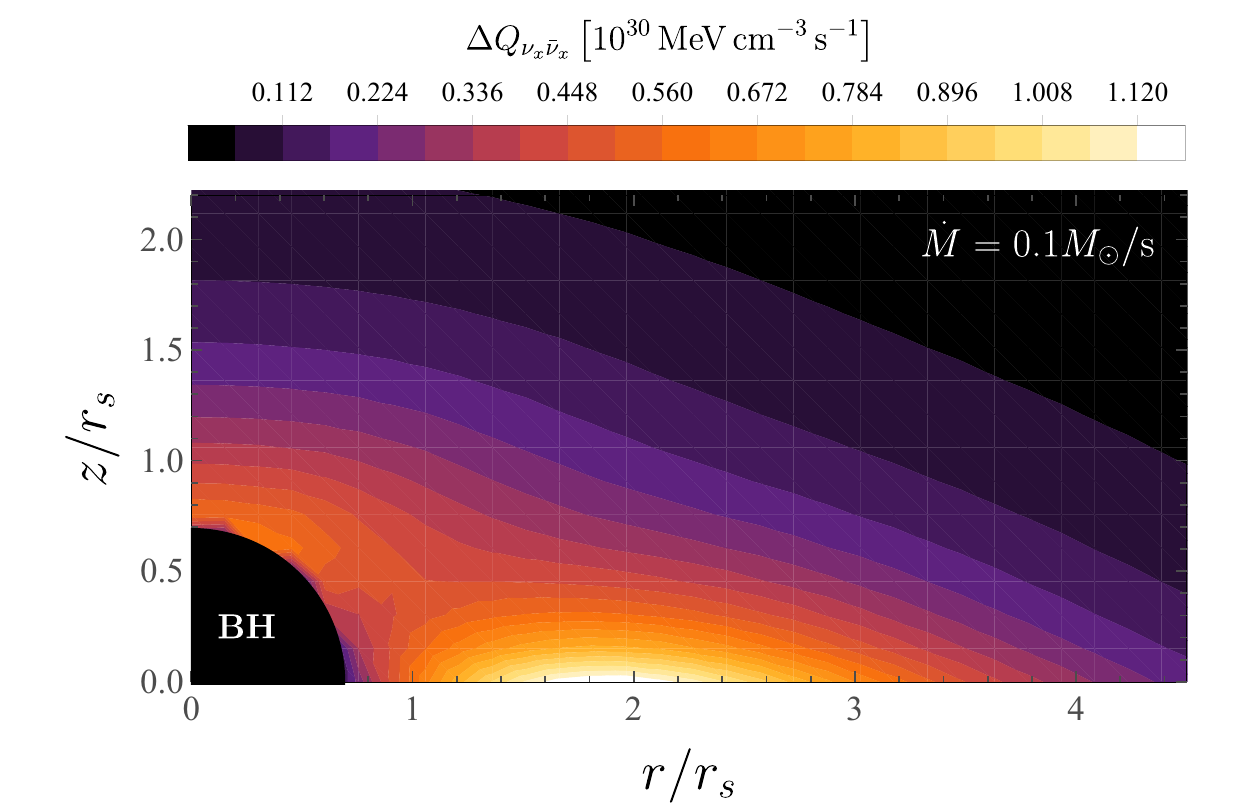}\includegraphics[width=0.5\hsize,clip]{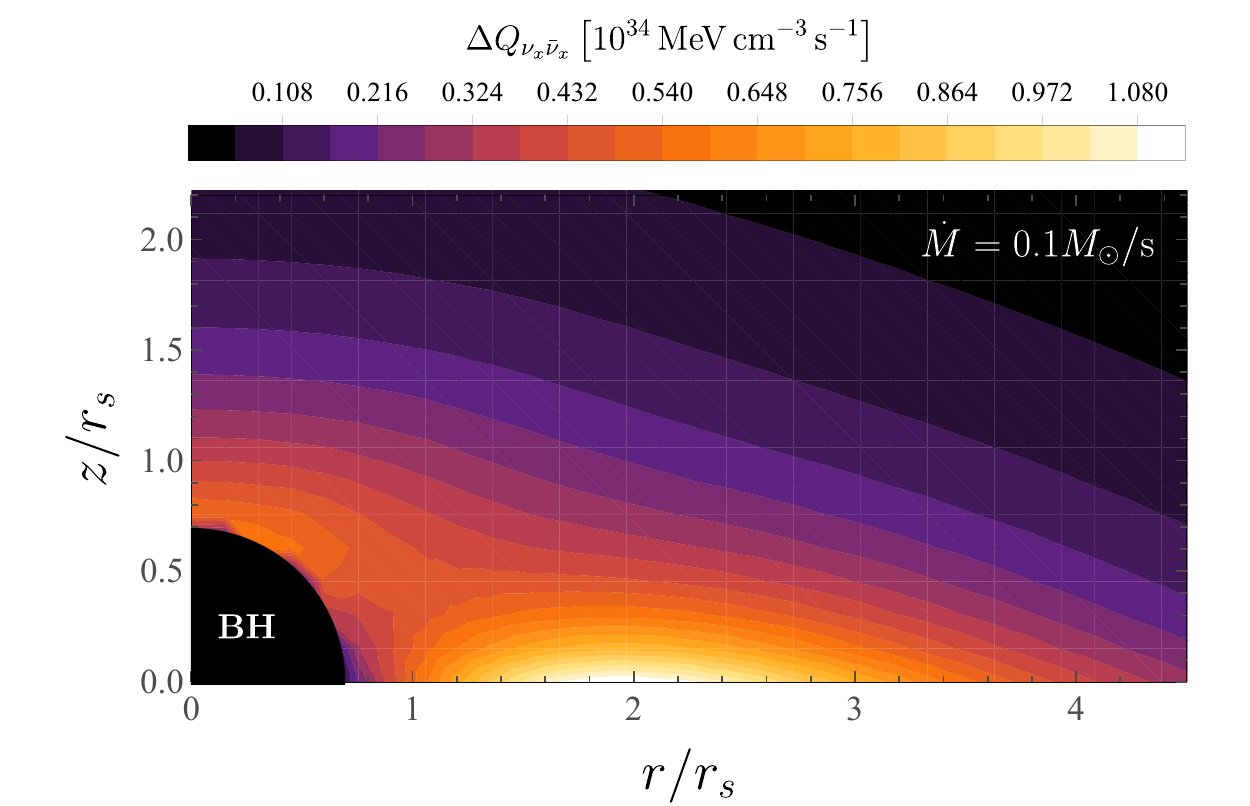}
\caption{Comparison of the neutrino annihilation luminosity per unit volume $\Delta Q_{\nu_{i}\bar{\nu}_{i}} =  \sum_{k,k^{\prime}} \Delta Q_{\nu_{i}\bar{\nu}_{i}kk^{\prime}}$ between disk without (\textit{left column}) and with (\textit{right column}) flavour equipartition for accretion rates $\dot{M} = 1M_{\odot}$ s$ ^{-1}$ and $\dot{M} = 0.01M_{\odot}$ s$ ^{-1}$.}
\label{fig:comp2}
\end{figure}

In Table~\ref{tab:Luminosities} we show the neutrino luminosities and the neutrino annihilation luminosities for disks with and without neutrino collective effects. In each case, flavour equipartition induces a loss in $L_{\nu_{e}}$ by a factor of $\sim$3, and a loss in $L_{\bar{\nu}_{e}}$ luminosity by a factor of $\sim$2. At the same time, $L_{\nu_{x}}$ and $L_{\bar{\nu}_{e}}$ are increased by a factor $\sim$10. This translates into a reduction of the energy deposition rate due to electron neutrino annihilation by a factor of $\sim$7 while the energy deposition rate due to non-electronic neutrinos goes from being negligible to be of the same order of the electronic energy deposition rate. The net effect is to reduce the total energy deposition rate of neutrino annihilation by a factor of $\sim(3-5)$ for the accretion rates considered. In particular, \textcolor{black}{we obtain a factor of} $3.03$ and $3.66$ for $\dot{M} = 1~M_{\odot}$~s$ ^{-1}$ and $\dot{M} = 0.01~M_{\odot}$ s$ ^{-1}$, respectively and a factor of 4.73 for $\dot{M} = 0.1~M_{\odot}$ s$ ^{-1}$. The highest value correspond to \textcolor{black}{an} intermediate value of the accretion rate because, for this case, there is a $\nu_{e}$ cooling suppression ($\tau_{\nu_{e}} > 1$) and the quotient $\tau_{\nu_{e}}/\tau_{\nu_{x}}$ is maximal. By Eq.~(\ref{eq:fluxcomp}), the difference between the respective cooling terms is also maximal. In Fig.~\ref{fig:comp2} we show the energy deposition rate per unit volume around the BH for each flavour with accretion rates $\dot{M} = 1~M_\odot$~s$^{-1}$ and $\dot{M} = 0.1~M_\odot$~s$ ^{-1}$. There we can see the drastic enhancement of the non-electronic neutrino energy deposition rate and the reduction of the electronic deposition rate. Due to the double peak in the neutrino density for $\dot{M} = 0.01~M_\odot$~s$^{-1}$ case (see Fig.~\ref{fig:Disks}), the deposition rate per unit volume also shows two peaks. One at $r_{s}<r<2r_{s}$ and the other at $10\,r_{s}<r<11\,r_{s}$. Even so, the behaviour is similar to the other cases.  

%

%%%%%%%%%%%%%%%%%%%%%%%%%%%%%%%%%%%%%%%%%%%%%%%%%%%%%%%%%%%%%%%%%%%%%%%%%%%%%%%%%%%%%%%%%%%%%%%%
%%%%%%%%%%%%%%%%%%%%%%%%%%%%%%%%%%%%%%%%%%%%%%%%%%%%%%%%%%%%%%%%%%%%%%%%%%%%%%%%%%%%%%%%%%%%%%%%
\section{Discussion}\label{sec:5}
%%%%%%%%%%%%%%%%%%%%%%%%%%%%%%%%%%%%%%%%%%%%%%%%%%%%%%%%%%%%%%%%%%%%%%%%%%%%%%%%%%%%%%%%%%%%%%%%%
%%%%%%%%%%%%%%%%%%%%%%%%%%%%%%%%%%%%%%%%%%%%%%%%%%%%%%%%%%%%%%%%%%%%%%%%%%%%%%%%%%%%%%%%%%%%%%%%%

The generation of a seed, energetic $e^{-}e^{+}$ plasma seems to be a general prerequisite of GRB theoretical models for the explanation of the prompt (MeV) gamma-ray emission. The $e^{-}e^{+}$ pair annihilation produce photons leading to an opaque pair-photon plasma that self-accelerates, expanding to ultrarelativistic Lorentz factors of the order of $10^2$--$10^3$ (see, e.g., \cite{1998A&A...338L..87P,1999A&AS..138..511R,2000A&A...359..855R}). The reaching of transparency of MeV-photons at large Lorentz factor and corresponding large radii is requested to solve the so-called \emph{compactness problem} posed by the observed non-thermal spectrum in the prompt emission~\cite{1990ApJ...365L..55S,1993MNRAS.263..861P,1993ApJ...415..181M}. There is a vast literature on this subject and we refer the reader to \cite{1999PhR...314..575P,2004RvMP...76.1143P,2002ARA26A..40..137M,2006RPPh...69.2259M,2014ARA&A..52...43B,2015PhR...561....1K}, and references therein, for further details. 
Neutrino-cooled accretion disks onto rotating BHs have been proposed as a possible way of producing the above-mentioned $e^{-}e^{+}$ plasma. The reason is that such disks emit a large amount of neutrino and antineutrinos that can undergo pair annihilation near the BH \cite{1999ApJ...518..356P,2001ApJ...557..949N,2002ApJ...577..311K,2002ApJ...579..706D,2005ApJ...629..341K,2005ApJ...632..421L,2006ApJ...643L..87G,2007ApJ...657..383C,2007ApJ...662.1156K,2010A&A...509A..55J,2013ApJ...766...31K,2013MNRAS.431.2362L,2013ApJS..207...23X}. The viability of this scenario clearly depends on the energy deposition rate of neutrino-antineutrinos into $e^{-}e^{+}$ and so on the local (anti)neutrino density and energy.
We have here shown that, inside these hyperaccreting disks, a rich neutrino oscillations phenomenology is present due to the high neutrino density. Consequently, the neutrino/antineutrino emission and the corresponding pair annihilation process around the BH leading to electron-positron pairs, are affected by neutrino flavour conversion. Using the thin disk and $\alpha$-viscosity approximations, we have built a simple stationary model of general relativistic neutrino-cooled accretion disks around a Kerr BH, that takes into account not only a wide range of neutrino emission processes and nucleosynthesis but also the dynamics of flavour oscillations. The main assumption relies on considering the neutrino oscillation behaviour within small neighbouring regions of the disk as independent from each other. This, albeit being a first approximation to a more detailed picture, has allowed us to set the main framework to analyze the neutrino oscillations phenomenology in inside neutrino-cooled disks.
In the absence of oscillations, a variety of neutrino-cooled accretion disks onto Kerr BHs, without neutrino flavour oscillations, have been modelled in the literature (see e.g. \cite{1998ApJ...498..313G,1999ApJ...518..356P,2007ApJ...657..383C,2013ApJS..207...23X} and \cite{LIU20171} for a recent review). The physical setting of our disk model follows closely the ones considered in \cite{2007ApJ...657..383C}, but with some extensions and differences in some aspects: 

\begin{enumerate}[leftmargin=*,labelsep=4.9mm]

    \item The equation of vertical hydrostatic equilibrium, Eq.~(\ref{eq:press}), can be derived in several ways \cite{1973blho.conf..343N,1998ApJ...498..313G,Abramowicz_1997}. We followed a particular approach consistent with the assumptions in \cite{1973blho.conf..343N}, in which we took the vertical average of a hydrostatic Euler equation in polar coordinates. \textcolor{black}{The result is an equation} that leads to smaller values of the disk pressure when compared with other models. It is expected that the pressure at the centre of the disk is smaller than the average density multiplied by the local tidal acceleration at the equatorial plane. Still, the choice between the assortment of pressure relations is tantamount to a fine-tuning of the model. Within the thin disk approximation, all these approaches are equivalent since they all assume vertical equilibrium and neglect self-gravity. 

    \item Following the BdHN scenario for the explanation of GRBs associated with Type Ic SNe (see Sec.~\ref{sec:1}), we considered a gas composed of $^{16}$O at the outermost radius of the disk and followed the evolution of the ion content using the Saha equation to fix the local NSE. In \cite{2007ApJ...657..383C}, only $^{4}$He is present and, in \cite{2013ApJS..207...23X}, ions up to $^{56}$Fe are introduced. The affinity between these cases implies that this particular model of disk accretion is insensible to the initial mass fraction distribution. This is explained by the fact that the average binding energy for most ions is very similar, hence any cooling or heating due to a redistribution of nucleons, given by the NSE, is negligible when compared to the energy consumed by direct photodisintegration of alpha particles. Additionally, once most ions are dissociated, the main cooling mechanism is neutrino emission that is similar for all models, modulo the supplementary neutrino emission processes included in addition to electron and positron capture. However, during our numerical calculations, we noticed that the inclusion of non-electron neutrino emission processes can reduce the electron fraction by up to $\sim 8\%$. This effect is observed again during the simulation of flavour equipartition alluding to the need for detailed calculations of neutrino emissivities when establishing NSE state. We obtain similar results to \cite{2007ApJ...657..383C} (see Fig.~\ref{fig:Disks}), but by varying the accretion rate and fixing the viscosity parameter. This suggests that a more natural differentiating set of variables in the hydrodynamic equations of an $\alpha$-viscosity disk is the combination of the quotient $\dot{M}/\alpha$ and either $\dot{M}$ or $\alpha$. This result is already evident in, for example, Fig.~11 and Fig.~12 of \cite{2007ApJ...657..383C}, but was not mentioned there.
\end{enumerate}

Concerning neutrino oscillations, we showed that the conditions inside the ignition radius, the oscillation potentials follow the relation $\langle \omega \rangle \ll \mu \ll \lambda$, as it is illustrated by Fig.~\ref{fig:Potentials}. We also showed that the within this region the number densities of electron neutrinos and antineutrinos are very similar. As a consequence of this particular environment very fast pair conversions $\nu_{e}\bar{\nu}_{e} \rightleftharpoons \nu_{x}\bar{\nu}_{x}$, induced by bipolar oscillations, are obtained for the inverted mass hierarchy case with oscillation frequencies between $10^{9}$ s$^{-1}$ and $10^{5}$ s$^{-1}$. For the normal hierarchy case no flavour changes are observed (see Fig.~\ref{fig:SurvProbI} and Fig.~\ref{fig:SurvSev}). Bearing in mind the magnitude of these frequencies and the low neutrino travel times through the disk, we conclude that an accretion disk under our main assumption cannot represent a steady-state. However, using numerical and algebraic results obtained in \cite{Raffelt:2007yz,Fogli:2007bk,EstebanPretel:2007ec}, and references therein, we were able to generalize our model to a more realistic picture of neutrino oscillations. The main consequence of the interaction between neighbouring regions of the disk is the onset of kinematic decoherence in a timescale of the order of the oscillation times. Kinematic decoherence induces fast flavour equipartition among electronic and non-electronic neutrinos throughout the disk. Therefore, the neutrino content emerging from the disk is very different from the one that is usually assumed (see e.g. \cite{2012PhRvD..86h5015M,2016PhRvD..93l3004L,2020arXiv200901843P}). The comparison between disks with and without flavour equipartition is summarized in Fig.~\ref{fig:comp} and Table~\ref{tab:Luminosities}. We found that flavour equipartition, while leaving antineutrino cooling practically unchanged, it enhances neutrino cooling by allowing the energy contained (and partially trapped inside the disk due to high opacity) within the $\nu_{e}$ gas to escape in the form of $\nu_{x}$, rendering the disk insensible to the electron neutrino opacity. We give in Eq.~(\ref{eq:fluxcomp}) a relation to estimate the change in $F_{\nu}$ as a function of $\tau_{\nu_{e}}\tau_{\nu_{x}}$ that describes correctly the behaviour of the disk under flavour equipartition. The variation of the flavour content in the emission flux implies a loss in $L_{\nu_{e}}$ and an increase in $L_{\nu_{x}}$ and $L_{\bar{\nu}_{e}}$. As a consequence, the total energy deposition rate of the process $\nu+\bar{\nu}\to e^{-}+e^{+}$ is reduced. We showed that this reduction can be as high 80\% and is maximal whenever the quotient $\tau_{\nu_{e}}/\tau_{\nu_{x}}$ is also maximal and the condition $\tau_{\nu_{e}}>1$ is obtained. 

At this point, we can identify several issues which must still to be investigated in view of the results we have presented:

First, throughout the accretion disk literature, several fits of the neutrino and neutrino annihilation luminosity can be found (see e.g. \cite{LIU20171} and references therein). However, all these fits were calculated without taking into account neutrino oscillations. Since we have shown that oscillations directly impact luminosity, these results need to be extended. 
    
Second, the calculations of the neutrino and antineutrino annihilation luminosities we have performed ignore general relativistic effects, \textcolor{black}{save for the correction given by the capture function,} and the possible neutrino oscillations from the disk surface to the annihilation point. In \cite{1999ApJ...517..859S}, it has been shown that general relativistic effects can enhance the neutrino annihilation luminosity in a neutron star binary merger by a factor of $10$. In \cite{1999ApJ...518..356P}, however, it is argued that in BHs this effect has to be mild since the energy gained by falling into the gravitational potential is lost by the electron-positron pairs when they climb back up. Nonetheless, this argument ignores the bending of neutrino trajectories and neutrino capture by the BH which can be significant for $r\lesssim 10r_{s}$. In \cite{2007A&A...463...51B}, the increment is calculated to be no more than a factor of $2$ and can be less depending on the geometry of the emitting surface. But, as before, these calculations assume a purely $\nu_{e}\bar{\nu}_{e}$ emission and ignore oscillations after the emission. Simultaneously, the literature on neutrino oscillation above accretion disks (see e.g. \cite{2012PhRvD..86h5015M,2020arXiv200901843P}) do not take int account oscillations inside the disk and assume only $\nu_{e}\bar{\nu}_{e}$ emission. A similar situation occurs in works studying the effect of neutrino emission on r-process nucleosynthesis in hot outflows (wind) ejected from the disk (see e.g. \cite{Caballero_2012}). 

It is still unclear how the complete picture (oscillations inside the disk $\to$ oscillations above the disk + relativistic effects) affect the final energy deposition. We are currently working on the numerical calculation of the annihilation energy deposition rate using a ray tracing code and including neutrino oscillations from the point of their creation until they are annihilated, i.e., within the accretion disk as well as after its emission from the surface of the disk and during its trajectory until reaching the annihilation point. These results and their consequences for the energy deposition annihilation rate will be the subject of a future publication.

\textcolor{black}{The knowledge of the final behavior of a neutrino-dominated accretion disk with neutrino oscillations requires time-dependent, multi-dimensional, neutrino-transport simulations coupled with the evolution of the disk.} These simulations are computationally costly even for systems with a high degree of symmetry, therefore a first approximation is needed to identify key theoretical and numerical features involved in the study of neutrino oscillations in neutrino-cooled accretion disks. This work serves as a platform for such a first approximation. Considering that kinematic decoherence is a general feature of anisotropic neutrino gases, with the simplified model presented here, we were able to obtain an analytical result that agrees with the physics understanding of accretion disks.
\textcolor{black}{In \cite{2016PhRvD..93l3004L} it is pointed out that for a total energy in $\bar{\nu}_{e}$ of $10^{52}$ erg and an average neutrino energy $\langle E_{\nu,\bar{\nu}} \rangle \sim 20$ MeV, the Hyper-Kamiokande neutrino-horizon is of the order of 1 Mpc. If we take a total energy carried out by $\bar{\nu}_{e}$ of the order of the gravitational gain by accretion ($E_{g}\sim 10^{52}-10^{53}$~erg) in the more energetic case of binary-driven hypernovae and the neutrino energies in Fig.~\ref{fig:Disks}, we should expect the neutrino-horizon distance
to be also of the order of 1 Mpc. However, if we adopt the local binary-driven hypernovae rate $\sim$ 1 Gpc$^{-3}$ yr$^{-1}$ \citep{2016ApJ...832..136R}, it is clear that the direct detection of this neutrino signal is quite unlikely. But we have shown that neutrino oscillation can have an effect on $e^{-}e^{+}$ plasma production above BHs in GRB models. Additionally, the unique conditions inside the disk and its geometry lend themselves to a variety of neutrino oscillations that can have an impact other astrophysical phenomena. Not only in plasma production, but also in r-process nucleosynthesis in disk winds. This, in particular, is a subject of a future publication. As such, this topic deserves appropriate attention since it paves the way for new, additional astrophysical scenarios for testing neutrino physics.}

%%%%%%%%%%%%%%%%%%%%%%%%%%%%%%%%%%%%%%%%%%
%%%%%%%%%%%%%%%%%%%%%%%%%%%%%%%%%%%%%%%%%%
\vspace{6pt} 

%%%%%%%%%%%%%%%%%%%%%%%%%%%%%%%%%%%%%%%%%%
%% optional
%\supplementary{The following are available online at \linksupplementary{s1}, Figure S1: title, Table S1: title, Video S1: title.}


%%%%%%%%%%%%%%%%%%%%%%%%%%%%%%%%%%%%%%%%%%
\authorcontributions{All authors have contributed equally to this work.}

%%%%%%%%%%%%%%%%%%%%%%%%%%%%%%%%%%%%%%%%%%
\funding{J.D.U. was supported by COLCIENCIAS under the program Becas Doctorados en el Exterior Convocatoria No. 728. E.A.B-V. was supported from COLCIENCIAS under the program Becas Doctorados Nacionales Convocatoria No. 727, the International Center for Relativistic Astrophysics Network (ICRANet), Universidad Industrial de Santander (UIS) and the International Relativistic Astrophysics Ph.D Program (IRAP-PhD).}

%%%%%%%%%%%%%%%%%%%%%%%%%%%%%%%%%%%%%%%%%%%
\acknowledgments{The authors thank Prof. F. D. Lora-Clavijo for the constant support in the development of the numerical code for the integration of the geodesic equations via the backward ray-tracing method, and Prof. C. L. Fryer for insightful discussions on astrophysical consequences of neutrino flavor oscillations in accretion disks.}

%%%%%%%%%%%%%%%%%%%%%%%%%%%%%%%%%%%%%%%%%%
\conflictsofinterest{The authors declare no conflict of interest.} 

%%%%%%%%%%%%%%%%%%%%%%%%%%%%%%%%%%%%%%%%%%
%% optional
\abbreviations{The following abbreviations are used in this manuscript:\\

\noindent 
\begin{tabular}{@{}ll}
BdHN & Binary-Driven Hypernova\\
BH & Black Hole\\
CF & Coordinate Frame\\
CO$_{\rm core}$ & Carbon-Oxygen Star\\
CRF & Co-rotating Frame\\
GRB & Gamma-Ray Burst\\
IGC & Induced Gravitational Collapse\\
ISCO & Innermost Stable Circular Orbit\\
LNRF & Locally Non-Rotating Frame\\
MSW & Mikheyev-Smirnov-Wolfenstein\\
NDAF & Neutrino-Dominated Accretion Flows\\
NS & Neutron Star\\
NSE & Nuclear Statistical Equilibrium\\
SN & Supernova\\
\end{tabular}}

%%%%%%%%%%%%%%%%%%%%%%%%%%%%%%%%%%%%%%%%%%
%% optional
\appendixtitles{yes} %Leave argument "no" if all appendix headings stay EMPTY (then no dot is printed after "Appendix A"). If the appendix sections contain a heading then change the argument to "yes".
\appendix

%%%%%%%%%%%%%%%%%%%%%%%%%%%%%%%%%%%%%%%%%%
%%%%%%%%%%%%%%%%%%%%%%%%%%%%%%%%%%%%%%%%%%
\section{Transformations and Christoffel symbols}\label{app:christoffel}
%%%%%%%%%%%%%%%%%%%%%%%%%%%%%%%%%%%%%%%%%%
%%%%%%%%%%%%%%%%%%%%%%%%%%%%%%%%%%%%%%%%%%

For the sake of completeness, here we give the explicitly the transformation used in Eq.~(\ref{eq:fourvel}) and the Christoffel symbols used during calculations. The coordinate transformation matrices between the CF and the LNRF on the tangent vector space is \cite{1972ApJ...178..347B}

\begin{equation}
e_{\hat{\nu}}^{\;\;\mu} = \begin{pmatrix}
    \frac{1}{\sqrt{\omega^2 g_{\phi\phi} - g_{tt}}} & 0 & 0 & 0 \\
    0 & \frac{1}{\sqrt{g_{rr}}} & 0 & 0 \\
    0 & 0 & \frac{1}{\sqrt{g_{\theta\theta}}} & 0 \\
    \frac{\omega}{\sqrt{\omega^2 g_{\phi\phi} - g_{tt}}} & 0 & 0 & \frac{1}{\sqrt{g_{\phi\phi}}}
  \end{pmatrix},\;\;\; e_{\;\;\mu}^{\hat{\nu}} = \begin{pmatrix}
  \sqrt{\omega^2 g_{\phi\phi} - g_{tt}} & 0 & 0 & 0 \\
  0 & \sqrt{g_{rr}} & 0 & 0 \\
  0 & 0 & \sqrt{g_{\theta\theta}} & 0 \\
  -\omega\sqrt{g_{\phi\phi}} & 0 & 0 & \sqrt{g_{\phi\phi}} \\
  \end{pmatrix},
    \label{eq:bardeen}
\end{equation}

so that the basis vectors transform as $\boldsymbol{\partial}_{\hat{\nu}} = e^{\mu}_{\;\; \tilde{\nu}}\boldsymbol{\partial}_{\mu}$, that is, with $\boldsymbol{e}^{T}$. For clarity, coordinates on the LNRF have a caret $(x^{\hat{\mu}})$, coordinates on the CRF have a tilde $(x^{\tilde{\mu}})$ and coordinates on the LRF have two $(x^{\tilde{\tilde{\mu}}})$. An observer on the LNRF sees the fluid elements move with an azimuthal velocity $\beta^{\hat{\phi}}$. This observer then can perform a Lorentz boost $L_{\beta^{\hat{\phi}}}$ to a new frame. On this new frame an observer sees the fluid elements falling radially with velocity $\beta^{\tilde{r}}$, so it can perform another Lorentz boost $L_{\beta^{\tilde{r}}}$ to the LRF. Finally, the transformation between the the LRF and the CF coordinates $x^{\mu} = e_{\hat{\rho}}^{\;\;\mu} (L_{\beta^{\hat{\phi}}})_{\tilde{\alpha}}^{\;\;\hat{\rho}} (L_{\beta^{\tilde{r}}})_{\tilde{\tilde{\nu}}}^{\;\;\tilde{\alpha}} x^{\tilde{\tilde{\nu}}} = A^{\;\;\mu}_{\tilde{\tilde{\nu}}} x^{\tilde{\tilde{\nu}}}$, where the components of $\boldsymbol{A}$ are

\begin{equation}
  A^{\tilde{{\tilde{\nu}}}}_{\;\;\mu} = \begin{pmatrix}
    \gamma_{\tilde{r}}\gamma_{\hat{\phi}}\left( \sqrt{\omega^2 g_{\phi\phi} - g_{tt}} + \beta^{\hat{\phi}}\omega\sqrt{g_{\phi\phi}} \right) & -\gamma_{\tilde{r}}\beta^{\tilde{r}}\sqrt{g_{rr}} & 0 & -\gamma_{\tilde{r}}\gamma_{\hat{\phi}}\beta^{\hat{\phi}}\sqrt{g_{\phi\phi}} \\ 
    -\gamma_{\hat{\phi}}\gamma_{\tilde{r}}\beta^{\tilde{r}}\left( \sqrt{\omega^2 g_{\phi\phi} - g_{tt}} + \beta^{\hat{\phi}}\omega\sqrt{g_{\phi\phi}} \right) & \gamma_{\tilde{r}}\sqrt{g_{rr}} & 0 & \gamma_{\tilde{r}}\gamma_{\hat{\phi}}\beta^{\tilde{r}}\beta^{\hat{\phi}}\sqrt{g_{\phi\phi}} \\
    0 & 0 & \sqrt{g_{\theta\theta}} & 0 \\
    -\gamma_{\hat{\phi}}\left( \beta^{\hat{\phi}}\sqrt{\omega^2 g_{\phi\phi} - g_{tt}} + \omega\sqrt{g_{\phi\phi}} \right) & 0 & 0 & \gamma_{\hat{\phi}}\sqrt{g_{\phi\phi}}
  \end{pmatrix}.
\label{eq:fulltransform}
\end{equation}

Since Lorentz transformations do not commute, the transformation $\boldsymbol{A}$ raises the question: what happens if we invert the order? In this case, we would not consider a co-rotating frame but a \emph{cofalling} frame on which observers see fluid elements, not falling, but rotating. The new transformation velocities $\beta^{r'}$, $\beta^{\phi'}$ are subject to the conditions $\beta^{\phi'} = \gamma_{r'}\beta^{\hat{\phi}}$, $\beta^{r'} = \beta^{\tilde{r}}/\gamma_{\hat{\phi}}$ and $\gamma_{r'}\gamma_{\phi'} = \gamma_{\tilde{r}}\gamma_{\hat{\phi}}$. Although both approaches are valid, considering that the radial velocity is an unknown, the first approach is clearly cleaner. To obtain the coordinate transformation between the CF and the CRF $A_{\tilde{\nu}}^{\;\;\mu}$ and $A^{\tilde{\nu}}_{\;\;\mu}$ we can simply set $\beta^{\tilde{r}} = 0$ in Eqs.~(\ref{eq:fulltransform}). With this, we can calculate

\begin{equation}
\frac{d\hat{\phi}}{d\hat{t}} = \beta^{\hat{\phi}} =  \frac{u^{\mu}e_{\;\;\mu}^{\hat{\phi}}}{u^{\nu}e^{\hat{t}}_{\;\;\nu}} = \sqrt{\frac{g_{\phi\phi}}{\omega^2 g_{\phi\phi} - g_{tt}}}\left(\Omega - \omega\right),
\label{eq:betaphi}
\end{equation}

and

\begin{equation}
d\tilde{r} = \sqrt{g_{rr}}dr,\;\,d\tilde{t}=\frac{\gamma_{\hat{\phi}}}{\sqrt{\omega^2 g_{\phi\phi} - g_{tt}}} dt = \frac{1}{\sqrt{-g_{tt}-2\Omega g_{t\phi} - \Omega^2 g_{\phi\phi}}}dt,\;\, d\tilde{\theta} = \sqrt{g_{\theta\theta}}d\theta.
\label{eq:deltas}
\end{equation}

The non-vanishing Christoffel symbols are

\begin{align}
\Gamma^{t}_{\;tr} &= \frac{M\left( r^2 - M^2a^2\cos^2\theta \right)\left(r^2 + M^2a^2\right)}{\Sigma^2\Delta},\;\; \Gamma^{t}_{\;t\theta} = -\frac{M^3a^2r\sin{2\theta}}{\Sigma^2}, \nonumber\\
\Gamma^{t}_{\;r\phi} &= -\frac{M^2a\left( 3r^4 + M^2a^2r^2 + M^2a^2\cos^2\theta\left( r^2 - M^2a^2  \right) \right)\sin^2\theta}{\Sigma^2\Delta}, \nonumber \\
\Gamma^{t}_{\;\theta\phi} &= \frac{2M^4a^3r\cos\theta\sin^3\theta}{\Sigma^2},\;\; \Gamma^{r}_{\;tt} = \frac{M\Delta\left( r^2 - M^2a^2\cos^2\theta \right)}{\Sigma^3}, \nonumber \\
\Gamma^{r}_{\;t\phi} &= -\frac{M^2 a \Delta\left( r^2 - M^2a^2\cos^2\theta \right)\sin^2\theta}{\Sigma^3}, \nonumber \\
\Gamma^{r}_{\;rr} &= \frac{r}{\Sigma} + \frac{M-r}{\Delta},\;\; \Gamma^{r}_{\;r\theta} = -\frac{M^2 a^2\sin\theta}{M^2 a^2\cos\theta + r^2\tan\theta},\;\; \Gamma^{r}_{\;\theta\theta} = -\frac{r\Delta}{\Sigma}, \nonumber \\
\Gamma^{r}_{\;\phi\phi} &= \left(Ma\Gamma^{r}_{\;t\phi} - \Gamma^{r}_{\;\theta\theta}\right) \sin^2\theta,\;\; \Gamma^{\theta}_{\;tt} = -\Gamma^{t}_{\;\theta\phi}\frac{\csc^2\theta}{Ma\Sigma},\;\; \Gamma^{\theta}_{\;t\phi} = \frac{M^2ar\left( r^2 + M^2a^2 \right)\sin{2\theta}}{\Sigma^3}, \nonumber \\
\Gamma^{\theta}_{\;rr} &= \frac{M^2a^2\sin\theta\cos\theta}{\Sigma\Delta},\;\; \Gamma^{\theta}_{\;t\theta} = \frac{r}{\Sigma},\;\; \Gamma^{\theta}_{\;\theta\theta} = \Gamma^{r}_{\;r\theta}, \nonumber \\
\Gamma^{\theta}_{\;\phi\phi} &= \left( \frac{\Delta}{\Sigma} + \frac{2 M r\left( r^2 + M^2a^2 \right)^2}{\Sigma^3} \right)\sin\theta\cos\theta,\;\;
\Gamma^{\phi}_{\;tr} = -\frac{M^2a\left(r^2 - M^2a^2\cos^2\theta\right)}{\Sigma^2\Delta}, \nonumber \\
\Gamma^{\phi}_{\;t\theta} &= -\frac{2M^2ar\cot\theta}{\Sigma^2},\;\; 
\Gamma^{\phi}_{\;r\phi} = \frac{r\left( \Sigma - 2Mr \right)}{\Sigma\Delta} + \frac{Ma\Sigma}{\Delta^2}\Gamma^{r}_{\;t\phi},\;\; \Gamma^{\phi}_{\;\theta\phi} = \cot\theta - \Gamma^{t}_{\;t\theta}.
    \label{eq:christoffel}
\end{align}

Using the connection coefficients and the metric, both evaluated at the equatorial plane we can collect several equations for averaged quantities. The expansion of the fluid world lines is

\begin{equation}
\boldsymbol{\theta}=\nabla_{\mu}u^{\mu} = \frac{2}{r}u^r+\partial_{r}u^r.
    \label{eq:expansion}
\end{equation}

There are several ways to obtain an approximate version of the shear tensor \cite[e.g.][]{1998ApJ...498..313G,2012arXiv1207.6455M,Moghaddas:2017itv} but by far the simplest one is proposed by \cite{1973blho.conf..343N}. On the CRF the fluid four-velocity can be approximated by $u^{\tilde{\mu}}=(1,0,0,0)$ by Eq.~(\ref{eq:appr1}). Both the fluid four-acceleration $a_{\nu}=u^{\mu}\nabla_{\mu}u_ {\nu}$ and expansion parameter, Eq.~(\ref{eq:expansion}), vanish so that the shear tensor reduces to $2\sigma_{\tilde{\mu}\tilde{\nu}} = \nabla_{\tilde{\mu}}u_{\tilde{\nu}} + \nabla_{\tilde{\nu}}u_{\tilde{\mu}} $. In particular, the $r$-$\phi$ component is

\begin{equation}
\sigma_{\tilde{r}\tilde{\phi}} = -\frac{1}{2}\left(\Gamma_{\;\tilde{\phi}\tilde{r}}^{\tilde{t}}+\Gamma_{\;\tilde{r}\tilde{\phi}}^{\tilde{t}}\right) = -\frac{1}{4}\left(2c_{\tilde{t}\tilde{\phi}}^{\;\;\;\tilde{r}}+2c_{\tilde{t}\tilde{r}}^{\;\;\;\tilde{\phi}}\right) = \frac{1}{2}c_{\tilde{r}\tilde{t}}^{\;\;\;\tilde{\phi}} = \frac{\gamma^{2}_{\hat{\phi}}}{2}\frac{\sqrt{g_{\phi\phi}}}{\sqrt{\omega^2g_{\phi\phi}-g_{tt}}\sqrt{g_{rr}}}\partial_{r}\Omega,
    \label{eq:shear}
\end{equation}

where $c_{\tilde{\mu}\tilde{\nu}}^{\;\;\;\tilde{\alpha}}$ are the commutation coefficients for the CRF. Finally, of particular interest is the $\tilde{\theta}$ component of the Riemann curvature tensor

\begin{equation}
\left. {R^{\,\tilde{\theta}}}_{\tilde{t}\tilde{\theta}\tilde{t}}\, \right|_{_{\theta=\pi/2}} =\frac{M}{r^{3}}\frac{r^{2} - 4aM^{3/2}r^{1/2} + 3M^{2}a^{2}}{r^{2} - 3Mr + 2aM^{3/2}r^{1/2}},
    \label{eq:riemanncurv}
\end{equation}

which gives a measurement of the relative acceleration in the $\tilde{\theta}$ direction of nearly equatorial geodesics. 

%%%%%%%%%%%%%%%%%%%%%%%%%%%%%%%%%%%%%%%%%%%%%%%%%%%%%%%%%%%%%%%%%%%%%%%%%%%%%%%%%%%%%%%%%%%%%%%%%%%%%%%%%%%%%%%%%% 
\section{Stress-Energy tensor}\label{app:stress-energy}
%%%%%%%%%%%%%%%%%%%%%%%%%%%%%%%%%%%%%%%%%%%%%%%%%%%%%%%%%%%%%%%%%%%%%%%%%%%%%%%%%%%%%%%%%%%%%%%%%%%%%%%%%%%%%%%%%%

Here we present some equations related to the stress-energy that we used in this paper. Eq.~(\ref{eq:stress-energy}) for a zero bulk viscosity fluid in components is

\begin{equation}
T^{\mu}_{\nu} = \Pi u^{\mu}u_{\nu} + P \delta^{\mu}_{\nu} - 2\eta\sigma^{\mu}_{\nu} + q^{\mu} u_{\nu} + q_{\nu} u^{\mu},
    \label{eq:s-e-comp}
\end{equation}

whose (vanishing) covariant derivative is

\begin{align}
\nabla_{\mu}T^{\mu}_{\nu} &= u^{\mu}u_{\nu}\partial_{\mu}\Pi + \Pi\boldsymbol{\theta}u_{\nu} + \Pi a_{\nu} + \partial_{\nu}P - 2\eta\nabla_{\mu}\sigma^{\mu}_{\nu} + q^{\mu}\nabla_{\mu}u_{\nu} + u_{\nu}\nabla_{\mu}q^{\mu} + q_{\nu}\boldsymbol{\theta} + u^{\mu}\nabla_{\mu}q_{\nu}   \nonumber \\
                          &= u^{\mu}\left[u_{\nu}\left(\partial_{\mu}\Pi-\frac{\Pi}{\rho}\partial_{\mu}\rho\right)-\frac{q_{\nu}}{\rho}\partial_{\mu}\rho\right] + \Pi a_{\nu} + \partial_{\nu}P - 2\eta\nabla_{\mu}\sigma^{\mu}_{\nu} + q^{\mu}\nabla_{\mu}u_{\nu} + u_{\nu}\nabla_{\mu}q^{\mu} + u^{\mu}\nabla_{\mu}q_{\nu},
    \label{eq:cov-s-e-comp}
\end{align}

where baryon conservation is used $\rho\boldsymbol{\theta} = -u^{\mu}\partial_{\mu}\rho$. To get an equation of motion for the
fluid, we project along the direction perpendicular to $u_{\nu}$

\begin{align}
&P^{\nu}_{\beta}\nabla_{\mu}T^{\mu}_{\nu}  = u^{\mu}\left[u_{\beta}\left(\partial_{\mu}\Pi-\frac{\Pi}{\rho}\partial_{\mu}\rho\right)-\frac{q_{\beta}}{\rho}\partial_{\mu}\rho\right] + \Pi a_{\beta} + \partial_{\beta}P - 2\eta\nabla_{\mu}\sigma^{\mu}_{\beta} + q^{\mu}\nabla_{\mu}u_{\beta} + u_{\beta}\nabla_{\mu}q^{\mu} \nonumber \\
                          & + u^{\mu}\nabla_{\mu}q_{\beta} - u^{\mu}u_{\beta}\left[\partial_{\mu}\Pi-\frac{\Pi}{\rho}\partial_{\mu}\rho\right] + u^{\nu}u_{\beta}\partial_{\nu}P - 2\eta u^{\nu}u_{\beta}\nabla_{\mu}\sigma^{\mu}_{\nu} - u_{\beta}\nabla_{\mu}q^{\mu} + u^{\nu}u_{\beta}u^{\mu}\nabla_{\mu}q_{\nu} \nonumber \\
                          &=-\frac{q_{\beta}}{\rho}u^{\mu}\partial_{\mu}\rho + \Pi a_{\beta} + \partial_{\beta}P - 2\eta \nabla_{\mu}\sigma^{\mu}_{\beta} + q^{\mu}\nabla_{\mu}u_{\beta}+ u^{\mu}\nabla_{\mu}q_{\beta} + u_{\beta}u^{\nu}\partial_{\nu}P - 2\eta u^{\nu}u_{\beta}\nabla_{\mu}\sigma^{\mu}_{\nu} + u^{\nu}u_{\beta}u^{\mu}\nabla_{\mu}q_{\nu} \nonumber \\
                          &\qquad\;\;=-\frac{q_{\beta}}{\rho}u^{\mu}\partial_{\mu}\rho + \Pi a_{\beta} + \partial_{\beta}P - 2\eta \nabla_{\mu}\sigma^{\mu}_{\beta} + q^{\mu}\nabla_{\mu}u_{\beta}+ u^{\mu}\nabla_{\mu}q_{\beta} + u_{\beta}\left(u^{\nu}\partial_{\nu}P  + 2\eta \sigma^{\mu\nu}\sigma_{\mu\nu} - q_{\nu}a^{\nu} \right),
    \label{eq:proj-cov-s-e-comp}
\end{align}

where the identities $q_{\mu}u^{\mu}=u^{\mu}a_{\mu}=\sigma^{\mu\nu}u_{\nu}=0$, $u_{\mu}u^{\nu}=-1$, $\sigma^{\mu\nu}\sigma_{\mu\nu} = \sigma^{\mu\nu}\nabla_{\mu}u_{\nu}$ are used. Combining the Eq.~(\ref{eq:cov-s-e-comp}) and Eq.~(\ref{eq:proj-cov-s-e-comp}) we get

\begin{equation}
u^{\mu}\left[\partial_{\mu}U-\frac{U + P}{\rho}\partial_{\mu}\rho\right] = 2\eta\sigma^{\mu\nu}\sigma_{\mu\nu} - q_{\mu}a^{\mu} - \nabla_{\mu}q^{\mu}.
    \label{eq:fineq}
\end{equation}

With Eq.~(\ref{eq:expansion}) we can obtain an equation for mass conservation

\begin{align}
0 =  &\nabla_{\mu} \left(\rho u^{\mu}\right) = u^{\mu}\partial_{\mu}\rho + \rho\boldsymbol{\theta} = u^{\mu}\partial_{\mu}\rho + \rho\left(\frac{2}{r}u^r+\partial_{r}u^r\right), \nonumber \\
&\Rightarrow \partial_{r}\left(r^2\rho u^{r}\right) + r^{2}u^{j}\partial_{j}\rho = 0,\;\; {\textrm for }\, j\in\left\{t,\theta,\phi\right\}.
    \label{eq:mass}
\end{align}

Finally, we reproduce the \emph{zero torque at the innermost stable circular orbit} condition that appears in \cite{1974ApJ...191..499P}. Using the killing vector fields $\boldsymbol{\partial}_{\phi}$, $\boldsymbol{\partial}_{t}$ and the approximation $\Pi\approx\rho$, we can calculate

\begin{align}
&0=\boldsymbol{\nabla}\cdot\left(\boldsymbol{T}\cdot\boldsymbol{\partial}_{\phi}\right) = \nabla_{\mu}T^{\mu}_{\phi} = \frac{1}{\sqrt{-g}}\partial_{\mu}\left(\sqrt{-g}T^{\mu}_{\phi}\right) \approx \frac{1}{r^2}\partial_{r}\left(\rho u^{r}u_{\phi}r^2  - 2\eta\sigma^{r}_{\phi}r^2 \right) + u_{\phi}\partial_{\theta}q^{\theta}, \nonumber \\
&\Rightarrow \partial_{r}\left(\rho u^{r}u_{\phi}r^2  - 2\eta\sigma^{r}_{\phi}r^2 \right) = -r^{2} u_{\phi}\partial_{\theta}q^{\theta},  \nonumber \\
&\Rightarrow \partial_{r}\left(\frac{\dot{M}}{2\pi}u_{\phi} + 4rH\eta\sigma^{r}_{\phi} \right) = 2Hu_{\phi}\epsilon\;\, {\textrm{ after integrating vertically and using Eq.~(\ref{eq:masscon})}}. \nonumber \\
&{\textrm{Analogously for $\boldsymbol{\partial}_{t}$},}\;\;\partial_{r}\left(\frac{\dot{M}}{2\pi}u_{t} - 4rH\Omega\eta\sigma^{r}_{\phi} \right) = 2Hu_{t}\epsilon\;\, {\textrm{ using Eq.~(\ref{eq:shear2})}}.
\label{eq:angmom}
\end{align}

The vertical integration of the divergence of the heat flux is as follows: Since, on average, $\boldsymbol{q} = q^{\theta}\boldsymbol{\partial}_{\theta}$, we have $\nabla_{\mu}q^{\mu} = \partial_{\theta}q^{\theta}$ and by Eq.~(\ref{eq:fulltransform}),  $q^{\theta}=rq^{\hat{\hat{\theta}}}$. Vertically integrating yields

\begin{equation}
\int_{\theta_{\textrm{min}}}^{\theta_{\textrm{max}}}\!\partial_{\theta}q^{\theta}rd\theta = r\! \left .q^{\theta}\right|_{_{\theta_{\textrm{min}}}}^{ ^{\theta_{\textrm{max}}}} = 2q^{\tilde{\tilde{\theta}}} = 2H\epsilon,
    \label{eq:vertflux}
\end{equation}

where $q^{\tilde{\tilde{\theta}}}$ is the averaged energy flux radiating out of a face of the disk, as measured by an observer on the LRF, which we approximate as the half-thickness of the disk $H$ times the average energy density per unit proper time $\epsilon$ lost by the disk. With the variable change $z=8\pi rH\eta\sigma^{r}_{\phi}/\dot{M}$ and $y=4\pi H\epsilon/\dot{M}$ the equations reduce to

\begin{subequations}
\begin{gather}
\partial_{r}\left(u_{\phi} + z \right) = yu_{\phi},\\
\partial_{r}\left(u_{t} - \Omega z \right) = yu_{t}.
\end{gather}\label{eq:enmom2}\end{subequations}

Using the relation $\partial_{r}u_{t}=-\Omega \partial_{r}u_{\phi}$ \cite[see Eq.~(10.7.29) in][]{1971reas.book.....Z} and $\partial_{r}\left(u_{t}+\Omega u_{\phi}\right)=u_{\phi}\partial_{r}\Omega$ we can combine the previous equations to obtain

\begin{subequations}
\begin{gather}
z = -\frac{y\left(u_{t}+\Omega u_{\phi}\right)}{\partial_{r}\Omega},\\
\partial_{r}\left(AB^2\right) = B\partial_{r}u_{\phi},
\end{gather}\label{eq:diffeq}\end{subequations}

with $A=y/\partial_{r}\Omega$ and $B=u_{t}+\Omega u_{\phi}$. To integrate these equations we use the zero torque condition $z(r=r^{*})=0$ where $r^{*}$ is the radius of the innermost stable circular orbit, which gives the relation 

\begin{equation}
y=\frac{\partial_{r}\Omega}{\left(u_{t}+\Omega u_{\phi}\right)^2}\int^{r}_{r^{*}}\!\left(u_{t}+\Omega u_{\phi}\right)\partial_{r}u_{\phi}dr=\frac{\partial_{r}\Omega}{\left(u_{t}+\Omega u_{\phi}\right)^2}\left(\left. u_{t}u_{\phi}\right|_{r^{*}}^{r}-2\!\int^{r}_{r^{*}}\!u_{\phi}\partial_{r}u_{t}dr\right),
\label{eq:soldiff}
\end{equation}

or, equivalently,

\begin{equation}
8\pi H r\rho\nu_{\textrm{turb}}\sigma^{r}_{\phi}\approx 8\pi H r\Pi\nu_{\textrm{turb}}\sigma^{r}_{\phi} = -\frac{\dot{M}}{\left(u_{t}+\Omega u_{\phi}\right)}\left(\left. u_{t}u_{\phi}\right|_{r^{*}}^{r}-2\!\int^{r}_{r^{*}}\!u_{\phi}\partial_{r}u_{t}dr\right).
\label{eq:reldiff}
\end{equation}

Using Eq.~(\ref{eq:fourvel}), the approximation $\gamma_{\tilde{r}} \approx 1$ and the variable change $r=x M^2$ the integral can be easily evaluated by partial fractions

\begin{subequations}
\begin{equation}
8\pi H r\rho\nu_{\textrm{turb}}\sigma^{r}_{\phi} =\dot{M}M f\left(x,x^{*}\right),
\end{equation}
\begin{equation}
f\left(x,x^{*}\right) = \frac{x^3+a}{x^{3/2}\sqrt{x^3-3x+2a}}\left[x-x^{*}-\frac{3}{2}a\ln\left(\frac{x}{x^{*}}\right)+\frac{1}{2}\sum_{i=1}^{3}\frac{ax_{i}^2-2x_{i}^{\phantom{2}}\!+a}{x_{i}^2-1}\ln\left(\frac{x-x_{i}}{x^{*}-x_{i}}\right)\right],
\end{equation}
\label{eq:pol}\end{subequations}

where $x_{1},x_{2},x_{3}$ are the roots of the polynomial $x^3-3x+2a$.

%%%%%%%%%%%%%%%%%%%%%%%%%%%%%%%%%%%%%%%%%%%%%%%%%%%%%%%%%%%%%%%%%%%%%%%%%%%%%%%%%%%%%%%%%%%%%%%%%%%%%%%%%%%%%%%%%% 
\section{Nuclear Statistical Equilibrium}\label{app:Coulomb}
%%%%%%%%%%%%%%%%%%%%%%%%%%%%%%%%%%%%%%%%%%%%%%%%%%%%%%%%%%%%%%%%%%%%%%%%%%%%%%%%%%%%%%%%%%%%%%%%%%%%%%%%%%%%%%%%%%

The results in this section appear in \cite{PhysRevE.62.8554}. We include them here since they are necessary to solve Eq.~(\ref{eq:saha}). Neutrino dominated accretion disks reach densities above $\sim 10^7$~g~cm$^{-3}$ and temperatures above $\sim 5\times 10^9$~K. For these temperatures, forward and reverse nuclear reactions are balanced and the abundances in the plasma are determined by the condition $\mu_{i} = Z_{i}\mu_{p} + N_{i}\mu_{n}$, that is, the Nuclear Statistical Equilibrium. However, for densities above $10^6$~g~cm$^{-3}$, the electron screening of charged particle reactions can affect the nuclear reaction rates. For this reason, to obtain an accurate NSE state it is necessary to include Coulomb corrections to the ion chemical potential. The Coulomb correction to the $i$-th chemical potential is given by

\begin{align}
\frac{\mu_{i}^{C}}{T} = K_{1}&\left[\Gamma_{i}\sqrt{\Gamma_{i}+K_{2}} - K_{2}\ln\left(\sqrt{\frac{\Gamma_{i}}{K_{2}}}+\sqrt{1 + \frac{\Gamma_{i}}{K_{2}}}\right)\right] \nonumber\\
&+ 2K_{3}\left[\sqrt{\Gamma_{i}} - \arctan\!\sqrt{\Gamma_{i}} \right] + Z_{1}\left[ \Gamma_{i} - Z_{2}\ln \left( 1 + \frac{\Gamma_{i}}{Z_{1}} \right) \right] + \frac{Z_{3}}{2}\ln\left( 1 + \frac{\Gamma_{i}^2}{Z_{4}} \right),
    \label{eq:coulombcorrection}
\end{align}

and the ion coupling parameter in terms of the electron coupling parameter is $\Gamma_{i}=\Gamma_{e}Z_{i}^{5/3}$ with

\begin{equation}
\Gamma_{e} = \frac{e^2}{T}\left(\frac{4\pi Y_{e}n_{B}}{3}\right)^{1/3}.
    \label{eq:eleccoup}
\end{equation}

where $e$ is the electron charge. The parameters $K_{i},C_{i}$ are given in table (\ref{tab:constcoulomb}).

\begin{table}
\caption{Constants appearing in Eq.~(\ref{eq:coulombcorrection}). See~\cite{PhysRevE.62.8554}.}
\centering
\begin{tabular}{c c c c c c c}
\toprule
  $K_{1}$ & $K_{2}$ & $K_{3}$ & $Z_{1}$ & $Z_{2}$ & $Z_{3}$ & $Z_{4}$ \\ 
 \midrule
    $-0.907347$ & $0.62849$ & $0.278497$ & $4.50\times 10^{-3}$ & $170.0$ & $-8.4\times 10^{-5}$ & $3.70\times 10^{-3}$ \\
 \bottomrule
\end{tabular}
\label{tab:constcoulomb}
\end{table}
% 

%%%%%%%%%%%%%%%%%%%%%%%%%%%%%%%%%%%%%%%%%%%%%%%%%%%%%%%%%%%%%%%%%%%%%%%%%%%%%%%%%%%%%%%%%%%%%%%%%%%%%%%%%%%%%%%%%%%%%%%%%%%%%%%%%%%%%%%%%
\section{Neutrino Interactions and cross-sections}\label{app:emisscross}
%%%%%%%%%%%%%%%%%%%%%%%%%%%%%%%%%%%%%%%%%%%%%%%%%%%%%%%%%%%%%%%%%%%%%%%%%%%%%%%%%%%%%%%%%%%%%%%%%%%%%%%%%%%%%%%%%%%%%%%%%%%%%%%%%%%%%%%%%

In this appendix we include the neutrino emission rates and neutrino cross-sections used in the accretion disk model. These expressions have been covered in \cite{Dicus:1972yr,1975ApJ...201..467T,1985ApJS...58..771B,1996A&A...311..532R,2001PhR...354....1Y,Burrows2004,BURROWS2006356}. We also include the expression energy emission rate for $\nu\bar{\nu}$ annihilation into electron-positron pairs. Whenever possible we write the rates in terms of generalized Fermi functions since some numerical calculations were done following \cite{1998ApJS..117..627A}. Before proceeding we list some useful expressions and constants in Planck units that will be used. The numerical values can be found in \cite{PhysRevD.98.030001}.

\begin{table}[H]
\caption{Constants used through this appendix to calculate emissivities and cross-sections. All quantities are reported in Planck units.}
\centering
\begin{tabular}{lll}
\toprule
\textbf{Symbol}	& \textbf{Value}	& \textbf{Name}\\
\midrule
$M_{w}$		& $6.584\times10^{-18}$			& W boson mass\\
$g_{w}$		& 0.653			& Weak coupling constant\\
$g_{a}$		& 1.26			& Axial-vector coupling constant\\
$\alpha^{*}$		& $\frac{1}{137}$			& Fine structure constant\\
$\sin^2 \theta_{\textrm W}$		& 0.231			& Weinberg angle\\
$\cos^{2}\theta_{c}$		& 0.947			& Cabibbo angle\\
$G_{F}$		& $1.738\times10^{33}$			& Fermi coupling constant\\
$C_{v,e}$		& $2\sin^2\theta_{\textrm W} + 1/2$			& Weak interaction vector constant for $\nu_{e}$\\
$C_{a,e}$ & $1/2$ & Weak interaction axial-vector constant for $\nu_{e}$ \\
$C_{v,e}$ & $C_{v,e} - 1$ & Weak interaction vector constant for $\nu_{x}$ \\
$C_{a,e}$ & $C_{a,e} - 1$ & Weak interaction axial-vector constant for $\nu_{x}$ \\
$\sigma_{0}$ & $6.546\times10^{21}$ & Weak interaction cross-section \\
\bottomrule
\end{tabular}
\end{table}

%%%%%%%%%%%%%%%%%%%%%%%%%%%%%%%%%%%%%%%%%%%%%%%%%%%%%%%%%%%%%%%%%%%%%%%%%%%%%%%%%%%%%%%%%%%%%%%%%%%%%%%%%%%%%%%%%%%%%%%%%%%%%%%%%%%%%%%%%
\subsection{Neutrino Emissivities}\label{app:subemissivity}
%%%%%%%%%%%%%%%%%%%%%%%%%%%%%%%%%%%%%%%%%%%%%%%%%%%%%%%%%%%%%%%%%%%%%%%%%%%%%%%%%%%%%%%%%%%%%%%%%%%%%%%%%%%%%%%%%%%%%%%%%%%%%%%%%%%%%%%%%

% 
\begin{itemize}
  \item Pair annihilation: $e^{-}\!+e^{+}\! \to \nu + \bar{\nu}$
\end{itemize}

This process generates neutrinos of all flavours but around 70$\%$ are electron neutrinos \cite{2018ApJ...852..120B}. This is due to the fact that the only charged leptons in the accretion systems we study are electrons and positrons, so creation of electron neutrinos occurs via either charged or neutral electroweak currents while creation of non-electronic neutrinos can only occur through neutral currents. Using the electron or positron four-momentum $p=(E,\boldsymbol{p})$, the Dicus cross-section for a particular flavour $i$ is \cite{Dicus:1972yr}

\begin{align}
\sigma_{D,i} &= \frac{G_{F}^{2}}{12\pi E_{e^{-}}E_{e^{+}}}\left[C_{+,i}\left(m_{e}^{4} + 3m_{e}^{2}p_{e^{-}}\! \cdot\! p_{e^{+}} + 2\left(p_{e^{-}}\! \cdot\! p_{e^{+}}\right)^{2} \right) + 3C_{-,i}\left(m_{e}^{4} + m_{e}^{2}p_{e^{-}}\! \cdot\! p_{e^{+}}\right) \right].
    \label{eq:dicus}
\end{align}

The factors $C_{\pm,i}$, are written in terms of the weak interaction vector and axial-vector constants: $C_{\pm,i}=C^{2}_{v,i} \pm C^{2}_{a,i}$ \cite{PhysRevD.98.030001}. Representing the Fermi-Dirac distribution for electrons (positrons) as $f_{e^{-}}(f_{e^{+}})$ with $\eta_{e^{\mp}}$ the electron (positron) degeneracy parameter including its rest mass. The number and energy emission rates can be calculated by replacing $\Lambda = 2$ and $\Lambda=E_{e^{-}}+E_{e^{+}}$ in the integral \cite{2001PhR...354....1Y}:

\begin{equation}
\frac{4}{\left(2\pi\right)^{6}}\int\! \Lambda \sigma_{D} f_{e^{-}}f_{e^{+}} d^{3}\!\boldsymbol{p}_{e^{-}} d^{3}\!\boldsymbol{p}_{e^{+}},
    \label{eq:dicus2}
\end{equation}

giving the expressions

\begin{subequations}
\begin{align}
    R_{\nu_{i}+\bar{\nu}_{i}} &= \frac{G_{F}^{2}m_{e}^{8}}{18\pi}\left[ C_{+,i}\left(8\mathbb{U}_{1}\mathbb{V}_{1} + 5\mathbb{U}_{-1}\mathbb{V}_{-1} + 9\mathbb{U}_{0}\mathbb{V}_{0} - 2\mathbb{U}_{-1}\mathbb{V}_{1} - 2\mathbb{U}_{1}\mathbb{V}_{-1}\right)\right. \nonumber\\
    &\,\,\, \left. + 9C_{-,i}\left(\mathbb{U}_{-1}\mathbb{V}_{-1} + \mathbb{U}_{0}\mathbb{V}_{0}\right) \right],  \\
    Q_{\nu_{i}+\bar{\nu}_{i}} &= \frac{G_{F}^{2}m_{e}^{9}}{36\pi}\left[ C_{+,i}\left( 8\left(\mathbb{U}_{2}\mathbb{V}_{1}+\mathbb{U}_{1}\mathbb{V}_{2}\right) + 7\left(\mathbb{U}_{1}\mathbb{V}_{0}+\mathbb{U}_{0}\mathbb{V}_{1}\right) + 5\left(\mathbb{U}_{-1}\mathbb{V}_{0}+\mathbb{U}_{0}\mathbb{V}_{-1}\right)\right.\right. \nonumber \\
    &\,\,\, \left.\left. - 2\left(\mathbb{U}_{2}\mathbb{V}_{-1}+\mathbb{U}_{-1}\mathbb{V}_{2}\right)\right) + 9C_{-,i}\left( \mathbb{U}_{0}\left(\mathbb{V}_{1}+\mathbb{V}_{-1}\right) + \mathbb{V}_{0}\left(\mathbb{U}_{1}+\mathbb{U}_{-1}\right)\right) \right].
\end{align}\label{eq:electronemissivity}\end{subequations}

The functions $\mathbb{U},\mathbb{V}$ can be written in terms of generalized Fermi functions

\begin{subequations}
\begin{align}
\mathbb{U}_{j} &= \sqrt{2}\xi^{3/2}\sum_{k=0}^{j+1}\binom{j+1}{k}\xi^{k}\mathcal{F}_{k+1/2,0}\left(\xi,\eta_{e^{-}}\right), \\
\mathbb{V}_{j} &= \sqrt{2}\xi^{3/2}\sum_{k=0}^{j+1}\binom{j+1}{k}\xi^{k}\mathcal{F}_{k+1/2,0}\left(\xi,\eta_{e^{+}}\right).
\end{align}
\label{eq:uvfunctions}\end{subequations}

It is often useful to define the functions

\begin{equation}
\varepsilon^{m}_{i} = \frac{2G^{2}_{F}\left(m_{e}\right)^{4}}{3\left(2\pi\right)^{7} } \int\!\! f_{e^{-}}f_{e^{+}}\left( E^{m}_{e^{-}} + E^{m}_{e^{+}} \right)\sigma_{D,i}\,d^{3}\mathbf{p}_{e^{-}}d^{3}\mathbf{p}_{e^{+}}.
\label{eq:emissionratenu}
\end{equation}

For $m=0$ and $m=1$ Eq.~(\ref{eq:emissionratenu}) gives the neutrino and antineutrino number emissivity (neutrino production rate), and the neutrino and antineutrino energy emissivity (energy per unit volume per unit time) for a certain flavour $i$, respectively (that is, Eq.~(\ref{eq:electronemissivity})). Hence, not only we are able to calculate the total number and energy emissivity, but we can also calculate the neutrino or antineutrino energy moments with

\begin{equation}
\langle E^{m}_{\nu_{i}\left(\bar{\nu}_{i}\right)} \rangle = \frac{\varepsilon^{m}_{i}}{\varepsilon^{0}_{i}},\,\, {\rm for }\,\, m\geq 1.
\label{eq:neutrinomoments}
\end{equation}
\begin{itemize}
  \item Electron capture and positron capture: $p + e^{-}\!\to n + \nu_{e}$, $n + e^{+}\!\to p + \bar{\nu}_{e}$ and $A + e^{-}\!\to A' + \nu_{e}$
\end{itemize}

Due to lepton number conservation this process generated only electron (anti)neutrinos. The number and energy emission rates for electron and positron capture by nucleons are

\begin{subequations}
\begin{align}
    R_{\nu_{e}} &= \frac{m_{e}^{5}G_{F}^{2}\cos^{2}\theta_{c}}{\sqrt{2}\pi^{3}}\left(1+3g_{A}^{2}\right)\Delta_{np}\xi^{3/2}\left[ \xi^{3}\mathcal{F}_{7/2,\chi}\left(\xi,\eta_{e^{-}}\right) \right. \nonumber \\
    &  \left. + \left(3-2\mathbb{Q}\right)\xi^{2}\mathcal{F}_{5/2,\chi}\left(\xi,\eta_{e^{-}}\right) + \left(1-\mathbb{Q}\right)\left(3-\mathbb{Q}\right)\xi\mathcal{F}_{3/2,\chi}\left(\xi,\eta_{e^{-}}\right) + \left(1-\mathbb{Q}\right)^{2}\mathcal{F}_{1/2,\chi}\left(\xi,\eta_{e^{-}}\right) \right], \label{eq:captN} \\
    Q_{\nu_{e}} &= \frac{m_{e}^{6}G_{F}^{2}\cos^{2}\theta_{c}}{\sqrt{2}\pi^{3}}\left(1+3g_{A}^{2}\right)\Delta_{np}\xi^{3/2}\left[ \xi^{4}\mathcal{F}_{9/2,\chi}\left(\xi,\eta_{e^{-}}\right) \right. \nonumber \\
    &\left. + \xi^{3}\left(4-3\mathbb{Q}\right)\mathcal{F}_{7/2,\chi}\left(\xi,\eta_{e^{-}}\right) + 3\left(\mathbb{Q}-1\right)\left(\mathbb{Q}-2\right)\xi^{2}\mathcal{F}_{5/2,\chi}\left(\xi,\eta_{e^{-}}\right) \right. \nonumber \\
    &\left. + \left(1-\mathbb{Q}\right)^{2}\left(4-\mathbb{Q}\right)\xi\mathcal{F}_{3/2,\chi}\left(\xi,\eta_{e^{-}}\right) + \left(1-\mathbb{Q}\right)^{3}\mathcal{F}_{1/2,\chi}\left(\xi,\eta_{e^{-}}\right) \right], \label{eq:captQ} \\
    R_{\bar{\nu}_{e}} &= \frac{m_{e}^{5}G_{F}^{2}\cos^{2}\theta_{c}}{\sqrt{2}\pi^{3}}\left(1+3g_{A}^{2}\right)\Delta_{pn}\xi^{3/2}\left[ \xi^{3}\mathcal{F}_{7/2,0}\left(\xi,\eta_{e^{+}}\right)  \right. \nonumber \\
    &\left. + \left(3+2\mathbb{Q}\right)\xi^{2}\mathcal{F}_{5/2,0}\left(\xi,\eta_{e^{+}}\right) + \left(1+\mathbb{Q}\right)\left(3+\mathbb{Q}\right)\xi\mathcal{F}_{3/2,0}\left(\xi,\eta_{e^{+}}\right)  + \left(1+\mathbb{Q}\right)^{2}\mathcal{F}_{1/2,0}\left(\xi,\eta_{e^{+}}\right) \right], \\
    Q_{\bar{\nu}_{e}} &= \frac{m_{e}^{6}G_{F}^{2}\cos^{2}\theta_{c}}{\sqrt{2}\pi^{3}}\left(1+3g_{A}^{2}\right)\Delta_{np}\xi^{3/2}\left[ \xi^{4}\mathcal{F}_{9/2,0}\left(\xi,\eta_{e^{+}}\right) \right. \nonumber \\
    &\left. + \xi^{3}\left(4+3\mathbb{Q}\right)\mathcal{F}_{7/2,0}\left(\xi,\eta_{e^{+}}\right) + 3\left(\mathbb{Q}+1\right)\left(\mathbb{Q}+2\right)\xi^{2}\mathcal{F}_{5/2,0}\left(\xi,\eta_{e^{+}}\right)  \right. \nonumber \\
    &\left. + \left(1+\mathbb{Q}\right)^{2}\left(4+\mathbb{Q}\right)\xi\mathcal{F}_{3/2,0}\left(\xi,\eta_{e^{+}}\right) + \left(1+\mathbb{Q}\right)^{3}\mathcal{F}_{1/2,0}\left(\xi,\eta_{e^{+}}\right) \right],
\end{align}\label{eq:captureelectronemissivity}\end{subequations}

where $\Delta_{ij} = \left(n_{i}-n_{j}\right)/\left(\exp\left(\eta_{i}-\eta_{j}\right)-1\right), \,\,i,j\in\left\{p,n\right\}$ are the Fermi blocking factors in the nucleon phase spaces and $\mathbb{Q} = (m_{n}-m_{p})m_{e} \approx 2.531$ is the nucleon mass difference. The number and energy emission rates for electron capture by an ion $i$ are

\begin{subequations}
\begin{align}
    &R_{\nu_{e},i} = \frac{\sqrt{2}m_{e}^{5}G_{F}^{2}\cos^{2}\theta_{c}}{7\pi^{3}}g_{A}^{2}n_{i}\kappa_{Z_{i}}\kappa_{N_{i}}\xi^{3/2}\left[ \xi^{3}\mathcal{F}_{7/2,\bar{\chi}}\left(\xi,\eta_{e^{-}}\right) \right. \nonumber \\
    &\left. + \left(3-2\mathbb{Q}\right)\xi^{2}\mathcal{F}_{5/2,\bar{\chi}}\left(\xi,\eta_{e^{-}}\right) + \left(1-\mathbb{Q}\right)\left(3-\mathbb{Q}\right)\xi\mathcal{F}_{3/2,\bar{\chi}}\left(\xi,\eta_{e^{-}}\right) \left(1-\mathbb{Q}\right)^{2}\mathcal{F}_{1/2,\bar{\chi}}\left(\xi,\eta_{e^{-}}\right) \right], \label{eq:captN2} \\
    &Q_{\nu_{e},i} = \frac{\sqrt{2}m_{e}^{6}G_{F}^{2}\cos^{2}\theta_{c}}{7\pi^{3}}g_{A}^{2}n_{i}\kappa_{Z_{i}}\kappa_{N_{i}}\xi^{3/2}\left[ \xi^{4}\mathcal{F}_{9/2,\bar{\chi}}\left(\xi,\eta_{e^{-}}\right) + \xi^{3}\left(4-3\mathbb{Q}\right)\mathcal{F}_{7/2,\bar{\chi}}\left(\xi,\eta_{e^{-}}\right) \right. \nonumber \\
    & \left.  + 3\left(\mathbb{Q}-1\right)\left(\mathbb{Q}-2\right)\xi^{2}\mathcal{F}_{5/2,\bar{\chi}}\left(\xi,\eta_{e^{-}}\right)  + \left(1-\mathbb{Q}\right)^{2}\left(4-\mathbb{Q}\right)\xi\mathcal{F}_{3/2,\bar{\chi}}\left(\xi,\eta_{e^{-}}\right) + \left(1-\mathbb{Q}\right)^{3}\mathcal{F}_{1/2,\bar{\chi}}\left(\xi,\eta_{e^{-}}\right) \right]. \label{eq:captQ2}
\end{align}\label{eq:ioncapt}\end{subequations}

The lower integration limits in these expressions are given by $\chi = (\mathbb{Q}-1)/\xi$ and $\bar{\chi} = (\mu_{n} - \mu_{p} + \Delta)/T - 1/\xi$ where $\Delta \approx 2.457\times10^{-22}$ is the energy of the neutron 1$f_{5/2}$ state above the ground state. The functions $\kappa_{Z_{i}},\kappa_{N_{i}}$ are

\begin{equation}
\kappa_{Z_{i}} = \begin{cases} 
      0 & {\textrm{if}}\;\, Z_{i} \leq 20. \\
      Z_{i}-20 & {\textrm{if}}\;\, 20 < Z_{i} \leq 28. \\
      8 & {\textrm{if}}\;\, Z_{i} > 28.
   \end{cases}, \quad
\kappa_{N_{i}} = \begin{cases} 
      6 & {\textrm{if}}\;\, N_{i} \leq 34. \\
      40 - N_{i} & {\textrm{if}}\;\, 34 < N_{i} \leq 40. \\
      0 & {\textrm{if}}\;\, N_{i} > 40.
   \end{cases}
\label{eq:ionemissfun}
\end{equation}
\begin{itemize}
  \item Plasmon decay: $\tilde{\gamma}\to \nu + \bar{\nu}$.
\end{itemize}
\begin{subequations}
\begin{align}
    R_{\nu_{e}+\bar{\nu}_{e}} &= \frac{C_{v,e}\sigma_{0}T^{8}}{96\pi^3 m_{e}^{2}\alpha^{*}}\tilde{\gamma}^{6}\left(\tilde{\gamma}+1\right)\exp\left(-\tilde{\gamma}\right), \\
    Q_{\nu_{e}+\bar{\nu}_{e}} &= \frac{C_{v,e}\sigma_{0}T^{9}}{192\pi^3 m_{e}^{2}\alpha^{*}}\tilde{\gamma}^{6}\left(\tilde{\gamma}^{2}+2\tilde{\gamma} +2\right)\exp\left(-\tilde{\gamma}\right), \\
    R_{\nu_{x}+\bar{\nu}_{x}} &= \frac{C_{v,x}\sigma_{0}T^{8}}{48\pi^3 m_{e}^{2}\alpha^{*}}\tilde{\gamma}^{6}\left(\tilde{\gamma}+1\right)\exp\left(-\tilde{\gamma}\right), \\
    Q_{\nu_{x}+\bar{\nu}_{x}} &= \frac{C_{v,x}\sigma_{0}T^{9}}{96\pi^3 m_{e}^{2}\alpha^{*}}\tilde{\gamma}^{6}\left(\tilde{\gamma}^{2}+2\tilde{\gamma} +2\right)\exp\left(-\tilde{\gamma}\right),
\end{align}\label{eq:plasmonemissivity}\end{subequations}

where $\tilde{\gamma} = \tilde{\gamma}_{0}\sqrt{\left(\pi^2+3\left(\eta_{e^{-}}+1/\xi\right)^{2}\right)/3}$ and $\tilde{\gamma}_{0} = 2\sqrt{\frac{\alpha^{*}}{3\pi}} \approx 5.565\times10^{-2}$.

\begin{itemize}
  \item Nucleon-nucleon bremsstrahlung $n_{1} + n_{2}\to n_{3} + n_{4} + \nu + \bar{\nu}$.
\end{itemize}

The nucleon-nucleon bremsstrahlung produces the same amount of neutrinos of all three flavours. The number and energy emission rates can be approximated by (see, e.g.,~\cite{BURROWS2006356})

\begin{subequations}
\begin{align}
    R_{\nu_{i}+\bar{\nu}_{i}} &= \left(2.59\times10^{13}\right)\left( X_{p}^{2} + X_{n}^{2} + \frac{28}{3}X_{p}X_{n} \right)n_{B}^{2} \xi^{9/2}, \\
    Q_{\nu_{i}+\bar{\nu}_{i}} &= \left(4.71\times10^{-9}\right)\left( X_{p}^{2} + X_{n}^{2} + \frac{28}{3}X_{p}X_{n} \right)n_{B}^{2} \xi^{10/2}.
\end{align}\label{eq:bremss}\end{subequations}
% 

%%%%%%%%%%%%%%%%%%%%%%%%%%%%%%%%%%%%%%%%%%%%%%%%%%%%%%%%%%%%%%%%%%%%%%%%%%%%%%%%%%%%%%%%%%%%%%%%%%%%%%%%%%%%%%%%%%%%%%%%%%%%%%%%
\subsection{Cross-Sections}\label{app:subemisscross1}
%%%%%%%%%%%%%%%%%%%%%%%%%%%%%%%%%%%%%%%%%%%%%%%%%%%%%%%%%%%%%%%%%%%%%%%%%%%%%%%%%%%%%%%%%%%%%%%%%%%%%%%%%%%%%%%%%%%%%%%%%%%%%%%%

We consider four interactions to describe the (anti)neutrino total cross-section.

\begin{itemize}
  \item Neutrino annihilation: $(\nu + \bar{\nu} \to e^{-}\!+e^{+})$.
\end{itemize}
\begin{subequations}
\begin{align}
\sigma_{\nu_{e}\bar{\nu}_{e}} &= \frac{4}{3}K_{\nu_{e}\bar{\nu}_{e}}\sigma_{0}\frac{\langle E_{\nu_{e}} \rangle\langle E_{\bar{\nu}_{e}} \rangle}{m_{e}^{2}} \;\; {\textrm{with}}\;\; K_{\nu_{e}\bar{\nu}_{e}} = \frac{1+4\sin^{2}\theta_{\textrm W}+8\sin^{4}\theta_{\textrm W}}{12}, \\
\sigma_{\nu_{x}\bar{\nu}_{x}} &= \frac{4}{3}K_{\nu_{x}\bar{\nu}_{x}}\sigma_{0}\frac{\langle E_{\nu_{x}} \rangle\langle E_{\bar{\nu}_{x}} \rangle}{m_{e}^{2}} \;\; {\textrm{with}}\;\; K_{\nu_{x}\bar{\nu}_{x}} = \frac{1-4\sin^{2}\theta_{\textrm W}+8\sin^{4}\theta_{\textrm W}}{12},
\label{eq:neutrinoannihilation}
\end{align}\end{subequations}

\begin{itemize}
  \item Electron (anti)neutrino absorption by nucleons: $(\nu_{e} + n \to e^{-} + p$ and $\bar{\nu}_{e} + p \to e^{+} + n)$.   
\end{itemize}
\begin{subequations}
\begin{align}
\sigma_{\nu_{e}n} &= \sigma_{0}\left(\frac{1 + 3 g_{a}^{2}}{4}\right)\left(\frac{\langle E_{\nu_{e}} \rangle}{m_{e}} + \mathbb{Q}\right)^{2}\sqrt{1-\frac{1}{\left(\frac{\langle E_{\nu_{e}} \rangle}{m_{e}} + \mathbb{Q}\right)^{2}}}, \\
\sigma_{\bar{\nu}_{e}p} &= 3.83\times 10^{22}\left(\frac{\wp\langle E_{\bar{\nu}_{e}} \rangle}{m_{e}} - \mathbb{Q}\right)^{2}\sqrt{1-\frac{1}{\left(\frac{\wp\langle E_{\bar{\nu}_{e}} \rangle}{m_{e}} - \mathbb{Q}\right)^{2}}}\left(\frac{\wp\langle E_{\bar{\nu}_{e}} \rangle}{m_{e}}\right)^{g(E_{\bar{\nu}_e})},\\
g(E_{\bar{\nu}_e}) &={-0.07056+0.02018\ln\left(\frac{\wp\langle E_{\bar{\nu}_{e}} \rangle}{m_{e}}\right)-0.001953\ln^{3}\left(\frac{\wp\langle E_{\bar{\nu}_{e}} \rangle}{m_{e}}\right)}.\label{eq:absorptioncon}
\end{align}\label{eq:neutrinoabsorption}\end{subequations}

where $\wp = 0.511$.

\begin{itemize}

  \item (anti)neutrino scattering by baryons: $(\nu + A_{i} \to \nu + A_{i}$ and $\bar{\nu} + A_{i} \to \bar{\nu} + A_{i})$.   
\end{itemize}
\begin{subequations}
\begin{align}
\sigma_{p} &= \frac{\sigma_{0}\langle E \rangle^{2}}{4m_{e}^{2}}\left(4\sin^{4}\theta_{W} - 2\sin^{2}\theta_{W} + \frac{1+3g^{2}_{a}}{4} \right), \\
\sigma_{n} &= \frac{\sigma_{0}\langle E \rangle^{2}}{4m_{e}^{2}}\frac{1+3g^{2}_{a}}{4}, \\
\sigma_{A_{i}} &= \frac{\sigma_{0}A_{i}^{2}\langle E \rangle^{2}}{16m_{e}^{2}}\left[\left( 4\sin^{2}\theta_{W} - 1 \right)\frac{Z_{i}}{A_{i}} + 1 - \frac{Z_{i}}{A_{i}} \right].
\label{eq:scatteringbaryon}
\end{align}\end{subequations}
\begin{itemize}
  \item (anti)neutrino scattering by electrons or positrons: $(\nu + e^{\pm} \to \nu + e^{\pm}$ and $\bar{\nu} + e^{\pm} \to \bar{\nu} + e^{\pm})$.   
\end{itemize}
\begin{align}
\sigma_{e} &= \frac{3}{8}\sigma_{0} \xi \frac{\langle E \rangle}{m_{e}} \left(1+\frac{\eta_{e} + 1/\xi}{4}\right)\left[\left(C_{v,i} + n_\ell C_{a,i}\right)^{2} + \frac{1}{3}\left(C_{v,i} - n_\ell C_{a,i}\right)^{2}\right].
\label{eq:celectronscattering}
\end{align}

Here, $n_\ell$ is the (anti)neutrino lepton number (that is, $1$ for neutrinos and $-1$ for antineutrinos, depending on the cross-section to be calculated), and, in the last four expressions, $\langle E \rangle$ is replaced by the average (anti)neutrino energy of the corresponding flavour. With these expressions, the total opacity for neutrinos or antineutrinos is

\begin{equation}
\kappa_{\nu_{i}\left(\bar{\nu}_{i}\right)} = \frac{\sum_{i} \sigma_{i} n_{i}}{\rho},
    \label{eq:totopcty}
\end{equation}

where $n_{i}$ is the number density of the target particle associated with the process corresponding to the cross-section $\sigma_{i}$. The (anti)neutrino optical depth appearing in Eq.~(\ref{eq:neutrinofluxpost}) can then be approximated as

\begin{equation}
\tau_{\nu_{i}\left(\bar{\nu}_{i}\right)} = \int\!\! \kappa_{\nu_{i}\left(\bar{\nu}_{i}\right)} \rho d\theta \approx \kappa_{\nu_{i}\left(\bar{\nu}_{i}\right)} \rho H.
    \label{eq:totoptdepth}
\end{equation}
%

%%%%%%%%%%%%%%%%%%%%%%%%%%%%%%%%%%%%%%%%%%%%%%%%%%%%%%%%%%%%%%%%%%%%%%%%%%%%%%%%%%%%%%%%%%%%%%%%%%%%%%%%%%%%%%%%%%%%%%%%%%%%%%%%
\subsection{Neutrino-antineutrino Pair Annihilation}\label{app:nu_barnu_ann}
%%%%%%%%%%%%%%%%%%%%%%%%%%%%%%%%%%%%%%%%%%%%%%%%%%%%%%%%%%%%%%%%%%%%%%%%%%%%%%%%%%%%%%%%%%%%%%%%%%%%%%%%%%%%%%%%%%%%%%%%%%%%%%%%

Since the main interaction between $\nu\bar{\nu}$ is the annihilation into $e^{-}e^{+}$, this process above neutrino-cooled disks has been proposed as the origin of the energetic plasma involved in the production of GRBs. Once the (anti)neutrino energy emissivity and average energies are calculated it is possible to calculate the energy deposition rate of the process $\nu_{i} + \bar{\nu}_{i} \to e^{-} + e^{+}$ for each flavour $i$. Ignoring Pauli blocking effects in the phase spaces of electron and positrons, the local energy deposition rate at a position $\mathbf{r}$ by $\nu\bar{\nu}$ annihilation can be written in terms of the neutrino and antineutrino distributions $f_{\nu_{i}}=f_{\nu_{i}}\left(\mathbf{r},E_{\nu}\right),f_{\bar{\nu}_{i}}=f_{\bar{\nu}_{i}}\left(\mathbf{r},E_{\bar{\nu}}\right)$ as \cite{1991A&A...244..378J}

\begin{align}
Q_{\nu_{i}\bar{\nu}_{i}} &= A_{1,i}\int^{\infty}_{0}\!\! dE_{\nu_{i}} \int^{\infty}_{0}\!\! dE_{\bar{\nu}_{i}} E_{\nu_{i}}^{3}E_{\bar{\nu}_{i}}^{3}\left(E_{\nu_{i}} + E_{\bar{\nu}_{i}}\right) \int_{S_{2}}d\Omega_{\nu_{i}} \int_{S_{2}}d\Omega_{\bar{\nu}_{i}} f_{\nu_{i}}f_{\bar{\nu}_{i}} \left(1 - \cos\theta\right)^{2},  \nonumber\\
& + A_{2,i} \int^{\infty}_{0}\!\! dE_{\nu_{i}} \int^{\infty}_{0}\!\! dE_{\bar{\nu}_{i}} E_{\nu_{i}}^{2}E_{\bar{\nu}_{i}}^{2}\left(E_{\nu_{i}} + E_{\bar{\nu}_{i}}\right) \int_{S_{2}}d\Omega_{\nu_{i}} \int_{S_{2}}d\Omega_{\bar{\nu}_{i}} f_{\nu_{i}}f_{\bar{\nu}_{i}} \left(1 - \cos\theta\right), 
    \label{eq:nunuan}
\end{align}

where we have introduced the constants appearing in Eq.~(\ref{eq:differentialdepp})

\begin{align}
A_{1,i} &= \frac{\sigma_{0}\left[\left(C_{v,i}-C_{a,i}\right)^{2}+\left(C_{v,i}+C_{a,i}\right)^{2}\right]}{12\pi^{2}m^{2}_{e}}, \nonumber \\
A_{2,i} &= \frac{\sigma_{0}\left[2C_{v,i}^{2}-C_{a,i}^{2}\right]}{6\pi^{2}m^{2}_{e}}.
    \label{eq:const_ann}
\end{align}

In Eq.~(\ref{eq:nunuan}), $\theta$ is the angle between the neutrino and antineutrino momentum and $d\Omega$ is the differential solid angle of the incident (anti)neutrino at $\mathbf{r}$. The integral can be re-written in terms of the total intensity (energy integrated intensity) $I_{\nu} = \int E^{3}_{\nu}f_{\nu} dE_{\nu}$ as \cite{1997A&A...319..122R}

\begin{align}
Q_{\nu_{i}\bar{\nu}_{i}} &=  A_{1,i} \int_{S_{2}}d\Omega_{\nu_{i}} I_{\nu_{i}} \int_{S_{2}}d\Omega_{\bar{\nu}_{i}} I_{\bar{\nu}_{i}} \left(\langle E_{\nu_{i}}\rangle + \langle E_{\bar{\nu}_{i}}\rangle\right) \left(1 - \cos\theta\right)^{2}  \nonumber\\
& + A_{2,i}  \int_{S_{2}}d\Omega_{\nu_{i}} I_{\nu_{i}}  \int_{S_{2}}d\Omega_{\bar{\nu}_{i}} I_{\bar{\nu}_{i}} \frac{ \langle E_{\nu_{i}}\rangle + \langle E_{\bar{\nu}_{i}}\rangle}{ \langle E_{\nu_{i}}\rangle \langle E_{\bar{\nu}_{i}}\rangle }  \left(1 - \cos\theta\right).
    \label{eq:nunuan2}
\end{align}

The incident radiation intensity passing through the solid differential angle $d\Omega$ at $\mathbf{r}$ is the intensity $I_{\mathbf{r}_d,\nu}$ emitted from the point on the disk $\mathbf{r}_d$ diluted by the inverse square distance $r_k = \vert \mathbf{r}-\mathbf{r}_{d} \vert$ between both points. Finally, assuming that each point $\mathbf{r}_d$ on the disk's surface acts as a half-isotropic radiator of (anti)neutrinos, the total flux emitted at $\mathbf{r}_d$ is $F_{\mathbf{r}_{d},\nu} = \int_{0}^{\pi/2}\int_{0}^{2\pi} I_{\mathbf{r}_{d},\nu} \cos\theta^{\prime}\sin\theta^{\prime} d\theta^{\prime} d\phi^{\prime} = \pi I_{\mathbf{r}_{d},\nu}$, with $\theta^{\prime},\phi^{\prime}$ the direction angles at $\mathbf{r}_d$. Collecting all obtains

\begin{align}
Q_{\nu_{i}\bar{\nu}_{i}} &= A_{1,i}\!\!\!\!\! \int\limits_{\mathbf{r}_{d,\nu_{i}}\in\text{disk}} \!\!\!\!\! d\mathbf{r}_{d,\nu_{i}} \!\!\! \int\limits_{\mathbf{r}_{d,\bar{\nu}_{i}}\in\text{disk}} \!\!\!\!\! d\mathbf{r}_{d,\bar{\nu}_{i}}\;\; \frac{F_{\mathbf{r}_d,\nu_{i}}}{r_{k,\nu_{i}}^{2}} \frac{F_{\mathbf{r}_d,\bar{\nu}_{i}}}{r_{k,\bar{\nu}_{i}}^{2}} \left(\langle E_{\nu_{i}}\rangle + \langle E_{\bar{\nu}_{i}}\rangle\right) \left(1 - \cos\theta\right)^{2}  \nonumber\\
&\;\;\;\;\;\;\; + A_{2,i} \!\!\!\!\!  \int\limits_{\mathbf{r}_{d,\nu_{i}}\in\text{disk}} \!\!\!\!\! d\mathbf{r}_{d,\nu_{i}}  \!\!\! \int\limits_{\mathbf{r}_{d,\bar{\nu}_{i}}\in\text{disk}} \!\!\!\!\! d\mathbf{r}_{d,\bar{\nu}_{i}} \;\; \frac{F_{\mathbf{r}_d,\nu_{i}}}{r_{k,\nu_{i}}^{2}} \frac{F_{\mathbf{r}_d,\bar{\nu}_{i}}}{r_{k,\bar{\nu}_{i}}^{2}} \frac{ \langle E_{\nu_{i}}\rangle + \langle E_{\bar{\nu}_{i}}\rangle}{ \langle E_{\nu_{i}}\rangle \langle E_{\bar{\nu}_{i}}\rangle }  \left(1 - \cos\theta\right).
    \label{eq:nunuan3}
\end{align}
% 

%%%%%%%%%%%%%%%%%%%%%%%%%%%%%%%%%%%%%%%%%%%%%%%%%%%%%%%%%%%%%%%%%%%%%%%%%%%%%%%%%%%%%%%%%%%%%%%%%%%%%%%%%%%%%%%%%%%%%%%%%%%%%%%%%%%%%%%%%


%=====================================
% References, variant B: external bibliography
%=====================================
\reftitle{References}
\externalbibliography{yes}
\bibliography{main.bib}

\begin{thebibliography}{-------}
\providecommand{\natexlab}[1]{#1}

\bibitem[de~Salas \em{et~al.}(2018)de~Salas, Forero, Ternes, Tortola, and
  Valle]{deSalas:2017kay}
de~Salas, P.F.; Forero, D.V.; Ternes, C.A.; Tortola, M.; Valle, J.W.F.
\newblock {Status of neutrino oscillations 2018: 3$\sigma$ hint for normal mass
  ordering and improved CP sensitivity}.
\newblock {\em Phys. Lett.} {\bf 2018}, {\em B782},~633--640,
  \href{http://xxx.lanl.gov/abs/1708.01186}{{\normalfont
  [arXiv:hep-ph/1708.01186]}}.
\newblock
  doi:{\changeurlcolor{black}\href{https://doi.org/10.1016/j.physletb.2018.06.019}{\detokenize{10.1016/j.physletb.2018.06.019}}}.

\bibitem[Wolfenstein(1978)]{1978PhRvD..17.2369W}
Wolfenstein, L.
\newblock {Neutrino Oscillations in Matter}.
\newblock {\em Phys. Rev.} {\bf 1978}, {\em D17},~2369--2374.
\newblock [,294(1977)],
  doi:{\changeurlcolor{black}\href{https://doi.org/10.1103/PhysRevD.17.2369}{\detokenize{10.1103/PhysRevD.17.2369}}}.

\bibitem[Mikheyev and Smirnov(1986)]{Mikheyev1986}
Mikheyev, S.P.; Smirnov, A.Y.
\newblock Resonant amplification of $\nu$ oscillations in matter and
  solar-neutrino spectroscopy.
\newblock {\em Il Nuovo Cimento C} {\bf 1986}, {\em 9},~17--26.
\newblock
  doi:{\changeurlcolor{black}\href{https://doi.org/10.1007/BF02508049}{\detokenize{10.1007/BF02508049}}}.

\bibitem[Barbieri and Dolgov(1991)]{BARBIERI1991743}
Barbieri, R.; Dolgov, A.
\newblock Neutrino oscillations in the early universe.
\newblock {\em Nuclear Physics B} {\bf 1991}, {\em 349},~743 -- 753.
\newblock
  doi:{\changeurlcolor{black}\href{https://doi.org/https://doi.org/10.1016/0550-3213(91)90396-F}{\detokenize{https://doi.org/10.1016/0550-3213(91)90396-F}}}.

\bibitem[Enqvist \em{et~al.}(1991)Enqvist, Kainulainen, and
  Maalampi]{Enqvist:1990ad}
Enqvist, K.; Kainulainen, K.; Maalampi, J.
\newblock {Refraction and Oscillations of Neutrinos in the Early Universe}.
\newblock {\em Nucl. Phys.} {\bf 1991}, {\em B349},~754--790.
\newblock
  doi:{\changeurlcolor{black}\href{https://doi.org/10.1016/0550-3213(91)90397-G}{\detokenize{10.1016/0550-3213(91)90397-G}}}.

\bibitem[Savage \em{et~al.}(1991)Savage, Malaney, and Fuller]{Savage:1990by}
Savage, M.J.; Malaney, R.A.; Fuller, G.M.
\newblock {Neutrino Oscillations and the Leptonic Charge of the Universe}.
\newblock {\em Astrophys. J.} {\bf 1991}, {\em 368},~1--11.
\newblock
  doi:{\changeurlcolor{black}\href{https://doi.org/10.1086/169665}{\detokenize{10.1086/169665}}}.

\bibitem[Kostelecky and Samuel(1993)]{Kostelecky:1993dm}
Kostelecky, V.A.; Samuel, S.
\newblock {Neutrino oscillations in the early universe with an inverted
  neutrino mass hierarchy}.
\newblock {\em Phys. Lett.} {\bf 1993}, {\em B318},~127--133.
\newblock
  doi:{\changeurlcolor{black}\href{https://doi.org/10.1016/0370-2693(93)91795-O}{\detokenize{10.1016/0370-2693(93)91795-O}}}.

\bibitem[Kostelecky and Samuel(1994)]{Kostelecky:1993ys}
Kostelecky, V.A.; Samuel, S.
\newblock {Nonlinear neutrino oscillations in the expanding universe}.
\newblock {\em Phys. Rev.} {\bf 1994}, {\em D49},~1740--1757.
\newblock
  doi:{\changeurlcolor{black}\href{https://doi.org/10.1103/PhysRevD.49.1740}{\detokenize{10.1103/PhysRevD.49.1740}}}.

\bibitem[Kostelecký \em{et~al.}(1993)Kostelecký, Pantaleone, and
  Samuel]{KOSTELECKY199346}
Kostelecký, V.; Pantaleone, J.; Samuel, S.
\newblock Neutrino oscillations in the early universe.
\newblock {\em Physics Letters B} {\bf 1993}, {\em 315},~46 -- 50.
\newblock
  doi:{\changeurlcolor{black}\href{https://doi.org/https://doi.org/10.1016/0370-2693(93)90156-C}{\detokenize{https://doi.org/10.1016/0370-2693(93)90156-C}}}.

\bibitem[McKellar and Thomson(1994)]{McKellar:1992ja}
McKellar, B.H.J.; Thomson, M.J.
\newblock {Oscillating doublet neutrinos in the early universe}.
\newblock {\em Phys. Rev.} {\bf 1994}, {\em D49},~2710--2728.
\newblock
  doi:{\changeurlcolor{black}\href{https://doi.org/10.1103/PhysRevD.49.2710}{\detokenize{10.1103/PhysRevD.49.2710}}}.

\bibitem[{Lunardini} and {Smirnov}(2001)]{2001PhRvD..64g3006L}
{Lunardini}, C.; {Smirnov}, A.Y.
\newblock {High-energy neutrino conversion and the lepton asymmetry in the
  universe}.
\newblock {\em Phys.~Rev.~D} {\bf 2001}, {\em 64},~073006,
  \href{http://xxx.lanl.gov/abs/hep-ph/0012056}{{\normalfont
  [arXiv:hep-ph/hep-ph/0012056]}}.
\newblock
  doi:{\changeurlcolor{black}\href{https://doi.org/10.1103/PhysRevD.64.073006}{\detokenize{10.1103/PhysRevD.64.073006}}}.

\bibitem[{Dolgov} \em{et~al.}(2002){Dolgov}, {Hansen}, {Pastor}, {Petcov},
  {Raffelt}, and {Semikoz}]{2002NuPhB.632..363D}
{Dolgov}, A.D.; {Hansen}, S.H.; {Pastor}, S.; {Petcov}, S.T.; {Raffelt}, G.G.;
  {Semikoz}, D.V.
\newblock {Cosmological bounds on neutrino degeneracy improved by flavor
  oscillations}.
\newblock {\em Nuclear Physics B} {\bf 2002}, {\em 632},~363--382,
  \href{http://xxx.lanl.gov/abs/hep-ph/0201287}{{\normalfont
  [arXiv:hep-ph/hep-ph/0201287]}}.
\newblock
  doi:{\changeurlcolor{black}\href{https://doi.org/10.1016/S0550-3213(02)00274-2}{\detokenize{10.1016/S0550-3213(02)00274-2}}}.

\bibitem[{Wong}(2002)]{2002PhRvD..66b5015W}
{Wong}, Y.Y.
\newblock {Analytical treatment of neutrino asymmetry equilibration from flavor
  oscillations in the early universe}.
\newblock {\em Phys.~Rev.~D} {\bf 2002}, {\em 66},~025015,
  \href{http://xxx.lanl.gov/abs/hep-ph/0203180}{{\normalfont
  [arXiv:hep-ph/hep-ph/0203180]}}.
\newblock
  doi:{\changeurlcolor{black}\href{https://doi.org/10.1103/PhysRevD.66.025015}{\detokenize{10.1103/PhysRevD.66.025015}}}.

\bibitem[{Abazajian} \em{et~al.}(2002){Abazajian}, {Beacom}, and
  {Bell}]{2002PhRvD..66a3008A}
{Abazajian}, K.N.; {Beacom}, J.F.; {Bell}, N.F.
\newblock {Stringent constraints on cosmological neutrino-antineutrino
  asymmetries from synchronized flavor transformation}.
\newblock {\em Phys.~Rev.~D} {\bf 2002}, {\em 66},~013008,
  \href{http://xxx.lanl.gov/abs/astro-ph/0203442}{{\normalfont
  [arXiv:astro-ph/astro-ph/0203442]}}.
\newblock
  doi:{\changeurlcolor{black}\href{https://doi.org/10.1103/PhysRevD.66.013008}{\detokenize{10.1103/PhysRevD.66.013008}}}.

\bibitem[{Kirilova}(2004)]{2004CEJPh...2..467K}
{Kirilova}, D.P.
\newblock {Neutrino oscillations and the early universe}.
\newblock {\em Central European Journal of Physics} {\bf 2004}, {\em
  2},~467--491,  \href{http://xxx.lanl.gov/abs/astro-ph/0312569}{{\normalfont
  [arXiv:astro-ph/astro-ph/0312569]}}.
\newblock
  doi:{\changeurlcolor{black}\href{https://doi.org/10.2478/BF02476426}{\detokenize{10.2478/BF02476426}}}.

\bibitem[{Bahcall} \em{et~al.}(2003){Bahcall}, {Concepcion Gonzalez-Garcia},
  and {na-Garay}]{2003JHEP...02..009B}
{Bahcall}, J.N.; {Concepcion Gonzalez-Garcia}, M.; {na-Garay}, C.P.
\newblock {Solar Neutrinos Before and After KamLAND}.
\newblock {\em Journal of High Energy Physics} {\bf 2003}, {\em 2003},~009,
  \href{http://xxx.lanl.gov/abs/hep-ph/0212147}{{\normalfont
  [arXiv:hep-ph/hep-ph/0212147]}}.
\newblock
  doi:{\changeurlcolor{black}\href{https://doi.org/10.1088/1126-6708/2003/02/009}{\detokenize{10.1088/1126-6708/2003/02/009}}}.

\bibitem[{Balantekin} and {Yuksel}(2003)]{2003hep.ph....1072B}
{Balantekin}, A.B.; {Yuksel}, H.
\newblock {Global Analysis of Solar Neutrino and KamLAND Data}.
\newblock {\em arXiv e-prints} {\bf 2003}, pp. hep--ph/0301072,
  \href{http://xxx.lanl.gov/abs/hep-ph/0301072}{{\normalfont
  [arXiv:hep-ph/hep-ph/0301072]}}.

\bibitem[{Fogli} \em{et~al.}(2003){Fogli}, {Lisi}, {Marrone}, {Montanino},
  {Palazzo}, and {Rotunno}]{2003hep.ph...10012F}
{Fogli}, G.L.; {Lisi}, E.; {Marrone}, A.; {Montanino}, D.; {Palazzo}, A.;
  {Rotunno}, A.M.
\newblock {Neutrino Oscillations: A Global Analysis}.
\newblock {\em arXiv e-prints} {\bf 2003}, pp. hep--ph/0310012,
  \href{http://xxx.lanl.gov/abs/hep-ph/0310012}{{\normalfont
  [arXiv:hep-ph/hep-ph/0310012]}}.

\bibitem[{de Holanda} and {Smirnov}(2004)]{2004APh....21..287D}
{de Holanda}, P.C.; {Smirnov}, A.Y.
\newblock {Solar neutrinos: the SNO salt phase results and physics of
  conversion}.
\newblock {\em Astroparticle Physics} {\bf 2004}, {\em 21},~287--301,
  \href{http://xxx.lanl.gov/abs/hep-ph/0309299}{{\normalfont
  [arXiv:hep-ph/hep-ph/0309299]}}.
\newblock
  doi:{\changeurlcolor{black}\href{https://doi.org/10.1016/j.astropartphys.2004.01.007}{\detokenize{10.1016/j.astropartphys.2004.01.007}}}.

\bibitem[{Giunti}(2004)]{2004EPJC...33S.852G}
{Giunti}, C.
\newblock {Status of neutrino masses and mixing}.
\newblock {\em European Physical Journal C} {\bf 2004}, {\em 33},~852--856,
  \href{http://xxx.lanl.gov/abs/hep-ph/0309024}{{\normalfont
  [arXiv:hep-ph/hep-ph/0309024]}}.
\newblock
  doi:{\changeurlcolor{black}\href{https://doi.org/10.1140/epjcd/s2003-03-917-2}{\detokenize{10.1140/epjcd/s2003-03-917-2}}}.

\bibitem[{Maltoni} \em{et~al.}(2003){Maltoni}, {Schwetz}, {T{\'o}rtola}, and
  {Valle}]{2003PhRvD..68k3010M}
{Maltoni}, M.; {Schwetz}, T.; {T{\'o}rtola}, M.A.; {Valle}, J.W.
\newblock {Status of three-neutrino oscillations after the SNO-salt data}.
\newblock {\em Phys.~Rev.~D} {\bf 2003}, {\em 68},~113010,
  \href{http://xxx.lanl.gov/abs/hep-ph/0309130}{{\normalfont
  [arXiv:hep-ph/hep-ph/0309130]}}.
\newblock
  doi:{\changeurlcolor{black}\href{https://doi.org/10.1103/PhysRevD.68.113010}{\detokenize{10.1103/PhysRevD.68.113010}}}.

\bibitem[{Dighe}(2010)]{2010JPhCS.203a2015D}
{Dighe}, A.
\newblock {Supernova neutrino oscillations: What do we understand?}
\newblock  Journal of Physics Conference Series,  2010, Vol. 203, {\em Journal
  of Physics Conference Series}, p. 012015,
  \href{http://xxx.lanl.gov/abs/0912.4167}{{\normalfont
  [arXiv:hep-ph/0912.4167]}}.
\newblock
  doi:{\changeurlcolor{black}\href{https://doi.org/10.1088/1742-6596/203/1/012015}{\detokenize{10.1088/1742-6596/203/1/012015}}}.

\bibitem[{Haxton} \em{et~al.}(2013){Haxton}, {Hamish Robertson}, and
  {Serenelli}]{2013ARA&A..51...21H}
{Haxton}, W.C.; {Hamish Robertson}, R.G.; {Serenelli}, A.M.
\newblock {Solar Neutrinos: Status and Prospects}.
\newblock {\em ARA\&A} {\bf 2013}, {\em 51},~21--61,
  \href{http://xxx.lanl.gov/abs/1208.5723}{{\normalfont
  [arXiv:astro-ph.SR/1208.5723]}}.
\newblock
  doi:{\changeurlcolor{black}\href{https://doi.org/10.1146/annurev-astro-081811-125539}{\detokenize{10.1146/annurev-astro-081811-125539}}}.

\bibitem[{Vissani}(2017)]{2017arXiv170605435V}
{Vissani}, F.
\newblock {Solar neutrino physics on the beginning of 2017}.
\newblock {\em Nuclear Physics and Atomic Energy} {\bf 2017}, {\em 18},~5--12,
  \href{http://xxx.lanl.gov/abs/1706.05435}{{\normalfont
  [arXiv:nucl-th/1706.05435]}}.
\newblock
  doi:{\changeurlcolor{black}\href{https://doi.org/10.15407/jnpae2017.01.005}{\detokenize{10.15407/jnpae2017.01.005}}}.

\bibitem[Notzold and Raffelt(1988)]{Notzold:1987ik}
Notzold, D.; Raffelt, G.
\newblock {Neutrino Dispersion at Finite Temperature and Density}.
\newblock {\em Nucl. Phys.} {\bf 1988}, {\em B307},~924--936.
\newblock
  doi:{\changeurlcolor{black}\href{https://doi.org/10.1016/0550-3213(88)90113-7}{\detokenize{10.1016/0550-3213(88)90113-7}}}.

\bibitem[{Pantaleone}(1992)]{1992PhLB..287..128P}
{Pantaleone}, J.
\newblock {Neutrino oscillations at high densities}.
\newblock {\em Physics Letters B} {\bf 1992}, {\em 287},~128--132.
\newblock
  doi:{\changeurlcolor{black}\href{https://doi.org/10.1016/0370-2693(92)91887-F}{\detokenize{10.1016/0370-2693(92)91887-F}}}.

\bibitem[Qian and Fuller(1995)]{Qian:1994wh}
Qian, Y.Z.; Fuller, G.M.
\newblock {Neutrino-neutrino scattering and matter enhanced neutrino flavor
  transformation in Supernovae}.
\newblock {\em Phys. Rev.} {\bf 1995}, {\em D51},~1479--1494,
  \href{http://xxx.lanl.gov/abs/astro-ph/9406073}{{\normalfont
  [arXiv:astro-ph/astro-ph/9406073]}}.
\newblock
  doi:{\changeurlcolor{black}\href{https://doi.org/10.1103/PhysRevD.51.1479}{\detokenize{10.1103/PhysRevD.51.1479}}}.

\bibitem[Pastor and Raffelt(2002)]{Pastor:2002we}
Pastor, S.; Raffelt, G.
\newblock {Flavor oscillations in the supernova hot bubble region: Nonlinear
  effects of neutrino background}.
\newblock {\em Phys. Rev. Lett.} {\bf 2002}, {\em 89},~191101,
  \href{http://xxx.lanl.gov/abs/astro-ph/0207281}{{\normalfont
  [arXiv:astro-ph/astro-ph/0207281]}}.
\newblock
  doi:{\changeurlcolor{black}\href{https://doi.org/10.1103/PhysRevLett.89.191101}{\detokenize{10.1103/PhysRevLett.89.191101}}}.

\bibitem[Duan \em{et~al.}(2006)Duan, Fuller, and Qian]{Duan:2005cp}
Duan, H.; Fuller, G.M.; Qian, Y.Z.
\newblock {Collective neutrino flavor transformation in supernovae}.
\newblock {\em Phys. Rev.} {\bf 2006}, {\em D74},~123004,
  \href{http://xxx.lanl.gov/abs/astro-ph/0511275}{{\normalfont
  [arXiv:astro-ph/astro-ph/0511275]}}.
\newblock
  doi:{\changeurlcolor{black}\href{https://doi.org/10.1103/PhysRevD.74.123004}{\detokenize{10.1103/PhysRevD.74.123004}}}.

\bibitem[{Sawyer}(2005)]{2005PhRvD..72d5003S}
{Sawyer}, R.F.
\newblock {Speed-up of neutrino transformations in a supernova environment}.
\newblock {\em Physical Review D} {\bf 2005}, {\em 72},~045003,
  \href{http://xxx.lanl.gov/abs/hep-ph/0503013}{{\normalfont
  [arXiv:astro-ph/hep-ph/0503013]}}.
\newblock
  doi:{\changeurlcolor{black}\href{https://doi.org/10.1103/PhysRevD.72.045003}{\detokenize{10.1103/PhysRevD.72.045003}}}.

\bibitem[Fuller and Qian(2006)]{Fuller:2005ae}
Fuller, G.M.; Qian, Y.Z.
\newblock {Simultaneous flavor transformation of neutrinos and antineutrinos
  with dominant potentials from neutrino-neutrino forward scattering}.
\newblock {\em Phys. Rev.} {\bf 2006}, {\em D73},~023004,
  \href{http://xxx.lanl.gov/abs/astro-ph/0505240}{{\normalfont
  [arXiv:astro-ph/astro-ph/0505240]}}.
\newblock
  doi:{\changeurlcolor{black}\href{https://doi.org/10.1103/PhysRevD.73.023004}{\detokenize{10.1103/PhysRevD.73.023004}}}.

\bibitem[{Duan} \em{et~al.}(2006){Duan}, {Fuller}, {Carlson}, and
  {Qian}]{2006PhRvL..97x1101D}
{Duan}, H.; {Fuller}, G.M.; {Carlson}, J.; {Qian}, Y.Z.
\newblock {Coherent Development of Neutrino Flavor in the Supernova
  Environment}.
\newblock {\em Physical Review Letters} {\bf 2006}, {\em 97},~241101,
  \href{http://xxx.lanl.gov/abs/astro-ph/0608050}{{\normalfont
  [arXiv:astro-ph/astro-ph/0608050]}}.
\newblock
  doi:{\changeurlcolor{black}\href{https://doi.org/10.1103/PhysRevLett.97.241101}{\detokenize{10.1103/PhysRevLett.97.241101}}}.

\bibitem[Fogli \em{et~al.}(2007)Fogli, Lisi, Marrone, and
  Mirizzi]{Fogli:2007bk}
Fogli, G.L.; Lisi, E.; Marrone, A.; Mirizzi, A.
\newblock {Collective neutrino flavor transitions in supernovae and the role of
  trajectory averaging}.
\newblock {\em JCAP} {\bf 2007}, {\em 0712},~010,
  \href{http://xxx.lanl.gov/abs/0707.1998}{{\normalfont
  [arXiv:hep-ph/0707.1998]}}.
\newblock
  doi:{\changeurlcolor{black}\href{https://doi.org/10.1088/1475-7516/2007/12/010}{\detokenize{10.1088/1475-7516/2007/12/010}}}.

\bibitem[Duan \em{et~al.}(2007)Duan, Fuller, and Qian]{Duan:2007fw}
Duan, H.; Fuller, G.M.; Qian, Y.Z.
\newblock {A Simple Picture for Neutrino Flavor Transformation in Supernovae}.
\newblock {\em Phys. Rev.} {\bf 2007}, {\em D76},~085013,
  \href{http://xxx.lanl.gov/abs/0706.4293}{{\normalfont
  [arXiv:astro-ph/0706.4293]}}.
\newblock
  doi:{\changeurlcolor{black}\href{https://doi.org/10.1103/PhysRevD.76.085013}{\detokenize{10.1103/PhysRevD.76.085013}}}.

\bibitem[Raffelt and Sigl(2007)]{Raffelt:2007yz}
Raffelt, G.G.; Sigl, G.
\newblock {Self-induced decoherence in dense neutrino gases}.
\newblock {\em Phys. Rev.} {\bf 2007}, {\em D75},~083002,
  \href{http://xxx.lanl.gov/abs/hep-ph/0701182}{{\normalfont
  [arXiv:hep-ph/hep-ph/0701182]}}.
\newblock
  doi:{\changeurlcolor{black}\href{https://doi.org/10.1103/PhysRevD.75.083002}{\detokenize{10.1103/PhysRevD.75.083002}}}.

\bibitem[Esteban-Pretel \em{et~al.}(2007)Esteban-Pretel, Pastor, Tomas,
  Raffelt, and Sigl]{EstebanPretel:2007ec}
Esteban-Pretel, A.; Pastor, S.; Tomas, R.; Raffelt, G.G.; Sigl, G.
\newblock {Decoherence in supernova neutrino transformations suppressed by
  deleptonization}.
\newblock {\em Phys. Rev.} {\bf 2007}, {\em D76},~125018,
  \href{http://xxx.lanl.gov/abs/0706.2498}{{\normalfont
  [arXiv:astro-ph/0706.2498]}}.
\newblock
  doi:{\changeurlcolor{black}\href{https://doi.org/10.1103/PhysRevD.76.125018}{\detokenize{10.1103/PhysRevD.76.125018}}}.

\bibitem[Esteban-Pretel \em{et~al.}(2008)Esteban-Pretel, Pastor, Tomas,
  Raffelt, and Sigl]{EstebanPretel:2007yq}
Esteban-Pretel, A.; Pastor, S.; Tomas, R.; Raffelt, G.G.; Sigl, G.
\newblock {Mu-tau neutrino refraction and collective three-flavor
  transformations in supernovae}.
\newblock {\em Phys. Rev.} {\bf 2008}, {\em D77},~065024,
  \href{http://xxx.lanl.gov/abs/0712.1137}{{\normalfont
  [arXiv:astro-ph/0712.1137]}}.
\newblock
  doi:{\changeurlcolor{black}\href{https://doi.org/10.1103/PhysRevD.77.065024}{\detokenize{10.1103/PhysRevD.77.065024}}}.

\bibitem[Chakraborty \em{et~al.}(2008)Chakraborty, Choubey, Dasgupta, and
  Kar]{Chakraborty:2008zp}
Chakraborty, S.; Choubey, S.; Dasgupta, B.; Kar, K.
\newblock {Effect of Collective Flavor Oscillations on the Diffuse Supernova
  Neutrino Background}.
\newblock {\em JCAP} {\bf 2008}, {\em 0809},~013,
  \href{http://xxx.lanl.gov/abs/0805.3131}{{\normalfont
  [arXiv:hep-ph/0805.3131]}}.
\newblock
  doi:{\changeurlcolor{black}\href{https://doi.org/10.1088/1475-7516/2008/09/013}{\detokenize{10.1088/1475-7516/2008/09/013}}}.

\bibitem[Duan \em{et~al.}(2008{\natexlab{a}})Duan, Fuller, Carlson, and
  Qian]{Duan:2007sh}
Duan, H.; Fuller, G.M.; Carlson, J.; Qian, Y.Z.
\newblock {Flavor Evolution of the Neutronization Neutrino Burst from an
  O-Ne-Mg Core-Collapse Supernova}.
\newblock {\em Phys. Rev. Lett.} {\bf 2008}, {\em 100},~021101,
  \href{http://xxx.lanl.gov/abs/0710.1271}{{\normalfont
  [arXiv:astro-ph/0710.1271]}}.
\newblock
  doi:{\changeurlcolor{black}\href{https://doi.org/10.1103/PhysRevLett.100.021101}{\detokenize{10.1103/PhysRevLett.100.021101}}}.

\bibitem[Duan \em{et~al.}(2008{\natexlab{b}})Duan, Fuller, and
  Carlson]{Duan:2008eb}
Duan, H.; Fuller, G.M.; Carlson, J.
\newblock {Simulating nonlinear neutrino flavor evolution}.
\newblock {\em Comput. Sci. Dis.} {\bf 2008}, {\em 1},~015007,
  \href{http://xxx.lanl.gov/abs/0803.3650}{{\normalfont
  [arXiv:astro-ph/0803.3650]}}.
\newblock
  doi:{\changeurlcolor{black}\href{https://doi.org/10.1088/1749-4699/1/1/015007}{\detokenize{10.1088/1749-4699/1/1/015007}}}.

\bibitem[Dasgupta \em{et~al.}(2008)Dasgupta, Dighe, and
  Mirizzi]{Dasgupta:2008my}
Dasgupta, B.; Dighe, A.; Mirizzi, A.
\newblock {Identifying neutrino mass hierarchy at extremely small theta(13)
  through Earth matter effects in a supernova signal}.
\newblock {\em Phys. Rev. Lett.} {\bf 2008}, {\em 101},~171801,
  \href{http://xxx.lanl.gov/abs/0802.1481}{{\normalfont
  [arXiv:hep-ph/0802.1481]}}.
\newblock
  doi:{\changeurlcolor{black}\href{https://doi.org/10.1103/PhysRevLett.101.171801}{\detokenize{10.1103/PhysRevLett.101.171801}}}.

\bibitem[Dasgupta and Dighe(2008)]{Dasgupta:2007ws}
Dasgupta, B.; Dighe, A.
\newblock {Collective three-flavor oscillations of supernova neutrinos}.
\newblock {\em Phys. Rev.} {\bf 2008}, {\em D77},~113002,
  \href{http://xxx.lanl.gov/abs/0712.3798}{{\normalfont
  [arXiv:hep-ph/0712.3798]}}.
\newblock
  doi:{\changeurlcolor{black}\href{https://doi.org/10.1103/PhysRevD.77.113002}{\detokenize{10.1103/PhysRevD.77.113002}}}.

\bibitem[Sawyer(2009)]{Sawyer:2008zs}
Sawyer, R.F.
\newblock {The multi-angle instability in dense neutrino systems}.
\newblock {\em Phys. Rev.} {\bf 2009}, {\em D79},~105003,
  \href{http://xxx.lanl.gov/abs/0803.4319}{{\normalfont
  [arXiv:astro-ph/0803.4319]}}.
\newblock
  doi:{\changeurlcolor{black}\href{https://doi.org/10.1103/PhysRevD.79.105003}{\detokenize{10.1103/PhysRevD.79.105003}}}.

\bibitem[Duan \em{et~al.}(2010)Duan, Fuller, and Qian]{Duan:2010bg}
Duan, H.; Fuller, G.M.; Qian, Y.Z.
\newblock {Collective Neutrino Oscillations}.
\newblock {\em Ann. Rev. Nucl. Part. Sci.} {\bf 2010}, {\em 60},~569--594,
  \href{http://xxx.lanl.gov/abs/1001.2799}{{\normalfont
  [arXiv:hep-ph/1001.2799]}}.
\newblock
  doi:{\changeurlcolor{black}\href{https://doi.org/10.1146/annurev.nucl.012809.104524}{\detokenize{10.1146/annurev.nucl.012809.104524}}}.

\bibitem[Wu and Qian(2011)]{Wu:2011yi}
Wu, M.R.; Qian, Y.Z.
\newblock {Resonances Driven by a Neutrino Gyroscope and Collective Neutrino
  Oscillations in Supernovae}.
\newblock {\em Phys. Rev.} {\bf 2011}, {\em D84},~045009,
  \href{http://xxx.lanl.gov/abs/1105.2068}{{\normalfont
  [arXiv:astro-ph.SR/1105.2068]}}.
\newblock
  doi:{\changeurlcolor{black}\href{https://doi.org/10.1103/PhysRevD.84.045009}{\detokenize{10.1103/PhysRevD.84.045009}}}.

\bibitem[{Bilenky}(2014)]{2014arXiv1408.2864B}
{Bilenky}, S.M.
\newblock {Neutrino oscillations: brief history and present status}.
\newblock {\em ArXiv e-prints} {\bf 2014},
  \href{http://xxx.lanl.gov/abs/1408.2864}{{\normalfont
  [arXiv:hep-ph/1408.2864]}}.

\bibitem[Kneller(2015)]{2015arXiv150701434K}
Kneller, J.P.
\newblock {The Physics Of Supernova Neutrino Oscillations}.
\newblock  {Proceedings, 12th Conference on the Intersections of Particle and
  Nuclear Physics (CIPANP 2015): Vail, Colorado, USA, May 19-24, 2015},  2015,
  \href{http://xxx.lanl.gov/abs/1507.01434}{{\normalfont
  [arXiv:hep-ph/1507.01434]}}.

\bibitem[{Volpe}(2016)]{2016JPhCS.718f2068V}
{Volpe}, C.
\newblock {Theoretical developments in supernova neutrino physics: mass
  corrections and pairing correlators}.
\newblock  Journal of Physics Conference Series,  2016, Vol. 718, {\em Journal
  of Physics Conference Series}, p. 062068,
  \href{http://xxx.lanl.gov/abs/1601.05018}{{\normalfont
  [arXiv:astro-ph.HE/1601.05018]}}.
\newblock
  doi:{\changeurlcolor{black}\href{https://doi.org/10.1088/1742-6596/718/6/062068}{\detokenize{10.1088/1742-6596/718/6/062068}}}.

\bibitem[{Mirizzi} \em{et~al.}(2016){Mirizzi}, {Tamborra}, {Janka}, {Saviano},
  {Scholberg}, {Bollig}, {H{\"u}depohl}, and
  {Chakraborty}]{2016NCimR..39....1M}
{Mirizzi}, A.; {Tamborra}, I.; {Janka}, H.T.; {Saviano}, N.; {Scholberg}, K.;
  {Bollig}, R.; {H{\"u}depohl}, L.; {Chakraborty}, S.
\newblock {Supernova neutrinos: production, oscillations and detection}.
\newblock {\em Nuovo Cimento Rivista Serie} {\bf 2016}, {\em 39},~1--112,
  \href{http://xxx.lanl.gov/abs/1508.00785}{{\normalfont
  [arXiv:astro-ph.HE/1508.00785]}}.
\newblock
  doi:{\changeurlcolor{black}\href{https://doi.org/10.1393/ncr/i2016-10120-8}{\detokenize{10.1393/ncr/i2016-10120-8}}}.

\bibitem[{Horiuchi} and {Kneller}(2018)]{2018JPhG...45d3002H}
{Horiuchi}, S.; {Kneller}, J.P.
\newblock {What can be learned from a future supernova neutrino detection?}
\newblock {\em Journal of Physics G Nuclear Physics} {\bf 2018}, {\em
  45},~043002,  \href{http://xxx.lanl.gov/abs/1709.01515}{{\normalfont
  [arXiv:astro-ph.HE/1709.01515]}}.
\newblock
  doi:{\changeurlcolor{black}\href{https://doi.org/10.1088/1361-6471/aaa90a}{\detokenize{10.1088/1361-6471/aaa90a}}}.

\bibitem[{Zaizen} \em{et~al.}(2018){Zaizen}, {Yoshida}, {Sumiyoshi}, and
  {Umeda}]{2018PhRvD..98j3020Z}
{Zaizen}, M.; {Yoshida}, T.; {Sumiyoshi}, K.; {Umeda}, H.
\newblock {Collective neutrino oscillations and detectabilities in failed
  supernovae}.
\newblock {\em Phys.~Rev.~D} {\bf 2018}, {\em 98},~103020,
  \href{http://xxx.lanl.gov/abs/1811.03320}{{\normalfont
  [arXiv:astro-ph.HE/1811.03320]}}.
\newblock
  doi:{\changeurlcolor{black}\href{https://doi.org/10.1103/PhysRevD.98.103020}{\detokenize{10.1103/PhysRevD.98.103020}}}.

\bibitem[{Zhang}(2018)]{2018pgrb.book.....Z}
{Zhang}, B.
\newblock {\em {The Physics of Gamma-Ray Bursts}}; Cambridge University Press,
  2018.
\newblock
  doi:{\changeurlcolor{black}\href{https://doi.org/10.1017/9781139226530}{\detokenize{10.1017/9781139226530}}}.

\bibitem[{Ruffini} \em{et~al.}(2006){Ruffini}, {Bernardini}, {Bianco},
  {Vitagliano}, {Xue}, {Chardonnet}, {Fraschetti}, and
  {Gurzadyan}]{2006tmgm.meet..369R}
{Ruffini}, R.; {Bernardini}, M.G.; {Bianco}, C.L.; {Vitagliano}, L.; {Xue},
  S.S.; {Chardonnet}, P.; {Fraschetti}, F.; {Gurzadyan}, V.
\newblock {Black Hole Physics and Astrophysics: The GRB-Supernova Connection
  and URCA-1 - URCA-2}.
\newblock  The Tenth Marcel Grossmann Meeting. Proceedings of the MG10 Meeting
  held at Brazilian Center for Research in Physics (CBPF), Rio de Janeiro,
  Brazil, 20-26 July 2003, Eds.: M{\'a}rio Novello; Santiago Perez Bergliaffa;
  Remo Ruffini. Singapore: World Scientific Publishing, in 3 volumes, ISBN
  981-256-667-8 (set), ISBN 981-256-980-4 (Part A), ISBN 981-256-979-0 (Part
  B), ISBN 981-256-978-2 (Part C), 2006, XLVIII + 2492 pp.: 2006, p.369;
  {Novello}, M.; {Perez Bergliaffa}, S.; {Ruffini}, R., Eds.,  2006, p. 369,
  \href{http://xxx.lanl.gov/abs/astro-ph/0503475}{{\normalfont
  [astro-ph/0503475]}}.
\newblock
  doi:{\changeurlcolor{black}\href{https://doi.org/10.1142/9789812704030_0026}{\detokenize{10.1142/9789812704030_0026}}}.

\bibitem[{Ruffini} \em{et~al.}(2008){Ruffini}, {Bernardini}, {Bianco}, {Caito},
  {Chardonnet}, {Cherubini}, {Dainotti}, {Fraschetti}, {Geralico}, {Guida},
  {Patricelli}, {Rotondo}, {Rueda Hernandez}, {Vereshchagin}, and
  {Xue}]{2008mgm..conf..368R}
{Ruffini}, R.; {Bernardini}, M.G.; {Bianco}, C.L.; {Caito}, L.; {Chardonnet},
  P.; {Cherubini}, C.; {Dainotti}, M.G.; {Fraschetti}, F.; {Geralico}, A.;
  {Guida}, R.; {Patricelli}, B.; {Rotondo}, M.; {Rueda Hernandez}, J.A.;
  {Vereshchagin}, G.; {Xue}, S.S.
\newblock {On Gamma-Ray Bursts}.
\newblock  The Eleventh Marcel Grossmann Meeting On Recent Developments in
  Theoretical and Experimental General Relativity, Gravitation and Relativistic
  Field Theories; {Kleinert}, H.; {Jantzen}, R.T.; {Ruffini}, R., Eds.,  2008,
  pp. 368--505,  \href{http://xxx.lanl.gov/abs/0804.2837}{{\normalfont
  [0804.2837]}}.
\newblock
  doi:{\changeurlcolor{black}\href{https://doi.org/10.1142/9789812834300_0019}{\detokenize{10.1142/9789812834300_0019}}}.

\bibitem[{Izzo} \em{et~al.}(2012){Izzo}, {Rueda}, and
  {Ruffini}]{2012A&A...548L...5I}
{Izzo}, L.; {Rueda}, J.A.; {Ruffini}, R.
\newblock {GRB 090618: a candidate for a neutron star gravitational collapse
  onto a black hole induced by a type Ib/c supernova}.
\newblock {\em A\&A} {\bf 2012}, {\em 548},~L5,
  \href{http://xxx.lanl.gov/abs/1206.2887}{{\normalfont
  [arXiv:astro-ph.HE/1206.2887]}}.
\newblock
  doi:{\changeurlcolor{black}\href{https://doi.org/10.1051/0004-6361/201219813}{\detokenize{10.1051/0004-6361/201219813}}}.

\bibitem[{Rueda} and {Ruffini}(2012)]{2012ApJ...758L...7R}
{Rueda}, J.A.; {Ruffini}, R.
\newblock {On the Induced Gravitational Collapse of a Neutron Star to a Black
  Hole by a Type Ib/c Supernova}.
\newblock {\em ApJl} {\bf 2012}, {\em 758},~L7,
  \href{http://xxx.lanl.gov/abs/1206.1684}{{\normalfont
  [arXiv:astro-ph.HE/1206.1684]}}.
\newblock
  doi:{\changeurlcolor{black}\href{https://doi.org/10.1088/2041-8205/758/1/L7}{\detokenize{10.1088/2041-8205/758/1/L7}}}.

\bibitem[Fryer \em{et~al.}(2014)Fryer, Rueda, and Ruffini]{2014ApJ...793L..36F}
Fryer, C.L.; Rueda, J.A.; Ruffini, R.
\newblock {Hypercritical Accretion, Induced Gravitational Collapse, and
  Binary-Driven Hypernovae}.
\newblock {\em Astrophys. J.} {\bf 2014}, {\em 793},~L36,
  \href{http://xxx.lanl.gov/abs/1409.1473}{{\normalfont
  [arXiv:astro-ph.HE/1409.1473]}}.
\newblock
  doi:{\changeurlcolor{black}\href{https://doi.org/10.1088/2041-8205/793/2/L36}{\detokenize{10.1088/2041-8205/793/2/L36}}}.

\bibitem[{Ruffini} \em{et~al.}(2015){Ruffini}, {Wang}, {Enderli}, {Muccino},
  {Kovacevic}, {Bianco}, {Penacchioni}, {Pisani}, and
  {Rueda}]{2015ApJ...798...10R}
{Ruffini}, R.; {Wang}, Y.; {Enderli}, M.; {Muccino}, M.; {Kovacevic}, M.;
  {Bianco}, C.L.; {Penacchioni}, A.V.; {Pisani}, G.B.; {Rueda}, J.A.
\newblock {GRB 130427A and SN 2013cq: A Multi-wavelength Analysis of An Induced
  Gravitational Collapse Event}.
\newblock {\em ApJ} {\bf 2015}, {\em 798},~10,
  \href{http://xxx.lanl.gov/abs/1405.5723}{{\normalfont
  [arXiv:astro-ph.HE/1405.5723]}}.
\newblock
  doi:{\changeurlcolor{black}\href{https://doi.org/10.1088/0004-637X/798/1/10}{\detokenize{10.1088/0004-637X/798/1/10}}}.

\bibitem[{Fryer} \em{et~al.}(2015){Fryer}, {Oliveira}, {Rueda}, and
  {Ruffini}]{2015PhRvL.115w1102F}
{Fryer}, C.L.; {Oliveira}, F.G.; {Rueda}, J.A.; {Ruffini}, R.
\newblock {Neutron-Star-Black-Hole Binaries Produced by Binary-Driven
  Hypernovae}.
\newblock {\em Physical Review Letters} {\bf 2015}, {\em 115},~231102,
  \href{http://xxx.lanl.gov/abs/1505.02809}{{\normalfont
  [arXiv:astro-ph.HE/1505.02809]}}.
\newblock
  doi:{\changeurlcolor{black}\href{https://doi.org/10.1103/PhysRevLett.115.231102}{\detokenize{10.1103/PhysRevLett.115.231102}}}.

\bibitem[{Wang} \em{et~al.}(2019){Wang}, {Rueda}, {Ruffini}, {Becerra},
  {Bianco}, {Becerra}, {Li}, and {Karlica}]{2019ApJ...874...39W}
{Wang}, Y.; {Rueda}, J.A.; {Ruffini}, R.; {Becerra}, L.; {Bianco}, C.;
  {Becerra}, L.; {Li}, L.; {Karlica}, M.
\newblock {Two Predictions of Supernova: GRB 130427A/SN 2013cq and GRB
  180728A/SN 2018fip}.
\newblock {\em ApJ} {\bf 2019}, {\em 874},~39,
  \href{http://xxx.lanl.gov/abs/1811.05433}{{\normalfont
  [arXiv:astro-ph.HE/1811.05433]}}.
\newblock
  doi:{\changeurlcolor{black}\href{https://doi.org/10.3847/1538-4357/ab04f8}{\detokenize{10.3847/1538-4357/ab04f8}}}.

\bibitem[{Rueda} \em{et~al.}(2020){Rueda}, {Ruffini}, {Karlica}, {Moradi}, and
  {Wang}]{2020ApJ...893..148R}
{Rueda}, J.A.; {Ruffini}, R.; {Karlica}, M.; {Moradi}, R.; {Wang}, Y.
\newblock {Magnetic Fields and Afterglows of BdHNe: Inferences from GRB
  130427A, GRB 160509A, GRB 160625B, GRB 180728A, and GRB 190114C}.
\newblock {\em ApJ} {\bf 2020}, {\em 893},~148,
  \href{http://xxx.lanl.gov/abs/1905.11339}{{\normalfont
  [arXiv:astro-ph.HE/1905.11339]}}.
\newblock
  doi:{\changeurlcolor{black}\href{https://doi.org/10.3847/1538-4357/ab80b9}{\detokenize{10.3847/1538-4357/ab80b9}}}.

\bibitem[{Rueda} \em{et~al.}(2019){Rueda}, {Ruffini}, and
  {Wang}]{2019Univ....5..110R}
{Rueda}, J.A.; {Ruffini}, R.; {Wang}, Y.
\newblock {Induced Gravitational Collapse, Binary-Driven Hypernovae, Long
  Gramma-ray Bursts and Their Connection with Short Gamma-ray Bursts}.
\newblock {\em Universe} {\bf 2019}, {\em 5},~110,
  \href{http://xxx.lanl.gov/abs/1905.06050}{{\normalfont
  [arXiv:astro-ph.HE/1905.06050]}}.
\newblock
  doi:{\changeurlcolor{black}\href{https://doi.org/10.3390/universe5050110}{\detokenize{10.3390/universe5050110}}}.

\bibitem[{Becerra} \em{et~al.}(2016){Becerra}, {Bianco}, {Fryer}, {Rueda}, and
  {Ruffini}]{2016ApJ...833..107B}
{Becerra}, L.; {Bianco}, C.L.; {Fryer}, C.L.; {Rueda}, J.A.; {Ruffini}, R.
\newblock {On the Induced Gravitational Collapse Scenario of Gamma-ray Bursts
  Associated with Supernovae}.
\newblock {\em ApJ} {\bf 2016}, {\em 833},~107,
  \href{http://xxx.lanl.gov/abs/1606.02523}{{\normalfont
  [arXiv:astro-ph.HE/1606.02523]}}.
\newblock
  doi:{\changeurlcolor{black}\href{https://doi.org/10.3847/1538-4357/833/1/107}{\detokenize{10.3847/1538-4357/833/1/107}}}.

\bibitem[{Bianco} \em{et~al.}(2001){Bianco}, {Ruffini}, and
  {Xue}]{2001A&A...368..377B}
{Bianco}, C.L.; {Ruffini}, R.; {Xue}, S.S.
\newblock {The elementary spike produced by a pure e$^{+}$e$^{-}$
  pair-electromagnetic pulse from a Black Hole: The PEM Pulse}.
\newblock {\em A\&A} {\bf 2001}, {\em 368},~377--390,
  \href{http://xxx.lanl.gov/abs/astro-ph/0102060}{{\normalfont
  [astro-ph/0102060]}}.
\newblock
  doi:{\changeurlcolor{black}\href{https://doi.org/10.1051/0004-6361:20000556}{\detokenize{10.1051/0004-6361:20000556}}}.

\bibitem[{Ruffini} \em{et~al.}(2019{\natexlab{a}}){Ruffini}, {Moradi}, {Rueda},
  {Becerra}, {Bianco}, {Cherubini}, {Filippi}, {Chen}, {Karlica}, {Sahakyan},
  {Wang}, and {Xue}]{2019ApJ...886...82R}
{Ruffini}, R.; {Moradi}, R.; {Rueda}, J.A.; {Becerra}, L.; {Bianco}, C.L.;
  {Cherubini}, C.; {Filippi}, S.; {Chen}, Y.C.; {Karlica}, M.; {Sahakyan}, N.;
  {Wang}, Y.; {Xue}, S.S.
\newblock {On the GeV Emission of the Type I BdHN GRB 130427A}.
\newblock {\em ApJ} {\bf 2019}, {\em 886},~82.
\newblock
  doi:{\changeurlcolor{black}\href{https://doi.org/10.3847/1538-4357/ab4ce6}{\detokenize{10.3847/1538-4357/ab4ce6}}}.

\bibitem[{Ruffini} \em{et~al.}(2019{\natexlab{b}}){Ruffini}, {Melon Fuksman},
  and {Vereshchagin}]{2019ApJ...883..191R}
{Ruffini}, R.; {Melon Fuksman}, J.D.; {Vereshchagin}, G.V.
\newblock {On the Role of a Cavity in the Hypernova Ejecta of GRB 190114C}.
\newblock {\em ApJ} {\bf 2019}, {\em 883},~191.
\newblock
  doi:{\changeurlcolor{black}\href{https://doi.org/10.3847/1538-4357/ab3c51}{\detokenize{10.3847/1538-4357/ab3c51}}}.

\bibitem[{Ruffini} \em{et~al.}(2018{\natexlab{a}}){Ruffini}, {Wang},
  {Aimuratov}, {Barres de Almeida}, {Becerra}, {Bianco}, {Chen}, {Karlica},
  {Kovacevic}, {Li}, {Melon Fuksman}, {Moradi}, {Muccino}, {Penacchioni},
  {Pisani}, {Primorac}, {Rueda}, {Shakeri}, {Vereshchagin}, and
  {Xue}]{2018ApJ...852...53R}
{Ruffini}, R.; {Wang}, Y.; {Aimuratov}, Y.; {Barres de Almeida}, U.; {Becerra},
  L.; {Bianco}, C.L.; {Chen}, Y.C.; {Karlica}, M.; {Kovacevic}, M.; {Li}, L.;
  {Melon Fuksman}, J.D.; {Moradi}, R.; {Muccino}, M.; {Penacchioni}, A.V.;
  {Pisani}, G.B.; {Primorac}, D.; {Rueda}, J.A.; {Shakeri}, S.; {Vereshchagin},
  G.V.; {Xue}, S.S.
\newblock {Early X-Ray Flares in GRBs}.
\newblock {\em ApJ} {\bf 2018}, {\em 852},~53,
  \href{http://xxx.lanl.gov/abs/1704.03821}{{\normalfont
  [arXiv:astro-ph.HE/1704.03821]}}.
\newblock
  doi:{\changeurlcolor{black}\href{https://doi.org/10.3847/1538-4357/aa9e8b}{\detokenize{10.3847/1538-4357/aa9e8b}}}.

\bibitem[{Ruffini} \em{et~al.}(2018{\natexlab{b}}){Ruffini}, {Karlica},
  {Sahakyan}, {Rueda}, {Wang}, {Mathews}, {Bianco}, and
  {Muccino}]{2018ApJ...869..101R}
{Ruffini}, R.; {Karlica}, M.; {Sahakyan}, N.; {Rueda}, J.A.; {Wang}, Y.;
  {Mathews}, G.J.; {Bianco}, C.L.; {Muccino}, M.
\newblock {A GRB Afterglow Model Consistent with Hypernova Observations}.
\newblock {\em ApJ} {\bf 2018}, {\em 869},~101,
  \href{http://xxx.lanl.gov/abs/1712.05000}{{\normalfont
  [arXiv:astro-ph.HE/1712.05000]}}.
\newblock
  doi:{\changeurlcolor{black}\href{https://doi.org/10.3847/1538-4357/aaeac8}{\detokenize{10.3847/1538-4357/aaeac8}}}.

\bibitem[{Becerra} \em{et~al.}(2015){Becerra}, {Cipolletta}, {Fryer}, {Rueda},
  and {Ruffini}]{2015ApJ...812..100B}
{Becerra}, L.; {Cipolletta}, F.; {Fryer}, C.L.; {Rueda}, J.A.; {Ruffini}, R.
\newblock {Angular Momentum Role in the Hypercritical Accretion of
  Binary-driven Hypernovae}.
\newblock {\em ApJ} {\bf 2015}, {\em 812},~100,
  \href{http://xxx.lanl.gov/abs/1505.07580}{{\normalfont
  [arXiv:astro-ph.HE/1505.07580]}}.
\newblock
  doi:{\changeurlcolor{black}\href{https://doi.org/10.1088/0004-637X/812/2/100}{\detokenize{10.1088/0004-637X/812/2/100}}}.

\bibitem[{Becerra} \em{et~al.}(2019){Becerra}, {Ellinger}, {Fryer}, {Rueda},
  and {Ruffini}]{2019ApJ...871...14B}
{Becerra}, L.; {Ellinger}, C.L.; {Fryer}, C.L.; {Rueda}, J.A.; {Ruffini}, R.
\newblock {SPH Simulations of the Induced Gravitational Collapse Scenario of
  Long Gamma-Ray Bursts Associated with Supernovae}.
\newblock {\em ApJ} {\bf 2019}, {\em 871},~14,
  \href{http://xxx.lanl.gov/abs/1803.04356}{{\normalfont
  [arXiv:astro-ph.HE/1803.04356]}}.
\newblock
  doi:{\changeurlcolor{black}\href{https://doi.org/10.3847/1538-4357/aaf6b3}{\detokenize{10.3847/1538-4357/aaf6b3}}}.

\bibitem[{Becerra} \em{et~al.}(2018){Becerra}, {Guzzo}, {Rossi-Torres},
  {Rueda}, {Ruffini}, and {Uribe}]{2018ApJ...852..120B}
{Becerra}, L.; {Guzzo}, M.M.; {Rossi-Torres}, F.; {Rueda}, J.A.; {Ruffini}, R.;
  {Uribe}, J.D.
\newblock {Neutrino Oscillations within the Induced Gravitational Collapse
  Paradigm of Long Gamma-Ray Bursts}.
\newblock {\em ApJ} {\bf 2018}, {\em 852},~120,
  \href{http://xxx.lanl.gov/abs/1712.07210}{{\normalfont
  [arXiv:astro-ph.HE/1712.07210]}}.
\newblock
  doi:{\changeurlcolor{black}\href{https://doi.org/10.3847/1538-4357/aaa296}{\detokenize{10.3847/1538-4357/aaa296}}}.

\bibitem[{Goodman}(1986)]{1986ApJ...308L..47G}
{Goodman}, J.
\newblock {Are gamma-ray bursts optically thick?}
\newblock {\em ApJl} {\bf 1986}, {\em 308},~L47--L50.
\newblock
  doi:{\changeurlcolor{black}\href{https://doi.org/10.1086/184741}{\detokenize{10.1086/184741}}}.

\bibitem[{Paczynski}(1986)]{1986ApJ...308L..43P}
{Paczynski}, B.
\newblock {Gamma-ray bursters at cosmological distances}.
\newblock {\em ApJl} {\bf 1986}, {\em 308},~L43--L46.
\newblock
  doi:{\changeurlcolor{black}\href{https://doi.org/10.1086/184740}{\detokenize{10.1086/184740}}}.

\bibitem[{Eichler} \em{et~al.}(1989){Eichler}, {Livio}, {Piran}, and
  {Schramm}]{1989Natur.340..126E}
{Eichler}, D.; {Livio}, M.; {Piran}, T.; {Schramm}, D.N.
\newblock {Nucleosynthesis, neutrino bursts and gamma-rays from coalescing
  neutron stars}.
\newblock {\em Nature} {\bf 1989}, {\em 340},~126--128.
\newblock
  doi:{\changeurlcolor{black}\href{https://doi.org/10.1038/340126a0}{\detokenize{10.1038/340126a0}}}.

\bibitem[{Narayan} \em{et~al.}(1991){Narayan}, {Piran}, and
  {Shemi}]{1991ApJ...379L..17N}
{Narayan}, R.; {Piran}, T.; {Shemi}, A.
\newblock {Neutron star and black hole binaries in the Galaxy}.
\newblock {\em ApJl} {\bf 1991}, {\em 379},~L17--L20.
\newblock
  doi:{\changeurlcolor{black}\href{https://doi.org/10.1086/186143}{\detokenize{10.1086/186143}}}.

\bibitem[{Balbus} and {Hawley}(1991)]{1991ApJ...376..214B}
{Balbus}, S.A.; {Hawley}, J.F.
\newblock {A powerful local shear instability in weakly magnetized disks. I -
  Linear analysis. II - Nonlinear evolution}.
\newblock {\em ApJ} {\bf 1991}, {\em 376},~214--233.
\newblock
  doi:{\changeurlcolor{black}\href{https://doi.org/10.1086/170270}{\detokenize{10.1086/170270}}}.

\bibitem[{Hawley} and {Balbus}(1991)]{1991ApJ...376..223H}
{Hawley}, J.F.; {Balbus}, S.A.
\newblock {A Powerful Local Shear Instability in Weakly Magnetized Disks. II.
  Nonlinear Evolution}.
\newblock {\em ApJ} {\bf 1991}, {\em 376},~223.
\newblock
  doi:{\changeurlcolor{black}\href{https://doi.org/10.1086/170271}{\detokenize{10.1086/170271}}}.

\bibitem[{Balbus} and {Hawley}(1998)]{1998RvMP...70....1B}
{Balbus}, S.A.; {Hawley}, J.F.
\newblock {Instability, turbulence, and enhanced transport in accretion disks}.
\newblock {\em Reviews of Modern Physics} {\bf 1998}, {\em 70},~1--53.
\newblock
  doi:{\changeurlcolor{black}\href{https://doi.org/10.1103/RevModPhys.70.1}{\detokenize{10.1103/RevModPhys.70.1}}}.

\bibitem[{Balbus}(2003)]{2003ARA&A..41..555B}
{Balbus}, S.A.
\newblock {Enhanced Angular Momentum Transport in Accretion Disks}.
\newblock {\em ARA\&A} {\bf 2003}, {\em 41},~555--597,
  \href{http://xxx.lanl.gov/abs/astro-ph/0306208}{{\normalfont
  [astro-ph/0306208]}}.
\newblock
  doi:{\changeurlcolor{black}\href{https://doi.org/10.1146/annurev.astro.41.081401.155207}{\detokenize{10.1146/annurev.astro.41.081401.155207}}}.

\bibitem[{Shakura} and {Sunyaev}(1973)]{1973A&A....24..337S}
{Shakura}, N.I.; {Sunyaev}, R.A.
\newblock {Black holes in binary systems. Observational appearance.}
\newblock {\em A\&A} {\bf 1973}, {\em 24},~337--355.

\bibitem[{King} \em{et~al.}(2007){King}, {Pringle}, and
  {Livio}]{2007MNRAS.376.1740K}
{King}, A.R.; {Pringle}, J.E.; {Livio}, M.
\newblock {Accretion disc viscosity: how big is alpha?}
\newblock {\em MNRAS} {\bf 2007}, {\em 376},~1740--1746,
  \href{http://xxx.lanl.gov/abs/astro-ph/0701803}{{\normalfont
  [astro-ph/0701803]}}.
\newblock
  doi:{\changeurlcolor{black}\href{https://doi.org/10.1111/j.1365-2966.2007.11556.x}{\detokenize{10.1111/j.1365-2966.2007.11556.x}}}.

\bibitem[{Pessah} \em{et~al.}(2008){Pessah}, {Chan}, and
  {Psaltis}]{2008MNRAS.383..683P}
{Pessah}, M.E.; {Chan}, C.K.; {Psaltis}, D.
\newblock {The fundamental difference between shear alpha viscosity and
  turbulent magnetorotational stresses}.
\newblock {\em MNRAS} {\bf 2008}, {\em 383},~683--690,
  \href{http://xxx.lanl.gov/abs/astro-ph/0612404}{{\normalfont
  [astro-ph/0612404]}}.
\newblock
  doi:{\changeurlcolor{black}\href{https://doi.org/10.1111/j.1365-2966.2007.12574.x}{\detokenize{10.1111/j.1365-2966.2007.12574.x}}}.

\bibitem[{King}(2012)]{2012MmSAI..83..466K}
{King}, A.
\newblock {Accretion disc theory since Shakura and Sunyaev}.
\newblock {\em Memorie della Societa Astronomica Italiana} {\bf 2012}, {\em
  83},~466,  \href{http://xxx.lanl.gov/abs/1201.2060}{{\normalfont
  [arXiv:astro-ph.HE/1201.2060]}}.

\bibitem[{Kotko} and {Lasota}(2012)]{2012A&A...545A.115K}
{Kotko}, I.; {Lasota}, J.P.
\newblock {The viscosity parameter {$\alpha$} and the properties of accretion
  disc outbursts in close binaries}.
\newblock {\em A\&A} {\bf 2012}, {\em 545},~A115,
  \href{http://xxx.lanl.gov/abs/1209.0017}{{\normalfont
  [arXiv:astro-ph.SR/1209.0017]}}.
\newblock
  doi:{\changeurlcolor{black}\href{https://doi.org/10.1051/0004-6361/201219618}{\detokenize{10.1051/0004-6361/201219618}}}.

\bibitem[{Pringle}(1981)]{1981ARA&A..19..137P}
{Pringle}, J.E.
\newblock {Accretion discs in astrophysics}.
\newblock {\em ARA\&A} {\bf 1981}, {\em 19},~137--162.
\newblock
  doi:{\changeurlcolor{black}\href{https://doi.org/10.1146/annurev.aa.19.090181.001033}{\detokenize{10.1146/annurev.aa.19.090181.001033}}}.

\bibitem[{Krolik}(1999)]{1999agnc.book.....K}
{Krolik}, J.H.
\newblock {\em {Active galactic nuclei : from the central black hole to the
  galactic environment}}; Princeton University Press,  1999.

\bibitem[{Abramowicz} \em{et~al.}(1999){Abramowicz}, {Bj{\"o}rnsson}, and
  {Pringle}]{1999tbha.book.....A}
{Abramowicz}, M.A.; {Bj{\"o}rnsson}, G.; {Pringle}, J.E.
\newblock {\em {Theory of Black Hole Accretion Discs}}; Cambridge University
  Press,  1999; p. 309.

\bibitem[{Manmoto}(2000)]{2000ApJ...534..734M}
{Manmoto}, T.
\newblock {Advection-dominated Accretion Flow around a Kerr Black Hole}.
\newblock {\em ApJ} {\bf 2000}, {\em 534},~734--746.
\newblock
  doi:{\changeurlcolor{black}\href{https://doi.org/10.1086/308768}{\detokenize{10.1086/308768}}}.

\bibitem[{Frank} \em{et~al.}(2002){Frank}, {King}, and
  {Raine}]{2002apa..book.....F}
{Frank}, J.; {King}, A.; {Raine}, D.J.
\newblock {\em {Accretion Power in Astrophysics: Third Edition}}; Cambridge
  University Press,  2002; p. 398.

\bibitem[{Blaes}(2004)]{2004adjh.conf..137B}
{Blaes}, O.M.
\newblock {Course 3: Physics Fundamentals of Luminous Accretion Disks around
  Black Holes}.
\newblock  Accretion Discs, Jets and High Energy Phenomena in Astrophysics;
  {Beskin}, V.; {Henri}, G.; {Menard}, F.; {et al.}., Eds.,  2004, pp.
  137--185,  \href{http://xxx.lanl.gov/abs/astro-ph/0211368}{{\normalfont
  [astro-ph/0211368]}}.

\bibitem[Narayan and McClintock(2008)]{NARAYAN2008733}
Narayan, R.; McClintock, J.E.
\newblock Advection-dominated accretion and the black hole event horizon.
\newblock {\em New Astronomy Reviews} {\bf 2008}, {\em 51},~733 -- 751.
\newblock Jean-Pierre Lasota, X-ray Binaries, Accretion Disks and Compact
  Stars,
  doi:{\changeurlcolor{black}\href{https://doi.org/https://doi.org/10.1016/j.newar.2008.03.002}{\detokenize{https://doi.org/10.1016/j.newar.2008.03.002}}}.

\bibitem[{Kato} \em{et~al.}(2008){Kato}, {Fukue}, and
  {Mineshige}]{2008bhad.book.....K}
{Kato}, S.; {Fukue}, J.; {Mineshige}, S.
\newblock {\em {Black-Hole Accretion Disks --- Towards a New Paradigm ---}};
  Kyoto University Press,  2008.

\bibitem[{Qian} \em{et~al.}(2009){Qian}, {Abramowicz}, {Fragile}, {Hor{\'a}k},
  {Machida}, and {Straub}]{2009A&A...498..471Q}
{Qian}, L.; {Abramowicz}, M.A.; {Fragile}, P.C.; {Hor{\'a}k}, J.; {Machida},
  M.; {Straub}, O.
\newblock {The Polish doughnuts revisited. I. The angular momentum distribution
  and equipressure surfaces}.
\newblock {\em A\&A} {\bf 2009}, {\em 498},~471--477,
  \href{http://xxx.lanl.gov/abs/0812.2467}{{\normalfont [0812.2467]}}.
\newblock
  doi:{\changeurlcolor{black}\href{https://doi.org/10.1051/0004-6361/200811518}{\detokenize{10.1051/0004-6361/200811518}}}.

\bibitem[{Montesinos}(2012)]{2012arXiv1203.6851M}
{Montesinos}, M.
\newblock {Review: Accretion Disk Theory}.
\newblock {\em ArXiv e-prints} {\bf 2012},
  \href{http://xxx.lanl.gov/abs/1203.6851}{{\normalfont
  [arXiv:astro-ph.HE/1203.6851]}}.

\bibitem[{Abramowicz} and {Fragile}(2013)]{2013LRR....16....1A}
{Abramowicz}, M.A.; {Fragile}, P.C.
\newblock {Foundations of Black Hole Accretion Disk Theory}.
\newblock {\em Living Reviews in Relativity} {\bf 2013}, {\em 16},~1,
  \href{http://xxx.lanl.gov/abs/1104.5499}{{\normalfont
  [arXiv:astro-ph.HE/1104.5499]}}.
\newblock
  doi:{\changeurlcolor{black}\href{https://doi.org/10.12942/lrr-2013-1}{\detokenize{10.12942/lrr-2013-1}}}.

\bibitem[{Yuan} and {Narayan}(2014)]{2014ARA&A..52..529Y}
{Yuan}, F.; {Narayan}, R.
\newblock {Hot Accretion Flows Around Black Holes}.
\newblock {\em ARA\&A} {\bf 2014}, {\em 52},~529--588,
  \href{http://xxx.lanl.gov/abs/1401.0586}{{\normalfont
  [arXiv:astro-ph.HE/1401.0586]}}.
\newblock
  doi:{\changeurlcolor{black}\href{https://doi.org/10.1146/annurev-astro-082812-141003}{\detokenize{10.1146/annurev-astro-082812-141003}}}.

\bibitem[{Blaes}(2014)]{2014SSRv..183...21B}
{Blaes}, O.
\newblock {General Overview of Black Hole Accretion Theory}.
\newblock {\em Space~Sci.~Rev.} {\bf 2014}, {\em 183},~21--41,
  \href{http://xxx.lanl.gov/abs/1304.4879}{{\normalfont
  [arXiv:astro-ph.HE/1304.4879]}}.
\newblock
  doi:{\changeurlcolor{black}\href{https://doi.org/10.1007/s11214-013-9985-6}{\detokenize{10.1007/s11214-013-9985-6}}}.

\bibitem[{Lasota}(2016)]{2016ASSL..440....1L}
{Lasota}, J.P.
\newblock {Black Hole Accretion Discs}.
\newblock  Astrophysics of Black Holes: From Fundamental Aspects to Latest
  Developments; {Bambi}, C., Ed.,  2016, Vol. 440, {\em Astrophysics and Space
  Science Library}, p.~1,
  \href{http://xxx.lanl.gov/abs/1505.02172}{{\normalfont
  [arXiv:astro-ph.HE/1505.02172]}}.
\newblock
  doi:{\changeurlcolor{black}\href{https://doi.org/10.1007/978-3-662-52859-4_1}{\detokenize{10.1007/978-3-662-52859-4_1}}}.

\bibitem[Liu \em{et~al.}(2017)Liu, Gu, and Zhang]{LIU20171}
Liu, T.; Gu, W.M.; Zhang, B.
\newblock Neutrino-dominated accretion flows as the central engine of gamma-ray
  bursts.
\newblock {\em New Astronomy Reviews} {\bf 2017}, {\em 79},~1 -- 25.
\newblock
  doi:{\changeurlcolor{black}\href{https://doi.org/https://doi.org/10.1016/j.newar.2017.07.001}{\detokenize{https://doi.org/10.1016/j.newar.2017.07.001}}}.

\bibitem[{Popham} \em{et~al.}(1999){Popham}, {Woosley}, and
  {Fryer}]{1999ApJ...518..356P}
{Popham}, R.; {Woosley}, S.E.; {Fryer}, C.
\newblock {Hyperaccreting Black Holes and Gamma-Ray Bursts}.
\newblock {\em ApJ} {\bf 1999}, {\em 518},~356--374,
  \href{http://xxx.lanl.gov/abs/astro-ph/9807028}{{\normalfont
  [astro-ph/9807028]}}.
\newblock
  doi:{\changeurlcolor{black}\href{https://doi.org/10.1086/307259}{\detokenize{10.1086/307259}}}.

\bibitem[{Narayan} \em{et~al.}(2001){Narayan}, {Piran}, and
  {Kumar}]{2001ApJ...557..949N}
{Narayan}, R.; {Piran}, T.; {Kumar}, P.
\newblock {Accretion Models of Gamma-Ray Bursts}.
\newblock {\em ApJ} {\bf 2001}, {\em 557},~949--957,
  \href{http://xxx.lanl.gov/abs/astro-ph/0103360}{{\normalfont
  [astro-ph/0103360]}}.
\newblock
  doi:{\changeurlcolor{black}\href{https://doi.org/10.1086/322267}{\detokenize{10.1086/322267}}}.

\bibitem[{Kohri} and {Mineshige}(2002)]{2002ApJ...577..311K}
{Kohri}, K.; {Mineshige}, S.
\newblock {Can Neutrino-cooled Accretion Disks Be an Origin of Gamma-Ray
  Bursts?}
\newblock {\em ApJ} {\bf 2002}, {\em 577},~311--321,
  \href{http://xxx.lanl.gov/abs/astro-ph/0203177}{{\normalfont
  [astro-ph/0203177]}}.
\newblock
  doi:{\changeurlcolor{black}\href{https://doi.org/10.1086/342166}{\detokenize{10.1086/342166}}}.

\bibitem[{Di Matteo} \em{et~al.}(2002){Di Matteo}, {Perna}, and
  {Narayan}]{2002ApJ...579..706D}
{Di Matteo}, T.; {Perna}, R.; {Narayan}, R.
\newblock {Neutrino Trapping and Accretion Models for Gamma-Ray Bursts}.
\newblock {\em ApJ} {\bf 2002}, {\em 579},~706--715,
  \href{http://xxx.lanl.gov/abs/astro-ph/0207319}{{\normalfont
  [astro-ph/0207319]}}.
\newblock
  doi:{\changeurlcolor{black}\href{https://doi.org/10.1086/342832}{\detokenize{10.1086/342832}}}.

\bibitem[{Kohri} \em{et~al.}(2005){Kohri}, {Narayan}, and
  {Piran}]{2005ApJ...629..341K}
{Kohri}, K.; {Narayan}, R.; {Piran}, T.
\newblock {Neutrino-dominated Accretion and Supernovae}.
\newblock {\em ApJ} {\bf 2005}, {\em 629},~341--361,
  \href{http://xxx.lanl.gov/abs/astro-ph/0502470}{{\normalfont
  [astro-ph/0502470]}}.
\newblock
  doi:{\changeurlcolor{black}\href{https://doi.org/10.1086/431354}{\detokenize{10.1086/431354}}}.

\bibitem[{Lee} \em{et~al.}(2005){Lee}, {Ramirez-Ruiz}, and
  {Page}]{2005ApJ...632..421L}
{Lee}, W.H.; {Ramirez-Ruiz}, E.; {Page}, D.
\newblock {Dynamical Evolution of Neutrino-cooled Accretion Disks: Detailed
  Microphysics, Lepton-driven Convection, and Global Energetics}.
\newblock {\em ApJ} {\bf 2005}, {\em 632},~421--437,
  \href{http://xxx.lanl.gov/abs/astro-ph/0506121}{{\normalfont
  [astro-ph/0506121]}}.
\newblock
  doi:{\changeurlcolor{black}\href{https://doi.org/10.1086/432373}{\detokenize{10.1086/432373}}}.

\bibitem[{Gu} \em{et~al.}(2006){Gu}, {Liu}, and {Lu}]{2006ApJ...643L..87G}
{Gu}, W.M.; {Liu}, T.; {Lu}, J.F.
\newblock {Neutrino-dominated Accretion Models for Gamma-Ray Bursts: Effects of
  General Relativity and Neutrino Opacity}.
\newblock {\em ApJl} {\bf 2006}, {\em 643},~L87--L90,
  \href{http://xxx.lanl.gov/abs/astro-ph/0604370}{{\normalfont
  [astro-ph/0604370]}}.
\newblock
  doi:{\changeurlcolor{black}\href{https://doi.org/10.1086/505140}{\detokenize{10.1086/505140}}}.

\bibitem[{Chen} and {Beloborodov}(2007)]{2007ApJ...657..383C}
{Chen}, W.X.; {Beloborodov}, A.M.
\newblock {Neutrino-cooled Accretion Disks around Spinning Black Holes}.
\newblock {\em ApJ} {\bf 2007}, {\em 657},~383--399,
  \href{http://xxx.lanl.gov/abs/astro-ph/0607145}{{\normalfont
  [astro-ph/0607145]}}.
\newblock
  doi:{\changeurlcolor{black}\href{https://doi.org/10.1086/508923}{\detokenize{10.1086/508923}}}.

\bibitem[{Kawanaka} and {Mineshige}(2007)]{2007ApJ...662.1156K}
{Kawanaka}, N.; {Mineshige}, S.
\newblock {Neutrino-cooled Accretion Disk and Its Stability}.
\newblock {\em ApJ} {\bf 2007}, {\em 662},~1156--1166,
  \href{http://xxx.lanl.gov/abs/astro-ph/0702630}{{\normalfont
  [astro-ph/0702630]}}.
\newblock
  doi:{\changeurlcolor{black}\href{https://doi.org/10.1086/517985}{\detokenize{10.1086/517985}}}.

\bibitem[{Janiuk} and {Yuan}(2010)]{2010A&A...509A..55J}
{Janiuk}, A.; {Yuan}, Y.F.
\newblock {The role of black hole spin and magnetic field threading the
  unstable neutrino disk in gamma ray bursts}.
\newblock {\em A\&A} {\bf 2010}, {\em 509},~A55,
  \href{http://xxx.lanl.gov/abs/0911.0395}{{\normalfont
  [arXiv:astro-ph.HE/0911.0395]}}.
\newblock
  doi:{\changeurlcolor{black}\href{https://doi.org/10.1051/0004-6361/200912725}{\detokenize{10.1051/0004-6361/200912725}}}.

\bibitem[{Kawanaka} \em{et~al.}(2013){Kawanaka}, {Piran}, and
  {Krolik}]{2013ApJ...766...31K}
{Kawanaka}, N.; {Piran}, T.; {Krolik}, J.H.
\newblock {Jet Luminosity from Neutrino-dominated Accretion Flows in Gamma-Ray
  Bursts}.
\newblock {\em ApJ} {\bf 2013}, {\em 766},~31,
  \href{http://xxx.lanl.gov/abs/1211.5110}{{\normalfont
  [arXiv:astro-ph.HE/1211.5110]}}.
\newblock
  doi:{\changeurlcolor{black}\href{https://doi.org/10.1088/0004-637X/766/1/31}{\detokenize{10.1088/0004-637X/766/1/31}}}.

\bibitem[{Luo} and {Yuan}(2013)]{2013MNRAS.431.2362L}
{Luo}, S.; {Yuan}, F.
\newblock {Global neutrino heating in hyperaccretion flows}.
\newblock {\em MNRAS} {\bf 2013}, {\em 431},~2362--2370,
  \href{http://xxx.lanl.gov/abs/1301.1102}{{\normalfont
  [arXiv:astro-ph.HE/1301.1102]}}.
\newblock
  doi:{\changeurlcolor{black}\href{https://doi.org/10.1093/mnras/stt337}{\detokenize{10.1093/mnras/stt337}}}.

\bibitem[{Xue} \em{et~al.}(2013){Xue}, {Liu}, {Gu}, and
  {Lu}]{2013ApJS..207...23X}
{Xue}, L.; {Liu}, T.; {Gu}, W.M.; {Lu}, J.F.
\newblock {Relativistic Global Solutions of Neutrino-dominated Accretion
  Flows}.
\newblock {\em ApJs} {\bf 2013}, {\em 207},~23,
  \href{http://xxx.lanl.gov/abs/1306.0655}{{\normalfont
  [arXiv:astro-ph.HE/1306.0655]}}.
\newblock
  doi:{\changeurlcolor{black}\href{https://doi.org/10.1088/0067-0049/207/2/23}{\detokenize{10.1088/0067-0049/207/2/23}}}.

\bibitem[{Malkus} \em{et~al.}(2012){Malkus}, {Kneller}, {McLaughlin}, and
  {Surman}]{2012PhRvD..86h5015M}
{Malkus}, A.; {Kneller}, J.P.; {McLaughlin}, G.C.; {Surman}, R.
\newblock {Neutrino oscillations above black hole accretion disks: Disks with
  electron-flavor emission}.
\newblock {\em Phys.~Rev.~D} {\bf 2012}, {\em 86},~085015,
  \href{http://xxx.lanl.gov/abs/1207.6648}{{\normalfont
  [arXiv:hep-ph/1207.6648]}}.
\newblock
  doi:{\changeurlcolor{black}\href{https://doi.org/10.1103/PhysRevD.86.085015}{\detokenize{10.1103/PhysRevD.86.085015}}}.

\bibitem[{Frensel} \em{et~al.}(2017){Frensel}, {Wu}, {Volpe}, and
  {Perego}]{2017PhRvD..95b3011F}
{Frensel}, M.; {Wu}, M.R.; {Volpe}, C.; {Perego}, A.
\newblock {Neutrino flavor evolution in binary neutron star merger remnants}.
\newblock {\em Phys.~Rev.~D} {\bf 2017}, {\em 95},~023011,
  \href{http://xxx.lanl.gov/abs/1607.05938}{{\normalfont
  [arXiv:astro-ph.HE/1607.05938]}}.
\newblock
  doi:{\changeurlcolor{black}\href{https://doi.org/10.1103/PhysRevD.95.023011}{\detokenize{10.1103/PhysRevD.95.023011}}}.

\bibitem[{Tian} \em{et~al.}(2017){Tian}, {Patwardhan}, and
  {Fuller}]{2017PhRvD..96d3001T}
{Tian}, J.Y.; {Patwardhan}, A.V.; {Fuller}, G.M.
\newblock {Neutrino flavor evolution in neutron star mergers}.
\newblock {\em Phys.~Rev.~D} {\bf 2017}, {\em 96},~043001,
  \href{http://xxx.lanl.gov/abs/1703.03039}{{\normalfont
  [arXiv:astro-ph.HE/1703.03039]}}.
\newblock
  doi:{\changeurlcolor{black}\href{https://doi.org/10.1103/PhysRevD.96.043001}{\detokenize{10.1103/PhysRevD.96.043001}}}.

\bibitem[Wu and Tamborra(2017)]{PhysRevD.95.103007}
Wu, M.R.; Tamborra, I.
\newblock Fast neutrino conversions: Ubiquitous in compact binary merger
  remnants.
\newblock {\em Phys. Rev. D} {\bf 2017}, {\em 95},~103007.
\newblock
  doi:{\changeurlcolor{black}\href{https://doi.org/10.1103/PhysRevD.95.103007}{\detokenize{10.1103/PhysRevD.95.103007}}}.

\bibitem[{Padilla-Gay} \em{et~al.}(2020){Padilla-Gay}, {Shalgar}, and
  {Tamborra}]{2020arXiv200901843P}
{Padilla-Gay}, I.; {Shalgar}, S.; {Tamborra}, I.
\newblock {Multi-Dimensional Solution of Fast Neutrino Conversions in Binary
  Neutron Star Merger Remnants}.
\newblock {\em arXiv e-prints} {\bf 2020}, p. arXiv:2009.01843,
  \href{http://xxx.lanl.gov/abs/2009.01843}{{\normalfont
  [arXiv:astro-ph.HE/2009.01843]}}.

\bibitem[Janiuk \em{et~al.}(2013)Janiuk, Mioduszewski, and
  Moscibrodzka]{Janiuk_2013}
Janiuk, A.; Mioduszewski, P.; Moscibrodzka, M.
\newblock {ACCRETION} {AND} {OUTFLOW} {FROM} A {MAGNETIZED}, {NEUTRINO}
  {COOLED} {TORUS} {AROUND} {THE} {GAMMA}-{RAY} {BURST} {CENTRAL} {ENGINE}.
\newblock {\em The Astrophysical Journal} {\bf 2013}, {\em 776},~105.
\newblock
  doi:{\changeurlcolor{black}\href{https://doi.org/10.1088/0004-637x/776/2/105}{\detokenize{10.1088/0004-637x/776/2/105}}}.

\bibitem[{Janiuk}(2017)]{2017ApJ...837...39J}
{Janiuk}, A.
\newblock {Microphysics in the Gamma-Ray Burst Central Engine}.
\newblock {\em ApJ} {\bf 2017}, {\em 837},~39,
  \href{http://xxx.lanl.gov/abs/1609.09361}{{\normalfont
  [arXiv:astro-ph.HE/1609.09361]}}.
\newblock
  doi:{\changeurlcolor{black}\href{https://doi.org/10.3847/1538-4357/aa5f16}{\detokenize{10.3847/1538-4357/aa5f16}}}.

\bibitem[Janiuk \em{et~al.}(2018)Janiuk, Sapountzis, Mortier, and
  Janiuk]{JSFI177}
Janiuk, A.; Sapountzis, K.; Mortier, J.; Janiuk, I.
\newblock Numerical Simulations of Black Hole Accretion Flows.
\newblock {\em Supercomputing Frontiers and Innovations} {\bf 2018}, {\em 5}.

\bibitem[{Janiuk}(2019)]{2019ApJ...882..163J}
{Janiuk}, A.
\newblock {The r-process Nucleosynthesis in the Outflows from Short GRB
  Accretion Disks}.
\newblock {\em ApJ} {\bf 2019}, {\em 882},~163,
  \href{http://xxx.lanl.gov/abs/1907.00809}{{\normalfont
  [arXiv:astro-ph.HE/1907.00809]}}.
\newblock
  doi:{\changeurlcolor{black}\href{https://doi.org/10.3847/1538-4357/ab3349}{\detokenize{10.3847/1538-4357/ab3349}}}.

\bibitem[{Bardeen}(1970)]{1970ApJ...162...71B}
{Bardeen}, J.M.
\newblock {A Variational Principle for Rotating Stars in General Relativity}.
\newblock {\em ApJ} {\bf 1970}, {\em 162},~71.
\newblock
  doi:{\changeurlcolor{black}\href{https://doi.org/10.1086/150635}{\detokenize{10.1086/150635}}}.

\bibitem[{Bardeen} \em{et~al.}(1972){Bardeen}, {Press}, and
  {Teukolsky}]{1972ApJ...178..347B}
{Bardeen}, J.M.; {Press}, W.H.; {Teukolsky}, S.A.
\newblock {Rotating Black Holes: Locally Nonrotating Frames, Energy Extraction,
  and Scalar Synchrotron Radiation}.
\newblock {\em ApJ} {\bf 1972}, {\em 178},~347--370.
\newblock
  doi:{\changeurlcolor{black}\href{https://doi.org/10.1086/151796}{\detokenize{10.1086/151796}}}.

\bibitem[{Gammie} and {Popham}(1998)]{1998ApJ...498..313G}
{Gammie}, C.F.; {Popham}, R.
\newblock {Advection-dominated Accretion Flows in the Kerr Metric. I. Basic
  Equations}.
\newblock {\em ApJ} {\bf 1998}, {\em 498},~313--326,
  \href{http://xxx.lanl.gov/abs/astro-ph/9705117}{{\normalfont
  [astro-ph/9705117]}}.
\newblock
  doi:{\changeurlcolor{black}\href{https://doi.org/10.1086/305521}{\detokenize{10.1086/305521}}}.

\bibitem[{Bardeen}(1970)]{1970Natur.226...64B}
{Bardeen}, J.M.
\newblock {Kerr Metric Black Holes}.
\newblock {\em Nature} {\bf 1970}, {\em 226},~64--65.
\newblock
  doi:{\changeurlcolor{black}\href{https://doi.org/10.1038/226064a0}{\detokenize{10.1038/226064a0}}}.

\bibitem[{Thorne}(1974)]{1974ApJ...191..507T}
{Thorne}, K.S.
\newblock {Disk-Accretion onto a Black Hole. II. Evolution of the Hole}.
\newblock {\em ApJ} {\bf 1974}, {\em 191},~507--520.
\newblock
  doi:{\changeurlcolor{black}\href{https://doi.org/10.1086/152991}{\detokenize{10.1086/152991}}}.

\bibitem[{Novikov} and {Thorne}(1973)]{1973blho.conf..343N}
{Novikov}, I.D.; {Thorne}, K.S.
\newblock {Astrophysics of black holes.}
\newblock  Black Holes (Les Astres Occlus); {Dewitt}, C.; {Dewitt}, B.S., Eds.,
   1973, pp. 343--450.

\bibitem[{Page} and {Thorne}(1974)]{1974ApJ...191..499P}
{Page}, D.N.; {Thorne}, K.S.
\newblock {Disk-Accretion onto a Black Hole. Time-Averaged Structure of
  Accretion Disk}.
\newblock {\em ApJ} {\bf 1974}, {\em 191},~499--506.
\newblock
  doi:{\changeurlcolor{black}\href{https://doi.org/10.1086/152990}{\detokenize{10.1086/152990}}}.

\bibitem[{Landau} and {Lifshitz}(1959)]{1959flme.book.....L}
{Landau}, L.D.; {Lifshitz}, E.M.
\newblock {\em {Fluid mechanics}}; Oxford: Pergamon Press, 1959,  1959.

\bibitem[Abramowicz \em{et~al.}(1996)Abramowicz, Chen, Granath, and
  Lasota]{Abramowicz_1996}
Abramowicz, M.A.; Chen, X.M.; Granath, M.; Lasota, J.P.
\newblock Advection-Dominated Accretion Flows Around Kerr Black Holes.
\newblock {\em The Astrophysical Journal} {\bf 1996}, {\em 471},~762--773.
\newblock
  doi:{\changeurlcolor{black}\href{https://doi.org/10.1086/178004}{\detokenize{10.1086/178004}}}.

\bibitem[Abramowicz \em{et~al.}(1997)Abramowicz, Lanza, and
  Percival]{Abramowicz_1997}
Abramowicz, M.A.; Lanza, A.; Percival, M.J.
\newblock Accretion Disks around Kerr Black Holes: Vertical Equilibrium
  Revisited.
\newblock {\em The Astrophysical Journal} {\bf 1997}, {\em 479},~179--183.
\newblock
  doi:{\changeurlcolor{black}\href{https://doi.org/10.1086/303869}{\detokenize{10.1086/303869}}}.

\bibitem[{Misner} \em{et~al.}(1973){Misner}, {Thorne}, and
  {Wheeler}]{1973grav.book.....M}
{Misner}, C.W.; {Thorne}, K.S.; {Wheeler}, J.A.
\newblock {\em {Gravitation}}; Princeton University Press,  1973.

\bibitem[{Mihalas} and {Mihalas}(1984)]{1984oup..book.....M}
{Mihalas}, D.; {Mihalas}, B.W.
\newblock {\em {Foundations of radiation hydrodynamics}}; Oxford University
  Press,  1984.

\bibitem[{Clifford} and {Tayler}(1965)]{1965MmRAS..69...21C}
{Clifford}, F.E.; {Tayler}, R.J.
\newblock {The equilibrium distribution of nuclides in matter at high
  temperatures.}
\newblock {\em MmRAS} {\bf 1965}, {\em 69},~21.

\bibitem[{Calder} \em{et~al.}(2007){Calder}, {Townsley}, {Seitenzahl}, {Peng},
  {Messer}, {Vladimirova}, {Brown}, {Truran}, and {Lamb}]{2007ApJ...656..313C}
{Calder}, A.C.; {Townsley}, D.M.; {Seitenzahl}, I.R.; {Peng}, F.; {Messer},
  O.E.B.; {Vladimirova}, N.; {Brown}, E.F.; {Truran}, J.W.; {Lamb}, D.Q.
\newblock {Capturing the Fire: Flame Energetics and Neutronization for Type Ia
  Supernova Simulations}.
\newblock {\em ApJ} {\bf 2007}, {\em 656},~313--332,
  \href{http://xxx.lanl.gov/abs/astro-ph/0611009}{{\normalfont
  [astro-ph/0611009]}}.
\newblock
  doi:{\changeurlcolor{black}\href{https://doi.org/10.1086/510709}{\detokenize{10.1086/510709}}}.

\bibitem[{Mavrodiev} and {Deliyergiyev}(2018)]{2018IJMPE..2750015M}
{Mavrodiev}, S.C.; {Deliyergiyev}, M.A.
\newblock {Modification of the nuclear landscape in the inverse problem
  framework using the generalized Bethe-Weizs{\"a}cker mass formula}.
\newblock {\em International Journal of Modern Physics E} {\bf 2018}, {\em
  27},~1850015--708.
\newblock
  doi:{\changeurlcolor{black}\href{https://doi.org/10.1142/S0218301318500155}{\detokenize{10.1142/S0218301318500155}}}.

\bibitem[{Rauscher} and {Thielemann}(2000)]{2000ADNDT..75....1R}
{Rauscher}, T.; {Thielemann}, F.K.
\newblock {Astrophysical Reaction Rates From Statistical Model Calculations}.
\newblock {\em Atomic Data and Nuclear Data Tables} {\bf 2000}, {\em
  75},~1--351,  \href{http://xxx.lanl.gov/abs/astro-ph/0004059}{{\normalfont
  [astro-ph/0004059]}}.
\newblock \url{http://nucastro.org/tables.html##partf},
  doi:{\changeurlcolor{black}\href{https://doi.org/10.1006/adnd.2000.0834}{\detokenize{10.1006/adnd.2000.0834}}}.

\bibitem[{Rauscher}(2003)]{2003ApJS..147..403R}
{Rauscher}, T.
\newblock {Nuclear Partition Functions at Temperatures Exceeding 10$^{10}$ K}.
\newblock {\em ApJs} {\bf 2003}, {\em 147},~403--408,
  \href{http://xxx.lanl.gov/abs/astro-ph/0304047}{{\normalfont
  [astro-ph/0304047]}}.
\newblock \url{http://nucastro.org/tables.html##partf},
  doi:{\changeurlcolor{black}\href{https://doi.org/10.1086/375733}{\detokenize{10.1086/375733}}}.

\bibitem[{Vincenti} and {Kruger}(1965)]{1965itpg.book.....V}
{Vincenti}, W.G.; {Kruger}, C.H.
\newblock {\em {Introduction to physical gas dynamics}}; Krieger Pub Co (June
  1, 1975),  1965.

\bibitem[Buresti(2015)]{Buresti2015}
Buresti, G.
\newblock A note on Stokes' hypothesis.
\newblock {\em Acta Mechanica} {\bf 2015}, {\em 226},~3555--3559.
\newblock
  doi:{\changeurlcolor{black}\href{https://doi.org/10.1007/s00707-015-1380-9}{\detokenize{10.1007/s00707-015-1380-9}}}.

\bibitem[{Particle Data Group}(2018)]{PhysRevD.98.030001}
{Particle Data Group}.
\newblock Review of Particle Physics.
\newblock {\em Phys. Rev. D} {\bf 2018}, {\em 98},~030001.
\newblock
  doi:{\changeurlcolor{black}\href{https://doi.org/10.1103/PhysRevD.98.030001}{\detokenize{10.1103/PhysRevD.98.030001}}}.

\bibitem[Dolgov(1981)]{Dolgov:1980cq}
Dolgov, A.D.
\newblock {Neutrinos in the Early Universe}.
\newblock {\em Sov. J. Nucl. Phys.} {\bf 1981}, {\em 33},~700--706.
\newblock [Yad. Fiz.33,1309(1981)].

\bibitem[Sigl and Raffelt(1993)]{Sigl:1992fn}
Sigl, G.; Raffelt, G.
\newblock {General kinetic description of relativistic mixed neutrinos}.
\newblock {\em Nucl. Phys.} {\bf 1993}, {\em B406},~423--451.
\newblock
  doi:{\changeurlcolor{black}\href{https://doi.org/10.1016/0550-3213(93)90175-O}{\detokenize{10.1016/0550-3213(93)90175-O}}}.

\bibitem[Hannestad \em{et~al.}(2006)Hannestad, Raffelt, Sigl, and
  Wong]{Hannestad:2006nj}
Hannestad, S.; Raffelt, G.G.; Sigl, G.; Wong, Y.Y.Y.
\newblock {Self-induced conversion in dense neutrino gases: Pendulum in flavour
  space}.
\newblock {\em Phys. Rev.} {\bf 2006}, {\em D74},~105010,
  \href{http://xxx.lanl.gov/abs/astro-ph/0608695}{{\normalfont
  [arXiv:astro-ph/astro-ph/0608695]}}.
\newblock [Erratum: Phys. Rev.D76,029901(2007)],
  doi:{\changeurlcolor{black}\href{https://doi.org/10.1103/PhysRevD.74.105010,
  10.1103/PhysRevD.76.029901}{\detokenize{10.1103/PhysRevD.74.105010,
  10.1103/PhysRevD.76.029901}}}.

\bibitem[Cardall(2008)]{Cardall:2007zw}
Cardall, C.Y.
\newblock {Liouville equations for neutrino distribution matrices}.
\newblock {\em Phys. Rev.} {\bf 2008}, {\em D78},~085017,
  \href{http://xxx.lanl.gov/abs/0712.1188}{{\normalfont
  [arXiv:astro-ph/0712.1188]}}.
\newblock
  doi:{\changeurlcolor{black}\href{https://doi.org/10.1103/PhysRevD.78.085017}{\detokenize{10.1103/PhysRevD.78.085017}}}.

\bibitem[Strack and Burrows(2005)]{Strack:2005ux}
Strack, P.; Burrows, A.
\newblock {Generalized Boltzmann formalism for oscillating neutrinos}.
\newblock {\em Phys. Rev.} {\bf 2005}, {\em D71},~093004,
  \href{http://xxx.lanl.gov/abs/hep-ph/0504035}{{\normalfont
  [arXiv:hep-ph/hep-ph/0504035]}}.
\newblock
  doi:{\changeurlcolor{black}\href{https://doi.org/10.1103/PhysRevD.71.093004}{\detokenize{10.1103/PhysRevD.71.093004}}}.

\bibitem[Dasgupta \em{et~al.}(2008)Dasgupta, Dighe, Mirizzi, and
  Raffelt]{Dasgupta:2008cu}
Dasgupta, B.; Dighe, A.; Mirizzi, A.; Raffelt, G.G.
\newblock {Collective neutrino oscillations in non-spherical geometry}.
\newblock {\em Phys. Rev.} {\bf 2008}, {\em D78},~033014,
  \href{http://xxx.lanl.gov/abs/0805.3300}{{\normalfont
  [arXiv:hep-ph/0805.3300]}}.
\newblock
  doi:{\changeurlcolor{black}\href{https://doi.org/10.1103/PhysRevD.78.033014}{\detokenize{10.1103/PhysRevD.78.033014}}}.

\bibitem[{Duan} \em{et~al.}(2006){Duan}, {Fuller}, and
  {Qian}]{2006PhRvD..74l3004D}
{Duan}, H.; {Fuller}, G.M.; {Qian}, Y.Z.
\newblock {Collective neutrino flavor transformation in supernovae}.
\newblock {\em Phys.~Rev.~D} {\bf 2006}, {\em 74},~123004,
  \href{http://xxx.lanl.gov/abs/astro-ph/0511275}{{\normalfont
  [arXiv:astro-ph/astro-ph/0511275]}}.
\newblock
  doi:{\changeurlcolor{black}\href{https://doi.org/10.1103/PhysRevD.74.123004}{\detokenize{10.1103/PhysRevD.74.123004}}}.

\bibitem[{Tolman}(1934)]{1934rtc..book.....T}
{Tolman}, R.C.
\newblock {\em {Relativity, Thermodynamics, and Cosmology}}; Clarendon Press,
  1934.

\bibitem[Klein(1949{\natexlab{a}})]{Klein1949}
Klein, O.
\newblock On the statistical derivation of the laws of chemical equilibrium.
\newblock {\em Il Nuovo Cimento (1943-1954)} {\bf 1949}, {\em 6},~171--180.
\newblock
  doi:{\changeurlcolor{black}\href{https://doi.org/10.1007/BF02780980}{\detokenize{10.1007/BF02780980}}}.

\bibitem[Klein(1949{\natexlab{b}})]{RevModPhys.21.531}
Klein, O.
\newblock On the Thermodynamical Equilibrium of Fluids in Gravitational Fields.
\newblock {\em Rev. Mod. Phys.} {\bf 1949}, {\em 21},~531--533.
\newblock
  doi:{\changeurlcolor{black}\href{https://doi.org/10.1103/RevModPhys.21.531}{\detokenize{10.1103/RevModPhys.21.531}}}.

\bibitem[{Paczynski}(1978)]{1978AcA....28...91P}
{Paczynski}, B.
\newblock {A model of selfgravitating accretion disk}.
\newblock {\em Acta Astron.} {\bf 1978}, {\em 28},~91--109.

\bibitem[Raffelt(1996)]{Raffelt:1996wa}
Raffelt, G.G.
\newblock {\em {Stars as laboratories for fundamental physics}}; University of
  Chicago Press,  1996.

\bibitem[{Harris} and {Stodolsky}(1982)]{1982PhLB..116..464H}
{Harris}, R.A.; {Stodolsky}, L.
\newblock {Two state systems in media and {\textquotedblleft}Turing's
  paradox{\textquotedblright}}.
\newblock {\em Physics Letters B} {\bf 1982}, {\em 116},~464--468.
\newblock
  doi:{\changeurlcolor{black}\href{https://doi.org/10.1016/0370-2693(82)90169-1}{\detokenize{10.1016/0370-2693(82)90169-1}}}.

\bibitem[Stodolsky(1987)]{PhysRevD.36.2273}
Stodolsky, L.
\newblock Treatment of neutrino oscillations in a thermal environment.
\newblock {\em Phys. Rev. D} {\bf 1987}, {\em 36},~2273--2277.
\newblock
  doi:{\changeurlcolor{black}\href{https://doi.org/10.1103/PhysRevD.36.2273}{\detokenize{10.1103/PhysRevD.36.2273}}}.

\bibitem[{Janka}(1991)]{1991A&A...244..378J}
{Janka}, H.T.
\newblock {Implications of detailed neutrino transport for the heating by
  neutrino-antineutrino annihilation in supernova explosions}.
\newblock {\em A\&A} {\bf 1991}, {\em 244},~378--382.

\bibitem[{Ruffert} \em{et~al.}(1997){Ruffert}, {Janka}, {Takahashi}, and
  {Schaefer}]{1997A&A...319..122R}
{Ruffert}, M.; {Janka}, H.T.; {Takahashi}, K.; {Schaefer}, G.
\newblock {Coalescing neutron stars - a step towards physical models. II.
  Neutrino emission, neutron tori, and gamma-ray bursts.}
\newblock {\em A\&A} {\bf 1997}, {\em 319},~122--153,
  \href{http://xxx.lanl.gov/abs/astro-ph/9606181}{{\normalfont
  [astro-ph/9606181]}}.

\bibitem[{Rosswog} \em{et~al.}(2003){Rosswog}, {Ramirez-Ruiz}, and
  {Davies}]{2003MNRAS.345.1077R}
{Rosswog}, S.; {Ramirez-Ruiz}, E.; {Davies}, M.B.
\newblock {High-resolution calculations of merging neutron stars - III.
  Gamma-ray bursts}.
\newblock {\em MNRAS} {\bf 2003}, {\em 345},~1077--1090,
  \href{http://xxx.lanl.gov/abs/astro-ph/0306418}{{\normalfont
  [astro-ph/0306418]}}.
\newblock
  doi:{\changeurlcolor{black}\href{https://doi.org/10.1046/j.1365-2966.2003.07032.x}{\detokenize{10.1046/j.1365-2966.2003.07032.x}}}.

\bibitem[{Kawanaka} and {Kohri}(2012)]{2012MNRAS.419..713K}
{Kawanaka}, N.; {Kohri}, K.
\newblock {A possible origin of the rapid variability of gamma-ray bursts due
  to convective energy transfer in hyperaccretion discs}.
\newblock {\em MNRAS} {\bf 2012}, {\em 419},~713--717,
  \href{http://xxx.lanl.gov/abs/1103.4713}{{\normalfont
  [arXiv:astro-ph.HE/1103.4713]}}.
\newblock
  doi:{\changeurlcolor{black}\href{https://doi.org/10.1111/j.1365-2966.2011.19733.x}{\detokenize{10.1111/j.1365-2966.2011.19733.x}}}.

\bibitem[{Preparata} \em{et~al.}(1998){Preparata}, {Ruffini}, and
  {Xue}]{1998A&A...338L..87P}
{Preparata}, G.; {Ruffini}, R.; {Xue}, S.S.
\newblock {The dyadosphere of black holes and gamma-ray bursts}.
\newblock {\em A\&A} {\bf 1998}, {\em 338},~L87--L90,
  \href{http://xxx.lanl.gov/abs/astro-ph/9810182}{{\normalfont
  [astro-ph/9810182]}}.

\bibitem[{Ruffini} \em{et~al.}(1999){Ruffini}, {Salmonson}, {Wilson}, and
  {Xue}]{1999A&AS..138..511R}
{Ruffini}, R.; {Salmonson}, J.D.; {Wilson}, J.R.; {Xue}, S.S.
\newblock {On evolution of the pair-electromagnetic pulse of a charged black
  hole}.
\newblock {\em A\&As} {\bf 1999}, {\em 138},~511--512,
  \href{http://xxx.lanl.gov/abs/astro-ph/9905021}{{\normalfont
  [astro-ph/9905021]}}.
\newblock
  doi:{\changeurlcolor{black}\href{https://doi.org/10.1051/aas:1999330}{\detokenize{10.1051/aas:1999330}}}.

\bibitem[{Ruffini} \em{et~al.}(2000){Ruffini}, {Salmonson}, {Wilson}, and
  {Xue}]{2000A&A...359..855R}
{Ruffini}, R.; {Salmonson}, J.D.; {Wilson}, J.R.; {Xue}, S.S.
\newblock {On the pair-electromagnetic pulse from an electromagnetic black hole
  surrounded by a baryonic remnant}.
\newblock {\em A\&A} {\bf 2000}, {\em 359},~855--864,
  \href{http://xxx.lanl.gov/abs/astro-ph/0004257}{{\normalfont
  [astro-ph/0004257]}}.

\bibitem[{Shemi} and {Piran}(1990)]{1990ApJ...365L..55S}
{Shemi}, A.; {Piran}, T.
\newblock {The appearance of cosmic fireballs}.
\newblock {\em ApJl} {\bf 1990}, {\em 365},~L55--L58.
\newblock
  doi:{\changeurlcolor{black}\href{https://doi.org/10.1086/185887}{\detokenize{10.1086/185887}}}.

\bibitem[{Piran} \em{et~al.}(1993){Piran}, {Shemi}, and
  {Narayan}]{1993MNRAS.263..861P}
{Piran}, T.; {Shemi}, A.; {Narayan}, R.
\newblock {Hydrodynamics of Relativistic Fireballs}.
\newblock {\em MNRAS} {\bf 1993}, {\em 263},~861,
  \href{http://xxx.lanl.gov/abs/astro-ph/9301004}{{\normalfont
  [astro-ph/9301004]}}.
\newblock
  doi:{\changeurlcolor{black}\href{https://doi.org/10.1093/mnras/263.4.861}{\detokenize{10.1093/mnras/263.4.861}}}.

\bibitem[{Meszaros} \em{et~al.}(1993){Meszaros}, {Laguna}, and
  {Rees}]{1993ApJ...415..181M}
{Meszaros}, P.; {Laguna}, P.; {Rees}, M.J.
\newblock {Gasdynamics of relativistically expanding gamma-ray burst sources -
  Kinematics, energetics, magnetic fields, and efficiency}.
\newblock {\em ApJ} {\bf 1993}, {\em 415},~181--190,
  \href{http://xxx.lanl.gov/abs/astro-ph/9301007}{{\normalfont
  [astro-ph/9301007]}}.
\newblock
  doi:{\changeurlcolor{black}\href{https://doi.org/10.1086/173154}{\detokenize{10.1086/173154}}}.

\bibitem[{Piran}(1999)]{1999PhR...314..575P}
{Piran}, T.
\newblock {Gamma-ray bursts and the fireball model}.
\newblock {\em Phys.~Rep.} {\bf 1999}, {\em 314},~575--667,
  \href{http://xxx.lanl.gov/abs/arXiv:astro-ph/9810256}{{\normalfont
  [arXiv:astro-ph/9810256]}}.
\newblock
  doi:{\changeurlcolor{black}\href{https://doi.org/10.1016/S0370-1573(98)00127-6}{\detokenize{10.1016/S0370-1573(98)00127-6}}}.

\bibitem[{Piran}(2004)]{2004RvMP...76.1143P}
{Piran}, T.
\newblock {The physics of gamma-ray bursts}.
\newblock {\em Reviews of Modern Physics} {\bf 2004}, {\em 76},~1143--1210,
  \href{http://xxx.lanl.gov/abs/astro-ph/0405503}{{\normalfont
  [astro-ph/0405503]}}.
\newblock
  doi:{\changeurlcolor{black}\href{https://doi.org/10.1103/RevModPhys.76.1143}{\detokenize{10.1103/RevModPhys.76.1143}}}.

\bibitem[{M{\'e}sz{\'a}ros}(2002)]{2002ARA26A..40..137M}
{M{\'e}sz{\'a}ros}, P.
\newblock {Theories of Gamma-Ray Bursts}.
\newblock {\em ARA\&A} {\bf 2002}, {\em 40},~137,
  \href{http://xxx.lanl.gov/abs/arXiv:astro-ph/0111170}{{\normalfont
  [arXiv:astro-ph/0111170]}}.
\newblock
  doi:{\changeurlcolor{black}\href{https://doi.org/10.1146/annurev.astro.40.060401.093821}{\detokenize{10.1146/annurev.astro.40.060401.093821}}}.

\bibitem[{M{\'e}sz{\'a}ros}(2006)]{2006RPPh...69.2259M}
{M{\'e}sz{\'a}ros}, P.
\newblock {Gamma-ray bursts}.
\newblock {\em Reports on Progress in Physics} {\bf 2006}, {\em
  69},~2259--2321,
  \href{http://xxx.lanl.gov/abs/astro-ph/0605208}{{\normalfont
  [astro-ph/0605208]}}.
\newblock
  doi:{\changeurlcolor{black}\href{https://doi.org/10.1088/0034-4885/69/8/R01}{\detokenize{10.1088/0034-4885/69/8/R01}}}.

\bibitem[{Berger}(2014)]{2014ARA&A..52...43B}
{Berger}, E.
\newblock {Short-Duration Gamma-Ray Bursts}.
\newblock {\em ARA\&A} {\bf 2014}, {\em 52},~43--105,
  \href{http://xxx.lanl.gov/abs/1311.2603}{{\normalfont
  [arXiv:astro-ph.HE/1311.2603]}}.
\newblock
  doi:{\changeurlcolor{black}\href{https://doi.org/10.1146/annurev-astro-081913-035926}{\detokenize{10.1146/annurev-astro-081913-035926}}}.

\bibitem[{Kumar} and {Zhang}(2015)]{2015PhR...561....1K}
{Kumar}, P.; {Zhang}, B.
\newblock {The physics of gamma-ray bursts and relativistic jets}.
\newblock {\em Phys.~Rep.} {\bf 2015}, {\em 561},~1--109,
  \href{http://xxx.lanl.gov/abs/1410.0679}{{\normalfont
  [arXiv:astro-ph.HE/1410.0679]}}.
\newblock
  doi:{\changeurlcolor{black}\href{https://doi.org/10.1016/j.physrep.2014.09.008}{\detokenize{10.1016/j.physrep.2014.09.008}}}.

\bibitem[{Liu} \em{et~al.}(2016){Liu}, {Zhang}, {Li}, {Ma}, and
  {Xue}]{2016PhRvD..93l3004L}
{Liu}, T.; {Zhang}, B.; {Li}, Y.; {Ma}, R.Y.; {Xue}, L.
\newblock {Detectable MeV neutrinos from black hole neutrino-dominated
  accretion flows}.
\newblock {\em Phys.~Rev.~D} {\bf 2016}, {\em 93},~123004,
  \href{http://xxx.lanl.gov/abs/1512.07203}{{\normalfont
  [arXiv:astro-ph.HE/1512.07203]}}.
\newblock
  doi:{\changeurlcolor{black}\href{https://doi.org/10.1103/PhysRevD.93.123004}{\detokenize{10.1103/PhysRevD.93.123004}}}.

\bibitem[{Salmonson} and {Wilson}(1999)]{1999ApJ...517..859S}
{Salmonson}, J.D.; {Wilson}, J.R.
\newblock {General Relativistic Augmentation of Neutrino Pair Annihilation
  Energy Deposition near Neutron Stars}.
\newblock {\em ApJ} {\bf 1999}, {\em 517},~859--865,
  \href{http://xxx.lanl.gov/abs/astro-ph/9908017}{{\normalfont
  [arXiv:astro-ph/astro-ph/9908017]}}.
\newblock
  doi:{\changeurlcolor{black}\href{https://doi.org/10.1086/307232}{\detokenize{10.1086/307232}}}.

\bibitem[{Birkl} \em{et~al.}(2007){Birkl}, {Aloy}, {Janka}, and
  {M{\"u}ller}]{2007A&A...463...51B}
{Birkl}, R.; {Aloy}, M.A.; {Janka}, H.T.; {M{\"u}ller}, E.
\newblock {Neutrino pair annihilation near accreting, stellar-mass black
  holes}.
\newblock {\em A\&A} {\bf 2007}, {\em 463},~51--67,
  \href{http://xxx.lanl.gov/abs/astro-ph/0608543}{{\normalfont
  [arXiv:astro-ph/astro-ph/0608543]}}.
\newblock
  doi:{\changeurlcolor{black}\href{https://doi.org/10.1051/0004-6361:20066293}{\detokenize{10.1051/0004-6361:20066293}}}.

\bibitem[{Caballero} \em{et~al.}(2012){Caballero}, {McLaughlin}, and
  {Surman}]{Caballero_2012}
{Caballero}, O.L.; {McLaughlin}, G.C.; {Surman}, R.
\newblock {Neutrino Spectra from Accretion Disks: Neutrino General Relativistic
  Effects and the Consequences for Nucleosynthesis}.
\newblock {\em ApJ} {\bf 2012}, {\em 745},~170,
  \href{http://xxx.lanl.gov/abs/1105.6371}{{\normalfont
  [arXiv:astro-ph.HE/1105.6371]}}.
\newblock
  doi:{\changeurlcolor{black}\href{https://doi.org/10.1088/0004-637X/745/2/170}{\detokenize{10.1088/0004-637X/745/2/170}}}.

\bibitem[{Ruffini} \em{et~al.}(2016){Ruffini}, {Rueda}, {Muccino}, {Aimuratov},
  {Becerra}, {Bianco}, {Kovacevic}, {Moradi}, {Oliveira}, {Pisani}, and
  {Wang}]{2016ApJ...832..136R}
{Ruffini}, R.; {Rueda}, J.A.; {Muccino}, M.; {Aimuratov}, Y.; {Becerra}, L.M.;
  {Bianco}, C.L.; {Kovacevic}, M.; {Moradi}, R.; {Oliveira}, F.G.; {Pisani},
  G.B.; {Wang}, Y.
\newblock {On the Classification of GRBs and Their Occurrence Rates}.
\newblock {\em ApJ} {\bf 2016}, {\em 832},~136,
  \href{http://xxx.lanl.gov/abs/1602.02732}{{\normalfont
  [arXiv:astro-ph.HE/1602.02732]}}.
\newblock
  doi:{\changeurlcolor{black}\href{https://doi.org/10.3847/0004-637X/832/2/136}{\detokenize{10.3847/0004-637X/832/2/136}}}.

\bibitem[Moghaddas \em{et~al.}(2012)Moghaddas, Ghanbari, and
  Ghodsi]{2012arXiv1207.6455M}
Moghaddas, M.; Ghanbari, J.; Ghodsi, A.
\newblock Shear Tensor and Dynamics of Relativistic Accretion Disks around
  Rotating Black Holes.
\newblock {\em Publications of the Astronomical Society of Japan} {\bf 2012},
  {\em 64}.
\newblock
  doi:{\changeurlcolor{black}\href{https://doi.org/10.1093/pasj/64.6.137}{\detokenize{10.1093/pasj/64.6.137}}}.

\bibitem[Moeen(2017)]{Moghaddas:2017itv}
Moeen, M.
\newblock Calculation of the relativistic bulk tensor and shear tensor of
  relativistic accretion flows in the Kerr metric.
\newblock {\em Iranian Journal of Astronomy and Astrophysics} {\bf 2017}, {\em
  4},~205--221,
  \href{http://xxx.lanl.gov/abs/http://ijaa.du.ac.ir/article\_122\_c850f8252ffbb5a2c5bbc939a691b85f.pdf}{{\normalfont
  [http://ijaa.du.ac.ir/article\_122\_c850f8252ffbb5a2c5bbc939a691b85f.pdf]}}.
\newblock
  doi:{\changeurlcolor{black}\href{https://doi.org/10.22128/ijaa.2017.122}{\detokenize{10.22128/ijaa.2017.122}}}.

\bibitem[{Zeldovich} and {Novikov}(1971)]{1971reas.book.....Z}
{Zeldovich}, Y.B.; {Novikov}, I.D.
\newblock {\em {Relativistic astrophysics. Vol.1: Stars and relativity}};
  University of Chicago Press,  1971.

\bibitem[Potekhin and Chabrier(2000)]{PhysRevE.62.8554}
Potekhin, A.Y.; Chabrier, G.
\newblock Equation of state of fully ionized electron-ion plasmas. II.
  Extension to relativistic densities and to the solid phase.
\newblock {\em Phys. Rev. E} {\bf 2000}, {\em 62},~8554--8563.
\newblock
  doi:{\changeurlcolor{black}\href{https://doi.org/10.1103/PhysRevE.62.8554}{\detokenize{10.1103/PhysRevE.62.8554}}}.

\bibitem[Dicus(1972)]{Dicus:1972yr}
Dicus, D.A.
\newblock {Stellar energy-loss rates in a convergent theory of weak and
  electromagnetic interactions}.
\newblock {\em Phys. Rev.} {\bf 1972}, {\em D6},~941--949.
\newblock
  doi:{\changeurlcolor{black}\href{https://doi.org/10.1103/PhysRevD.6.941}{\detokenize{10.1103/PhysRevD.6.941}}}.

\bibitem[{Tubbs} and {Schramm}(1975)]{1975ApJ...201..467T}
{Tubbs}, D.L.; {Schramm}, D.N.
\newblock {Neutrino Opacities at High Temperatures and Densities}.
\newblock {\em ApJ} {\bf 1975}, {\em 201},~467--488.
\newblock
  doi:{\changeurlcolor{black}\href{https://doi.org/10.1086/153909}{\detokenize{10.1086/153909}}}.

\bibitem[{Bruenn}(1985)]{1985ApJS...58..771B}
{Bruenn}, S.W.
\newblock {Stellar core collapse - Numerical model and infall epoch}.
\newblock {\em ApJs} {\bf 1985}, {\em 58},~771--841.
\newblock
  doi:{\changeurlcolor{black}\href{https://doi.org/10.1086/191056}{\detokenize{10.1086/191056}}}.

\bibitem[{Ruffert} \em{et~al.}(1996){Ruffert}, {Janka}, and
  {Schaefer}]{1996A&A...311..532R}
{Ruffert}, M.; {Janka}, H.T.; {Schaefer}, G.
\newblock {Coalescing neutron stars - a step towards physical models. I.
  Hydrodynamic evolution and gravitational-wave emission.}
\newblock {\em A\&A} {\bf 1996}, {\em 311},~532--566,
  \href{http://xxx.lanl.gov/abs/astro-ph/9509006}{{\normalfont
  [astro-ph/9509006]}}.

\bibitem[{Yakovlev} \em{et~al.}(2001){Yakovlev}, {Kaminker}, {Gnedin}, and
  {Haensel}]{2001PhR...354....1Y}
{Yakovlev}, D.G.; {Kaminker}, A.D.; {Gnedin}, O.Y.; {Haensel}, P.
\newblock {Neutrino emission from neutron stars}.
\newblock {\em Phys.~Rep.} {\bf 2001}, {\em 354},~1--155,
  \href{http://xxx.lanl.gov/abs/astro-ph/0012122}{{\normalfont
  [astro-ph/0012122]}}.
\newblock
  doi:{\changeurlcolor{black}\href{https://doi.org/10.1016/S0370-1573(00)00131-9}{\detokenize{10.1016/S0370-1573(00)00131-9}}}.

\bibitem[{Burrows} and {Thompson}(2004)]{Burrows2004}
{Burrows}, A.; {Thompson}, T.A.
\newblock {Neutrino-Matter Interaction Rates in Supernovae}.
\newblock  Astrophysics and Space Science Library; {Fryer}, C.L., Ed.,  2004,
  Vol. 302, pp. 133--174.
\newblock
  doi:{\changeurlcolor{black}\href{https://doi.org/10.1007/978-0-306-48599-2_5}{\detokenize{10.1007/978-0-306-48599-2_5}}}.

\bibitem[Burrows \em{et~al.}(2006)Burrows, Reddy, and Thompson]{BURROWS2006356}
Burrows, A.; Reddy, S.; Thompson, T.A.
\newblock Neutrino opacities in nuclear matter.
\newblock {\em Nuclear Physics A} {\bf 2006}, {\em 777},~356 -- 394.
\newblock Special Issue on Nuclear Astrophysics,
  doi:{\changeurlcolor{black}\href{https://doi.org/https://doi.org/10.1016/j.nuclphysa.2004.06.012}{\detokenize{https://doi.org/10.1016/j.nuclphysa.2004.06.012}}}.

\bibitem[{Aparicio}(1998)]{1998ApJS..117..627A}
{Aparicio}, J.M.
\newblock {A Simple and Accurate Method for the Calculation of Generalized
  Fermi Functions}.
\newblock {\em ApJs} {\bf 1998}, {\em 117},~627--632.
\newblock
  doi:{\changeurlcolor{black}\href{https://doi.org/10.1086/313121}{\detokenize{10.1086/313121}}}.

\end{thebibliography}

\end{document}